\documentclass[11pt]{cernrep}
\usepackage{graphicx}

\usepackage{epsf}
\usepackage{epsfig}

\renewcommand{\theequation}{\arabic{section}.\arabic{equation}}
\renewcommand{\thesection}{\arabic{section}}
\renewcommand{\thesubsection}{\arabic{section}.\arabic{subsection}}
\renewcommand{\thesubsubsection}{\arabic{section}.\arabic{subsection}.\arabic{subsubsection}}

\newcommand{\lsim}{
\mathrel{\hbox{\rlap{\hbox{\lower4pt\hbox{$\sim$}}}\hbox{$<$}}}}

\newcommand{\gsim}{
\mathrel{\hbox{\rlap{\hbox{\lower4pt\hbox{$\sim$}}}\hbox{$>$}}}}

\begin{document}


\thispagestyle{empty}

\begin{flushright}
CERN-PH-TH/2006-152\\
hep-ph/0608010
\end{flushright}

\vspace{1.6truecm}
\begin{center}
\boldmath
\large\bf  Flavour Physics and CP Violation
\unboldmath
\end{center}

\vspace{0.9truecm}
\begin{center}
Robert Fleischer\\[0.1cm]
{\sl CERN, Department of Physics, Theory Division\\
CH-1211 Geneva 23, Switzerland}
\end{center}

\vspace{1.3truecm}

\begin{center}
{\bf Abstract}
\end{center}

{\small
\vspace{0.2cm}\noindent
The starting point of these lectures is an introduction to the weak 
interactions of quarks and the Standard-Model description of CP violation, 
where the central r\^ole is played by the Cabibbo--Kobayashi--Maskawa matrix 
and the corresponding unitarity triangles. Since the $B$-meson system will govern 
the stage of (quark) flavour physics and CP violation in this decade, 
it will be our main focus. We  shall classify $B$-meson decays, introduce the 
theoretical tools to deal with them, investigate the requirements for non-vanishing 
CP-violating asymmetries, and discuss the main strategies to explore
CP violation and the preferred avenues for physics beyond the Standard Model
to enter. This formalism is then applied to discuss the status of important 
$B$-factory benchmark modes, where we focus on puzzling patterns in the
data that may indicate new-physics effects, as well as the prospects for 
$B$-decay studies at the LHC.
}

\vspace{1.5truecm}

\begin{center}
Lectures given at the {\it 2005 European School of High-Energy Physics,}\\
Kitzb\"uhel, Austria, 21 August -- 3 September 2005\\
To appear in the Proceedings (CERN Report)
\end{center}

\vfill
\noindent
CERN-PH-TH/2006-152\\
July 2006

\newpage
\thispagestyle{empty}
\vbox{}
\newpage
 
\setcounter{page}{1}


\pagestyle{plain}

\setcounter{page}{1}
\pagenumbering{roman}

\tableofcontents

\newpage

\setcounter{page}{1}
\pagenumbering{arabic}

\title{FLAVOUR PHYSICS AND CP VIOLATION}
\author{Robert Fleischer}
\institute{CERN, Geneva, Switzerland}
\maketitle
\begin{abstract}
The starting point of these lectures is an introduction to the weak 
interactions of quarks and the Standard-Model description of CP violation, 
where the central r\^ole is played by the Cabibbo--Kobayashi--Maskawa matrix 
and the corresponding unitarity triangles. Since the $B$-meson system will govern 
the stage of (quark) flavour physics and CP violation in this decade, 
it will be our main focus. We  shall classify $B$-meson decays, introduce the 
theoretical tools to deal with them, investigate the requirements for non-vanishing 
CP-violating asymmetries, and discuss the main strategies to explore
CP violation and the preferred avenues for physics beyond the Standard Model
to enter. This formalism is then applied to discuss the status of important 
$B$-factory benchmark modes, where we focus on puzzling patterns in the
data that may indicate new-physics effects, as well as the prospects for 
$B$-decay studies at the LHC.
\end{abstract}

\pagestyle{plain}

\pagestyle{plain}

\section{INTRODUCTION}\label{sec:intro}
\setcounter{equation}{0}
The history of CP violation, i.e.\ the non-invariance of the weak interactions
with respect to a combined charge-conjugation (C) and parity (P) 
transformation, goes back to the year 1964, where this phenomenon was
discovered through the observation of $K_{\rm L}\to\pi^+\pi^-$ decays
\cite{CP-obs}, which exhibit a branching ratio at the $10^{-3}$ level. This 
surprising effect is a manifestation of {\it indirect} CP violation, which arises 
from the fact that the mass eigenstates $K_{\rm L,S}$ of the neutral kaon 
system, which shows $K^0$--$\bar K^0$ mixing, are not eigenstates of the 
CP operator. In particular, the $K_{\rm L}$ state is governed by the CP-odd 
eigenstate, but has also a tiny admixture of the CP-even eigenstate, which 
may decay through CP-conserving interactions into the $\pi^+\pi^-$ final state. 
These CP-violating effects are described by the following observable:
\begin{equation}\label{epsK}
\varepsilon_K=(2.280\pm0.013)\times10^{-3}\times e^{i\pi/4}.
\end{equation}
On the other hand, CP-violating effects may also arise directly at the decay-amplitude
level, thereby yielding {\it direct} CP violation. This phenomenon, which leads to a
non-vanishing value of a quantity Re$(\varepsilon_K'/\varepsilon_K)$, could 
eventually be established in 1999 through the NA48 (CERN) and KTeV 
(FNAL) collaborations \cite{eps-prime}; the final results of the corresponding 
measurements are given by
\begin{equation}\label{epsp-eps-final}
\mbox{Re}(\varepsilon_K'/\varepsilon_K)=\left\{\begin{array}{ll}
(14.7\pm2.2)\times10^{-4}&\mbox{(NA48 \cite{NA48-final})}\\
(20.7\pm2.8)\times10^{-4}&\mbox{(KTeV \cite{KTeV-final}).}
\end{array}
\right.
\end{equation}

In this decade, there are huge experimental efforts to further 
explore CP violation and the quark-flavour sector of the Standard Model 
(SM). In these studies, the main actor is
the $B$-meson system, where we distinguish between charged and neutral 
$B$ mesons, which are characterized by the following valence-quark contents:
\begin{equation}\label{B-valence}
\begin{array}{c}
B^+\sim u \bar b, \quad B^+_c\sim c \bar b, \quad B^0_d\sim d \bar b, \quad
B^0_s\sim s \bar b, \\
B^-\sim \bar u b, \quad B^-_c\sim \bar c b, \quad \bar B^0_d\sim \bar d  b, \quad
\bar B^0_s\sim \bar s  b.
\end{array}
\end{equation}
In contrast to the charged $B$ mesons, their neutral counterparts
$B_q$ ($q\in \{d,s\}$) show
-- in analogy to $K^0$--$\bar K^0$ mixing -- the phenomenon of 
$B_q^0$--$\bar B_q^0$ mixing.
The asymmetric $e^+e^-$ $B$ factories at SLAC and KEK with their detectors 
BaBar and Belle, respectively, can only produce $B^+$ and $B^0_d$ mesons 
(and their anti-particles) since they operate at the $\Upsilon(4S)$ resonance, 
and have already collected ${\cal O}(10^8)$ $B\bar B$ pairs of this kind.
Moreover, first $B$-physics results from run II of the Tevatron were 
reported from the CDF and D0 collaborations, including also $B^+_c$
and $B^0_s$ studies, and second-generation $B$-decay
studies will become possible at the Large Hadron Collider (LHC)
at CERN, in particular thanks to the LHCb experiment, starting in the 
autumn of 2007. For the more distant future,
an $e^+$--$e^-$ ``super-$B$ factory'' is under consideration, with an
increase of luminosity by up to two orders of magnitude with respect to the 
currently operating machines. Moreover, there are plans to measure the 
very ``rare" kaon decays $K^+\to\pi^+\nu\bar\nu$ and $K_{\rm L}\to\pi^0\nu\bar\nu$, which are absent at the tree level in the SM, at CERN and KEK/J-PARC. 

In 2001, CP-violating effects were discovered in $B$ decays 
with the help of $B_d\to J/\psi K_{\rm S}$ modes by the BaBar and Belle 
collaborations \cite{CP-B-obs}, representing the first observation of CP violation 
outside the kaon system. This particular kind of CP violation, which is by
now well established, originates from the interference between 
$B^0_d$--$\bar B^0_d$ mixing and $B^0_d\to J/\psi K_{\rm S}$, 
$\bar B^0_d\to J/\psi K_{\rm S}$ decay processes, and is referred to as 
``mixing-induced" CP violation. In the summer of 2004, also direct CP 
violation could be detected in $B_d\to\pi^\mp K^\pm$ decays \cite{CP-B-dir}, 
thereby complementing the measurement of a non-zero
value of $\mbox{Re}(\varepsilon_K'/\varepsilon_K)$.

Studies of CP violation and flavour physics are particularly interesting since
``new physics" (NP), i.e.\ physics lying beyond the SM, typically leads to
new sources of flavour and CP violation. Furthermore, the origin of the
fermion masses, flavour mixing, CP violation etc.\ lies completely in the
dark and is expected to involve NP, too. Interestingly, CP violation
offers also a link to cosmology. One of the key features of our Universe is the
cosmological baryon asymmetry of ${\cal O}(10^{-10})$. As was pointed
out by Sakharov \cite{sach}, the necessary conditions for the generation of 
such an asymmetry include also the requirement that elementary interactions
violate CP (and C). Model calculations of the baryon asymmetry indicate, however,
that the CP violation present in the SM seems to be too small to generate
the observed asymmetry  \cite{shapos}. On the one hand, the required new sources 
of CP violation could be associated with very high energy scales, as in 
``leptogenesis", where new CP-violating effects appear in decays of heavy
Majorana neutrinos \cite{LG-rev}. On the other hand, new sources of
CP violation could also be accessible in the laboratory, as they arise 
naturally when going beyond the SM. 

Before searching for NP, it is essential to understand first the picture of
flavour physics and CP violation arising in the framework of the SM,
where the Cabibbo--Kobayashi--Maskawa (CKM) matrix -- the
quark-mixing matrix -- plays the key r\^ole \cite{cab,KM}. The 
corresponding phenomenology is extremely rich \cite{CKM-book}. In general,
the key problem for the theoretical interpretation is related to strong
interactions, i.e.\ to ``hadronic" uncertainties. A famous example is
$\mbox{Re}(\varepsilon_K'/\varepsilon_K)$, where we
have to deal with a subtle interplay between different contributions
which largely cancel \cite{epsp-rev}. Although the non-vanishing value of this
quantity has unambiguously ruled out ``superweak" models of
CP violation \cite{superweak}, it does currently not allow a stringent
test of the SM. 

In the $B$-meson system, there are various strategies to eliminate
the hadronic uncertainties in the exploration of CP violation (simply
speaking, there are many $B$ decays). Moreover, we may also search
for relations and/or correlations that hold in the SM but could well be
spoiled by NP. These topics will be the focus of this lecture, which is
complemented by the dedicated lectures on the experimental aspects 
of $K$- and $B$-meson decays in Refs.~\cite{jeitler} and \cite{widhalm}, respectively. 
The outline is as follows: in Section~\ref{sec:CP-SM}, we discuss the quark 
mixing in the SM by having a closer look at the CKM matrix and the
associated unitarity triangles. The main actors of this lecture -- the
$B$ mesons and their weak decays -- will then be introduced in 
Section~\ref{sec:Bdecays}. There we will also move towards studies
of CP violation and shall classify the main strategies for its exploration,
using amplitude relations and the phenomenon of $B^0_q$--$\bar B^0_q$ 
mixing ($q\in\{d,s\}$). In Section~\ref{sec:A-REL}, we illustrate the former 
kind of methods by having a closer look at clean amplitude relations between
$B^\pm\to K^\pm D$ and $B_c^\pm\to D_s^\pm D$ decays, whereas we discuss 
features of neutral
$B_q$ mesons in Section~\ref{sec:mix}. In Section~\ref{sec:NP}, we address the question of how NP could enter, and then apply these considerations in 
Section~\ref{sec:bench} to the $B$-factory benchmark modes 
$B^0_d\to J/\psi K_{\rm S}$, $B^0_d\to \phi K_{\rm S}$ and $B^0_d\to\pi^+\pi^-$.
Since the data for certain $B\to\pi K$
decays show a puzzling pattern for several years, we have devoted 
Section~\ref{sec:BpiK-puzzle} to a detailed discussion of this ``$B\to\pi K$
puzzle" and its interplay with rare $K$ and $B$ decays. In Section~\ref{sec:bd-pengs}, 
we focus on $b\to d$ penguin processes, which are now coming within experimental 
reach at the $B$ factories, thereby offering an exciting new playground. Finally, 
in Section~\ref{sec:LHC}, we discuss $B$-decay studies at the  LHC, where the 
physics potential of the $B^0_s$-meson system can be fully exploited. 
The conclusions and a brief outlook are given in Section~\ref{sec:concl}.

For detailed discussions and textbooks dealing with flavour physics and CP violation, 
the reader is referred to Refs.~\cite{BF-rev}--\cite{mannel-book}, alternative lecture 
notes can be found in Refs.~\cite{nir-argentina, buras-spain}, and a selection 
of more compact recent reviews is given in Refs.~\cite{ali-rev}--\cite{HoLi-rev}.
The data used in these lectures refer to the situation in the spring of 2006.

\begin{figure}
\centerline{
\epsfysize=3.3truecm
\epsffile{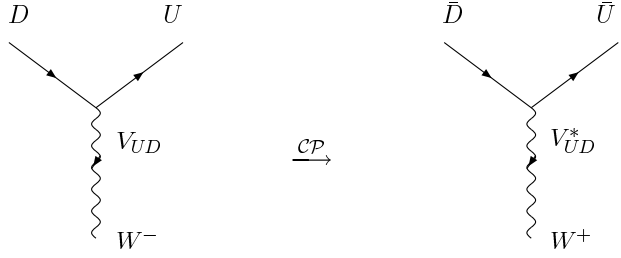}}
\caption{CP-conjugate charged-current quark-level interaction processes
in the SM.}\label{fig:CC} 
\end{figure}

\section{CP VIOLATION IN THE STANDARD MODEL}\label{sec:CP-SM}
\setcounter{equation}{0}
\subsection{Weak Interactions of Quarks and the Quark-Mixing Matrix}
In the framework of the Standard Model of electroweak interactions 
\cite{buchmueller,SM}, which is based on the spontaneously broken gauge group
\begin{equation}
SU(2)_{\rm L}\times U(1)_{\rm Y}
\stackrel{{\rm SSB}}
{\longrightarrow}U(1)_{\rm em},
\end{equation}
CP-violating effects may originate from the charged-current 
interactions of quarks, having the structure
\begin{equation}\label{cc-int}
D\to U W^-.
\end{equation}
Here $D\in\{d,s,b\}$ and $U\in\{u,c,t\}$ denote down- and up-type quark 
flavours, respectively, whereas the $W^-$ is the usual $SU(2)_{\rm L}$ 
gauge boson. From a phenomenological point of view, it is convenient to 
collect the generic ``coupling strengths'' $V_{UD}$ of the charged-current 
processes in (\ref{cc-int}) in the form of the following matrix:
\begin{equation}\label{ckm0}
\hat V_{\rm CKM}=
\left(\begin{array}{ccc}
V_{ud}&V_{us}&V_{ub}\\
V_{cd}&V_{cs}&V_{cb}\\
V_{td}&V_{ts}&V_{tb}
\end{array}\right),
\end{equation}
which is referred to as the Cabibbo--Kobayashi--Maskawa (CKM) matrix
\cite{cab,KM}. 

From a theoretical point of view, this matrix connects the electroweak 
states $(d',s',b')$ of the down, strange and bottom quarks with their 
mass eigenstates $(d,s,b)$ through the following unitary transformation 
\cite{buchmueller}:
\begin{equation}\label{ckm}
\left(\begin{array}{c}
d'\\
s'\\
b'
\end{array}\right)=\left(\begin{array}{ccc}
V_{ud}&V_{us}&V_{ub}\\
V_{cd}&V_{cs}&V_{cb}\\
V_{td}&V_{ts}&V_{tb}
\end{array}\right)\cdot
\left(\begin{array}{c}
d\\
s\\
b
\end{array}\right).
\end{equation}
Consequently, $\hat V_{\rm CKM}$ is actually a {\it unitary} matrix.
This feature ensures the absence of flavour-changing neutral-current 
(FCNC) processes at the tree level in the SM, and is hence at the basis 
of the famous Glashow--Iliopoulos--Maiani (GIM) mechanism \cite{GIM}. 
We shall return to the unitarity of the CKM matrix in
Subsection~\ref{ssec:UT}, discussing the ``unitarity triangles''.
If we express the non-leptonic charged-current interaction Lagrangian 
in terms of the mass eigenstates appearing in (\ref{ckm}), we arrive at 
\begin{equation}\label{cc-lag2}
{\cal L}_{\mbox{{\scriptsize int}}}^{\mbox{{\scriptsize CC}}}=
-\frac{g_2}{\sqrt{2}}\left(\begin{array}{ccc}
\bar u_{\mbox{{\scriptsize L}}},& \bar c_{\mbox{{\scriptsize L}}},
&\bar t_{\mbox{{\scriptsize L}}}\end{array}\right)\gamma^\mu\,\hat
V_{\mbox{{\scriptsize CKM}}}
\left(\begin{array}{c}
d_{\mbox{{\scriptsize L}}}\\
s_{\mbox{{\scriptsize L}}}\\
b_{\mbox{{\scriptsize L}}}
\end{array}\right)W_\mu^\dagger\,\,+\,\,\mbox{h.c.,}
\end{equation}
where the gauge coupling $g_2$ is related to the gauge group 
$SU(2)_{\mbox{{\scriptsize L}}}$, and the $W_\mu^{(\dagger)}$ field 
corresponds to the charged $W$ bosons. Looking at the
interaction vertices following from (\ref{cc-lag2}), we observe 
that the elements of the CKM matrix describe in fact the generic
strengths of the associated charged-current processes, as we have 
noted above. 

In Fig.~\ref{fig:CC}, we show the $D\to U W^-$ vertex and its CP 
conjugate. Since the corresponding CP transformation involves the 
replacement
\begin{equation}\label{CKM-CP}
V_{UD}\stackrel{{\cal CP}}{\longrightarrow}V_{UD}^\ast,
\end{equation}
CP violation could -- in principle -- be accommodated in the SM through 
complex phases in the CKM matrix. The crucial question in this context
is, of course, whether we may actually have physical complex phases in 
that matrix.

\subsection{Phase Structure of the CKM Matrix}
We have the freedom to redefine the up- and down-type quark fields
in the following manner:
\begin{equation}
U\to \exp(i\xi_U)U,\quad D\to \exp(i\xi_D)D. 
\end{equation}
If we perform such transformations in (\ref{cc-lag2}), the invariance 
of the charged-current interaction Lagrangian implies the 
following phase transformations of the CKM matrix elements:
\begin{equation}\label{CKM-trafo}
V_{UD}\to\exp(i\xi_U)V_{UD}\exp(-i\xi_D).
\end{equation}
Using these transformations to eliminate unphysical phases, it can be shown 
that the parametrization of the general $N\times N$ quark-mixing matrix, 
where $N$ denotes the number of fermion generations, involves the following 
parameters:
\begin{equation}
\underbrace{\frac{1}{2}N(N-1)}_{\mbox{Euler angles}} \, + \,
\underbrace{\frac{1}{2}(N-1)(N-2)}_{\mbox{complex phases}}=
(N-1)^2.
\end{equation}

If we apply this expression to the case of $N=2$ generations, we observe
that only one rotation angle -- the Cabibbo angle $\theta_{\rm C}$
\cite{cab} -- is required for the parametrization of the $2\times2$
quark-mixing matrix, which can be written in the following form:
\begin{equation}\label{Cmatrix}
\hat V_{\rm C}=\left(\begin{array}{cc}
\cos\theta_{\rm C}&\sin\theta_{\rm C}\\
-\sin\theta_{\rm C}&\cos\theta_{\rm C}
\end{array}\right),
\end{equation}
where $\sin\theta_{\rm C}=0.22$ can be determined from $K\to\pi\ell\bar\nu$ 
decays. On the other hand, in the case of $N=3$ generations, the 
parametrization of the corresponding $3\times3$ quark-mixing matrix involves 
three Euler-type angles and a single {\it complex} phase. This complex phase 
allows us to accommodate CP violation in the SM, as was pointed out by 
Kobayashi and Maskawa in 1973 \cite{KM}. The corresponding picture
is referred to as the Kobayashi--Maskawa (KM) mechanism of CP violation.

In the ``standard parametrization'' advocated by the Particle Data Group
(PDG) \cite{PDG}, the three-generation CKM matrix takes the following 
form:
\begin{equation}\label{standard}
\hat V_{\rm CKM}=\left(\begin{array}{ccc}
c_{12}c_{13}&s_{12}c_{13}&s_{13}e^{-i\delta_{13}}\\ -s_{12}c_{23}
-c_{12}s_{23}s_{13}e^{i\delta_{13}}&c_{12}c_{23}-
s_{12}s_{23}s_{13}e^{i\delta_{13}}&
s_{23}c_{13}\\ s_{12}s_{23}-c_{12}c_{23}s_{13}e^{i\delta_{13}}&-c_{12}s_{23}
-s_{12}c_{23}s_{13}e^{i\delta_{13}}&c_{23}c_{13}
\end{array}\right),
\end{equation}
where $c_{ij}\equiv\cos\theta_{ij}$ and $s_{ij}\equiv\sin\theta_{ij}$. 
Performing appropriate redefinitions of the quark-field phases, the real 
angles $\theta_{12}$, $\theta_{23}$ and $\theta_{13}$ can all be made to
lie in the first quadrant. The advantage of this parametrization is that
the generation labels $i,j=1,2,3$ are introduced in such a manner that
the mixing between two chosen generations vanishes if the corresponding
mixing angle $\theta_{ij}$ is set to zero. In particular, for 
$\theta_{23}=\theta_{13}=0$, the third generation decouples, and the
$2\times2$ submatrix describing the mixing between the first and 
second generations takes the same form as (\ref{Cmatrix}).

Another interesting parametrization of the CKM matrix was proposed by 
Fritzsch and Xing \cite{FX}:
\begin{equation}
\hat V_{\rm CKM}=\left(\begin{array}{ccc}
s_{\rm u} s_{\rm d} c + c_{\rm u} c_{\rm d} e^{-i\varphi} & 
s_{\rm u} c_{\rm d} c - c_{\rm u} s_{\rm d} e^{-i\varphi} &s_{\rm u} s\\
c_{\rm u} s_{\rm d} c - s_{\rm u} c_{\rm d} e^{-i\varphi} & 
c_{\rm u} c_{\rm d} c + s_{\rm u} s_{\rm d} e^{-i\varphi} &c_{\rm u} s\\
-s_{\rm d}s & -c_{\rm d}s & c
\end{array}\right).
\end{equation}
It is inspired by the hierarchical structure of the quark-mass spectrum
and is particularly useful in the context of models for fermion masses and
mixings. The characteristic feature of this parametrization is that
the complex phase arises only in the $2\times2$ submatrix involving
the up, down, strange and charm quarks. 

Let us finally note that physical observables, for instance CP-violating
asymmetries, {\it cannot} depend on the chosen parametrization of the CKM 
matrix, i.e.\ have to be invariant under the phase transformations specified 
in (\ref{CKM-trafo}).

\subsection{Further Requirements for CP Violation}
As we have just seen, in order to be able to accommodate CP violation within 
the framework of the SM through a complex phase in the CKM matrix, at least 
three generations are required. However, this feature is not sufficient for 
observable CP-violating effects. To this end, further conditions have to 
be satisfied, which can be summarized as follows \cite{jarlskog,BBG}:
\begin{equation}\label{CP-req}
(m_t^2-m_c^2)(m_t^2-m_u^2)(m_c^2-m_u^2)
(m_b^2-m_s^2)(m_b^2-m_d^2)(m_s^2-m_d^2)\times J_{\rm CP}\,\not=\,0,
\end{equation}
where
\begin{equation}\label{JCP}
J_{\rm CP}=|\mbox{Im}(V_{i\alpha}V_{j\beta}V_{i\beta}^\ast 
V_{j\alpha}^\ast)|\quad(i\not=j,\,\alpha\not=\beta)\,.
\end{equation}

The mass factors in (\ref{CP-req}) are related to the fact that the 
CP-violating phase of the CKM matrix could be eliminated through an 
appropriate unitary transformation of the quark fields if any two quarks 
with the same charge had the same mass. Consequently, the origin 
of CP violation is closely related to the ``flavour problem'' in
elementary particle physics, and cannot be understood in a deeper 
way, unless we have fundamental insights into the hierarchy of quark 
masses and the number of fermion generations.

The second element of (\ref{CP-req}), the ``Jarlskog parameter'' 
$J_{\rm CP}$ \cite{jarlskog}, can be interpreted as a measure of the 
strength of CP violation in the SM. It does not depend on the chosen 
quark-field parametrization, i.e.\ it is invariant under (\ref{CKM-trafo}), 
and the unitarity of the CKM matrix implies that all combinations 
$|\mbox{Im}(V_{i\alpha}V_{j\beta}V_{i\beta}^\ast V_{j\alpha}^\ast)|$ 
are equal to one another. Using the standard parametrization of the
CKM matrix introduced in (\ref{standard}), we obtain
\begin{equation}\label{JCP-PDG}
J_{\rm CP}=s_{12}s_{13}s_{23}c_{12}c_{23}c_{13}^2\sin\delta_{13}.
\end{equation}
The  experimental information on the CKM parameters implies 
$J_{\rm CP}={\cal O}(10^{-5})$, so that
CP-violating phenomena are hard to observe.  However, new complex couplings 
are typically present in scenarios for NP \cite{ellis}. Such additional sources
for CP violation could be detected through flavour experiments.

\boldmath\subsection{Experimental Information on $|V_{\rm CKM}|$}
\begin{figure}
\centerline{
\epsfysize=3.8truecm
\epsffile{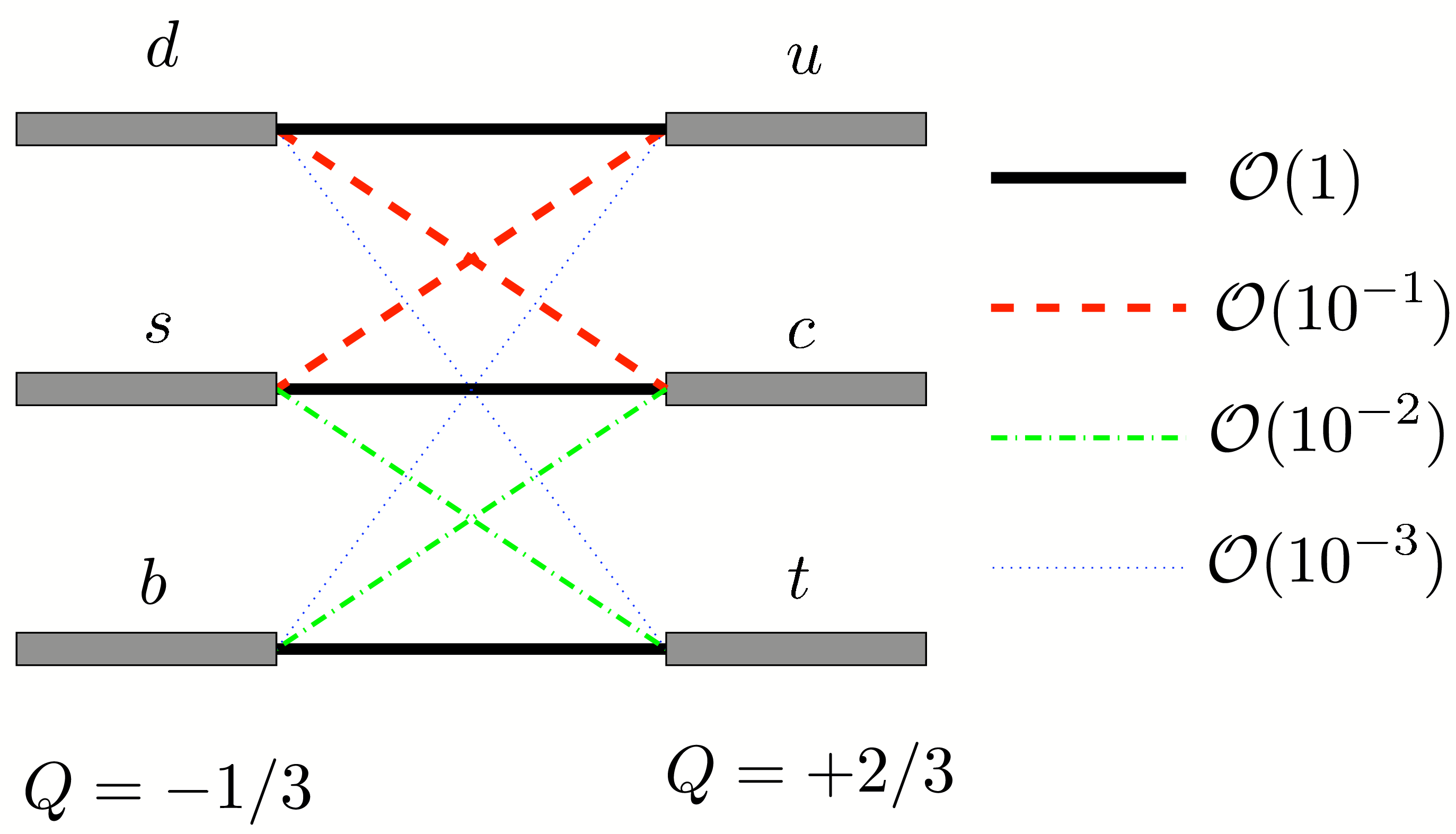}}
\vspace*{-0.2truecm}
\caption{Hierarchy of the quark transitions mediated through 
charged-current processes.}\label{fig:term}
\end{figure}

\unboldmath
In order to determine the magnitudes $|V_{ij}|$ of the elements of the
CKM matrix, we may use the following tree-level processes:
\begin{itemize}
\item Nuclear beta decays, neutron decays $\Rightarrow$ $|V_{ud}|$.
\item $K\to\pi\ell\bar\nu$ decays $\Rightarrow$ $|V_{us}|$.
\item $\nu$ production of charm off valence $d$ quarks
$\Rightarrow$ $|V_{cd}|$.
\item Charm-tagged $W$ decays (as well as $\nu$ production and 
semileptonic $D$ decays)  $\Rightarrow$ $|V_{cs}|$.
\item Exclusive and inclusive $b\to c \ell \bar\nu$ decays 
$\Rightarrow$ $|V_{cb}|$.
\item Exclusive and inclusive 
$b\to u \ell \bar \nu$ decays $\Rightarrow$ $|V_{ub}|$.
\item $\bar t\to \bar b \ell \bar\nu$ processes $\Rightarrow$ (crude direct 
determination of) $|V_{tb}|$.
\end{itemize}
If we use the corresponding experimental information, together with the 
CKM unitarity condition, and assume that there are only three generations, 
we arrive at the following 90\% C.L. limits for the $|V_{ij}|$ \cite{PDG}:
\begin{equation}\label{CKM-mag}
|\hat V_{\rm CKM}|=
\left(\begin{array}{ccc}
$0.9739\mbox{--}0.9751$ & $0.221\mbox{--}0.227$ & $0.0029\mbox{--}0.0045$\\ 
$0.221\mbox{--}0.227$ & $0.9730\mbox{--}0.9744$ & $0.039\mbox{--}0.044$\\ 
$0.0048\mbox{--}0.014$ & $0.037\mbox{--}0.043$ & $0.9990\mbox{--}0.9992$\\ 
\end{array}\right).
\end{equation}
In Fig.~\ref{fig:term}, we have illustrated the resulting hierarchy 
of the strengths of the charged-current quark-level processes:
transitions within the same generation are governed by 
CKM matrix elements of ${\cal O}(1)$, those between the first and the second 
generation are suppressed by CKM factors of ${\cal O}(10^{-1})$, those 
between the second and the third generation are suppressed by 
${\cal O}(10^{-2})$, and the transitions between the first and the third 
generation are even suppressed by CKM factors of ${\cal O}(10^{-3})$. 
In the standard parametrization (\ref{standard}), this hierarchy is 
reflected by 
\begin{equation}
s_{12}=0.22 \,\gg\, s_{23}={\cal O}(10^{-2}) \,\gg\, 
s_{13}={\cal O}(10^{-3}). 
\end{equation}

\subsection{Wolfenstein Parametrization of the CKM Matrix}
For phenomenological applications, it would be useful to have a 
parametrization of the CKM matrix available that makes the 
hierarchy arising in (\ref{CKM-mag}) -- and illustrated in
Fig.~\ref{fig:term} -- explicit \cite{wolf}. In order to derive such 
a parametrization, we introduce a set of new parameters, 
$\lambda$, $A$, $\rho$ and $\eta$, by imposing the following 
relations \cite{blo}:
\begin{equation}\label{set-rel}
s_{12}\equiv\lambda=0.22,\quad s_{23}\equiv A\lambda^2,\quad 
s_{13}e^{-i\delta_{13}}\equiv A\lambda^3(\rho-i\eta).
\end{equation}
If we now go back to the standard parametrization (\ref{standard}), we 
obtain an {\it exact} parametrization of the CKM matrix as a function of 
$\lambda$ (and $A$, $\rho$, $\eta$), allowing us to expand each CKM 
element in powers of the small parameter $\lambda$. If we neglect terms of 
${\cal O}(\lambda^4)$, we arrive at the famous ``Wolfenstein 
parametrization'' \cite{wolf}:
\begin{equation}\label{W-par}
\hat V_{\mbox{{\scriptsize CKM}}} =\left(\begin{array}{ccc}
1-\frac{1}{2}\lambda^2 & \lambda & A\lambda^3(\rho-i\eta) \\
-\lambda & 1-\frac{1}{2}\lambda^2 & A\lambda^2\\
A\lambda^3(1-\rho-i\eta) & -A\lambda^2 & 1
\end{array}\right)+{\cal O}(\lambda^4),
\end{equation}
which makes the hierarchical structure of the CKM matrix very transparent 
and is an important tool for phenomenological considerations, as we 
will see throughout these lectures. 

For several applications, next-to-leading order corrections in $\lambda$ 
play an important r\^ole. Using the exact parametrization following from 
(\ref{standard}) and (\ref{set-rel}), they can be calculated straightforwardly 
by expanding each CKM element to the desired accuracy in 
$\lambda$ \cite{blo,Brev01}:
\begin{displaymath}
V_{ud}=1-\frac{1}{2}\lambda^2-\frac{1}{8}\lambda^4+{\cal O}(\lambda^6),\quad
V_{us}=\lambda+{\cal O}(\lambda^7),\quad
V_{ub}=A\lambda^3(\rho-i\,\eta),
\end{displaymath}
\begin{displaymath}
V_{cd}=-\lambda+\frac{1}{2}A^2\lambda^5\left[1-2(\rho+i\eta)\right]+
{\cal O}(\lambda^7),
\end{displaymath}
\begin{equation}\label{NLO-wolf}
V_{cs}=1-\frac{1}{2}\lambda^2-\frac{1}{8}\lambda^4(1+4A^2)+
{\cal O}(\lambda^6),
\end{equation}
\begin{displaymath}
V_{cb}=A\lambda^2+{\cal O}(\lambda^8),\quad
V_{td}=A\lambda^3\left[1-(\rho+i\eta)\left(1-\frac{1}{2}\lambda^2\right)
\right]+{\cal O}(\lambda^7),
\end{displaymath}
\begin{displaymath}
V_{ts}=-A\lambda^2+\frac{1}{2}A(1-2\rho)\lambda^4-i\eta A\lambda^4
+{\cal O}(\lambda^6),\quad
V_{tb}=1-\frac{1}{2}A^2\lambda^4+{\cal O}(\lambda^6).
\end{displaymath}
It should be noted that 
\begin{equation}
V_{ub}\equiv A\lambda^3(\rho-i\eta)
\end{equation}
receives {\it by definition} no power corrections in $\lambda$ within
this prescription. If we follow \cite{blo} and introduce the generalized
Wolfenstein parameters
\begin{equation}\label{rho-eta-bar}
\bar\rho\equiv\rho\left(1-\frac{1}{2}\lambda^2\right),\quad
\bar\eta\equiv\eta\left(1-\frac{1}{2}\lambda^2\right),
\end{equation}
we may simply write, up to corrections of ${\cal O}(\lambda^7)$,
\begin{equation}\label{Vtd-expr}
V_{td}=A\lambda^3(1-\bar\rho-i\,\bar\eta).
\end{equation}
Moreover, we have to an excellent accuracy
\begin{equation}\label{Def-A}
V_{us}=\lambda\quad \mbox{and}\quad 
V_{cb}=A\lambda^2,
\end{equation}
as these quantities receive only corrections at the $\lambda^7$ and
$\lambda^8$ levels, respectively. In comparison with other generalizations
of the Wolfenstein parametrization found in the literature, the advantage
of (\ref{NLO-wolf}) is the absence of relevant corrections to $V_{us}$
and $V_{cb}$, and that $V_{ub}$ and $V_{td}$ take forms similar to those 
in (\ref{W-par}). As far as the Jarlskog parameter introduced in
(\ref{JCP}) is concerned, we obtain the simple expression
\begin{equation}
J_{\rm CP}=\lambda^6A^2\eta,
\end{equation}
which should be compared with (\ref{JCP-PDG}).

\subsection{Unitarity Triangles of the CKM Matrix}\label{ssec:UT}
The unitarity of the CKM matrix, which is described by
\begin{equation}
\hat V_{\mbox{{\scriptsize CKM}}}^{\,\,\dagger}\cdot\hat 
V_{\mbox{{\scriptsize CKM}}}=
\hat 1=\hat V_{\mbox{{\scriptsize CKM}}}\cdot\hat V_{\mbox{{\scriptsize 
CKM}}}^{\,\,\dagger},
\end{equation}
leads to a set of 12 equations, consisting of 6 normalization 
and 6 orthogonality relations. The latter can be represented as 6 
triangles in the complex plane \cite{AKL}, all having the same area, 
$2 A_{\Delta}=J_{\rm CP}$ \cite{JS}. Let us now have a closer look at 
these relations: those describing the orthogonality of different columns 
of the CKM matrix are given by
\begin{eqnarray}
\underbrace{V_{ud}V_{us}^\ast}_{{\cal O}(\lambda)}+
\underbrace{V_{cd}V_{cs}^\ast}_{{\cal O}(\lambda)}+
\underbrace{V_{td}V_{ts}^\ast}_{{\cal O}(\lambda^5)} & = &
0\\
\underbrace{V_{us}V_{ub}^\ast}_{{\cal O}(\lambda^4)}+
\underbrace{V_{cs}V_{cb}^\ast}_{{\cal O}(\lambda^2)}+
\underbrace{V_{ts}V_{tb}^\ast}_{{\cal O}(\lambda^2)} & = &
0\\
\underbrace{V_{ud}V_{ub}^\ast}_{(\rho+i\eta)A\lambda^3}+
\underbrace{V_{cd}V_{cb}^\ast}_{-A\lambda^3}+
\underbrace{V_{td}V_{tb}^\ast}_{(1-\rho-i\eta)A\lambda^3} & = &
0,\label{UT1}
\end{eqnarray}
whereas those associated with the orthogonality of different rows 
take the following form:
\begin{eqnarray}
\underbrace{V_{ud}^\ast V_{cd}}_{{\cal O}(\lambda)}+
\underbrace{V_{us}^\ast V_{cs}}_{{\cal O}(\lambda)}+
\underbrace{V_{ub}^\ast V_{cb}}_{{\cal O}(\lambda^5)} & = &
0\\
\underbrace{V_{cd}^\ast V_{td}}_{{\cal O}(\lambda^4)}+
\underbrace{V_{cs}^\ast V_{ts}}_{{\cal O}(\lambda^2)}+
\underbrace{V_{cb}^\ast V_{tb}}_{{\cal O}(\lambda^2)} & = & 0\\
\underbrace{V_{ud}^\ast V_{td}}_{(1-\rho-i\eta)A\lambda^3}+
\underbrace{V_{us}^\ast V_{ts}}_{-A\lambda^3}+
\underbrace{V_{ub}^\ast V_{tb}}_{(\rho+i\eta)A\lambda^3}
& = & 0.\label{UT2}
\end{eqnarray}
Here we have also indicated the structures that arise if we apply the 
Wolfenstein parametrization by keeping just the leading, non-vanishing 
terms. We observe that only in (\ref{UT1}) and (\ref{UT2}), which 
describe the orthogonality of the first and third columns 
and of the first and third rows, respectively, all three sides are 
of comparable magnitude, ${\cal O}(\lambda^3)$, while in the 
remaining relations, one side is suppressed with respect to the others 
by factors of ${\cal O}(\lambda^2)$ or ${\cal O}(\lambda^4)$. Consequently,
we have to deal with only {\it two} non-squashed unitarity triangles 
in the complex plane. However, as we have already indicated in (\ref{UT1}) 
and (\ref{UT2}), the corresponding orthogonality relations agree with each 
other at the $\lambda^3$ level, yielding
\begin{equation}\label{UTLO}
\left[(\rho+i\eta)+(-1)+(1-\rho-i\eta)\right]A\lambda^3=0.
\end{equation}
Consequently, they describe the same triangle, which is usually referred 
to as {\it the} unitarity triangle of the CKM matrix 
\cite{JS,ut}.

\begin{figure}[t]
\centerline{
\begin{tabular}{ll}
   {\small(a)} & {\small(b)} \\
   \qquad \includegraphics[width=6.3truecm]{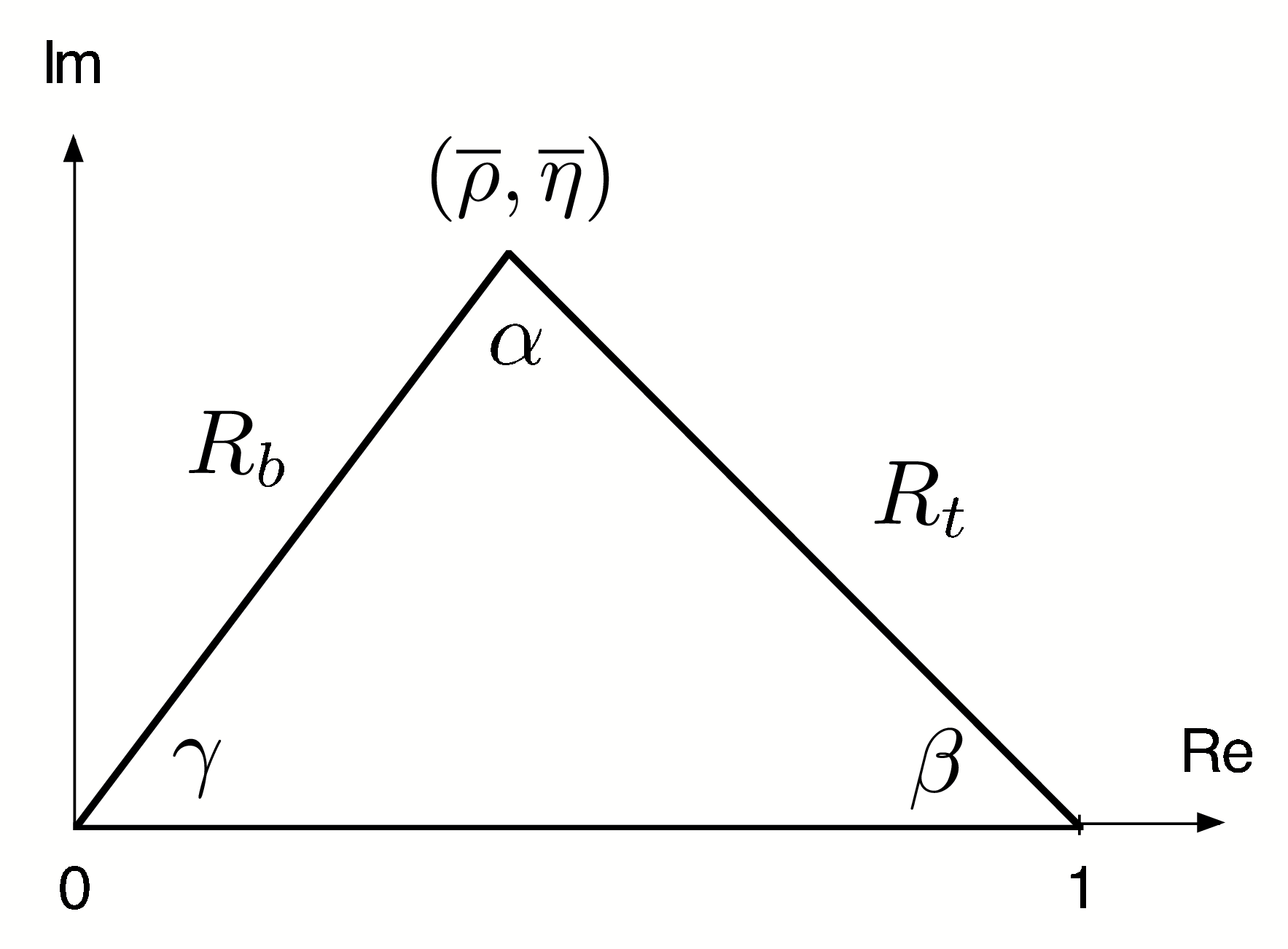}
&
\qquad \includegraphics[width=6.3truecm]{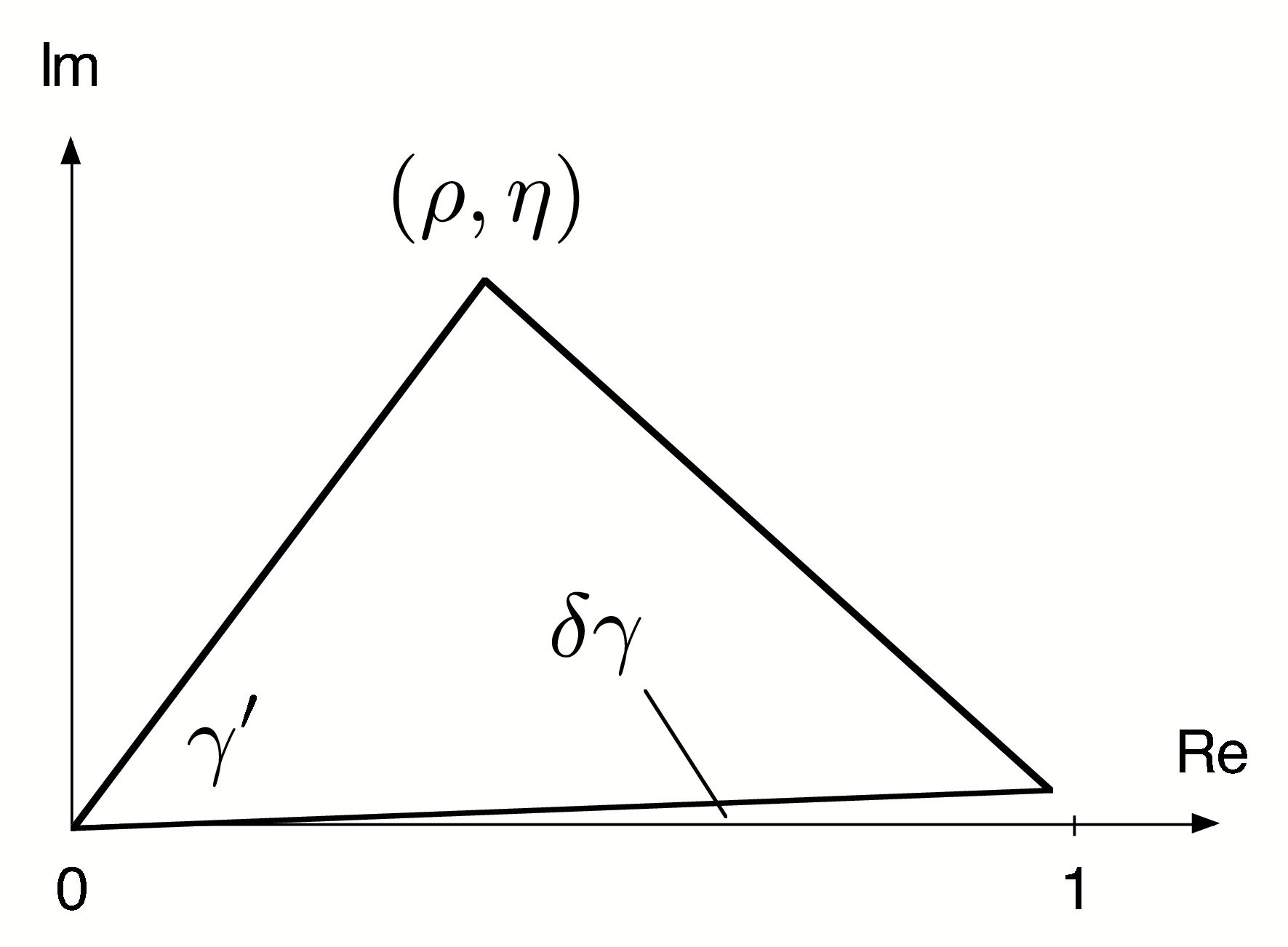}
 \end{tabular}}
 \vspace*{-0.2truecm}
\caption{The two non-squashed unitarity triangles of the CKM matrix, as
explained in the text: (a) and (b) correspond to the orthogonality 
relations (\ref{UT1}) and (\ref{UT2}), respectively. In Asia, the notation 
$\phi_1\equiv\beta$,
$\phi_2\equiv\alpha$ and $\phi_3\equiv\gamma$ is used for the angles of the
triangle shown in (a).}
\label{fig:UT}
\end{figure}

Concerning second-generation $B$-decay studies in the LHC era, 
the experimental accuracy will be so tremendous that we will also have 
to take the next-to-leading order terms of the Wolfenstein expansion
into account, and will have to distinguish between the unitarity triangles 
following from (\ref{UT1}) and (\ref{UT2}). Let us first have a closer
look at the former relation. Including terms of ${\cal O}(\lambda^5)$, 
we obtain the following generalization of (\ref{UTLO}):
\begin{equation}\label{UT1-NLO}
\left[(\bar\rho+i\bar\eta)+(-1)+(1-\bar\rho-
i\bar\eta)\right]A\lambda^3 +{\cal O}(\lambda^7)=0, 
\end{equation}
where $\bar\rho$ and $\bar\eta$ are as defined in (\ref{rho-eta-bar}). 
If we divide this relation by the overall normalization factor $A\lambda^3$, 
and introduce
\begin{equation}\label{Rb-def}
R_b\equiv\sqrt{\overline{\rho}^2+\overline{\eta}^2}=\left(1-\frac{\lambda^2}{2}
\right)\frac{1}{\lambda}\left|\frac{V_{ub}}{V_{cb}}\right|
\end{equation}
\begin{equation}\label{Rt-def}
R_t\equiv\sqrt{(1-\overline{\rho})^2+\overline{\eta}^2}=
\frac{1}{\lambda}\left|\frac{V_{td}}{V_{cb}}\right|,
\end{equation}
we arrive at the unitarity triangle illustrated in Fig.\ \ref{fig:UT} (a). 
It is a straightforward generalization of the leading-order
case described by (\ref{UTLO}): instead of $(\rho,\eta)$, the apex
is now simply given by $(\bar\rho,\bar\eta)$ \cite{blo}. The two sides 
$R_b$ and $R_t$, as well as the three angles $\alpha$, $\beta$ and $\gamma$, 
will show up at several places throughout these lectures. Moreover, the 
relations
\begin{equation}
V_{ub}=A\lambda^3\left(\frac{R_b}{1-\lambda^2/2}\right)e^{-i\gamma},\quad 
V_{td}=A\lambda^3 R_t e^{-i\beta}
\end{equation}
are also useful for phenomenological applications, since they make the
dependences of $\gamma$ and $\beta$ explicit; they correspond to the
phase convention chosen both in the standard parametrization 
(\ref{standard}) and in the generalized Wolfenstein parametrization 
(\ref{NLO-wolf}). Finally, if we take also (\ref{set-rel}) into account, 
we obtain
\begin{equation}
\delta_{13}=\gamma.
\end{equation}

Let us now turn to (\ref{UT2}). Here we arrive at an expression 
that is more complicated than (\ref{UT1-NLO}): 
\begin{equation}
\left[\left\{1-\frac{\lambda^2}{2}-
(1-\lambda^2)\rho-i(1-\lambda^2)\eta
\right\}\!+\!\left\{-1+\left(\frac{1}{2}-\rho\right)
\lambda^2-i\eta\lambda^2\right\}
\!+\!\left\{\rho+i\eta\right\}\right]A\lambda^3+{\cal O}(\lambda^7)=0.
\end{equation}
If we divide again by $A\lambda^3$, we obtain the unitarity triangle
sketched in Fig.\ \ref{fig:UT} (b), where the apex is given by $(\rho,\eta)$ 
and {\it not} by $(\bar\rho,\bar\eta)$. On the other hand, we encounter a 
tiny angle 
\begin{equation}
\delta\gamma\equiv\lambda^2\eta={\cal O}(1^\circ)
\end{equation}
between real axis and basis of the triangle, which satisfies 
\begin{equation}
\gamma=\gamma'+\delta\gamma, 
\end{equation}
where $\gamma$ coincides with the corresponding angle in  
Fig.\ \ref{fig:UT} (a).

Whenever we will refer to a ``unitarity triangle'' (UT) in the following 
discussion, we mean the one illustrated in Fig.\ \ref{fig:UT} (a), which
is the generic generalization of the leading-order case described
by (\ref{UTLO}). As we will see below, the UT is the central target 
of the experimental tests of the SM description of CP violation. 
Interestingly, also the tiny angle $\delta\gamma$ can be probed directly 
through certain CP-violating effects that can be explored at 
hadron colliders, in particular at the LHC.

\begin{figure}[t]
\centerline{
\begin{tabular}{ll}
  \includegraphics[width=7.0truecm]{CKMfitter-LP05.epsf} &
\includegraphics[width=9.0truecm]{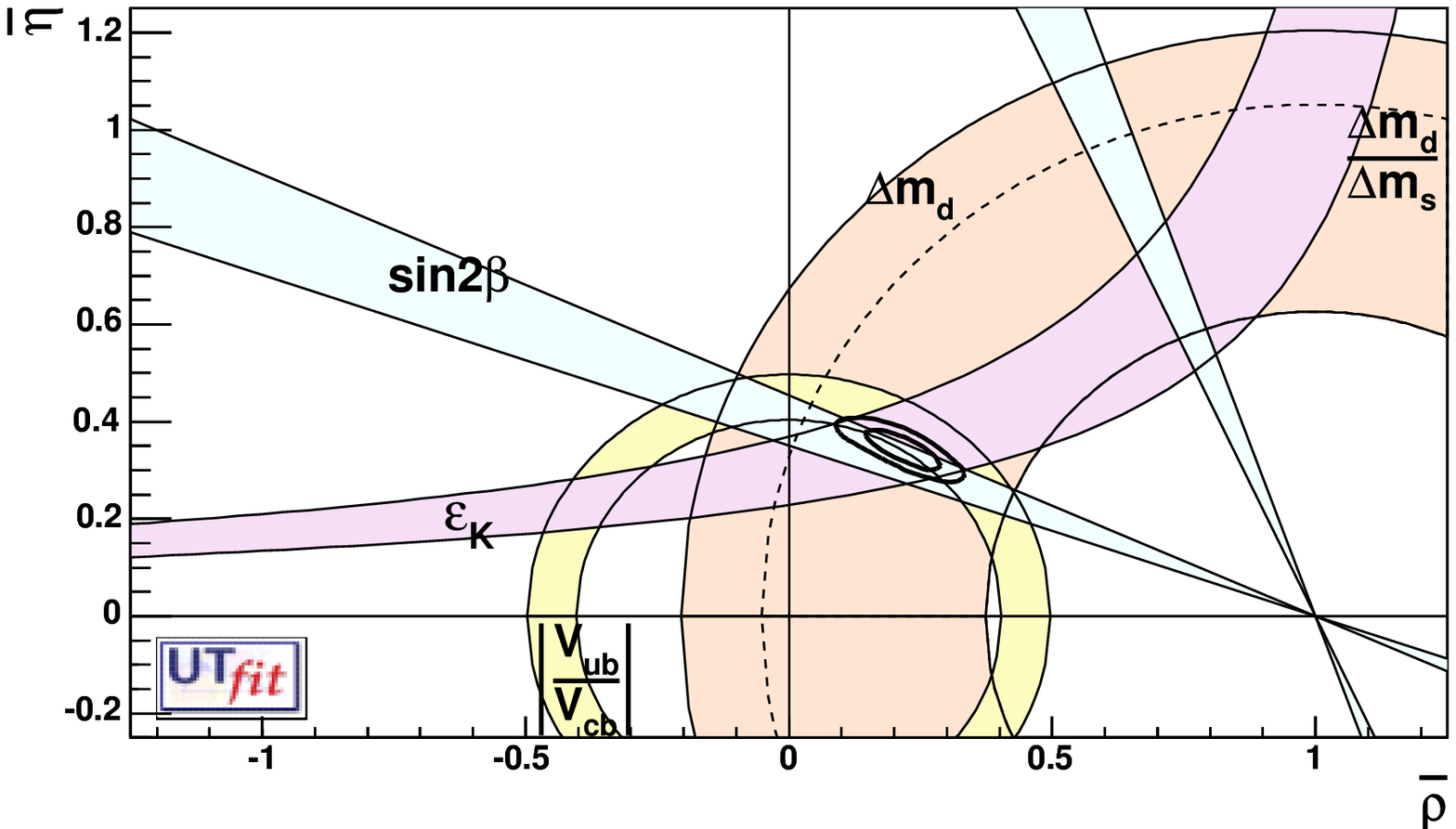}
 \end{tabular}}
\caption{Analyses of the CKMfitter and 
UTfit  collaborations \cite{CKMfitter,UTfit}.}\label{fig:UTfits}
\end{figure}

\subsection{The Determination of the Unitarity Triangle}\label{subsec:CKM-fits}
The next obvious question is how to determine the UT. There are two 
conceptually different avenues that we may follow to this end:
\begin{itemize}
\item[(i)] In the ``CKM fits'', theory is used to convert 
experimental data into contours in the $\bar\rho$--$\bar\eta$ plane. In particular, 
semi-leptonic $b\to u \ell \bar\nu_\ell$, $c \ell \bar\nu_\ell$ decays and 
$B^0_q$--$\bar B^0_q$ mixing ($q\in\{d,s\}$) allow us to determine the UT sides 
$R_b$ and $R_t$, respectively, i.e.\ to fix two circles in the $\bar\rho$--$\bar\eta$ 
plane. Furthermore, the indirect CP violation in the neutral kaon system
described by $\varepsilon_K$ can be transformed into a hyperbola. 
\item[(ii)] Theoretical considerations allow us to convert measurements of 
CP-violating effects in $B$-meson decays into direct information on the UT angles. 
The most prominent example is the determination of $\sin2\beta$ through 
CP violation in $B^0_d\to J/\psi K_{\rm S}$ decays, but several other strategies 
were proposed.
\end{itemize}
The goal is to ``overconstrain'' the UT as much as possible. In the future, 
additional contours can be fixed in the $\bar\rho$--$\bar\eta$ plane through 
the measurement of rare decays. 

In Fig.~\ref{fig:UTfits}, we show examples of the comprehensive
analyses of the UT that are performed (and continuously updated)
by the ``CKM Fitter Group'' \cite{CKMfitter}
and the ``UTfit collaboration''~\cite{UTfit}. In these figures, we can nicely see the
circles that are determined through the semi-leptonic $B$ decays and the 
$\varepsilon_K$ hyperbolas. Moreover, also the straight lines following from the 
direct measurement of $\sin 2\beta$ with the help of $B^0_d\to J/\psi K_{\rm S}$ 
modes are shown. We observe that the global consistency is very good. However,
looking closer, we also see that the most recent average for 
$(\sin 2\beta)_{\psi K_{\rm S}}$ is now on the lower side, so that the situation in 
the $\bar\rho$--$\bar\eta$ plane is no longer ``perfect". Moreover, as we
shall discuss in detail in the course of these lectures, there are certain puzzles in the
$B$-factory data, and several important aspects could not yet be addressed
experimentally and are hence still essentially unexplored. Consequently, we may hope
that flavour studies will eventually establish deviations from the SM description 
of CP violation. Since $B$ mesons play a key r\^ole in these explorations, let us 
next have a closer look at them.

\section{DECAYS OF {\boldmath$B$\unboldmath} MESONS}\label{sec:Bdecays}
\setcounter{equation}{0}
The $B$-meson system consists of charged and neutral $B$ mesons, which are 
characterized by the valence quark contents in (\ref{B-valence}).
The characteristic feature
of the neutral $B_q$ ($q\in \{d,s\}$) mesons is the phenomenon
of $B_q^0$--$\bar B_q^0$ mixing, which will be discussed in 
Section~\ref{sec:mix}. As far as the weak decays of 
$B$ mesons are concerned, we distinguish between leptonic, 
semileptonic and non-leptonic transitions.

\begin{figure}
\vskip -0.2truein
\begin{center}
\leavevmode
\epsfysize=3.8truecm 
\epsffile{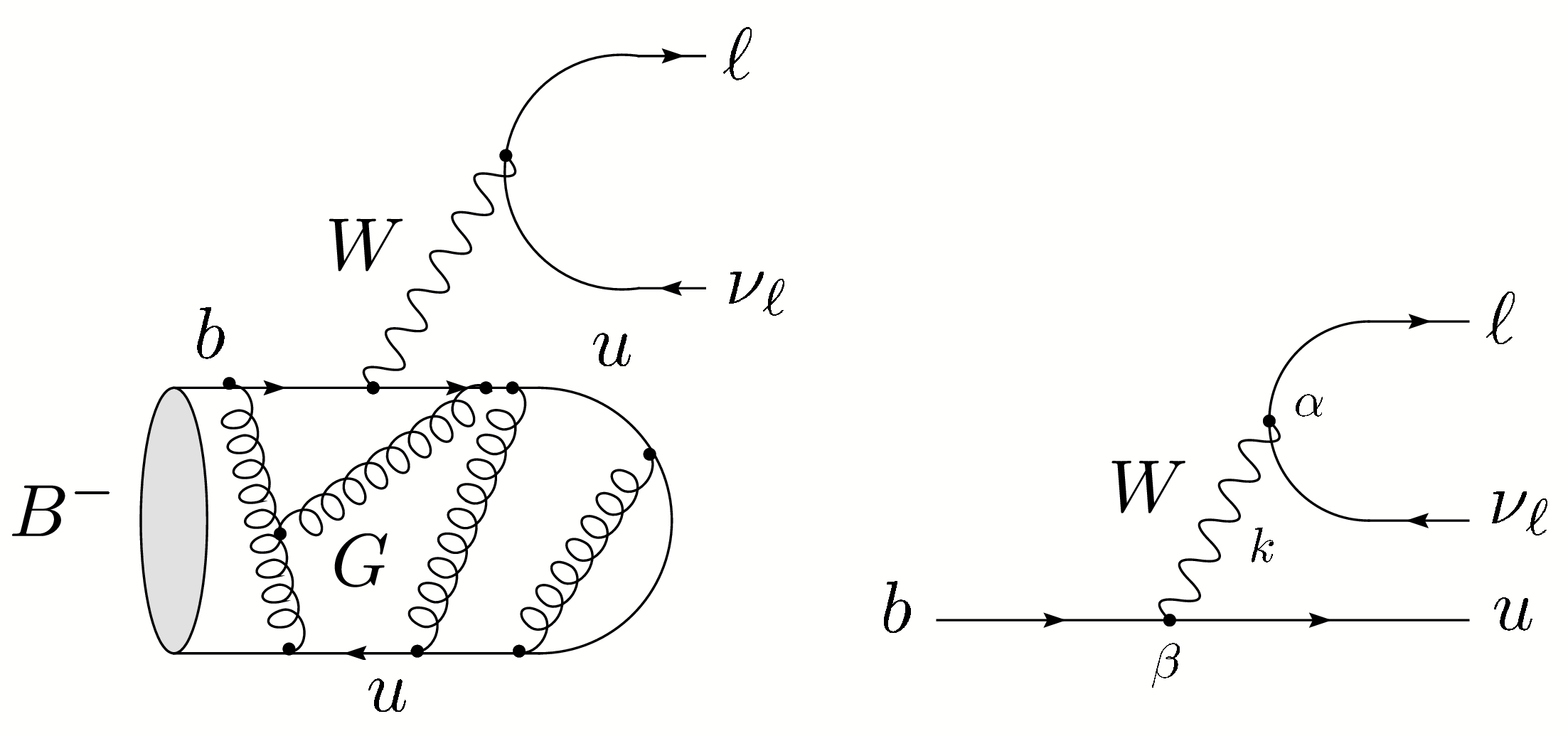} 
\end{center}
\vspace*{-0.8truecm}
\caption{Feynman diagrams contributing to the leptonic decay
$B^-\to \ell\bar\nu_\ell$.}\label{fig:lep}
\end{figure}

\subsection{Leptonic Decays}
The simplest $B$-meson decay class is given by leptonic decays 
of the kind $B^-\to \ell\bar\nu_\ell$, as illustrated in Fig.~\ref{fig:lep}.
If we evaluate the corresponding Feynman diagram, we arrive at the 
following transition amplitude:
\begin{equation}\label{Tfi-lept}
T_{fi}=-\,\frac{g_2^2}{8} V_{ub}
\underbrace{\left[\bar u_\ell\gamma^\alpha(1-\gamma_5)v_\nu
\right]}_{\mbox{Dirac spinors}}
\left[\frac{g_{\alpha\beta}}{k^2-M_W^2}\right]
\underbrace{\langle 0|\bar u\gamma^\beta
(1-\gamma_5)b|B^-\rangle}_{\mbox{hadronic ME}},
\end{equation}
where $g_2$ is the $SU(2)_{\rm L}$ gauge coupling, $V_{ub}$ the corresponding
element of the CKM matrix, $\alpha$ and $\beta$ are Lorentz indices,
and $M_W$ denotes the mass of the $W$ gauge boson. Since the four-momentum
$k$ that is carried by the $W$ satisfies $k^2=M_B^2\ll M_W^2$, we may
write
\begin{equation}\label{W-int-out}
\frac{g_{\alpha\beta}}{k^2-M_W^2}\quad\longrightarrow\quad
-\,\frac{g_{\alpha\beta}}{M_W^2}\equiv-\left(\frac{8G_{\rm F}}{\sqrt{2}g_2^2}
\right)g_{\alpha\beta},
\end{equation}
where $G_{\rm F}$ is Fermi's constant. Consequently, we may ``integrate out'' 
the $W$ boson in (\ref{Tfi-lept}), which yields
\begin{equation}\label{Tfi-lept-2}
T_{fi}=\frac{G_{\rm F}}{\sqrt{2}}V_{ub}\left[\bar u_\ell\gamma^\alpha
(1-\gamma_5)v_\nu\right]\langle 0|\bar u\gamma_\alpha(1-\gamma_5)b
|B^-\rangle.
\end{equation}
In this simple expression, {\it all} the hadronic physics is encoded 
in the {\it hadronic matrix element} 
\begin{displaymath}
\langle 0|\bar u\gamma_\alpha(1-\gamma_5)b
|B^-\rangle,
\end{displaymath}
i.e.\ there are no other strong-interaction QCD effects (for a detailed discussion
of QCD, see Ref.~\cite{ecker}). Since the $B^-$ meson is a pseudoscalar particle, 
we have
\begin{equation}\label{ME-rel1}
\langle 0|\overline{u}\gamma_\alpha b|B^-\rangle=0,
\end{equation}
and may write
\begin{equation}\label{ME-rel2}
\langle 0|\bar u\gamma_\alpha\gamma_5 b|B^-(q)\rangle =
i f_B q_\alpha,
\end{equation}
where $f_B$ is the $B$-meson {\it decay constant}, which is an important
input for phenomenological studies. In order to determine this
quantity, which is a very challenging task, non-perturbative 
techniques, such as QCD sum-rule analyses \cite{khodjamirian} or 
lattice studies, where a numerical evaluation of the QCD path integral is 
performed with the help of a space-time lattice, \cite{luscher}--\cite{luscher-rev}, 
are required.  If we use (\ref{Tfi-lept-2}) with (\ref{ME-rel1}) and 
(\ref{ME-rel2}), and perform the corresponding phase-space integrations, 
we obtain the following decay rate:
\begin{equation}
\Gamma(B^-\to\ell \bar \nu_\ell)=\frac{G_{\rm F}^2}{8\pi}
M_Bm_\ell^2\left(1-\frac{m_\ell^2}{M_B^2}\right)^2f_B^2|V_{ub}|^2 ,
\end{equation}
where $M_B$ and $m_\ell$ denote the masses of the $B^-$ and $\ell$,
respectively. Because of the tiny value of $|V_{ub}|\propto\lambda^3$ 
and a helicity-suppression mechanism, we obtain unfortunately very small 
branching ratios of ${\cal O}(10^{-10})$ and ${\cal O}(10^{-7})$ for 
$\ell=e$ and $\ell=\mu$, respectively \cite{fulvia}. The helicity 
suppression is not effective for $\ell=\tau$, but -- because of the 
required $\tau$ reconstruction -- these modes are also very challenging from 
an experimental point of view. Nevertheless, the Belle experiment has recently
reported the first evidence for the purely leptonic decay 
$B^- \to \tau^- \bar\nu_\tau$, with the following branching ratio \cite{Belle-leptonic}:
\begin{equation}
\mbox{BR}(B^- \to \tau^- \bar\nu_\tau) = \left[1.06 ^{+0.34}_{-0.28} \,
\mbox{(stat)} \,  ^{+0.18}_{-0.16} \, \mbox{(syst)}\right]\times 10^{-4},
\end{equation}
which corresponds to a significance of 4.2 standard deviations. Using the
SM expression for this branching ratio and the measured values of
$G_{\rm F}$, $M_B$, $m_\tau$ and the $B$-meson lifetime, the 
product of the $B$-meson decay constant $f_B$ and the magnitude of the
CKM matrix element $|V_{ub}|$ is obtained as
\begin{equation}
f_B|V_{ub}|=\left[7.73^{+1.24}_{-1.02}\,
\mbox{(stat)} \,  ^{+0.66}_{-0.58} \, \mbox{(syst)}\right]\times 10^{-4} \, \mbox{GeV}.
\end{equation}
The determination of this quantity is very interesting, as knowledge of $|V_{ub}|$
allows us to extract $f_B$, thereby providing tests of non-perturbative calculations 
of this important parameter. 

Before discussing the determination of  $|V_{ub}|$ from semileptonic $B$ decays 
in the next subsection, let us have a look at the leptonic $D$-meson decay
$D^+\to \mu^+\nu$. It is governed by the CKM factor 
\begin{equation}
|V_{cd}|=|V_{us}|+{\cal O}(\lambda^5)=
\lambda[1+{\cal O}(\lambda^4)],
\end{equation}
whereas $B^-\to \mu^-\bar \nu$ involves $|V_{ub}|=\lambda^3 R_b$.
Consequently, we win a factor of ${\cal O}(\lambda^4)$ in the decay rate,
so that $D^+\to \mu^+\nu$ is accessible at the CLEO-c experiment \cite{CLEO-c}. 
Since the corresponding CKM factor is well known, the decay constant  $f_{D^+}$
defined in analogy to (\ref{ME-rel2}) can be extracted,  
allowing another interesting testing ground for lattice calculations. Thanks to
recent progress in these techniques \cite{davies-eps}, the ``quenched" 
approximation, which had to be applied for many many years and ingnores 
quark loops, is no longer required for the calculation of $f_{D^+}$. In the 
summer of 2005, there was a first show down between the corresponding 
theoretical prediction and experiment:
the lattice result of $f_{D^+}=(201\pm3\pm17)\mbox{MeV}$ was reported
\cite{fD-lat}, while CLEO-c announced the measurement of
$f_{D^+}=(222.6 \pm 16.7 ^{+2.8}_{-3.4})\,\mbox{MeV}$ \cite{fD-cleoc}.
Both numbers agree well within the uncertainties, and it will be interesting to
stay tuned for future results.

\begin{figure}
\vskip -0.2truein
\begin{center}
\leavevmode
\epsfysize=3.8truecm 
\epsffile{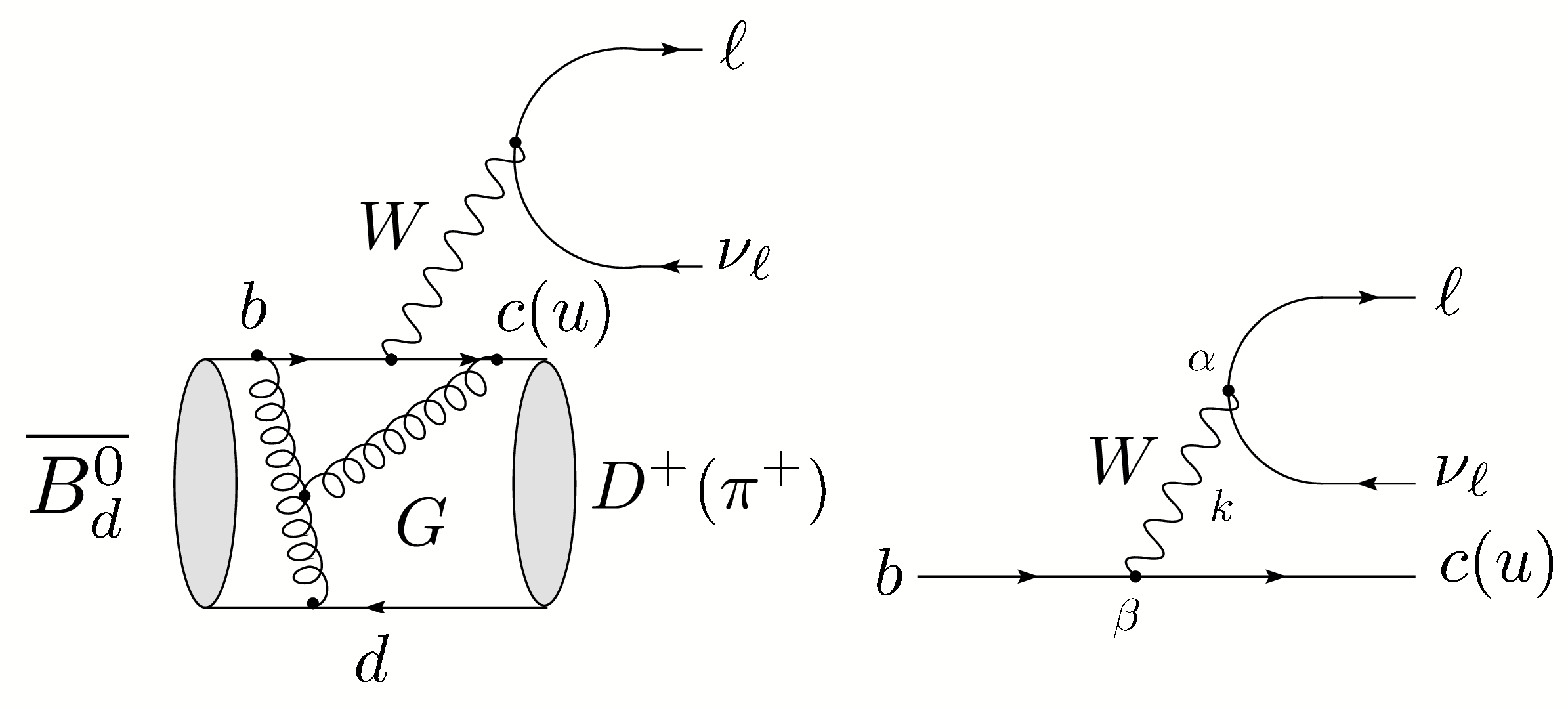} 
\end{center}
\vspace*{-0.8truecm}
\caption{Feynman diagrams contributing to semileptonic 
$\bar B^0_d\to D^+ (\pi^+) \ell \bar \nu_\ell$ decays.}\label{fig:semi}
\end{figure}

\subsection{Semileptonic Decays}\label{subsec:semi-lept}
\subsubsection{General Structure}
Semileptonic $B$-meson decays of the kind shown in Fig.~\ref{fig:semi}
have a structure that is more complicated than the one of the 
leptonic transitions. If we evaluate the corresponding Feynman diagram
for the $b\to c$ case, we obtain
\begin{equation}\label{Tfi-semi-full}
T_{fi}=-\,\frac{g_2^2}{8} V_{cb}
\underbrace{\left[\bar u_\ell\gamma^\alpha(1-\gamma_5)v_\nu
\right]}_{\mbox{Dirac spinors}}
\left[\frac{g_{\alpha\beta}}{k^2-M_W^2}\right]
\underbrace{\langle D^+|\bar c\gamma^\beta
(1-\gamma_5)b|\bar B^0_d\rangle}_{\mbox{hadronic ME}}.
\end{equation}
Because of $k^2\sim M_B^2\ll M_W^2$, we may again -- as in (\ref{Tfi-lept}) --
integrate out the $W$ boson with the help of (\ref{W-int-out}), which
yields
\begin{equation}\label{Tfi-semi}
T_{fi}=\frac{G_{\rm F}}{\sqrt{2}}V_{cb}\left[\bar u_\ell\gamma^\alpha
(1-\gamma_5)v_\nu\right]\langle  D^+|\bar c\gamma_\alpha(1-\gamma_5)b
|\bar B^0_d\rangle,
\end{equation}
where {\it all} the hadronic physics is encoded in the hadronic
matrix element
\begin{displaymath}
\langle D^+|\bar c\gamma_\alpha
(1-\gamma_5)b|\bar B^0_d\rangle,
\end{displaymath}
i.e.\ there are {\it no} other QCD effects.
Since the $\bar B^0_d$ and $D^+$ are pseudoscalar mesons, we have
\begin{equation}
\langle D^+|\bar c\gamma_\alpha\gamma_5b|
\bar B^0_d\rangle=0,
\end{equation}
and may write
\begin{equation}\label{BD-ME}
\langle D^+(k)|\bar c\gamma_\alpha b|\bar B^0_d(p)
\rangle=F_1(q^2)\left[(p+k)_\alpha -
\left(\frac{M_B^2-M_D^2}{q^2}\right)q_\alpha\right]
+F_0(q^2)\left(\frac{M_B^2-M_D^2}{q^2}\right)q_\alpha, 
\end{equation}
where $q\equiv p-k$, and the $F_{1,0}(q^2)$ denote the {\it form factors}
of the $\bar B\to D$ transitions. Consequently, in contrast to the simple 
case of the leptonic transitions, semileptonic decays involve {\it two} 
hadronic form factors instead of the decay constant $f_B$. In order to 
calculate these parameters, which depend on the momentum transfer $q$, 
again non-perturbative techniques (QCD sum rules, lattice,  etc.) are 
required.

\subsubsection{Aspects of the Heavy-Quark Effective Theory}
If the mass $m_Q$ of a quark $Q$ is much larger than the QCD scale parameter
$\Lambda_{\rm QCD}={\cal O}(100\,\mbox{MeV})$ \cite{ecker}, it is referred to as 
a ``heavy'' quark. Since the bottom and charm quarks have masses at the 
level of $5\,\mbox{GeV}$ and $1\,\mbox{GeV}$, respectively, they belong 
to this important category. As far as the extremely heavy top quark, 
with $m_t\sim 170\,\mbox{GeV}$ is concerned, it decays unfortunately 
through weak interactions before a hadron can be formed. Let us now 
consider a heavy quark that is bound inside a hadron, i.e.\ a bottom 
or a charm quark. The heavy quark then moves almost with the 
hadron's four velocity $v$ and is almost on-shell, so that
\begin{equation}
p_Q^\mu=m_Qv^\mu + k^\mu,
\end{equation}
where $v^2=1$ and $k\ll m_Q$ is the ``residual'' momentum. Owing to 
the interactions of the heavy quark with the light degrees of freedom of 
the hadron, the residual momentum may only change by 
$\Delta k\sim\Lambda_{\rm QCD}$, and $\Delta v \to 0$ for 
$\Lambda_{\rm QCD}/m_Q\to 0$. 

It is now instructive to have a look at the elastic scattering process 
$\bar B(v)\to \bar B(v')$ in the limit of $\Lambda_{\rm QCD}/m_b \to 0$, 
which is characterized by the following matrix element:
\begin{equation}\label{BB-ME}
\frac{1}{M_B}\langle\bar B(v')|\bar b_{v'}\gamma_\alpha b_v
|\bar B(v)\rangle=\xi(v'\cdot v)(v+v')_\alpha.
\end{equation}
Since the contraction of this matrix element with $(v-v')^\alpha$ has to 
vanish because of $\not \hspace*{-0.1truecm}v b_v= b_v$ and 
$\overline{b}_{v'} \hspace*{-0.2truecm}\not \hspace*{-0.1truecm}v' = 
\overline{b}_{v'}$, no $(v-v')_\alpha$ term arises in the parametrization
in (\ref{BB-ME}). On the other hand, the $1/M_B$ factor is related
to the normalization of states, i.e.\ the right-hand side of
\begin{equation}
\left(\frac{1}{\sqrt{M_B}}\langle\bar B(p')|\right)
\left(|\bar B(p)\rangle\frac{1}{\sqrt{M_B}}\right)=2v^0(2\pi)^3
\delta^3(\vec p-\vec p')
\end{equation}
does not depend on $M_B$. Finally, current conservation implies 
the following normalization condition:
\begin{equation}
\xi(v'\cdot v=1)=1,
\end{equation}
where  the ``Isgur--Wise'' function $\xi(v'\cdot v)$ does 
{\it not} depend on the flavour of the heavy quark (heavy-quark symmetry)
\cite{IW}. Consequently, for $\Lambda_{\rm QCD}/m_{b,c}\to 0$, we may write
\begin{equation}\label{BD-ME-HQ}
\frac{1}{\sqrt{M_D M_B}}\langle D(v')|\bar c_{v'}\gamma_\alpha b_v
|\bar B(v)\rangle=\xi(v'\cdot v)(v+v')_\alpha,
\end{equation}
and observe that this transition amplitude is governed -- in the 
heavy-quark limit -- by {\it one} hadronic form factor $\xi(v'\cdot v)$, 
which satisfies $\xi(1)=1$. If we now compare (\ref{BD-ME-HQ})
with (\ref{BD-ME}), we obtain 
\begin{equation}
F_1(q^2)=\frac{M_D+M_B}{2\sqrt{M_DM_B}}\xi(w)
\end{equation}
\begin{equation}
F_0(q^2)=\frac{2\sqrt{M_DM_B}}{M_D+M_B}\left[\frac{1+w}{2}\right]\xi(w),
\end{equation}
with
\begin{equation}
w\equiv v_D\cdot v_B=\frac{M_D^2+M_B^2-q^2}{2M_DM_B}.
\end{equation}
Similar relations hold  for the $\bar B\to D^\ast$ form factors
because of the heavy-quark spin symmetry, since the $D^\ast$ is 
related to the $D$ by a rotation of the heavy-quark spin. A detailed 
discussion of these interesting features and the associated ``heavy-quark 
effective theory'' (HQET) is beyond the scope of these lectures. For
a detailed overview, we refer the reader to Ref.~\cite{neubert-rev}, where
also a comprehensive list of original references can be found. 
For a more phenomenological discussion, also Ref.~\cite{BaBar-book} is very
useful.

\subsubsection{Applications}
An important application of the formalism sketched above is the
extraction of the CKM element $|V_{cb}|$. To this end, 
$\bar B\to D^*\ell\bar \nu$ decays are particularly promising. 
The corresponding rate can be written as 
\begin{equation}\label{BD-rate}
\frac{{\rm d}\Gamma}{{\rm d}w}=G_{\rm F}^2 K(M_B,M_{D^\ast},w) 
F(w)^2 |V_{cb}|^2,
\end{equation}
where $K(M_B,M_{D^\ast},w)$ is a known kinematic function, and 
$F(w)$ agrees with the Isgur--Wise function, up to perturbative
QCD corrections and $\Lambda_{\rm QCD}/m_{b,c}$ terms. The form 
factor $F(w)$ is a non-perturbative quantity. However, it satisfies 
the following normalization condition:
\begin{equation}\label{F1-norm}
F(1)=\eta_A(\alpha_s)\left[1+\frac{0}{m_c}+
\frac{0}{m_b}+
{\cal O}(\Lambda_{\rm QCD}^2/m_{b,c}^2)\right],
\end{equation} 
where $\eta_A(\alpha_s)$ is a perturbatively calculable short-distance
QCD factor, and the $\Lambda_{\rm QCD}/m_{b,c}$ corrections {\it vanish}
\cite{neubert-rev,neu-BDast}. The important latter feature is an 
implication of Luke's theorem \cite{luke}. Consequently, 
if we extract $F(w)|V_{cb}|$ from a measurement of (\ref{BD-rate}) 
as a function of $w$ and extrapolate to the ``zero-recoil point'' $w=1$
(where the rate vanishes), we may determine $|V_{cb}|$. In the case of 
$\bar B\to D\ell\bar \nu$ decays, we have 
${\cal O}(\Lambda_{\rm QCD}/m_{b,c})$ corrections to the 
corresponding rate ${\rm d}\Gamma/{\rm d}w$ at $w=1$. 
In order to determine $|V_{cb}|$, inclusive $B\to X_c\ell\bar \nu$ decays 
offer also very attractive avenues. As becomes obvious from (\ref{Def-A})
and the considerations in Subsection~\ref{ssec:UT}, $|V_{cb}|$ fixes 
the normalization of the UT. Moreover, this quantity is an important 
input parameter for various theoretical calculations. The CKM matrix
element $|V_{cb}|$ is currently known with $2\%$ precision;  performing an analysis
of leptonic and hadronic moments in inclusive $b\to c \ell \bar \nu$ processes 
\cite{Gambino}, the following value was extracted from the $B$-factory 
data \cite{OBuchmuller}:
\begin{equation}\label{Vcb}
|V_{cb}| = (42.0\pm 0.7)\times 10^{-3},
\end{equation}
which agrees with that from exclusive decays.

Let us now turn to $\bar B\to \pi\ell\bar\nu, \rho\ell\bar\nu$ decays, 
which originate from $b\to u\ell \bar\nu$ quark-level processes, as
can be seen in Fig.~\ref{fig:semi}, and provide access to $|V_{ub}|$. 
If we complement this CKM matrix element with $|V_{cb}|$, we may determine 
the side $R_b$ of the UT with the help of (\ref{Rb-def}). The 
determination of $|V_{ub}|$ is hence a very important aspect of
flavour physics. Since the $\pi$ and $\rho$ are ``light'' mesons, 
the HQET symmetry relations 
cannot be applied to the $\bar B\to \pi\ell\bar\nu, \rho\ell\bar\nu$ modes. 
Consequently, in order to determine $|V_{ub}|$ from these exclusive 
channels, the corresponding heavy-to-light form factors have to be
described by models. An important alternative is provided by inclusive 
decays. The corresponding decay rate takes the following form:
\begin{equation}\label{inclusive-rate}
\Gamma(\bar B\to X_u \ell \bar \nu)=
\frac{G_{\rm F}^2|V_{ub}|^2}{192\pi^3}m_b^5
\left[1-2.41\frac{\alpha_s}{\pi}+\frac{\lambda_1-9\lambda_2}{2m_b^2}
+\ldots\right],
\end{equation}
where $\lambda_1$ and $\lambda_2$ are non-perturbative parameters, 
which describe the hadronic matrix elements of certain ``kinetic'' 
and ``chromomagnetic'' operators appearing within the framework of
the HQET. Using the heavy-quark expansions
\begin{equation}\label{mass-exp}
M_B=m_b+\bar\Lambda-\frac{\lambda_1+3\lambda_2}{2m_b}+\ldots, \quad
M_{B^\ast}=m_b+\bar\Lambda-\frac{\lambda_1-\lambda_2}{2m_b}+\ldots
\end{equation}
for the $B^{(\ast)}$-meson masses, where $\bar\Lambda\sim\Lambda_{\rm QCD}$ 
is another non-perturbative parameter that is related to the light degrees
of freedom, the parameter $\lambda_2$ can be determined from the measured
values of the $M_{B^{(\ast)}}$. The strong dependence of 
(\ref{inclusive-rate}) on $m_b$ is a significant
source of uncertainty. On the other hand, the $1/m_b^2$ corrections
can be better controlled than in the exclusive case (\ref{F1-norm}), 
where we have, moreover, to deal with $1/m_c^2$ corrections. From an 
experimental point of view, we have to struggle with large backgrounds, 
which originate from $b\to c \ell \bar\nu$ processes and require also 
a model-dependent treatment. The determination of $|V_{ub}|$ from 
$B$-meson decays caused by $b\to u\ell \bar\nu$ 
quark-level processes is therefore a very challenging issue, and the situation 
is less favourable than with $|V_{cb}|$: there is a 1$\,\sigma$ discrepancy 
between the values from inclusive and exclusive transitions \cite{HFAG}:
\begin{equation}
|V_{ub}|_{\rm incl} = (4.4\pm 0.3)\times 10^{-3}\,,\quad 
|V_{ub}|_{\rm excl} = (3.8\pm 0.6)\times 10^{-3}\,,
\end{equation}
which has to be settled in the future. 
The error on $|V_{ub}|_{\rm excl}$ is dominated by the theoretical
uncertainty of lattice and light-cone sum rule calculations of $B\to\pi$ and
$B\to\rho$ transition form factors \cite{Vublatt,LCSR}, whereas for
$|V_{ub}|_{\rm incl}$ experimental and theoretical errors are at par.
Using the values of $|V_{cb}|$ and $|V_{ub}|$ given above
and $\lambda =0.225\pm0.001$ \cite{blucher}, we obtain
\begin{equation}\label{Rb}
R_b^{\rm incl} = 0.45\pm 0.03\,,\qquad R_b^{\rm excl} = 0.39\pm
0.06\,,
\end{equation}
where the labels ``incl" and ``excl" refer to the determinations of $|V_{ub}|$
through inclusive and exclusive $b\to u\ell\bar \nu_\ell$ transitions, 
respectively.

For a much more detailed discussion of the determinations of $|V_{cb}|$ 
and $|V_{ub}|$, addressing also various recent developments 
and the future prospects, we refer the reader to Ref.~\cite{CKM-book}, where 
also the references to the vast original literature can be found. Another 
excellent presentation is given in Ref.~\cite{BaBar-book}.

\begin{figure}
   \centerline{
   \begin{tabular}{lc}
     {\small(a)} & \\
    &     \includegraphics[width=3.3truecm]{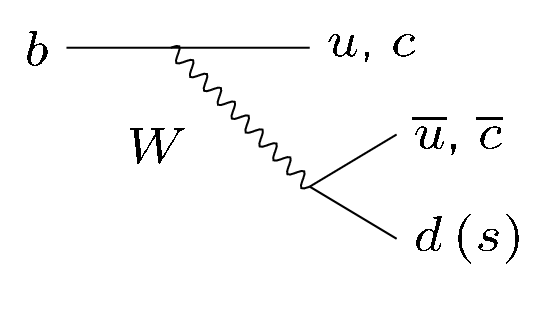}\\
     {\small(b)} & \\
    &  \includegraphics[width=5.0truecm]{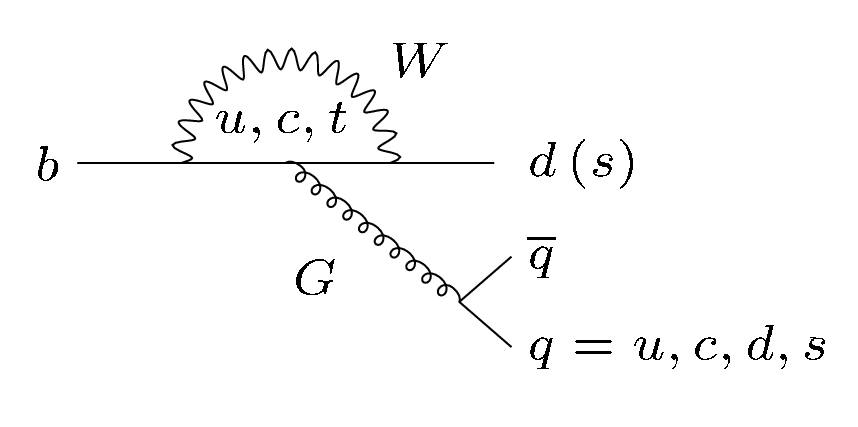}\\
     {\small(c)} & \\
    &     \includegraphics[width=8.3truecm]{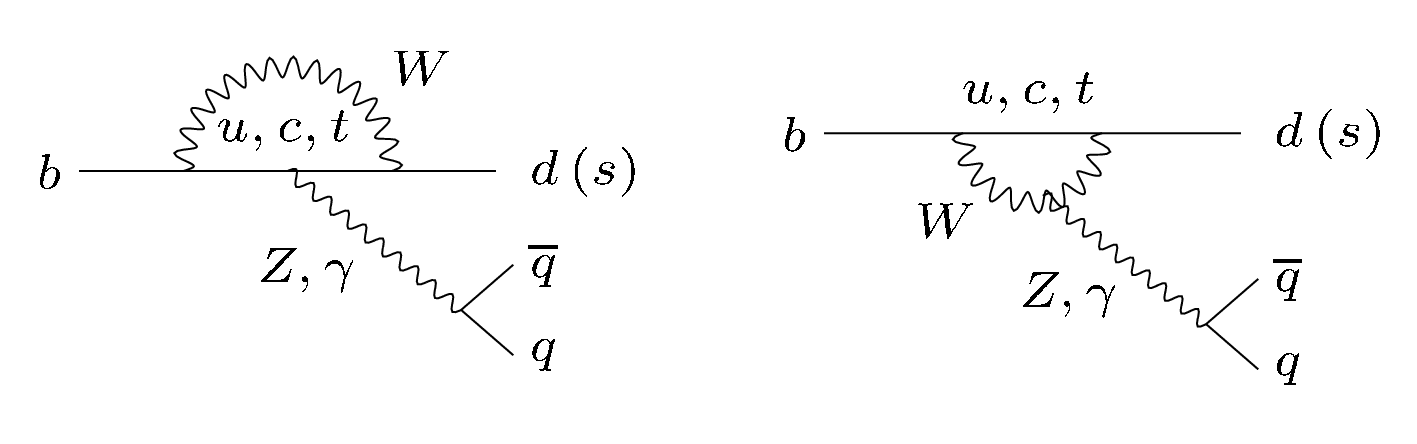} 
     \end{tabular}}
     \caption{Feynman diagrams of the topologies characterizing non-leptonic 
     $B$ decays: trees (a), QCD penguins (b), and electroweak penguins 
     (c).}\label{fig:topol}
\end{figure}

\subsection{Non-Leptonic Decays}\label{subsec:non-lept}
\subsubsection{Classification}\label{sec:class}
The most complicated $B$ decays are the non-leptonic transitions, 
which are mediated by 
$b\to q_1\,\bar q_2\,d\,(s)$ quark-level processes, with 
$q_1,q_2\in\{u,d,c,s\}$. There are two kinds of 
topologies contributing to such decays: tree-diagram-like and ``penguin'' 
topologies. The latter consist of gluonic (QCD) and electroweak (EW) 
penguins. In Fig.~\ref{fig:topol}, the corresponding 
leading-order Feynman diagrams are shown. Depending
on the flavour content of their final states, we may classify 
$b\to q_1\,\bar q_2\,d\,(s)$ decays as follows:
\begin{itemize}
\item $q_1\not=q_2\in\{u,c\}$: {\it only} tree diagrams contribute.
\item $q_1=q_2\in\{u,c\}$: tree {\it and} penguin diagrams contribute.
\item $q_1=q_2\in\{d,s\}$: {\it only} penguin diagrams contribute.
\end{itemize}

\begin{figure}
\begin{center}
\leavevmode
\begin{tabular}{cc}
\epsfysize=4.0truecm 
\epsffile{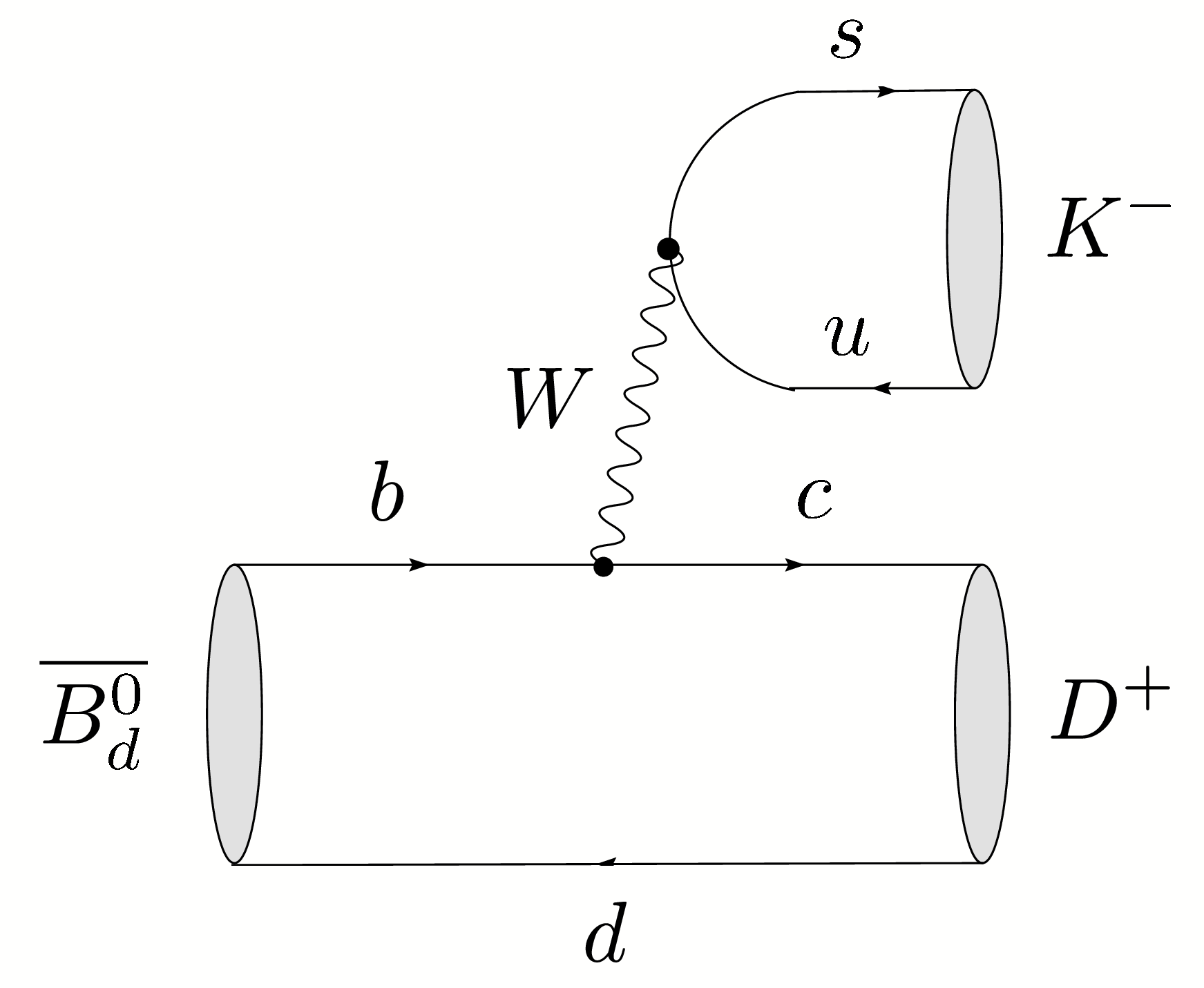} &
\epsfysize=2.0truecm 
\epsffile{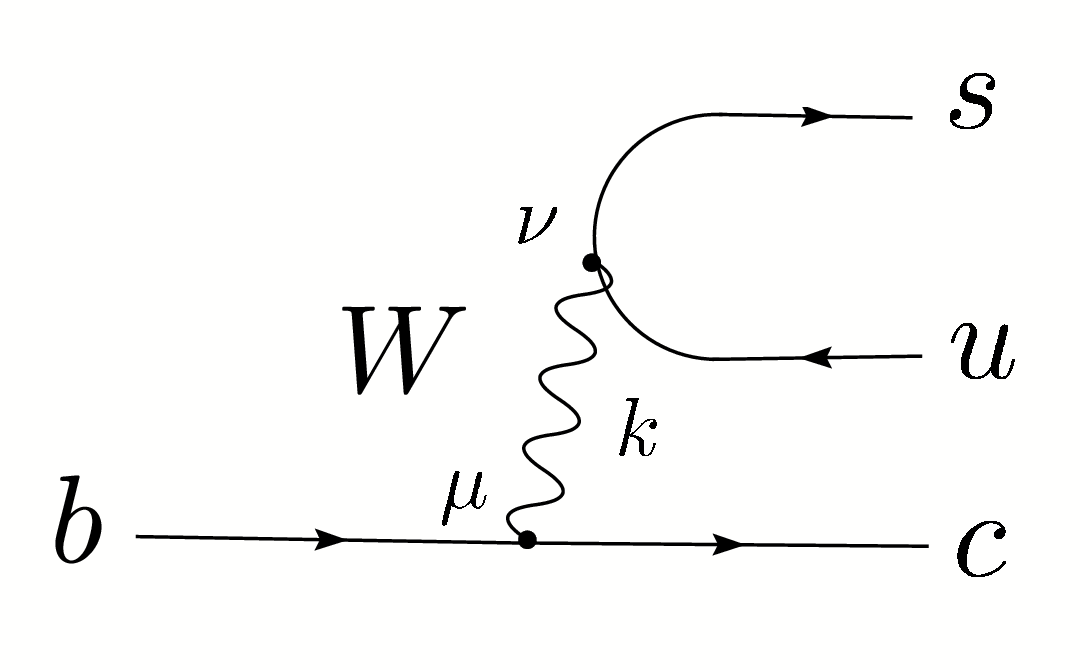} 
\end{tabular}
\end{center}
\vspace*{-0.8truecm}
\caption{Feynman diagrams contributing to the non-leptonic 
$\bar B^0_d\to D^+K^-$ decay.}\label{fig:non-lept-ex}
\end{figure}

\begin{figure}
\begin{center}
\leavevmode
\epsfysize=2.0truecm 
\epsffile{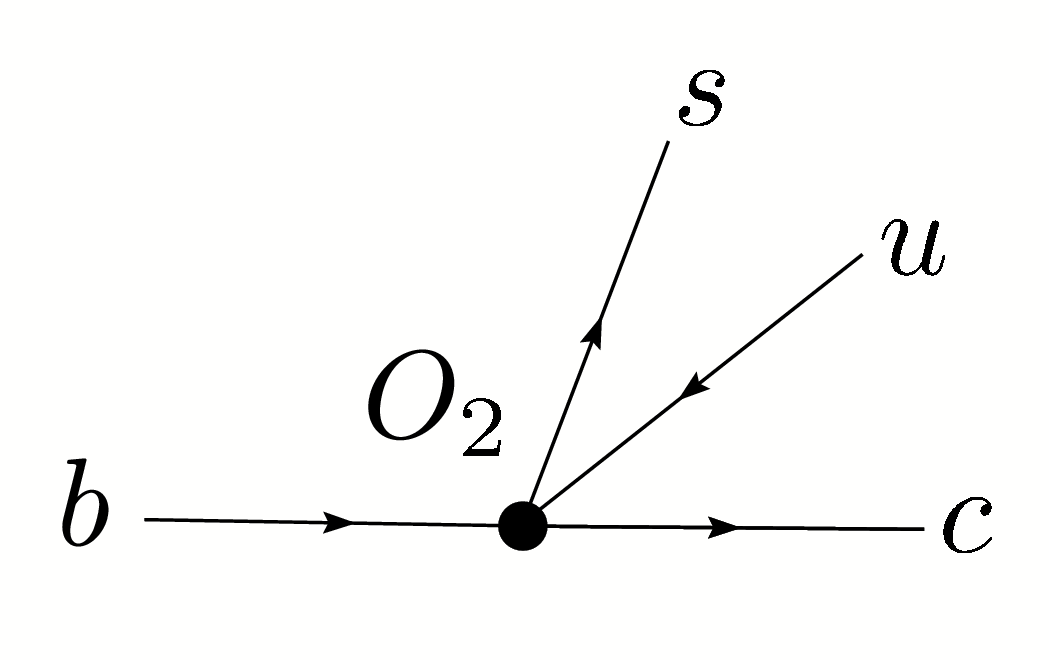} 
\end{center}
\vspace*{-0.8truecm}
\caption{The description of the $b\to d \bar u s$ process through the four-quark
operator $O_2$ in the effective theory after the $W$ boson has been integrated 
out.}\label{fig:non-lept-eff}
\end{figure}

\subsubsection{Low-Energy Effective Hamiltonians}\label{subsec:ham}
In order to analyse non-leptonic $B$ decays theoretically, one uses 
low-energy effective Hamiltonians, which are calculated by making use 
of the ``operator product expansion'', yielding transition 
matrix elements of the following structure:
\begin{equation}\label{ee2}
\langle f|{\cal H}_{\rm eff}|i\rangle=\frac{G_{\rm F}}{\sqrt{2}}
\lambda_{\rm CKM}\sum_k C_k(\mu)\langle f|Q_k(\mu)|i\rangle\,.
\end{equation}
The technique of the operator product expansion allows us to separate 
the short-distance contributions to this transition amplitude from the 
long-distance ones, which are described by perturbative quantities 
$C_k(\mu)$ (``Wilson coefficient functions'') and non-perturbative 
quantities $\langle f|Q_k(\mu)|i\rangle$ (``hadronic matrix elements''), 
respectively. As before, $G_{\rm F}$ is the Fermi constant, whereas
$\lambda_{\rm CKM}$ is a CKM factor and $\mu$ denotes an appropriate 
renormalization scale. The $Q_k$ are local operators, which 
are generated by electroweak interactions and QCD, and govern ``effectively'' 
the decay in question. The Wilson coefficients $C_k(\mu)$ can be 
considered as scale-dependent couplings related to the vertices described
by the $Q_k$.

In order to illustrate this rather abstract formalism, let us consider
the decay $\bar B^0_d\to D^+K^-$, which allows a transparent discussion 
of the evaluation of the corresponding low-energy effective Hamiltonian.
Since this transition originates from a $b\to c \bar u s$ quark-level 
process, it is -- as we have seen in our classification in 
Subsection~\ref{sec:class} -- a pure ``tree'' decay, i.e.\ we do not have
to deal with penguin topologies, which simplifies the analysis
considerably. The leading-order Feynman diagram contributing to
$\bar B^0_d\to D^+K^-$ can straightforwardly be obtained from 
Fig.~\ref{fig:semi} by substituting $\ell$ and $\nu_\ell$ by $s$ and $u$, 
respectively, as can be seen in Fig.~\ref{fig:non-lept-ex}. Consequently, 
the lepton current is simply replaced by a 
quark current, which will have important implications shown below. 
Evaluating the corresponding Feynman diagram yields
\begin{equation}\label{trans-ampl}
-\,\frac{g_2^2}{8}V_{us}^\ast V_{cb}
\left[\bar s\gamma^\nu(1-\gamma_5)u\right]
\left[\frac{g_{\nu\mu}}{k^2-M_W^2}\right]
\left[\bar c\gamma^\mu(1-\gamma_5)b\right].
\end{equation}
Because of $k^2\sim m_b^2\ll M_W^2$, we may -- as in (\ref{Tfi-semi-full}) -- 
``integrate out'' the $W$ boson with the help of (\ref{W-int-out}),
and arrive at
\begin{eqnarray}
\lefteqn{{\cal H}_{\rm eff}=\frac{G_{\rm F}}{\sqrt{2}}V_{us}^\ast V_{cb}
\left[\bar s_\alpha\gamma_\mu(1-\gamma_5)u_\alpha\right]
\left[\bar c_\beta\gamma^\mu(1-\gamma_5)b_\beta\right]}\nonumber\\
&&=\frac{G_{\rm F}}{\sqrt{2}}V_{us}^\ast V_{cb}
(\bar s_\alpha u_\alpha)_{\mbox{{\scriptsize 
V--A}}}(\bar c_\beta b_\beta)_{\mbox{{\scriptsize V--A}}}
\equiv\frac{G_{\rm F}}{\sqrt{2}}V_{us}^\ast V_{cb}O_2\,,
\end{eqnarray}
where $\alpha$ and $\beta$ denote the colour indices of the $SU(3)_{\rm C}$
gauge group of QCD. Effectively, our $b\to c \bar u s$ decay process 
is now described by the ``current--current'' operator $O_2$, as is illustrated
in Fig.~\ref{fig:non-lept-eff}.

\begin{figure}
\begin{center}
\leavevmode
\begin{tabular}{ccc}
\epsfysize=2.7truecm 
\epsffile{DIAG-fact-full.epsf} & \mbox{} &
\epsfysize=2.7truecm 
\epsffile{DIAG-fact-eff.epsf}
\end{tabular}
\end{center}
\vspace*{-0.5truecm}
\caption{Factorizable QCD corrections in the full and effective 
theories.}\label{fig:QCD-fact}
\end{figure}

\begin{figure}
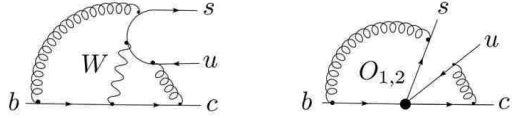

\begin{center}
\leavevmode
\begin{tabular}{ccc}
\epsfysize=1.8truecm 
\epsffile{DIAG-non-fact-full.epsf} &   \mbox{} &
\epsfysize=1.8truecm 
\epsffile{DIAG-non-fact-eff.epsf}
\end{tabular}
\end{center}
\vspace*{-0.5truecm}
\caption{Non-factorizable QCD corrections in the full and effective 
theories.}\label{fig:QCD-nonfact}
\end{figure}

So far, we neglected QCD corrections. Their important impact is
twofold: thanks to {\it factorizable} QCD corrections as shown in Fig.~\ref{fig:QCD-fact}, 
the Wilson coefficient $C_2$ acquires a renormalization-scale dependence,
i.e.\ $C_2(\mu)\not=1$. On the other hand,  {\it non-factorizable} QCD corrections as
illustrated in Fig.~\ref{fig:QCD-nonfact} generate a second current--current operator
through ``operator mixing", which is given by
\begin{equation}
O_1\equiv\left[\bar s_\alpha\gamma_\mu(1-\gamma_5)u_\beta\right]
\left[\bar c_\beta\gamma^\mu(1-\gamma_5)b_\alpha\right].
\end{equation}
Consequently, we eventually arrive at  a low-energy effective Hamiltonian of the
following structure:
\begin{equation}\label{Heff-example}
{\cal H}_{\rm eff}=\frac{G_{\rm F}}{\sqrt{2}}V_{us}^\ast V_{cb}
\left[C_1(\mu)O_1+C_2(\mu)O_2\right].
\end{equation}
In order to evaluate the Wilson  
coefficients $C_1(\mu)\not=0$ and $C_2(\mu)\not=1$ \cite{HEFF-TREE}, 
we must first calculate the QCD corrections to the decay processes 
both in the full theory, i.e. with $W$ exchange, and in the effective 
theory, where the $W$ is integrated out (see Figs.~\ref{fig:QCD-fact} and 
\ref{fig:QCD-nonfact}), and have then to express 
the QCD-corrected transition amplitude in terms of QCD-corrected matrix 
elements and Wilson coefficients as in (\ref{ee2}). This procedure is 
called ``matching'' between the full and the effective theory. 
The results for the $C_k(\mu)$ thus obtained contain 
terms of $\mbox{log}(\mu/M_W)$, which become large for $\mu={\cal O}(m_b)$, 
the scale governing the hadronic matrix elements of the $O_k$. Making use of 
the renormalization group, which exploits the fact that the transition 
amplitude (\ref{ee2}) cannot depend on the chosen renormalization scale 
$\mu$, we may sum up the following terms of the Wilson coefficients:
\begin{equation}
\alpha_s^n\left[\log\left(\frac{\mu}{M_W}\right)\right]^n 
\,\,\mbox{(LO)},\quad\,\,\alpha_s^n\left[\log\left(\frac{\mu}{M_W}\right)
\right]^{n-1}\,\,\mbox{(NLO)},\quad ...\quad ;
\end{equation}
detailed discussions of these rather technical aspects can be found in
Refs.~\cite{BBL-rev,B-LH98}.

For the exploration of CP violation, the class of non-leptonic $B$ decays 
that receives contributions both from tree and from penguin topologies plays 
a key r\^ole. In this important case, the operator basis is much larger 
than in our example (\ref{Heff-example}), where we considered a pure 
``tree'' decay. If we apply the relation
\begin{equation}\label{CKM-UT-Rel}
V_{ur}^\ast V_{ub}+V_{cr}^\ast V_{cb}+V_{tr}^\ast V_{tb}=0
\quad (r\in\{d,s\}),
\end{equation}
which follows from the unitarity of the CKM matrix, and ``integrate out''
the top quark (which enters through the penguin loop processes) and 
the $W$ boson, we may write 
\begin{equation}\label{e4}
{\cal H}_{\mbox{{\scriptsize eff}}}=\frac{G_{\mbox{{\scriptsize 
F}}}}{\sqrt{2}}\left[\sum\limits_{j=u,c}V_{jr}^\ast V_{jb}\left\{
\sum\limits_{k=1}^2C_k(\mu)\,Q_k^{jr}
+\sum\limits_{k=3}^{10}C_k(\mu)\,Q_k^{r}\right\}\right].
\end{equation}
Here we have introduced another quark-flavour label $j\in\{u,c\}$,
and the $Q_k^{jr}$ can be divided as follows:
\begin{itemize}
\item Current--current operators:
\begin{equation}
\begin{array}{rcl}
Q_{1}^{jr}&=&(\bar r_{\alpha}j_{\beta})_{\mbox{{\scriptsize V--A}}}
(\bar j_{\beta}b_{\alpha})_{\mbox{{\scriptsize V--A}}}\\
Q_{2}^{jr}&=&(\bar r_\alpha j_\alpha)_{\mbox{{\scriptsize 
V--A}}}(\bar j_\beta b_\beta)_{\mbox{{\scriptsize V--A}}}.
\end{array}
\end{equation}
\item QCD penguin operators:
\begin{equation}\label{qcd-penguins}
\begin{array}{rcl}
Q_{3}^r&=&(\bar r_\alpha b_\alpha)_{\mbox{{\scriptsize V--A}}}\sum_{q'}
(\bar q'_\beta q'_\beta)_{\mbox{{\scriptsize V--A}}}\\
Q_{4}^r&=&(\bar r_{\alpha}b_{\beta})_{\mbox{{\scriptsize V--A}}}
\sum_{q'}(\bar q'_{\beta}q'_{\alpha})_{\mbox{{\scriptsize V--A}}}\\
Q_{5}^r&=&(\bar r_\alpha b_\alpha)_{\mbox{{\scriptsize V--A}}}\sum_{q'}
(\bar q'_\beta q'_\beta)_{\mbox{{\scriptsize V+A}}}\\
Q_{6}^r&=&(\bar r_{\alpha}b_{\beta})_{\mbox{{\scriptsize V--A}}}
\sum_{q'}(\bar q'_{\beta}q'_{\alpha})_{\mbox{{\scriptsize V+A}}}.
\end{array}
\end{equation}
\item EW penguin operators (the $e_{q'}$ denote the
electrical quark charges):
\begin{equation}
\begin{array}{rcl}
Q_{7}^r&=&\frac{3}{2}(\bar r_\alpha b_\alpha)_{\mbox{{\scriptsize V--A}}}
\sum_{q'}e_{q'}(\bar q'_\beta q'_\beta)_{\mbox{{\scriptsize V+A}}}\\
Q_{8}^r&=&
\frac{3}{2}(\bar r_{\alpha}b_{\beta})_{\mbox{{\scriptsize V--A}}}
\sum_{q'}e_{q'}(\bar q_{\beta}'q'_{\alpha})_{\mbox{{\scriptsize V+A}}}\\
Q_{9}^r&=&\frac{3}{2}(\bar r_\alpha b_\alpha)_{\mbox{{\scriptsize V--A}}}
\sum_{q'}e_{q'}(\bar q'_\beta q'_\beta)_{\mbox{{\scriptsize V--A}}}\\
Q_{10}^r&=&
\frac{3}{2}(\bar r_{\alpha}b_{\beta})_{\mbox{{\scriptsize V--A}}}
\sum_{q'}e_{q'}(\bar q'_{\beta}q'_{\alpha})_{\mbox{{\scriptsize V--A}}}.
\end{array}
\end{equation}
\end{itemize}
The current--current, QCD and EW penguin operators are related to the tree, 
QCD and EW penguin processes shown in 
Fig.~\ref{fig:topol}. At a renormalization scale
$\mu={\cal O}(m_b)$, the Wilson coefficients of the current--current operators
are $C_1(\mu)={\cal O}(10^{-1})$ and $C_2(\mu)={\cal O}(1)$, whereas those
of the penguin operators are ${\cal O}(10^{-2})$ \cite{B-LH98,BBL-rev}. 
Note that penguin 
topologies with internal charm- and up-quark exchanges \cite{BSS}
are described in this framework by penguin-like matrix elements of 
the corresponding current--current operators \cite{RF-DIPL}, and 
may also have important phenomenological consequences \cite{BF-PEN,CHARM-PEN}.

Since the ratio $\alpha/\alpha_s={\cal O}(10^{-2})$ of the QED and QCD 
couplings is very small, we would expect na\"\i vely that EW penguins 
should play a minor r\^ole in comparison with QCD penguins. This would 
actually be the case if the top quark was not ``heavy''. However, since 
the Wilson coefficient $C_9$ increases strongly with $m_t$, we obtain 
interesting EW penguin effects in several $B$ decays: $B\to K\phi$ 
modes are affected significantly by EW penguins, whereas $B\to\pi\phi$ 
and $B_s\to\pi^0\phi$ transitions are even {\it dominated} by such 
topologies \cite{RF-EWP,RF-rev}. EW penguins also have an important 
impact on the $B\to\pi K$ system \cite{EWP-BpiK}.

The low-energy effective Hamiltonians discussed above apply to all 
$B$ decays that are caused by the same quark-level transition, i.e.\ 
they are ``universal''. Consequently, the differences between the 
various exclusive modes of a given decay class arise within this formalism 
only through the hadronic matrix elements of the relevant four-quark 
operators. Unfortunately, the evaluation of such matrix elements is 
associated with large uncertainties and is a very challenging task. In this 
context, ``factorization'' is a widely used concept, which is our next 
topic.

\subsubsection{Factorization of Hadronic Matrix Elements}\label{ssec:ME-fact}
In order to discuss ``factorization'', let us consider once more 
the decay $\bar B^0_d\to D^+K^-$. Evaluating the corresponding 
transition amplitude, we encounter the hadronic matrix elements of the 
$O_{1,2}$ operators between the $\langle K^-D^+|$ final and the 
$|\bar B^0_d\rangle$ initial states. If we use the well-known 
$SU(N_{\rm C})$ colour-algebra relation
\begin{equation}
T^a_{\alpha\beta}T^a_{\gamma\delta}=\frac{1}{2}\left(\delta_{\alpha\delta}
\delta_{\beta\gamma}-\frac{1}{N_{\rm C}}\delta_{\alpha\beta}
\delta_{\gamma\delta}\right)
\end{equation}
to rewrite the operator $O_1$, we obtain
\begin{displaymath}
\langle K^-D^+|{\cal H}_{\rm eff}|\bar B^0_d\rangle=
\frac{G_{\rm F}}{\sqrt{2}}V_{us}^\ast V_{cb}\Bigl[a_1\langle K^-D^+|
(\bar s_\alpha u_\alpha)_{\mbox{{\scriptsize V--A}}}
(\bar c_\beta b_\beta)_{\mbox{{\scriptsize V--A}}}
|\bar B^0_d\rangle
\end{displaymath}
\vspace*{-0.3truecm}
\begin{equation}\label{ME-rewritten}
+2\,C_1\langle K^-D^+|
(\bar s_\alpha\, T^a_{\alpha\beta}\,u_\beta)_{\mbox{{\scriptsize 
V--A}}}(\bar c_\gamma 
\,T^a_{\gamma\delta}\,b_\delta)_{\mbox{{\scriptsize V--A}}}
|\bar B^0_d\rangle\Bigr],\nonumber
\end{equation}
with
\begin{equation}\label{a1-def}
a_1=C_1/N_{\rm C}+C_2 \sim 1.
\end{equation}
It is now straightforward to ``factorize'' the hadronic matrix elements
in (\ref{ME-rewritten}):
\begin{eqnarray}
\lefteqn{\left.\langle K^-D^+|
(\bar s_\alpha u_\alpha)_{\mbox{{\scriptsize 
V--A}}}(\bar c_\beta b_\beta)_{\mbox{{\scriptsize V--A}}}
|\bar B^0_d\rangle\right|_{\rm fact}}\nonumber\\
&&=\langle K^-|\left[\bar s_\alpha\gamma_\mu(1-\gamma_5)u_\alpha\right]
|0\rangle\langle D^+|\left[\bar c_\beta\gamma^\mu
(1-\gamma_5)b_\beta\right]|\bar B^0_d\rangle\nonumber\\
&&=\underbrace{i f_K}_{\mbox{decay constant}} \, \times \, 
\underbrace{F^{(BD)}_0(M_K^2)}_{\mbox{$B\to D$ form factor}} 
\, \times \,\underbrace{(M_B^2-M_D^2),}_{\mbox{kinematical factor}}
\end{eqnarray}
\begin{equation}
\left.\langle K^-D^+|
(\bar s_\alpha\, T^a_{\alpha\beta}\,u_\beta)_{\mbox{{\scriptsize 
V--A}}}(\bar c_\gamma 
\,T^a_{\gamma\delta}\,b_\delta)_{\mbox{{\scriptsize V--A}}}
|\bar B^0_d\rangle\right|_{\rm fact}=0.
\end{equation}
The quantity $a_1$ is a phenomenological ``colour factor'', 
which governs ``colour-allowed'' decays; the
decay $\bar B^0_d\to D^+K^-$ belongs to this category, since the 
colour indices of the $K^-$ meson and the $\bar B^0_d$--$D^+$ system 
run independently from each other in the corresponding leading-order 
diagram shown in Fig.~\ref{fig:non-lept-ex}. On the other hand, in the case 
of ``colour-suppressed'' modes, for instance $\bar B^0_d\to \pi^0D^0$, where only 
one colour index runs through the whole diagram, we have to deal with the combination
\begin{equation}\label{a2-def}
a_2=C_1+C_2/N_{\rm C}\sim0.25.
\end{equation}

The concept of factorizing the hadronic matrix elements of four-quark 
operators into the product of hadronic matrix elements of quark currents 
has a long history \cite{Neu-Ste}, and can be justified, for example, 
in the large-$N_{\rm C}$ limit \cite{largeN}. Interesting recent 
developments are the following:
\begin{itemize}
\item ``QCD factorization'' \cite{BBNS}, which is in 
accordance with the old picture that factorization should 
hold for certain decays in the limit of $m_b\gg\Lambda_{\rm QCD}$ 
\cite{QCDF-old}, provides a formalism to calculate the 
relevant amplitudes at the leading order of a $\Lambda_{\rm QCD}/m_b$ 
expansion. The resulting expression for the transition amplitudes 
incorporates elements both of the na\"\i ve factorization approach 
sketched above and of the hard-scattering picture. Let us consider a 
decay $\bar B\to M_1M_2$, where $M_1$ picks up the spectator quark. 
If $M_1$ is either a heavy ($D$) or a light ($\pi$, $K$) meson, and 
$M_2$ a light ($\pi$, $K$) meson, QCD factorization gives a transition 
amplitude of the following structure:
\begin{equation}
A(\bar B\to M_1M_2)=\left[\mbox{``na\"\i ve factorization''}\right]
\times\left[1+{\cal O}(\alpha_s)+{\cal O}(\Lambda_{\rm QCD}/m_b)\right].
\end{equation}
While the ${\cal O}(\alpha_s)$ terms, i.e.\ the radiative
non-factorizable corrections, can be calculated systematically, 
the main limitation of the theoretical accuracy originates from 
the ${\cal O}(\Lambda_{\rm QCD}/m_b)$ terms. 

\item Another QCD approach to deal with non-leptonic $B$-meson decays -- 
the ``perturbative hard-scattering approach '' (PQCD) -- was developed 
independently in \cite{PQCD}, and differs from the QCD factorization 
formalism in some technical aspects.

\item An interesting technique for ``factorization proofs'' is provided 
by the framework of the ``soft collinear effective theory'' (SCET) 
\cite{SCET}, which has received a lot of attention in the recent literature 
and led to various applications.

\item Non-leptonic $B$ decays can also be studied within 
QCD light-cone sum-rule approaches \cite{sum-rules}.
\end{itemize}
A detailed presentation of these topics would be very technical and is
beyond the scope of these lectures. However, for the discussion of the
CP-violating effects in the $B$-meson system, we must only be familiar 
with the general structure of the non-leptonic $B$ decay amplitudes and 
not enter the details of the techniques to deal with the
corresponding hadronic matrix elements. Let us finally note that the 
$B$-factory data will eventually decide how well factorization and 
the new concepts sketched above are actually working. For example, 
data on the $B\to\pi\pi$ system point towards 
large non-factorizable corrections \cite{BFRS2,BFRS3}, to which we 
shall return in Subsection~\ref{ssec:Bpipi-hadr}.

\subsection{Towards Studies of CP Violation}\label{To-CP}
As we have seen above, leptonic and semileptonic $B$-meson decays
involve only a single weak (CKM) amplitude. On the other hand, the 
structure of non-leptonic transitions is considerably more complicated. 
Let us consider a non-leptonic decay $\bar B\to\bar f$ that is described by
the low-energy effective Hamiltonian  in (\ref{e4}). The corresponding
decay amplitude is then given as follows:
\begin{eqnarray}
\lefteqn{\hspace*{-1.3truecm}A(\bar B\to \bar f)=\langle \bar f\vert
{\cal H}_{\mbox{{\scriptsize eff}}}\vert\bar B\rangle}\nonumber\\
&&\hspace*{-1.3truecm}=\frac{G_{\mbox{{\scriptsize F}}}}{\sqrt{2}}\left[
\sum\limits_{j=u,c}V_{jr}^\ast V_{jb}\left\{\sum\limits_{k=1}^2
C_{k}(\mu)\langle \bar f\vert Q_{k}^{jr}(\mu)\vert\bar B\rangle
+\sum\limits_{k=3}^{10}C_{k}(\mu)\langle \bar f\vert Q_{k}^r(\mu)
\vert\bar B\rangle\right\}\right].~~~\mbox{}\label{Bbarfbar-ampl}
\end{eqnarray}
Concerning the CP-conjugate process $B\to\ f$, we have
\begin{eqnarray}
\lefteqn{\hspace*{-1.3truecm}A(B \to f)=\langle f|
{\cal H}_{\mbox{{\scriptsize 
eff}}}^\dagger|B\rangle}\nonumber\\
&&\hspace*{-1.3truecm}=\frac{G_{\mbox{{\scriptsize F}}}}{\sqrt{2}}
\left[\sum\limits_{j=u,c}V_{jr}V_{jb}^\ast \left\{\sum\limits_{k=1}^2
C_{k}(\mu)\langle f\vert Q_{k}^{jr\dagger}(\mu)\vert B\rangle
+\sum\limits_{k=3}^{10}C_{k}(\mu)\langle f\vert Q_k^{r\dagger}(\mu)
\vert B\rangle\right\}\right].~~~\mbox{}\label{Bf-ampl}
\end{eqnarray}
If we use now that strong interactions are invariant under CP transformations, 
insert $({\cal CP})^\dagger({\cal CP})=\hat 1$ both after the $\langle f|$ and in 
front of the $|B\rangle$, and take the relation 
\begin{equation}
({\cal CP})Q_k^{jr\dagger}({\cal CP})^\dagger=Q_k^{jr}
\end{equation}
into account, we arrive at
\begin{eqnarray}
\lefteqn{\hspace*{-1.3truecm}A(B \to f)=
e^{i[\phi_{\mbox{{\scriptsize CP}}}(B)-\phi_{\mbox{{\scriptsize CP}}}(f)]}}\nonumber\\
&&\hspace*{-1.3truecm}\times\frac{G_{\mbox{{\scriptsize F}}}}{\sqrt{2}}
\left[\sum\limits_{j=u,c}V_{jr}V_{jb}^\ast\left\{\sum\limits_{k=1}^2
C_{k}(\mu)\langle \bar f\vert Q_{k}^{jr}(\mu)\vert\bar B\rangle
+\sum\limits_{k=3}^{10}C_{k}(\mu)
\langle \bar f\vert Q_{k}^r(\mu)\vert\bar B\rangle\right\}\right],
\end{eqnarray}
where the convention-dependent phases $\phi_{\mbox{{\scriptsize CP}}}(B)$ 
and $\phi_{\mbox{{\scriptsize CP}}}(f)$ are defined through
\begin{equation}\label{CP-phase-def}
({\cal CP})\vert B\rangle=
e^{i\phi_{\mbox{{\scriptsize CP}}}(B)}
\vert\bar B\rangle, \quad 
({\cal CP})\vert f\rangle=
e^{i\phi_{\mbox{{\scriptsize CP}}}(f)}
\vert\bar f\rangle.
\end{equation}
Consequently, we may write
\begin{eqnarray}
A(\bar B\to\bar f)&=&e^{+i\varphi_1}
|A_1|e^{i\delta_1}+e^{+i\varphi_2}|A_2|e^{i\delta_2}\label{par-ampl}\\
A(B\to f)&=&e^{i[\phi_{\mbox{{\scriptsize CP}}}(B)-\phi_{\mbox{{\scriptsize CP}}}(f)]}
\left[e^{-i\varphi_1}|A_1|e^{i\delta_1}+e^{-i\varphi_2}|A_2|e^{i\delta_2}
\right].\label{par-ampl-CP}
\end{eqnarray}
Here the CP-violating phases $\varphi_{1,2}$ originate from the CKM factors 
$V_{jr}^\ast V_{jb}$, and the CP-conserving ``strong'' amplitudes
$|A_{1,2}|e^{i\delta_{1,2}}$ involve the hadronic matrix elements of the 
four-quark operators. In fact, these expressions are the most general forms
of any non-leptonic $B$-decay amplitude in the SM, i.e.\ they do not only
refer to the $\Delta C=\Delta U=0$ case described by (\ref{e4}). 
Using (\ref{par-ampl}) and (\ref{par-ampl-CP}), we obtain
the following CP asymmetry:
\begin{eqnarray}
{\cal A}_{\rm CP}&\equiv&\frac{\Gamma(B\to f)-
\Gamma(\bar B\to\bar f)}{\Gamma(B\to f)+\Gamma(\bar B
\to \bar f)}=\frac{|A(B\to f)|^2-|A(\bar B\to \bar f)|^2}{|A(B\to f)|^2+
|A(\bar B\to \bar f)|^2}\nonumber\\
&=&\frac{2|A_1||A_2|\sin(\delta_1-\delta_2)
\sin(\varphi_1-\varphi_2)}{|A_1|^2+2|A_1||A_2|\cos(\delta_1-\delta_2)
\cos(\varphi_1-\varphi_2)+|A_2|^2}.\label{direct-CPV}
\end{eqnarray}
We observe that a non-vanishing value can be generated through the 
interference between the two weak amplitudes, provided both a non-trivial 
weak phase difference $\varphi_1-\varphi_2$ and a non-trivial strong phase 
difference $\delta_1-\delta_2$ are present. This kind of
CP violation is referred to as ``direct'' CP violation, as it originates 
directly at the amplitude level of the considered decay. It is the 
$B$-meson counterpart of the effects that are probed through 
$\mbox{Re}(\varepsilon'/\varepsilon)$ in the neutral kaon 
system,\footnote{In order to calculate this quantity, an approriate 
low-energy effective Hamiltonian having the same structure as (\ref{e4}) 
is used. The large theoretical uncertainties mentioned in Section~\ref{sec:intro} 
originate from a strong cancellation 
between the contributions of the QCD and EW penguins (caused by the
large top-quark mass) and the associated
hadronic matrix elements.} and could recently be established with the help of 
$B_d\to\pi^\mp K^\pm$ decays \cite{CP-B-dir}, as we will see in
Subsection~\ref{ssec:Bpi+pi-}.

Since $\varphi_1-\varphi_2$ is in general given by one of the UT angles  -- 
usually $\gamma$ -- the goal is to extract this quantity from the measured 
value of ${\cal A}_{\rm CP}$. Unfortunately, hadronic uncertainties affect this
determination through the poorly known hadronic matrix elements in 
(\ref{Bbarfbar-ampl}). In order to deal with this problem, we may proceed along 
one of the following two avenues:
\begin{itemize}
\item[(i)] Amplitude relations can be used to eliminate the 
hadronic matrix elements. We distinguish between exact relations, 
using pure ``tree'' decays  of the kind $B^\pm\to K^\pm D$ \cite{gw,ADS} or 
$B_c^\pm\to D^\pm_s D$ \cite{fw}, and relations, which follow from the flavour symmetries 
of strong interactions, i.e.\ isospin or $SU(3)_{\rm F}$, and involve 
$B_{(s)}\to\pi\pi,\pi K,KK$ modes~\cite{GHLR}. 
\item[(ii)] In decays of neutral $B_q$ mesons, interference effects 
between $B^0_q$--$\bar B^0_q$ mixing and decay processes may induce
 ``mixing-induced CP violation''. If a single CKM amplitude governs the decay, 
 the hadronic matrix elements cancel in the corresponding
CP asymmetries; otherwise we have to use again amplitude relations.
The most important example is the decay $B^0_d\to J/\psi K_{\rm S}$ \cite{bisa}.
\end{itemize}
Before discussing the features of neutral $B_q$ mesons and  
$B^0_q$--$\bar B^0_q$ mixing in detail in Section~\ref{sec:mix}, let us illustrate
the use of amplitude relations for clean extractions of the UT angle $\gamma$
from decays of charged $B_u$ and $B_c$ mesons.

\section{AMPLITUDE RELATIONS}\label{sec:A-REL}
\setcounter{equation}{0}
\boldmath
\subsection{$B^\pm\to K^\pm D$}
\begin{figure}
\begin{center}
\leavevmode
\epsfysize=3.9truecm 
\epsffile{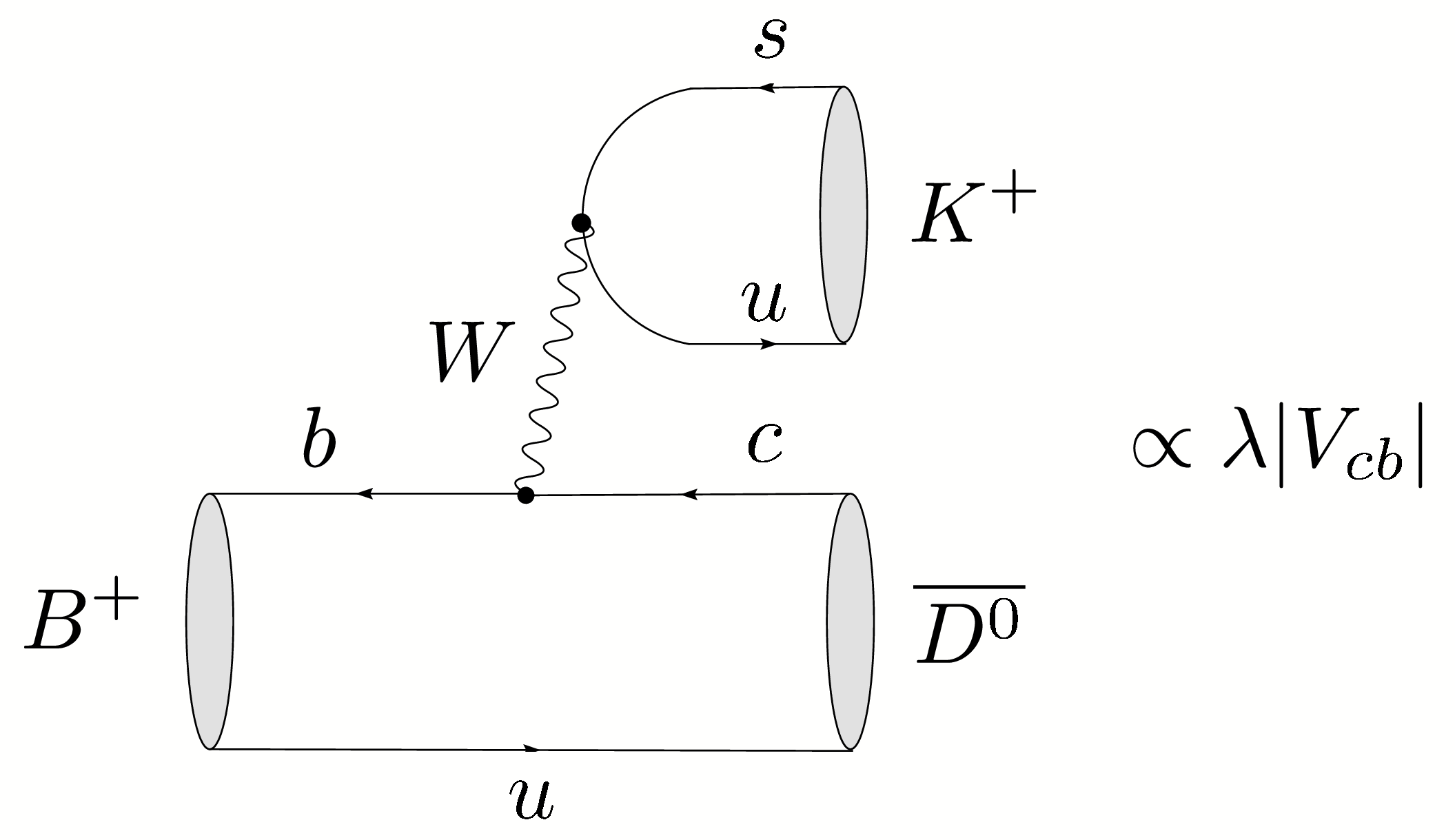} \hspace*{1truecm}
\epsfysize=4.3truecm 
\epsffile{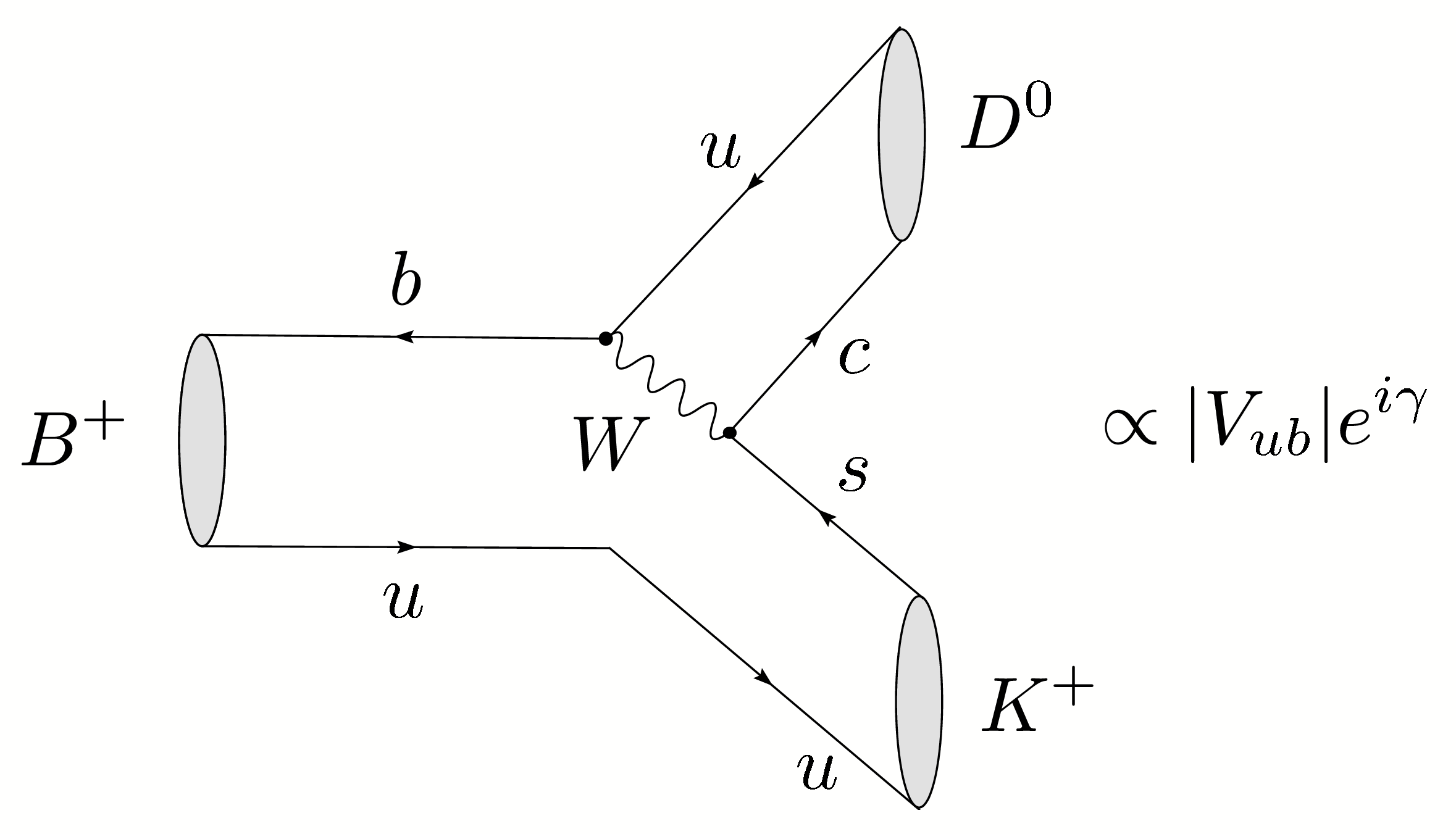}
\end{center}
\vspace*{-0.4truecm}
\caption{Feynman diagrams contributing to $B^+\to K^+\bar D^0$ and 
$B^+\to K^+D^0$. }\label{fig:BDK}
\end{figure}

\begin{figure}
\vspace*{0.3truecm}
\begin{center}
\leavevmode
\epsfysize=2.7truecm 
\epsffile{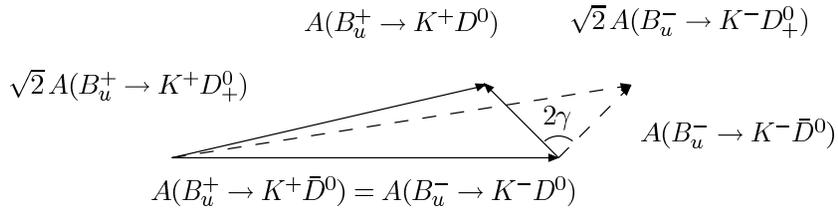} 
\end{center}
\vspace*{-0.6truecm}
\caption{The extraction of $\gamma$ from 
$B^\pm\to K^\pm\{D^0,\bar D^0,D^0_+\}$ 
decays.}\label{fig:BDK-triangle}
\end{figure}

\unboldmath
The prototype of the strategies using theoretically clean amplitude 
relations is provided by $B^\pm \to K^\pm D$ decays \cite{gw}. Looking at 
Fig.~\ref{fig:BDK}, we observe that $B^+\to K^+\bar D^0$ and $B^+\to K^+D^0$ 
are pure ``tree'' decays. If we consider, in addition, the transition 
$B^+\to D^0_+K^+$, where $D^0_+$ denotes the CP 
eigenstate of the neutral $D$-meson system with eigenvalue $+1$,
\begin{equation}\label{ED85}
|D^0_+\rangle=\frac{1}{\sqrt{2}}\left[|D^0\rangle+
|\bar D^0\rangle\right],
\end{equation}
we obtain interference effects, which are described by
\begin{eqnarray}
\sqrt{2}A(B^+\to K^+D^0_+)&=&A(B^+\to K^+D^0)+
A(B^+\to K^+\bar D^0)\\
\sqrt{2}A(B^-\to K^-D^0_+)&=&A(B^-\to K^-\bar D^0)+
A(B^-\to K^-D^0).
\end{eqnarray}
These relations can be represented as two triangles in 
the complex plane. Since we have only to deal with tree-diagram-like 
topologies, we have moreover
\begin{eqnarray}
A(B^+\to K^+\bar D^0)&=&A(B^-\to K^-D^0)\\
A(B^+\to K^+D^0)&=&A(B^-\to K^-\bar D^0)\times e^{2i\gamma},
\end{eqnarray}
allowing a {\it theoretically clean} extraction of $\gamma$, as shown 
in Fig.~\ref{fig:BDK-triangle}. Unfortunately, these triangles are 
very squashed, since $B^+\to K^+D^0$ is colour-suppressed 
with respect to $B^+\to K^+\bar D^0$:
\begin{equation}\label{BDK-suppr}
\left|\frac{A(B^+\to K^+D^0)}{A(B^+\to K^+\bar D^0}\right|=
\left|\frac{A(B^-\to K^-\bar D^0)}{A(B^-\to K^-D^0}\right|\approx
\frac{1}{\lambda}\frac{|V_{ub}|}{|V_{cb}|}\times\frac{a_2}{a_1}
\approx 0.4\times0.3={\cal O}(0.1),
\end{equation}
where the phenomenological ``colour'' factors were introduced in
Subsection~\ref{ssec:ME-fact}. 

Another -- more subtle -- problem is related to the measurement of
$\mbox{BR}(B^+\to K^+D^0)$. From the theoretical point of view, 
$D^0\to K^-\ell^+\nu$ would be ideal to measure this tiny 
branching ratio. However, because of the huge background from 
semileptonic $B$ decays, we must rely on Cabibbo-allowed hadronic 
$D^0\to f_{\rm NE}$ decays, such as $f_{\rm NE}=\pi^+K^-$, $\rho^+K^-$,
$\ldots$, i.e.\ have to measure 
\begin{equation}\label{chain1}
B^+\to K^+D^0 \,[\to f_{\rm NE}].
\end{equation}
Unfortunately, we then encounter another decay path into the {\it same} 
final state $K^+ f_{\rm NE}$ through 
\begin{equation}\label{chain2}
B^+\to K^+\bar D^0 \,[\to f_{\rm NE}], 
\end{equation}
where BR$(B^+\to K^+\bar D^0)$ is {\it larger} than BR$(B^+\to K^+D^0)$
by a factor of ${\cal O}(10^2)$, while $\bar D^0\to f_{\rm NE}$ is doubly 
Cabibbo-suppressed, i.e.\ the corresponding branching ratio is suppressed
with respect to the one of $D^0\to f_{\rm NE}$ by a factor of 
${\cal O}(10^{-2})$. Consequently, we obtain interference effects of 
${\cal O}(1)$ between the decay chains in (\ref{chain1}) and (\ref{chain2}). 
However, if two different final states $f_{\rm NE}$ are considered, 
$\gamma$  can be extracted \cite{ADS}, although this determination is  
then more involved than the original triangle approach presented in 
\cite{gw}. 

The angle $\gamma$ can be determined in a variety of ways
through CP-violating effects in pure
tree decays of type $B\to D^{(*)} K^{(*)}$ \cite{WG-sum}. Using the
present $B$-factory data, the following results were obtained through a
combination of various methods:
\begin{equation}\label{gam-DK}
\left.\gamma\right|_{D^{(*)} K^{(*)}} = \left\{
\begin{array}{ll}
(62^{+35}_{-25})^\circ & \mbox{(CKMfitter collaboration
    \cite{CKMfitter}),}\\[5pt]
(65\pm 20)^\circ & \mbox{(UTfit collaboration \cite{UTfit})}.
\end{array}
\right.
\end{equation}
Here we have discarded a second solution given by $180^\circ+\left.
\gamma\right|_{D^{(*)} K^{(*)}}$
in the third quadrant of the $\bar\rho$--$\bar\eta$ plane, as it is disfavoured
by the global fits of
the UT, and by the data for mixing-induced CP violation in pure tree decays
of type
$B_d\to D^{\pm}\pi^\mp, D^{\ast\pm}\pi^\mp, ...$ \cite{RF-gam-ca}. A similar
comment applies
to the information from $B\to\pi\pi, \pi K$ modes \cite{BFRS-5}.

\boldmath
\subsection{$B_c^\pm\to D_s^\pm D$}
\unboldmath
In addition to the ``conventional'' $B_u^\pm$ mesons, there is yet another 
species of charged $B$ mesons, the $B_c$-meson system, which consists of
$B_c^+\sim c\overline{b}$ and $B_c^-\sim b\overline{c}$. These mesons were 
observed by the CDF collaboration through their decay 
$B_c^+\to J/\psi \ell^+ \nu$, with the following mass and lifetime 
\cite{CDF-Bc}:
\begin{equation}
M_{B_c}=(6.40\pm0.39\pm0.13)\,\mbox{GeV}, \quad
\tau_{B_c}=(0.46^{+0.18}_{-0.16}\pm 0.03)\,\mbox{ps}.
\end{equation}
Meanwhile, the D0 collaboration observed the $B_c^+\to J/\psi\,\mu^+ X$ mode
\cite{D0-Bc}, which led to the following $B_c$ mass and lifetime determinations:
\begin{equation}
M_{B_c}=(5.95^{+0.14}_{-0.13}\pm0.34)\,\mbox{GeV}, \quad
\tau_{B_c}=(0.448^{+0.123}_{-0.096}\pm 0.121)\,\mbox{ps},
\end{equation}
and CDF reported evidence for the $B_c^+\to J/\psi \pi^+$ channel \cite{CDF-Bc-nl},
implying
\begin{equation}
M_{B_c}= (6.2870 \pm 0.0048  \pm 0.0011)\,\mbox{GeV}.
\end{equation}

\begin{figure}
\begin{center}
\leavevmode
\epsfysize=4.0truecm 
\epsffile{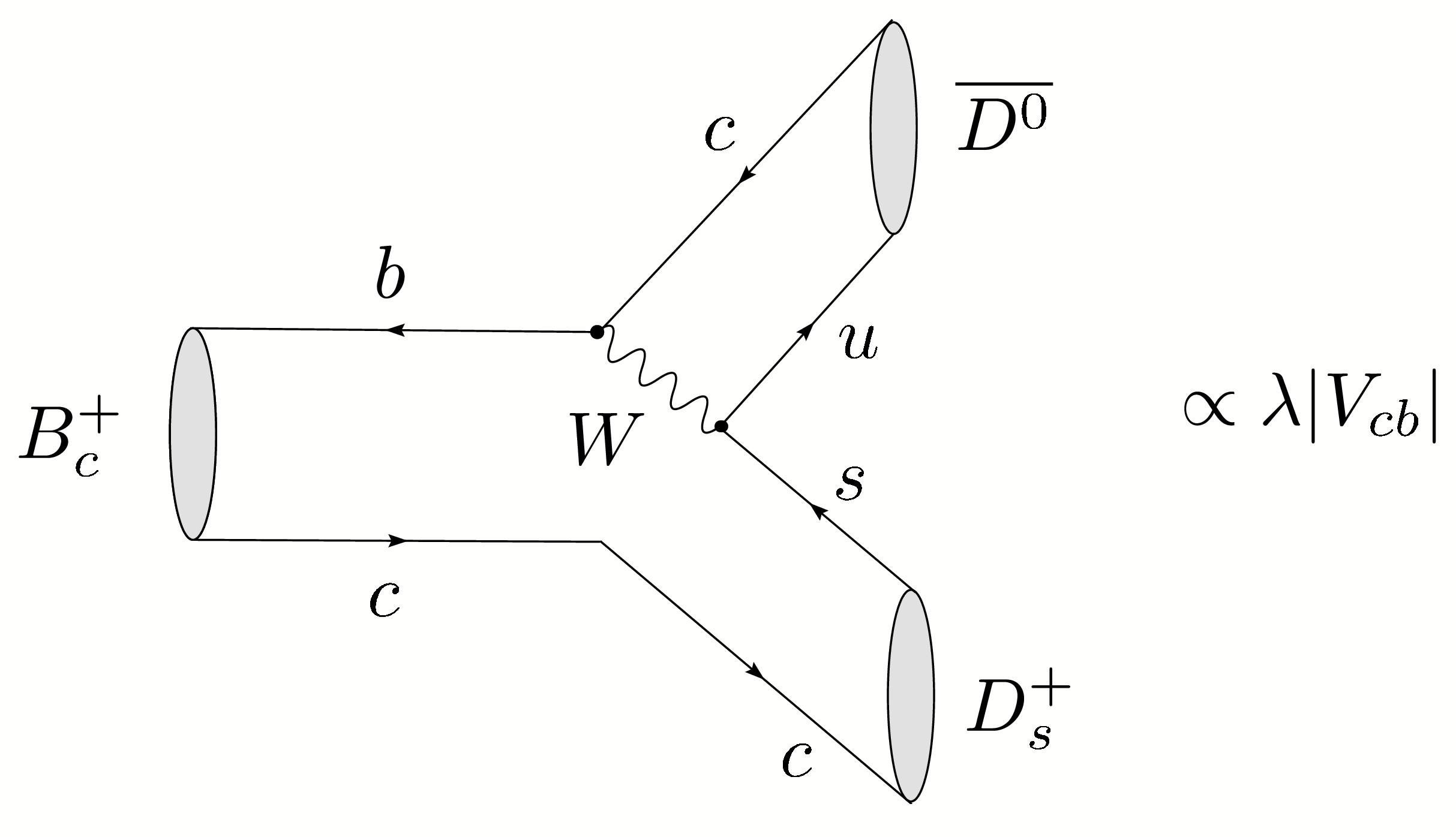} \hspace*{1truecm}
\epsfysize=3.8truecm 
\epsffile{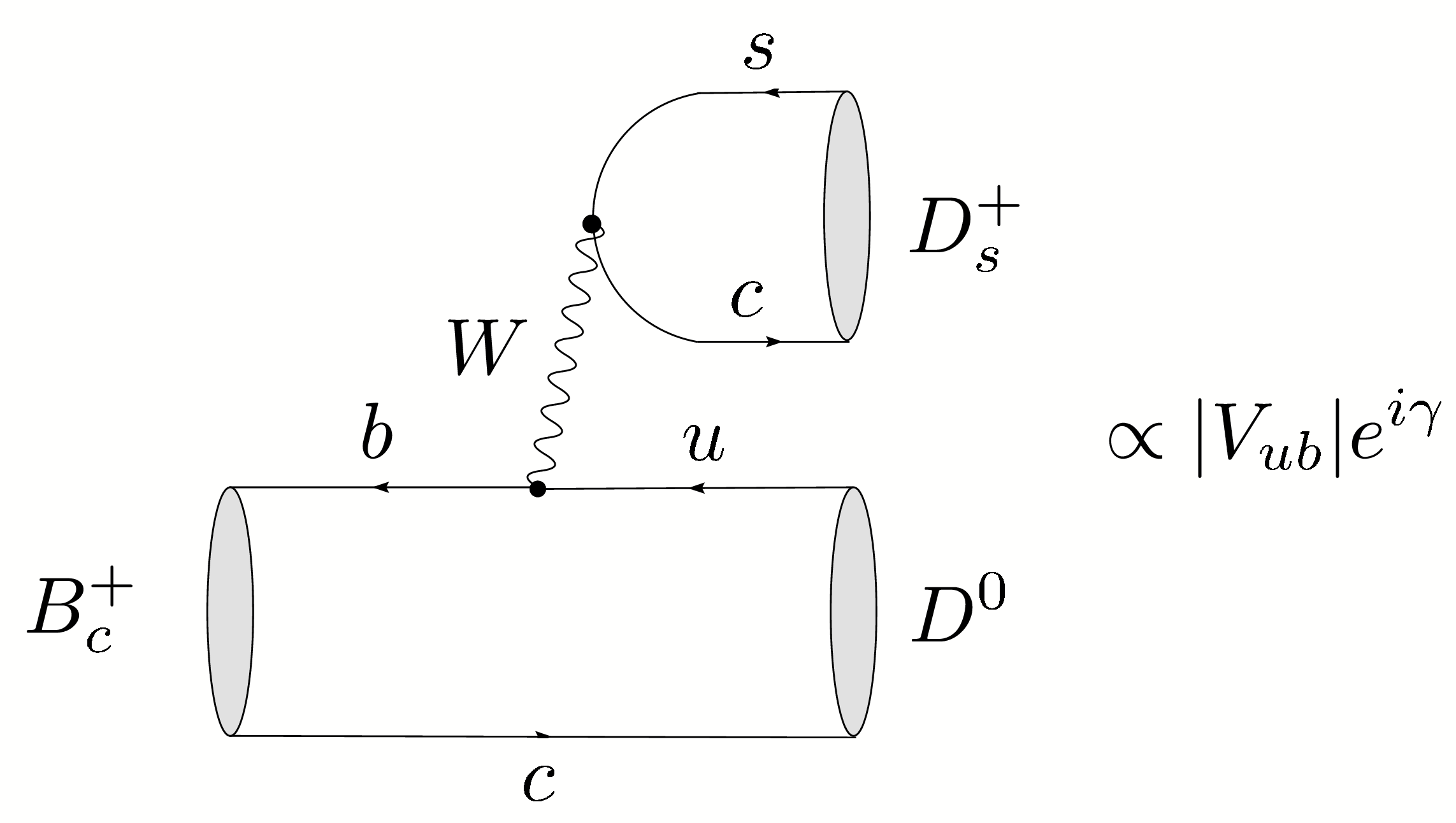}
\end{center}
\vspace*{-0.4truecm}
\caption{Feynman diagrams contributing to $B^+_c\to D_s^+\bar D^0$ and 
$B^+\to D_s^+D^0$. }\label{fig:BcDsD}
\end{figure}

\begin{figure}
\vspace*{0.3truecm}
\begin{center}
\leavevmode
\epsfysize=4.7truecm 
\epsffile{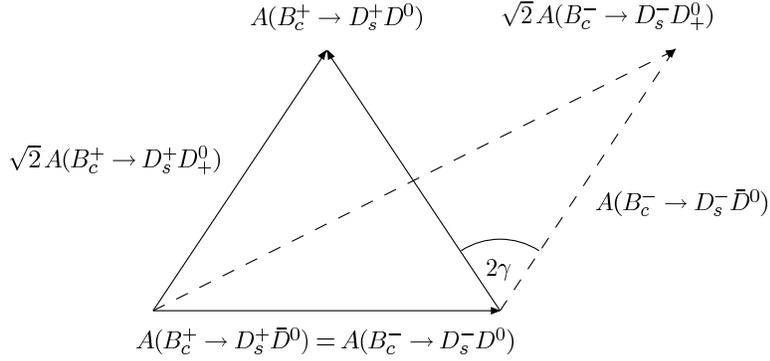} 
\end{center}
\vspace*{-0.6truecm}
\caption{The extraction of $\gamma$ from 
$B_c^\pm\to D^\pm_s\{D^0,\bar D^0,D^0_+\}$ decays.}\label{fig:triangles}
\end{figure}

Since run II of the Tevatron will provide further insights into $B_c$ physics and 
a huge number of $B_c$ mesons will be produced at LHCb, the 
natural question of how to explore CP violation with charged $B_c$ decays arises,
in particular whether an extraction of $\gamma$ with the help of the
triangle approach is possible.  Such a determination is actually offered by 
$B_c^\pm\to D_s^\pm D$ decays, which are the $B_c$ counterparts 
of the $B_u^\pm\to K^\pm D$ modes (see Fig.\ \ref{fig:BcDsD}), 
and satisfy the following amplitude relations \cite{masetti}:
\begin{eqnarray}
\sqrt{2}A(B_c^+\to D_s^+D^0_+)&=&A(B_c^+\to D_s^+D^0)+
A(B_c^+\to D_s^+\bar D^0)\\
\sqrt{2}A(B_c^-\to D_s^-D^0_+)&=&A(B_c^-\to D_s^-\bar D^0)+
A(B_c^-\to D_s^-D^0),
\end{eqnarray}
with
\begin{eqnarray}
A(B^+_c\to D_s^+\bar D^0)&=&A(B^-_c\to D_s^-D^0)\\
A(B_c^+\to D_s^+D^0)&=&A(B_c^-\to D_s^-\bar D^0)\times e^{2i\gamma}.
\end{eqnarray}
At first sight, everything is completely analogous to the $B_u^\pm\to K^\pm D$
case. However, there is an important difference \cite{fw}, 
which becomes obvious by comparing the Feynman diagrams shown in 
Figs.~\ref{fig:BDK} and \ref{fig:BcDsD}: in the $B_c^\pm\to D_s^\pm D$ 
system, the amplitude with the rather small CKM matrix element $V_{ub}$ 
is not colour-suppressed, while the larger element $V_{cb}$ comes with 
a colour-suppression factor. Therefore, we obtain
\begin{equation}\label{Bc-ratio1}
\left|\frac{A(B^+_c\to D_s^+ D^0)}{A(B^+_c\to D_s^+ 
\bar D^0)}\right|=\left|\frac{A(B^-_c\to D_s^-\bar D^0)}{A(B^-_c\to D_s^- 
D^0)}\right|\approx\frac{1}{\lambda}\frac{|V_{ub}|}{|V_{cb}|}
\times\frac{a_1}{a_2}\approx0.4\times 3 = {\cal O}(1),
\end{equation}
and conclude that the two amplitudes are similar in size. In contrast 
to this favourable situation, in the decays $B_u^{\pm}\to K^{\pm}D$, 
the matrix element $V_{ub}$ comes with the colour-suppression factor, 
resulting in a very stretched triangle. The extraction of $\gamma$ from 
the $B_c^\pm\to D_s^\pm D$ triangles is illustrated in 
Fig.~\ref{fig:triangles}, which should be compared with the
squashed $B^\pm_u\to K^\pm D$ triangles shown in 
Fig.\ \ref{fig:BDK-triangle}. Another important advantage is that 
the interference effects arising from $D^0,\bar D^0\to\pi^+K^-$ are 
practically unimportant for the measurement of BR$(B^+_c\to D_s^+ D^0)$ 
and BR$(B^+_c\to D_s^+ \bar D^0)$ since the $B_c$-decay amplitudes are 
of the same order of magnitude. Consequently, the $B_c^\pm\to D_s^\pm D$
decays provide -- from the theoretical point of view -- the ideal
realization of the ``triangle'' approach to determine $\gamma$. On
the other hand, the practical implementation still appears to 
be challenging, although detailed experimental feasibility studies for
LHCb are strongly encouraged. The corresponding branching ratios
were estimated in Ref.~\cite{IKP}, with a pattern in accordance 
with (\ref{Bc-ratio1}).

\begin{figure}
\centerline{
 \includegraphics[width=5.9truecm]{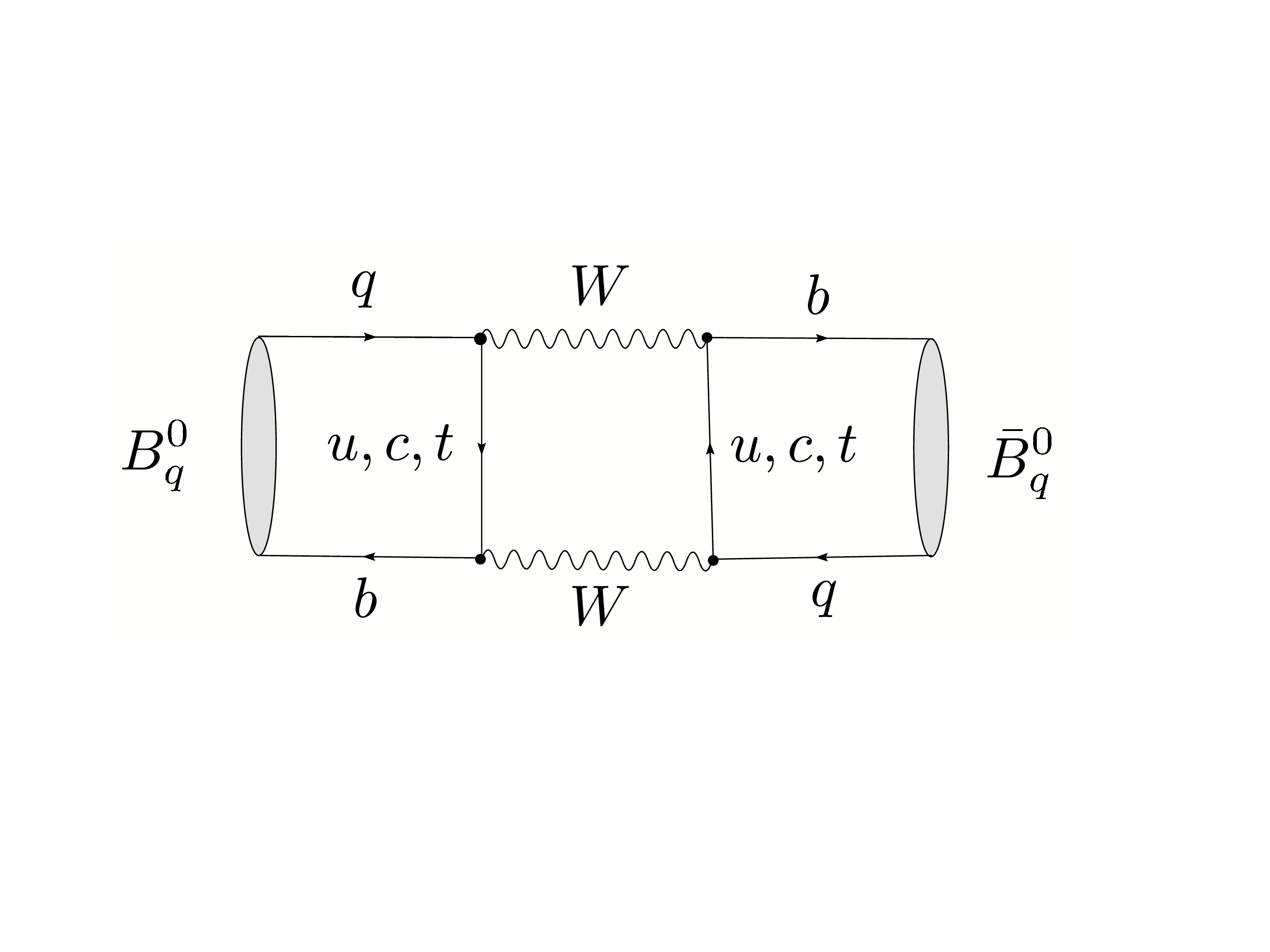}
 \hspace*{0.5truecm}
 \includegraphics[width=5.9truecm]{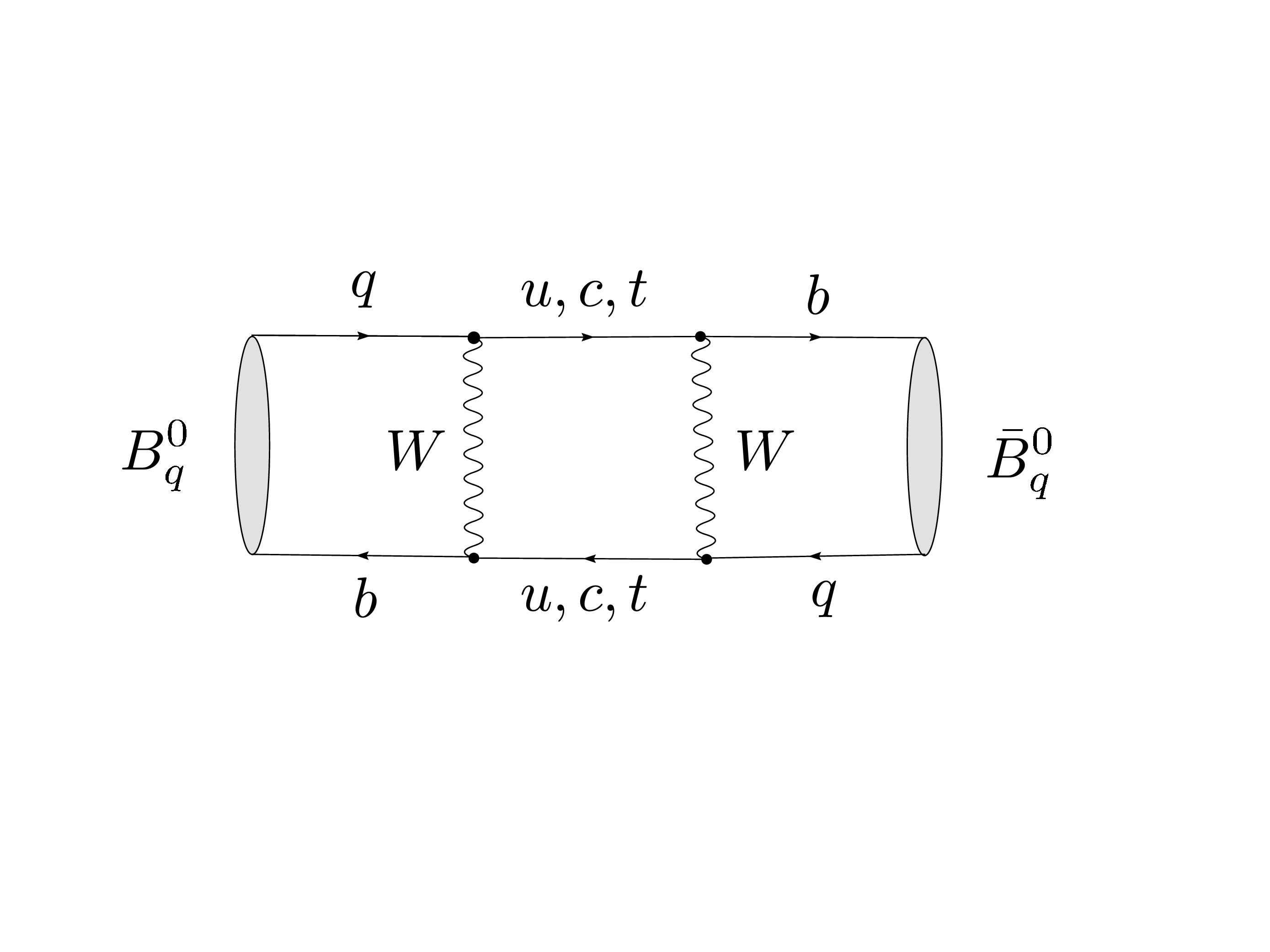}  
 }
 \vspace*{-0.4truecm}
 \caption{Box diagrams contributing to $B^0_q$--$\bar B^0_q$ mixing in the
 SM ($q\in\{d,s\}$).}
   \label{fig:boxes}
\end{figure}

\section{FEATURES OF NEUTRAL {\boldmath$B$\unboldmath} 
MESONS}\label{sec:mix}
\setcounter{equation}{0}
\boldmath\subsection{Schr\"odinger Equation for $B^0_q$--$\bar B^0_q$ 
Mixing}\unboldmath\label{ssec:BBbar-mix}
Within the SM, $B^0_q$--$\bar B^0_q$ mixing ($q\in\{d,s\}$) arises from
the box diagrams shown in Fig.~\ref{fig:boxes}. Because of
this phenomenon, an initially, i.e.\ at time $t=0$, present 
$B^0_q$-meson state evolves into a time-dependent linear combination of 
$B^0_q$ and $\bar B^0_q$ states:
\begin{equation}
|B_q(t)\rangle=a(t)|B^0_q\rangle + b(t)|\bar B^0_q\rangle,
\end{equation}
where $a(t)$ and $b(t)$ are governed by a Schr\"odinger equation of 
the following form:
\begin{equation}\label{SG-OSZ}
i\,\frac{{\rm d}}{{\rm d} t}\left(\begin{array}{c} a(t)\\ b(t)
\end{array}
\right)= H \cdot\left(\begin{array}{c}
a(t)\\ b(t)\nonumber
\end{array}
\right) \equiv
\Biggl[\underbrace{\left(\begin{array}{cc}
M_{0}^{(q)} & M_{12}^{(q)}\\ M_{12}^{(q)\ast} & M_{0}^{(q)}
\end{array}\right)}_{\mbox{mass matrix}}-
\frac{i}{2}\underbrace{\left(\begin{array}{cc}
\Gamma_{0}^{(q)} & \Gamma_{12}^{(q)}\\
\Gamma_{12}^{(q)\ast} & \Gamma_{0}^{(q)}
\end{array}\right)}_{\mbox{decay matrix}}\Biggr]
\cdot\left(\begin{array}{c}
a(t)\\ b(t)\nonumber
\end{array}
\right).
\end{equation}
The special form $H_{11}=H_{22}$ of the Hamiltonian $H$ is an implication 
of the CPT theorem, i.e.\ of the invariance under combined CP and 
time-reversal (T) transformations. 

It is straightforward to calculate the eigenstates 
$\vert B_{\pm}^{(q)}\rangle$ and eigenvalues $\lambda_{\pm}^{(q)}$ 
of (\ref{SG-OSZ}):
\begin{equation}
\vert B_{\pm}^{(q)} \rangle  =
\frac{1}{\sqrt{1+\vert \alpha_q\vert^{2}}}
\left(\vert B^{0}_q\rangle\pm\alpha_q\vert\bar B^{0}_q\rangle\right)
\end{equation}
\begin{equation}\label{lam-pm}
\lambda_{\pm}^{(q)}  =
\left(M_{0}^{(q)}-\frac{i}{2}\Gamma_{0}^{(q)}\right)\pm
\left(M_{12}^{(q)}-\frac{i}{2}\Gamma_{12}^{(q)}\right)\alpha_q,
\end{equation}
where
\begin{equation}\label{alpha-q-expr}
\alpha_q e^{+i\left(\Theta_{\Gamma_{12}}^{(q)}+n'
\pi\right)}=
\sqrt{\frac{4\vert M_{12}^{(q)}\vert^{2}
e^{-i2\delta\Theta_{M/\Gamma}^{(q)}}+\vert\Gamma_{12}^{(q)}\vert^{2}}{4\vert 
M_{12}^{(q)}\vert^{2}+\vert\Gamma_{12}^{(q)}\vert^{2}- 
4\vert M_{12}^{(q)}\vert\vert\Gamma_{12}^{(q)}\vert
\sin\delta\Theta_{M/\Gamma}^{(q)}}}.
\end{equation}
Here we have written
\begin{equation}
M_{12}^{(q)}\equiv e^{i\Theta_{M_{12}}^{(q)}}\vert
M_{12}^{(q)}\vert,\quad \Gamma_{12}^{(q)}\equiv
e^{i\Theta_{\Gamma_{12}}^{(q)}}\vert\Gamma_{12}^{(q)}\vert,\quad
\delta\Theta_{M/\Gamma}^{(q)}\equiv
\Theta_{M_{12}}^{(q)}-\Theta_{\Gamma_{12}}^{(q)},
\end{equation}
and have introduced the quantity $n'\in\{0,1\}$ to parametrize the 
sign of the square root in (\ref{alpha-q-expr}). 

Evaluating the dispersive parts of the box diagrams shown in
Fig.~\ref{fig:boxes}, which are dominated by internal top-quark 
exchanges, yields (for a more detailed discussion, see Ref.~\cite{BF-rev}):
\begin{equation}\label{M12-calc}
M_{12}^{(q)}=
\frac{G_{\rm F}^2 M_{W}^{2}}{12\pi^{2}}
\eta_B M_{B_q}f_{B_q}^{2}\hat B_{B_q}\left(V_{tq}^\ast V_{tb}\right)^2 
S_0(x_{t}) e^{i(\pi-\phi_{\mbox{{\scriptsize CP}}}(B_q))},
\end{equation}
where $\phi_{\mbox{{\scriptsize CP}}}(B_q)$ is a convention-dependent
phase, which is defined in analogy to (\ref{CP-phase-def}).
The short-distance physics is encoded in the ``Inami--Lim'' function 
$S_0(x_{t}\equiv m_t^2/M_W^2)$ \cite{IL}, which can be written -- to a good 
approximation -- in the SM  as \cite{Buras-Schladming}
\begin{equation}
S_0(x_t)=2.40\times\left[\frac{m_t}{167\,\mbox{GeV}}\right]^{1.52},
\end{equation}
and in the perturbative QCD correction factor  $\eta_B=0.55\pm0.01$ \cite{eta-B},
which does {\it not} depend on $q\in\{d,s\}$, i.e.\ is the same for $B_d$ and $B_s$ 
mesons. On the other hand, the non-perturbative physics is described by the 
quantities $f_{B_q}\hat B_{B_q}^{1/2}$, involving -- in addition to the $B_q$ decay 
constant $f_{B_q}$ -- the ``bag'' parameter $\hat B_{B_q}$, which is related to the hadronic matrix element $\langle \bar B^0_q|(\bar bq)_{{\rm V}-{\rm A}}
(\bar bq)_{{\rm V}-{\rm A}}|B^0_q\rangle$. These non-perturbative parameters
can be determined through QCD sum-rule calculations \cite{SR-calc}
or lattice studies. Concerning the latter analyses, the front runners are 
now unquenched calculations with 2 or 3 dynamical quarks. 
Despite tremendous progress, the results still suffer from several uncertainties. For the
analysis of the mixing parameters discussed below \cite{BF-DMs}, we use 
two sets of parameters from the JLQCD \cite{JLQCD} and 
HPQCD \cite{HPQCD} lattice collaborations:
\begin{equation}\label{JLQCD}
\begin{array}{rcl}
\left.f_{B_d}\hat{B}_{B_d}^{1/2}\right|_{\rm JLQCD} &=& (0.215\pm
0.019^{+0}_{-0.023})\,{\rm GeV}\\
\left.f_{B_s}\hat{B}_{B_s}^{1/2}\right|_{\rm JLQCD} &=& (0.245\pm
0.021^{+0.003}_{-0.002})\,{\rm GeV},
\end{array}
\end{equation}
which were obtained for two flavours of dynamical light (``Wilson") quarks, and
\begin{equation}\label{HPQCD}
\begin{array}{rcl}
\left.f_{B_d}\hat{B}_{B_d}^{1/2}\right|_{\rm (HP+JL)QCD}& =& (0.244\pm
0.026)\,{\rm GeV}\\
\left.f_{B_s}\hat{B}_{B_s}^{1/2}\right|_{\rm (HP+JL)QCD} &=&
(0.295\pm0.036)\,{\rm GeV},
\end{array}
\end{equation}
where $f_{B_q}$ comes from HPQCD (3 dynamical
flavours) and $\hat B_{B_q}$ from JLQCD as no value for this parameter
is available from the former collaboration \cite{Okamoto}. 

If we calculate also the absorptive parts of the box diagrams in
Fig~\ref{fig:boxes}, we obtain
\begin{equation}
\frac{\Gamma_{12}^{(q)}}{M_{12}^{(q)}}\approx
-\frac{3\pi}{2S_0(x_{t})}\left(\frac{m_b^2}{M_W^2}\right)
={\cal O}(m_b^2/m_t^2)\ll 1.
\end{equation}
Consequently, we may expand (\ref{alpha-q-expr}) in 
$\Gamma_{12}^{(q)}/M_{12}^{(q)}$. Neglecting second-order 
terms, we arrive at
\begin{equation}
\alpha_q=\left[1+\frac{1}{2}\left|
\frac{\Gamma_{12}^{(q)}}{M_{12}^{(q)}}\right|\sin\delta
\Theta_{M/\Gamma}^{(q)}\right]e^{-i\left(\Theta_{M_{12}}^{(q)}+n'\pi\right)}.
\end{equation}

The deviation of $|\alpha_q|$ from 1 measures CP violation in 
$B^0_q$--$\bar B^0_q$ oscillations, and can be probed through
the following ``wrong-charge'' lepton asymmetries:
\begin{equation}
{\cal A}^{(q)}_{\mbox{{\scriptsize SL}}}\equiv
\frac{\Gamma(B^0_q(t)\to \ell^-\bar\nu X)-\Gamma(\bar B^0_q(t)\to
\ell^+\nu X)}{\Gamma(B^0_q(t)\to \ell^-\bar \nu X)+
\Gamma(\bar B^0_q(t)\to \ell^+\nu X)}
=\frac{|\alpha_q|^4-1}{|\alpha_q|^4+1}\approx\left|
\frac{\Gamma_{12}^{(q)}}{M_{12}^{(q)}}\right|
\sin\delta\Theta^{(q)}_{M/\Gamma}.
\end{equation}
Because of $|\Gamma_{12}^{(q)}|/|M_{12}^{(q)}|\propto
m_b^2/m_t^2$ and $\sin\delta\Theta^{(q)}_{M/\Gamma}\propto m_c^2/m_b^2$,
the asymmetry ${\cal A}^{(q)}_{\mbox{{\scriptsize SL}}}$ is suppressed by 
a factor of $m_c^2/m_t^2={\cal O}(10^{-4})$ and is hence tiny in the SM.
However, this observable may be enhanced through NP effects, thereby
representing an interesing probe for physics beyond the SM 
\cite{LLNP,BBLN-CFLMT}.  The current experimental average for the $B_d$-meson
system compiled by the ``Heavy Flavour Averaging Group" \cite{HFAG} reads
as follows:
\begin{equation}
{\cal A}^{(d)}_{\mbox{{\scriptsize SL}}}=0.0030\pm0.0078,
\end{equation}
and does not indicate any non-vanishing effect.

\subsection{Mixing Parameters}\label{ssec:Mix-Par}
Let us denote the masses of the eigenstates of (\ref{SG-OSZ}) by 
$M^{(q)}_{\rm H}$ (``heavy'') and $M^{(q)}_{\rm L}$ (``light''). 
It is then useful to introduce 
\begin{equation}
M_q\equiv\frac{M^{(q)}_{\rm H}+M^{(q)}_{\rm L}}{2}=
M^{(q)}_0,
\end{equation}
as well as the mass difference
\begin{equation}\label{DeltaMq-def}
\Delta M_q\equiv M_{\rm H}^{(q)}-M_{\rm L}^{(q)}=2|M_{12}^{(q)}|>0,
\end{equation}
which is by definition {\it positive}. While $B^0_d$--$\bar B^0_d$ mixing is well established and
\begin{equation}\label{DMd-exp}
\Delta M_d = (0.507\pm 0.004)\,{\rm ps}^{-1}
\end{equation}
known with impressive
experimental accuracy \cite{HFAG},  only lower bounds on $\Delta M_s$ 
were available,  for many years, from the LEP (CERN) 
experiments and SLD (SLAC) \cite{LEPBOSC}. In the spring of 2006, 
$\Delta M_s$ could eventually be pinned down at the Tevatron: the D0 
collaboration reported a two-sided bound 
\begin{equation}\label{D0-range}
17 \,{\rm ps}^{-1}< \Delta M_s < 21\,{\rm ps}^{-1} \quad \mbox{(90\% C.L.)},
\end{equation}
corresponding to a 2.5\,$\sigma$ signal at $\Delta M_s=19\,{\rm ps}^{-1}$ 
\cite{D0}, and CDF announced the following result \cite{CDF}:
\begin{equation}\label{CDF-DMs}
\Delta M_s = \left[17.31^{+0.33}_{-0.18}({\rm
  stat})\pm 0.07({\rm syst})\right]{\rm ps}^{-1}.
\end{equation}

The decay widths $\Gamma_{\rm H}^{(q)}$ and 
$\Gamma_{\rm L}^{(q)}$ of the mass eigenstates, which correspond to 
$M^{(q)}_{\rm H}$ and $M^{(q)}_{\rm L}$, respectively, satisfy 
\begin{equation}
\Delta\Gamma_q\equiv\Gamma_{\rm H}^{(q)}-\Gamma_{\rm L}^{(q)}=
\frac{4\mbox{\,Re}\left[M_{12}^{(q)}\Gamma_{12}^{(q)\ast}\right]}{\Delta M_q},
\end{equation}
whereas 
\begin{equation}
\Gamma_q\equiv\frac{\Gamma^{(q)}_{\rm H}+\Gamma^{(q)}_{\rm L}}{2}=
\Gamma^{(q)}_0.
\end{equation}
There is the following interesting relation:
\begin{equation}\label{DGoG}
\frac{\Delta\Gamma_q}{\Gamma_q}\approx-\frac{3\pi}{2S_0(x_t)}
\left(\frac{m_b^2}{M_W^2}\right)x_q=-{\cal O}(10^{-2})\times x_q,
\end{equation}
where
\begin{equation}\label{mix-par}
x_q\equiv\frac{\Delta M_q}{\Gamma_q}=\left\{\begin{array}{cc}
0.771\pm0.012&(q=d)\\
{\cal O}(20)& (q=s)
\end{array}\right.
\end{equation}
is often referred to as {\it the} 
$B^0_q$--$\bar B^0_q$ ``mixing parameter''.\footnote{Note that
$\Delta\Gamma_q/\Gamma_q$ is negative in the SM because of the
minus sign in (\ref{DGoG}).}
Consequently, we observe that $\Delta\Gamma_d/\Gamma_d\sim 10^{-2}$ is 
negligibly small, while $\Delta\Gamma_s/\Gamma_s\sim 10^{-1}$ may
be sizeable. In fact, as was reviewed in Ref.~\cite{lenz}, the 
state of the art of calculations of these quantities is given as follows:
\begin{equation}\label{DGam-numbers}
\frac{|\Delta\Gamma_d|}{\Gamma_d}=(3\pm1.2)\times 10^{-3}, \quad
\frac{|\Delta\Gamma_s|}{\Gamma_s}=0.12\pm0.05.
\end{equation}
Recently, the first results for 
$\Delta\Gamma_s$ were reported from the Tevatron, using the 
$B^0_s\to J/\psi\phi$ channel \cite{DDF}:
\begin{equation}
\frac{|\Delta\Gamma_s|}{\Gamma_s}=\left\{
\begin{array}{ll}
0.65^{+0.25}_{-0.33}\pm0.01 & \mbox{(CDF \cite{CDF-DG})}\\
0.24^{+0.28+0.03}_{-0.38-0.04} & \mbox{(D0 \cite{D0-DG})}.
\end{array}
\right.
\end{equation}
It will be interesting to follow the evolution of the data for this quantity. 

In Subsections~\ref{ssec:BpsiK} and \ref{ssec:Bs-prelim}, we will give detailed
discussions of the theoretical interpretation of the data for the $B^0_q$--$\bar B^0_q$
mixing parameters.

\subsection{Time-Dependent Decay Rates}
The time evolution of initially, i.e.\ at $t=0$, pure $B^0_q$- and 
$\bar B^0_q$-meson states is given by
\begin{equation}
|B^0_q(t)\rangle=f_+^{(q)}(t)|B^{0}_q\rangle
+\alpha_qf_-^{(q)}(t)|\bar B^{0}_q\rangle
\end{equation}
and
\begin{equation}
|\bar B^0_q(t)\rangle=\frac{1}{\alpha_q}f_-^{(q)}(t)
|B^{0}_q\rangle+f_+^{(q)}(t)|\bar B^{0}_q\rangle,
\end{equation}
respectively, with
\begin{equation}\label{f-functions}
f_{\pm}^{(q)}(t)=\frac{1}{2}\left[e^{-i\lambda_+^{(q)}t}\pm
e^{-i\lambda_-^{(q)}t}\right].
\end{equation}
These time-dependent state vectors allow the calculation of the 
corresponding transition rates. To this end, it is useful to introduce
\begin{equation}\label{g-funct-1}
|g^{(q)}_{\pm}(t)|^2=\frac{1}{4}\left[e^{-\Gamma_{\rm L}^{(q)}t}+
e^{-\Gamma_{\rm H}^{(q)}t}\pm2\,e^{-\Gamma_q t}\cos(\Delta M_qt)\right]
\end{equation}
\begin{equation}\label{g-funct-2}
g_-^{(q)}(t)\,g_+^{(q)}(t)^\ast=\frac{1}{4}\left[e^{-\Gamma_{\rm L}^{(q)}t}-
e^{-\Gamma_{\rm H}^{(q)}t}+2\,i\,e^{-\Gamma_q t}\sin(\Delta M_qt)\right],
\end{equation}
as well as
\begin{equation}\label{xi-def}
\xi_f^{(q)}=e^{-i\Theta_{M_{12}}^{(q)}}
\frac{A(\bar B_q^0\to f)}{A(B_q^0\to f)},\quad
\xi_{\bar f}^{(q)}=e^{-i\Theta_{M_{12}}^{(q)}}
\frac{A(\bar B_q^0\to \bar f)}{A(B_q^0\to \bar f)}.
\end{equation}
Looking at (\ref{M12-calc}), we find
\begin{equation}\label{theta-def}
\Theta_{M_{12}}^{(q)}=\pi+2\mbox{arg}(V_{tq}^\ast V_{tb})-
\phi_{\mbox{{\scriptsize CP}}}(B_q),
\end{equation}
and observe that this phase depends on the chosen CKM and 
CP phase conventions specified in (\ref{CKM-trafo}) and (\ref{CP-phase-def}), 
respectively. However, these dependences are cancelled through the 
amplitude ratios in (\ref{xi-def}), so that $\xi_f^{(q)}$ and 
$\xi_{\bar f}^{(q)}$ are {\it convention-independent} observables. 
Whereas $n'$ enters the functions in (\ref{f-functions}) through 
(\ref{lam-pm}), the dependence on this parameter is cancelled in 
(\ref{g-funct-1}) and (\ref{g-funct-2}) through the introduction of 
the {\it positive} mass difference $\Delta M_q$ (see (\ref{DeltaMq-def})). 
Combining the formulae listed above, we eventually arrive at the 
following transition rates for decays of initially, i.e.\ at $t=0$, 
present $B^0_q$ or $\bar B^0_q$ mesons:
\begin{equation}\label{rates}
\Gamma(\stackrel{{\mbox{\tiny (--)}}}{B^0_q}(t)\to f)
=\left[|g_\mp^{(q)}(t)|^2+|\xi_f^{(q)}|^2|g_\pm^{(q)}(t)|^2-
2\mbox{\,Re}\left\{\xi_f^{(q)}
g_\pm^{(q)}(t)g_\mp^{(q)}(t)^\ast\right\}
\right]\tilde\Gamma_f,
\end{equation}
where the time-independent rate $\tilde\Gamma_f$ corresponds to the 
``unevolved'' decay amplitude $A(B^0_q\to f)$, and can be calculated by 
performing the usual phase-space integrations. The rates into the 
CP-conjugate final state $\bar f$ can straightforwardly be obtained from 
(\ref{rates}) by making the substitutions
\begin{equation}
\tilde\Gamma_f  \,\,\,\to\,\,\, 
\tilde\Gamma_{\bar f},
\quad\,\,\xi_f^{(q)} \,\,\,\to\,\,\, 
\xi_{\bar f}^{(q)}.
\end{equation}

\subsection{``Untagged'' Rates}
The expected sizeable width difference $\Delta\Gamma_s$ may provide interesting studies of CP 
violation through ``untagged'' $B_s$ rates 
(see Ref.~\cite{DDF} and \cite{dun}--\cite{DFN}), which are defined as 
\begin{equation}
\langle\Gamma(B_s(t)\to f)\rangle
\equiv\Gamma(B^0_s(t)\to f)+\Gamma(\bar B^0_s(t)\to f),
\end{equation}
and are characterized by the feature that we do not distinguish between
initially, i.e.\ at time $t=0$, present $B^0_s$ or $\bar B^0_s$ mesons. 
If we consider a final state $f$ to which both a $B^0_s$ and a $\bar B^0_s$ 
may decay, and use the expressions in (\ref{rates}), we find
\begin{equation}\label{untagged-rate}
\hspace*{-0.7truecm}\langle\Gamma(B_s(t)\to f)\rangle
\propto \left[\cosh(\Delta\Gamma_st/2)-{\cal A}_{\Delta\Gamma}(B_s\to f)
\sinh(\Delta\Gamma_st/2)\right]e^{-\Gamma_s t},
\end{equation}
with
\begin{equation}\label{ADGam}
{\cal A}_{\rm \Delta\Gamma}(B_s\to f)\equiv
\frac{2\,\mbox{Re}\,\xi^{(s)}_f}{1+\bigl|\xi^{(s)}_f
\bigr|^2}.
\end{equation}
We observe that the rapidly oscillating 
$\Delta M_st$ terms cancel, and that we may obtain information about the 
phase structure of the observable $\xi_f^{(s)}$, thereby providing valuable
insights into CP violation. 

Following these lines, for instance,  the untagged observables offered by 
the angular distribution of the $B_s\to K^{*+}K^{*-}, K^{*0}\bar K^{*0}$ decay 
products allow a determination of the UT angle $\gamma$, 
provided $\Delta\Gamma_s$ is 
actually sizeable \cite{FD-CP}. Untagged $B_s$-decay rates are interesting in 
terms of efficiency, acceptance and purity, and are already applied for the physics
analyses at the Tevatron. Later on, they will help to fully exploit the physics 
potential of the $B_s$-meson system at the LHC.

\subsection{CP Asymmetries}\label{subsec:CPasym}
A particularly simple -- but also very interesting  -- situation arises 
if we restrict ourselves to decays of neutral $B_q$ mesons 
into final states $f$ that are eigenstates of the CP operator, i.e.\
satisfy the relation 
\begin{equation}\label{CP-eigen}
({\cal CP})|f\rangle=\pm |f\rangle. 
\end{equation}
Consequently, we have $\xi_f^{(q)}=\xi_{\bar f}^{(q)}$ in this case, 
as can be seen in (\ref{xi-def}). Using the decay rates in (\ref{rates}), 
we find that the corresponding time-dependent CP asymmetry is given by
\begin{eqnarray}
{\cal A}_{\rm CP}(t)&\equiv&\frac{\Gamma(B^0_q(t)\to f)-
\Gamma(\bar B^0_q(t)\to f)}{\Gamma(B^0_q(t)\to f)+
\Gamma(\bar B^0_q(t)\to f)}\nonumber\\
&=&\left[\frac{{\cal A}_{\rm CP}^{\rm dir}(B_q\to f)\,\cos(\Delta M_q t)+
{\cal A}_{\rm CP}^{\rm mix}(B_q\to f)\,\sin(\Delta 
M_q t)}{\cosh(\Delta\Gamma_qt/2)-{\cal A}_{\rm 
\Delta\Gamma}(B_q\to f)\,\sinh(\Delta\Gamma_qt/2)}\right],\label{ee6}
\end{eqnarray}
with
\begin{equation}\label{CPV-OBS}
{\cal A}^{\mbox{{\scriptsize dir}}}_{\mbox{{\scriptsize CP}}}(B_q\to f)\equiv
\frac{1-\bigl|\xi_f^{(q)}\bigr|^2}{1+\bigl|\xi_f^{(q)}\bigr|^2},\qquad
{\cal A}^{\mbox{{\scriptsize mix}}}_{\mbox{{\scriptsize
CP}}}(B_q\to f)\equiv\frac{2\,\mbox{Im}\,\xi^{(q)}_f}{1+
\bigl|\xi^{(q)}_f\bigr|^2}.
\end{equation}
Because of the relation
\begin{equation}
{\cal A}^{\mbox{{\scriptsize dir}}}_{\mbox{{\scriptsize CP}}}(B_q\to f)=
\frac{|A(B^0_q\to f)|^2-|A(\bar B^0_q\to \bar f)|^2}{|A(B^0_q\to f)|^2+
|A(\bar B^0_q\to \bar f)|^2},
\end{equation}
this observable measures the direct CP violation in the decay
$B_q\to f$, which originates from the interference between different
weak amplitudes, as we have seen in (\ref{direct-CPV}). On the other
hand, the interesting {\it new} aspect of (\ref{ee6}) is due to 
${\cal A}^{\mbox{{\scriptsize mix}}}_{\mbox{{\scriptsize
CP}}}(B_q\to f)$, which originates from interference effects between 
$B_q^0$--$\bar B_q^0$ mixing and decay processes, and describes
``mixing-induced'' CP violation. Finally, the width difference 
$\Delta\Gamma_q$, which may be sizeable in the $B_s$-meson system, 
provides access to ${\cal A}_{\Delta\Gamma}(B_q\to f)$  introduced 
in (\ref{ADGam}). However, this observable is not independent from 
${\cal A}^{\mbox{{\scriptsize dir}}}_{\mbox{{\scriptsize CP}}}(B_q\to f)$ and 
${\cal A}^{\mbox{{\scriptsize mix}}}_{\mbox{{\scriptsize CP}}}(B_q\to f)$,
satisfying 
\begin{equation}\label{Obs-rel}
\Bigl[{\cal A}_{\rm CP}^{\rm dir}(B_q\to f)\Bigr]^2+
\Bigl[{\cal A}_{\rm CP}^{\rm mix}(B_q\to f)\Bigr]^2+
\Bigl[{\cal A}_{\Delta\Gamma}(B_q\to f)\Bigr]^2=1.
\end{equation}

In order to calculate $\xi_f^{(q)}$, we use the general expressions
(\ref{par-ampl}) and (\ref{par-ampl-CP}), where 
$e^{-i\phi_{\mbox{{\scriptsize CP}}}(f)}=\pm1$ because of (\ref{CP-eigen}), and 
$\phi_{\mbox{{\scriptsize CP}}}(B)=\phi_{\mbox{{\scriptsize CP}}}(B_q)$. If we insert
these amplitude parametrizations into (\ref{xi-def}) and take (\ref{theta-def}) into 
account, we observe that the phase-convention-dependent 
quantity $\phi_{\mbox{{\scriptsize CP}}}(B_q)$ cancels, and finally 
arrive at
\begin{equation}\label{xi-re}
\xi_f^{(q)}=\mp\, e^{-i\phi_q}\left[
\frac{e^{+i\varphi_1}|A_1|e^{i\delta_1}+
e^{+i\varphi_2}|A_2|e^{i\delta_2}}{
e^{-i\varphi_1}|A_1|e^{i\delta_1}+
e^{-i\varphi_2}|A_2|e^{i\delta_2}}\right],
\end{equation}
where
\begin{equation}\label{phiq-def}
\phi_q\equiv 2\,\mbox{arg} (V_{tq}^\ast V_{tb})=\left\{\begin{array}{cl}
+2\beta&\mbox{($q=d$)}\\
-2\delta\gamma&\mbox{($q=s$)}\end{array}\right.
\end{equation}
is associated with the CP-violating weak $B_q^0$--$\bar B_q^0$ mixing
phase arising in the SM; $\beta$ and $\delta\gamma$ refer to the corresponding
angles in the unitarity triangles shown in Fig.\ \ref{fig:UT}.

In analogy to (\ref{direct-CPV}), the caclulation
of $\xi_f^{(q)}$ is -- in general -- also affected by large hadronic uncertainties. 
However, if one CKM amplitude plays the dominant r\^ole in the $B_q\to f$
transition, we obtain
\begin{equation}\label{xi-si}
\xi_f^{(q)}=\mp\, e^{-i\phi_q}\left[
\frac{e^{+i\phi_f/2}|M_f|e^{i\delta_f}}{e^{-i\phi_f/2}|M_f|e^{i\delta_f}}
\right]=\mp\, e^{-i(\phi_q-\phi_f)},
\end{equation}
and observe that the hadronic matrix element $|M_f|e^{i\delta_f}$ 
cancels in this expression. Since the requirements for 
direct CP violation discussed above are no longer satisfied, direct CP violation 
vanishes in this important special case, i.e.\ 
${\cal A}^{\mbox{{\scriptsize dir}}}_{\mbox{{\scriptsize CP}}}
(B_q\to f)=0$. On the other hand, this is {\it not} the case for the mixing-induced 
CP asymmetry. In particular, 
\begin{equation}\label{Amix-simple}
{\cal A}^{\rm mix}_{\rm CP}(B_q\to f)=\pm\sin\phi
\end{equation}
is now governed by the CP-violating weak phase difference 
$\phi\equiv\phi_q-\phi_f$ and is not affected by hadronic 
uncertainties. The corresponding time-dependent CP asymmetry
takes then the simple form
\begin{equation}\label{Amix-t-simple}
\left.\frac{\Gamma(B^0_q(t)\to f)-
\Gamma(\bar B^0_q(t)\to \bar f)}{\Gamma(B^0_q(t)\to f)+
\Gamma(\bar B^0_q(t)\to \bar f)}\right|_{\Delta\Gamma_q=0}
=\pm\sin\phi\,\sin(\Delta M_q t),
\end{equation}
and allows an elegant determination of $\sin\phi$.

\section{HOW COULD NEW PHYSICS ENTER?}\label{sec:NP}
\begin{figure}
   \centering
   \includegraphics[width=9.0truecm]{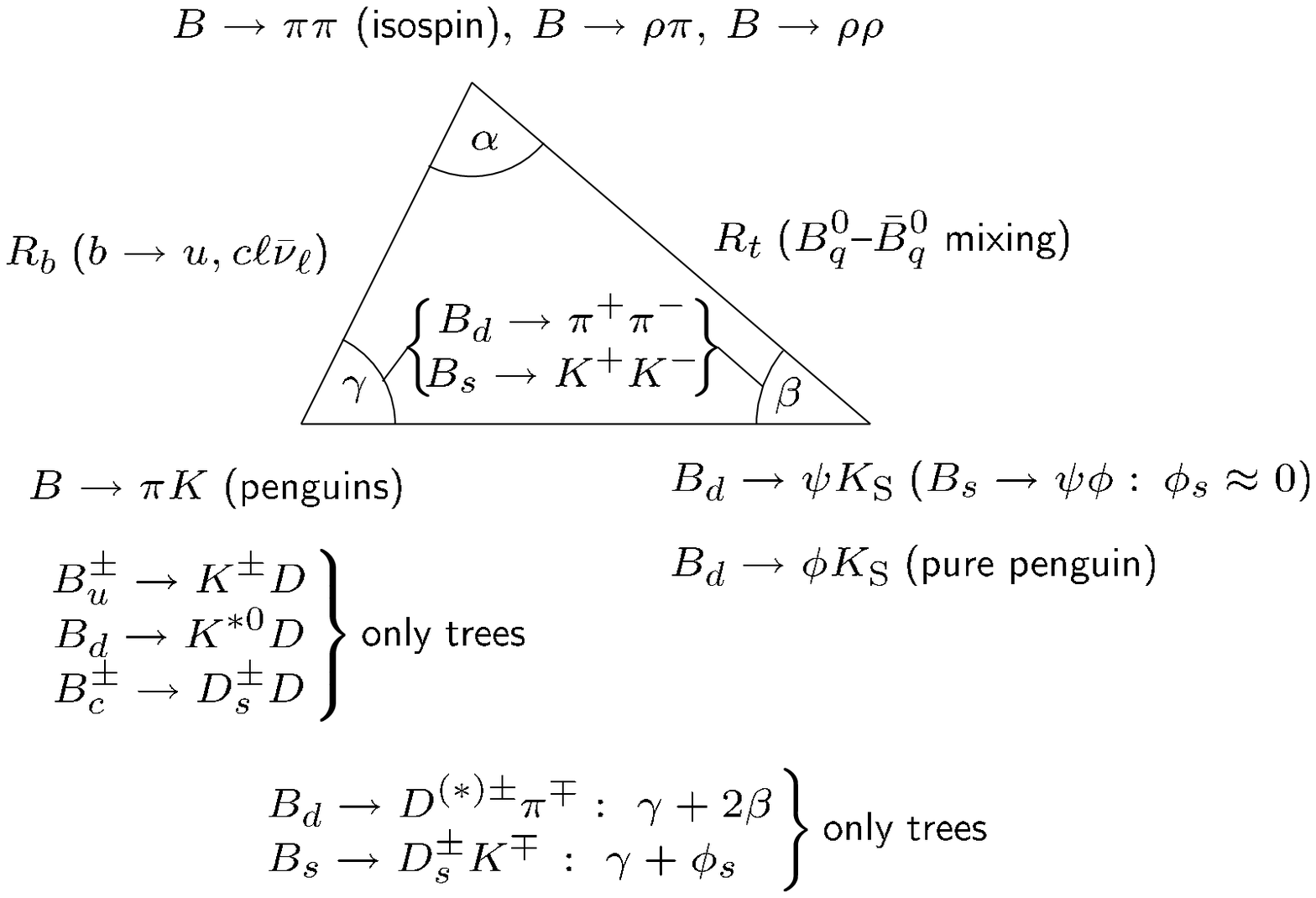} 
   \caption{A brief roadmap of $B$-decay strategies for the exploration of
   CP violation.}
   \label{fig:flavour-map}
\end{figure}

\setcounter{equation}{0}
Using the concept of the low-energy effective Hamiltonians introduced
in Subsection~\ref{subsec:ham}, we may address this important question
in a systematic manner \cite{buras-NP}:
\begin{itemize}
\item NP may modify the ``strength" of the SM operators through new
short-distance functions which depend on the NP parameters, such as the masses 
of charginos, squarks, charged Higgs particles and $\tan\bar\beta\equiv v_2/v_1$ 
in the ``minimal supersymmetric SM'' (MSSM). The NP particles may enter in 
box and penguin topologies, and are ``integrated out'' as the $W$ boson and 
top quark in the SM. Consequently, the initial conditions for the
renormalization-group evolution take the following form:
\begin{equation}\label{WC-NP}
C_k \to C_k^{\rm SM} + C_k^{\rm NP}.
\end{equation}
It should be emphasized that the NP pieces $C_k^{\rm NP}$ may also involve 
new CP-violating phases which are {\it not} related to the CKM matrix.
\item NP may enhance the operator basis:
\begin{equation}
\{Q_k\} \to \{Q_k^{\rm SM}, Q_l^{\rm NP}\},
\end{equation}
so that operators which are not present (or strongly suppressed) in the 
SM may actually play an important r\^ole. In this case, we encounter, 
in general, also new sources for flavour and CP violation.
\end{itemize}
The $B$-meson system offers a variety of processes and strategies for the
exploration of CP violation \cite{CKM-book,RF-Phys-Rep}, as we have illustrated in 
Fig.~\ref{fig:flavour-map} through a collection of prominent examples. 
We see that there are processes with a very {\it different} dynamics that 
are -- in the SM -- sensitive to the {\it same} angles of the UT. 
Moreover, rare $B$- and $K$-meson decays \cite{rare}, 
which originate from loop effects in the SM, provide complementary insights 
into flavour physics and interesting correlations with the CP-B sector; key 
examples are $B\to X_s\gamma$ and the exclusive modes
$B\to K^\ast \gamma$, $B\to\rho\gamma$, as well as $B_{s,d}\to \mu^+\mu^-$ 
and $K^+\to\pi^+\nu\bar\nu$, $K_{\rm L}\to\pi^0\nu\bar\nu$.

\begin{figure}
\centerline{
\epsfxsize=0.33\textwidth\epsffile{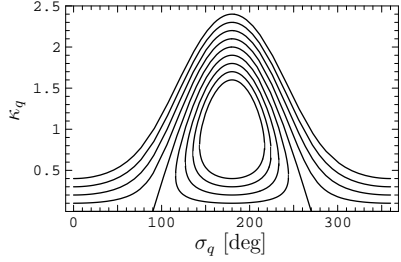}
 }
 \vspace*{-0.3truecm}
\caption[]{The dependence of $\kappa_q$ on $\sigma_q$ for values of 
$\rho_q$ varied between 1.4 (most upper curve) and 0.6 (most inner curve),
in steps of 0.1.}\label{fig:kappa-rho}
\end{figure}

\begin{figure}
\centerline{
\epsfxsize=0.33\textwidth\epsffile{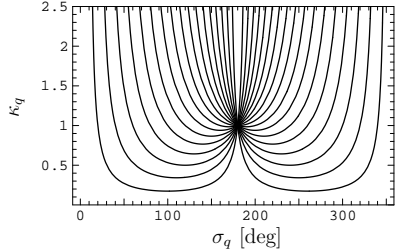}
 }
 \vspace*{-0.3truecm}
\caption[]{The dependence of $\kappa_q$ on $\sigma_q$ for values of 
$\phi_q^{\rm NP}$ varied between $\pm10^\circ$ (lower curves) and 
$\pm170^\circ$ in steps of $10^\circ$: the curves for $0^\circ<\sigma_q<180^\circ$
and $180^\circ<\sigma_q<360^\circ$ correspond to positive and negative values
of $\phi_q^{\rm NP}$, respectively.}\label{fig:kappa-phi}
\end{figure}

In the presence of NP contributions, the subtle interplay between the different 
processes could well be disturbed. There are two popular avenues for NP to 
enter the roadmap of quark-flavour physics:
\begin{itemize}
\item {\it $B^0_q$--$\bar B^0_q$ mixing:} NP could enter through the exchange
of new particles in the box diagrams, or through new contributions at the
tree level. In general, we may write
\begin{equation}
M_{12}^{(q)} =M_{12}^{q,{\rm SM}} \left(1 + \kappa_q e^{i\sigma_q}\right),
\end{equation}
where the expression for $M_{12}^{q,{\rm SM}}$ can be found 
in (\ref{M12-calc}). Consequently, we obtain
\begin{eqnarray}
\Delta M_q & = &\Delta M_q^{\rm SM}+\Delta M_q^{\rm NP} =
\Delta M_q^{\rm SM}\left| 1 + \kappa_q
  e^{i\sigma_q}\right|,\label{DMq-NP}\\
\phi_q & = & \phi_q^{\rm SM}+\phi_q^{\rm NP}=
\phi_q^{\rm SM} + \arg (1+\kappa_q e^{i\sigma_q}),\label{phiq-NP}
\end{eqnarray}
with $\Delta M_q^{\rm SM}$ and $\phi_q^{\rm SM}$ given in (\ref{DeltaMq-def}) and
(\ref{phiq-def}), respectively. 
Using dimensional arguments borrowed from effective field 
theory \cite{FM-BpsiK,FIM}, it can be shown that 
$\Delta M_q^{\rm NP}/\Delta M_q^{\rm SM}\sim1$ and
$\phi_q^{\rm NP}/\phi_q^{\rm SM}\sim1$ could -- in principle -- be possible
for a NP scale $\Lambda_{\rm NP}$ in the TeV regime; such a pattern may 
also arise in specific NP scenarios. Introducing 
\begin{equation}\label{rhoq-def}
\rho_q\equiv
\left|\frac{\Delta M_q}{\Delta M_q^{\rm SM}}\right|=
\sqrt{1+2\kappa_q\cos\sigma_q+\kappa_q^2}\,,
\end{equation}
the measured values of the mass differences $\Delta M_q$ can be converted
into constraints in NP parameter space through the contours shown in
Fig.~\ref{fig:kappa-rho}. Further constraints are implied by the NP
phases $\phi_q^{\rm NP}$, which can be probed through mixing-induced
CP asymmetries, through the curves in the $\sigma_q$--$\kappa_q$
plane shown in Fig.~\ref{fig:kappa-phi}. Interestingly, $\kappa_q$ is bounded
from below for any value of $\phi_q^{\rm NP}\not=0$. For example, even a
small phase $|\phi_q^{\rm NP}|=10^\circ$ implies a clean lower bound of
$\kappa_q\geq0.17$, i.e.\ NP contributions of at most 17\% \cite{BF-DMs}. 
\item {\it Decay amplitudes:} NP has typically a small effect if SM tree processes
play the dominant r\^ole. However, NP could well have a significant impact on 
the FCNC sector: new particles may enter in penguin or box diagrams, or new 
FCNC contributions may even be generated at the tree level. In fact, sizeable 
contributions arise generically in field-theoretical estimates with 
$\Lambda_{\rm NP}\sim\mbox{TeV}$ \cite{FM-BphiK}, as well as in specific 
NP models. 
\end{itemize}
Concerning model-dependent NP analyses, in particular SUSY
scenarios have received a lot of attention; for a selection of recent studies, see
Refs.~\cite{GOSST}--\cite{GHK}. Examples of other fashionable NP scenarios 
are left--right-symmetric models \cite{LR-sym}, scenarios with extra dimensions
\cite{extra-dim}, models with an extra $Z'$ \cite{Z-prime}, ``little Higgs'' 
scenarios \cite{little-higgs}, and models with a fourth generation \cite{hou-4}.

The simplest extension of the SM is given by models with ``minimal flavour violation'' (MFV). Following the characterization given in Ref.~\cite{MFV-1}, 
the flavour-changing processes are here still governed by the CKM matrix -- in 
particular there are no new sources for CP violation --  and the only relevant 
operators are those present in the SM (for an alternative definition, see 
Ref.~\cite{MFV-2}). Specific examples are the Two-Higgs Doublet Model II,
the MSSM without new sources of flavour violation and $\tan\bar\beta$ not
too large, models with one extra universal dimension and the simplest
little Higgs models. Due to their simplicity, the extensions of the SM with
MFV show several correlations between various observables, 
thereby allowing for powerful tests of this scenario \cite{buras-MFV}. A 
systematic discussion of  models with ``next-to-minimal flavour violation" was 
recently given in Ref.~\cite{NMFV}.

There are other fascinating probes for the search of NP. Important examples are 
the $D$-meson system \cite{petrov}, electric dipole moments \cite{PR}, or 
flavour-violating charged lepton decays \cite{CEPRT}. Since a discussion of
these topics is beyond the scope of these lectures, the interested reader should consult 
the corresponding references. Let us next have a closer look at prominent $B$ 
decays, with a particular emphasis of the impact of NP.

\begin{figure}[t]
\centerline{
 \includegraphics[width=5.7truecm]{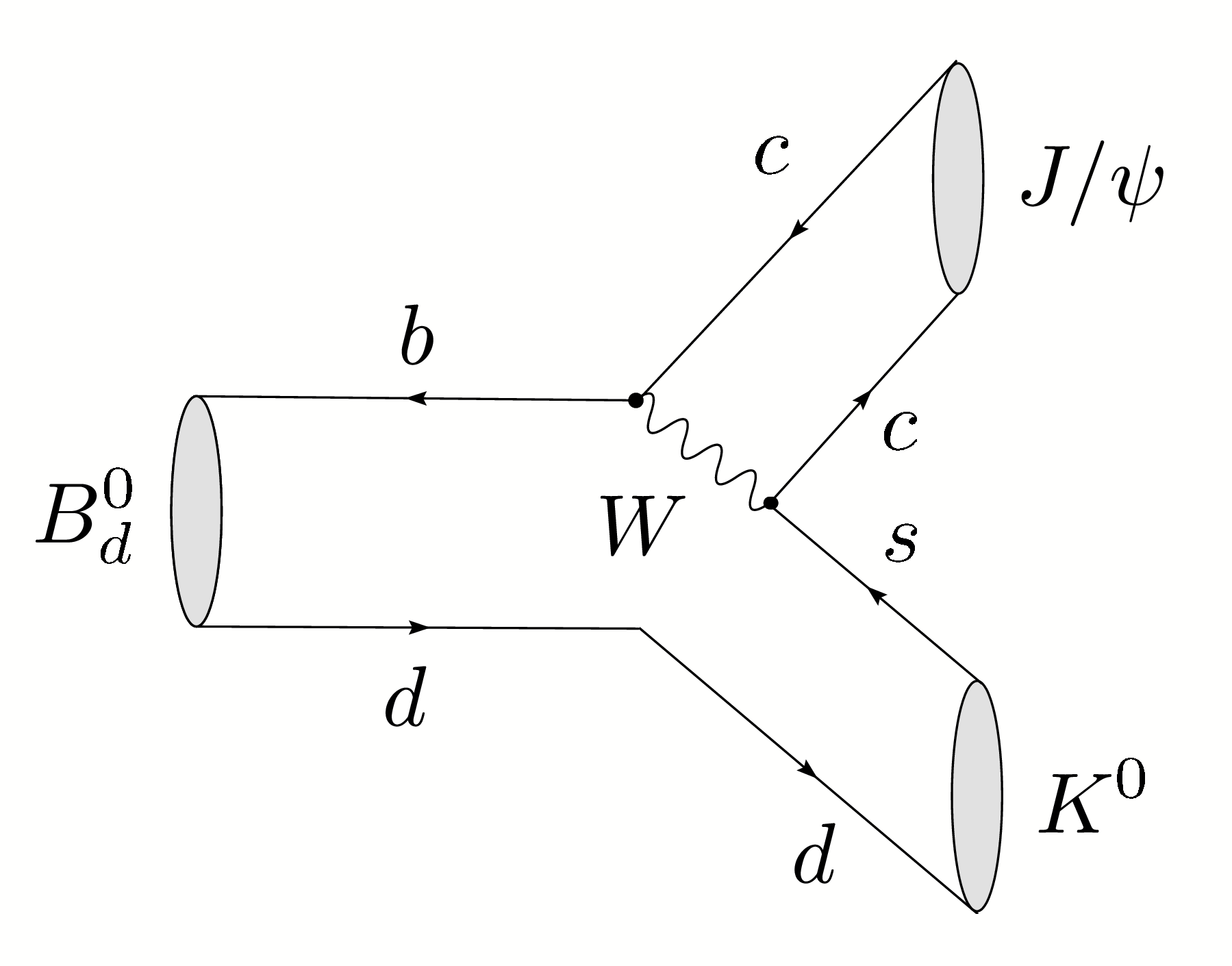}
 \hspace*{0.5truecm}
 \includegraphics[width=5.7truecm]{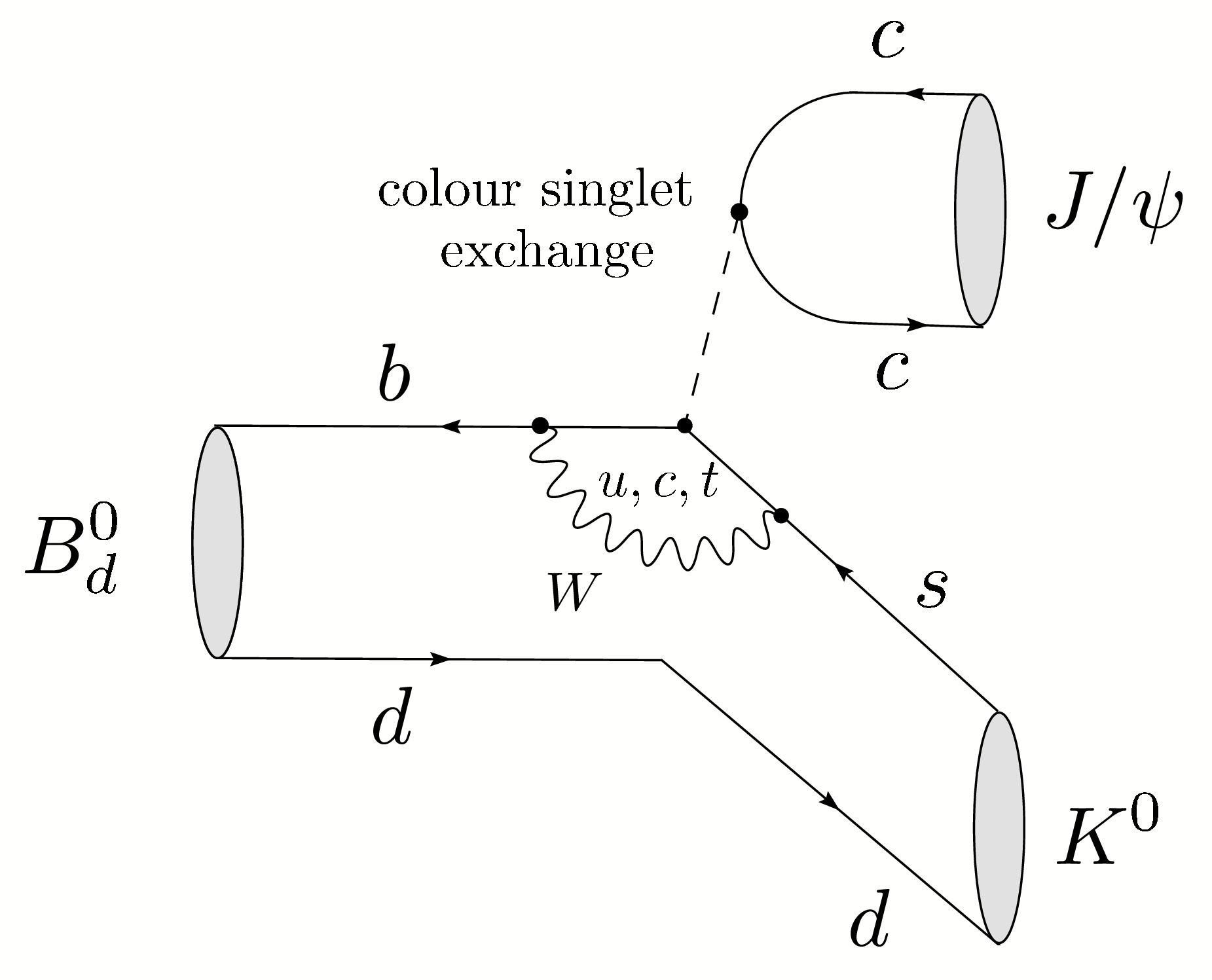}  
 }
 \vspace*{-0.3truecm}
\caption{Feynman diagrams contributing to $B^0_d\to J/\psi K^0$ 
decays.}\label{fig:BpsiK-diag}
\end{figure}

\section{STATUS OF IMPORTANT \boldmath$B$\unboldmath-FACTORY BENCHMARK 
MODES}\label{sec:bench}
\setcounter{equation}{0}
\boldmath
\subsection{$B^0_d\to J/\psi K_{\rm S}$}\label{ssec:BpsiK}
\unboldmath
\subsubsection{Basic Formulae}
This decay has a CP-odd final state, and originates from 
$\bar b\to\bar c c \bar s$ quark-level transitions. Consequently, as we
discussed in Subsection~\ref{sec:class},
it receives contributions both from tree and from penguin topologies, 
as can be seen in Fig.~\ref{fig:BpsiK-diag}. In the SM, the decay 
amplitude can hence be written as follows \cite{RF-BdsPsiK}:
\begin{equation}\label{Bd-ampl1}
A(B_d^0\to J/\psi K_{\rm S})=\lambda_c^{(s)}\left(A_{\rm T}^{c'}+
A_{\rm P}^{c'}\right)+\lambda_u^{(s)}A_{\rm P}^{u'}
+\lambda_t^{(s)}A_{\rm P}^{t'}.
\end{equation}
Here the
\begin{equation}\label{lamqs-def}
\lambda_q^{(s)}\equiv V_{qs}V_{qb}^\ast
\end{equation}
are CKM factors, $A_{\rm T}^{c'}$ is the CP-conserving strong tree amplitude, 
while the $A_{\rm P}^{q'}$ describe the penguin topologies with internal 
$q$ quarks ($q\in\{u,c,t\})$, including QCD and EW penguins; 
the primes remind us that we are dealing with a $\bar b\to\bar s$ 
transition. If we eliminate now $\lambda_t^{(s)}$ through (\ref{CKM-UT-Rel}) 
and apply the Wolfenstein parametrization, we obtain
\begin{equation}\label{BdpsiK-ampl2}
A(B_d^0\to J/\psi K_{\rm S})\propto\left[1+\lambda^2 a e^{i\theta}
e^{i\gamma}\right],
\end{equation}
where
\begin{equation}
a e^{i\vartheta}\equiv\left(\frac{R_b}{1-\lambda^2}\right)
\left[\frac{A_{\rm P}^{u'}-A_{\rm P}^{t'}}{A_{\rm T}^{c'}+
A_{\rm P}^{c'}-A_{\rm P}^{t'}}\right]
\end{equation}
is a hadronic parameter. Using now the formalism of 
Subsection~\ref{subsec:CPasym} yields
\begin{equation}\label{xi-BdpsiKS}
\xi_{\psi K_{\rm S}}^{(d)}=+e^{-i\phi_d}\left[\frac{1+
\lambda^2a e^{i\vartheta}e^{-i\gamma}}{1+\lambda^2a e^{i\vartheta}
e^{+i\gamma}}\right].
\end{equation}
Unfortunately, $a e^{i\vartheta}$, which is a measure for the ratio of the
$B_d^0\to J/\psi K_{\rm S}$ penguin to tree contributions,
can only be estimated with large hadronic uncertainties. However, since 
this parameter enters (\ref{xi-BdpsiKS}) in a doubly Cabibbo-suppressed way, its 
impact on the CP-violating observables is practically negligible. We can put 
this important statement on a more quantitative basis by making the plausible
assumption that $a={\cal O}(\bar\lambda)={\cal O}(0.2)={\cal O}(\lambda)$,
where $\bar\lambda$ is a ``generic'' expansion parameter:
\begin{eqnarray}
{\cal A}^{\mbox{{\scriptsize dir}}}_{\mbox{{\scriptsize
CP}}}(B_d\to J/\psi K_{\mbox{{\scriptsize S}}})&=&0+
{\cal O}(\overline{\lambda}^3)\label{Adir-BdpsiKS}\\
{\cal A}^{\mbox{{\scriptsize mix}}}_{\mbox{{\scriptsize
CP}}}(B_d\to J/\psi K_{\mbox{{\scriptsize S}}})&=&-\sin\phi_d +
{\cal O}(\overline{\lambda}^3) \, \stackrel{\rm SM}{=} \, -\sin2\beta+
{\cal O}(\overline{\lambda}^3).\label{Amix-BdpsiKS}
\end{eqnarray}
Consequently, (\ref{Amix-BdpsiKS}) allows an essentially {\it clean}
determination of $\sin2\beta$ \cite{bisa}.

\subsubsection{Experimental Status}
Since the CKM fits performed within the SM pointed to a large value of 
$\sin2\beta$, $B^0_d\to J/\psi K_{\rm S}$ offered the exciting perspective 
of exhibiting {\it large} mixing-induced CP violation. In 2001, the measurement of  
${\cal A}^{\mbox{{\scriptsize mix}}}_{\mbox{{\scriptsize CP}}}
(B_d\to J/\psi K_{\mbox{{\scriptsize S}}})$
allowed indeed the first observation of CP violation {\it outside} the 
$K$-meson system \cite{CP-B-obs}.
The most recent data are still not showing any signal for {\it direct} CP violation
in $B^0_d\to J/\psi K_{\rm S}$ within the current uncertainties, as is expected from 
(\ref{Adir-BdpsiKS}). The current world average reads \cite{HFAG}
\begin{equation}
{\cal A}_{\rm CP}^{\rm dir}(B_d\to J/\psi K_{\rm S})=0.026\pm0.041.
\end{equation}
As far as (\ref{Amix-BdpsiKS}) is concerned, we have
\begin{equation}\label{s2b-psiK-exp}
\hspace*{-2.0truecm}(\sin2\beta)_{\psi K_{\rm S}}\equiv 
-{\cal A}^{\mbox{{\scriptsize mix}}}_{\mbox{{\scriptsize
CP}}}(B_d\to J/\psi K_{\mbox{{\scriptsize S}}})
=\left\{
\begin{array}{ll}
0.722\pm0.040\pm0.023 & \mbox{(BaBar \cite{s2b-babar})}\\
0.652\pm0.039\pm0.020 & \mbox{(Belle \cite{s2b-belle}),}
\end{array}
\right.
\end{equation}
which gives the following world average \cite{HFAG}:
\begin{equation}\label{s2b-average}
(\sin 2\beta)_{\psi K_{\rm S}}=0.687\pm0.032.
\end{equation}
In the SM, the theoretical uncertainties are generically expected to be
below the 0.01 level; significantly smaller effects are found in \cite{BMR}, 
whereas a fit performed in \cite{CPS} yields a theoretical penguin uncertainty 
comparable to the present experimental systematic error. A possibility
to control these uncertainties is provided by the $B^0_s\to J/\psi K_{\rm S}$ 
channel \cite{RF-BdsPsiK}, which can be explored at the LHC \cite{LHC-Book}.

In Ref.~\cite{FM-BpsiK}, a set of observables to search for NP contributions to 
the $B\to J/\psi K $ decay amplitudes was introduced. It
uses also the charged $B^\pm\to J/\psi K^\pm$ decay, and is given by
\begin{equation}\label{BpsiK}
{\cal B}_{\psi K}\equiv \frac{1-{\cal A}_{\psi K}}{1+{\cal A}_{\psi K}}, 
\end{equation}
with
\begin{equation}\label{ApsiK-def}
{\cal A}_{\psi K}\equiv\left[\frac{\mbox{BR}(B^+\to J/\psi K^+)+
\mbox{BR}(B^-\to J/\psi K^-)}{\mbox{BR}(B^0_d\to J/\psi K^0)+
\mbox{BR}(\bar B^0_d\to J/\psi\bar K^0)}
\right]\left[\frac{\tau_{B^0_d}}{\tau_{B^+}}\right],
\end{equation}
and
\begin{equation}\label{DpmPsiK}
{\cal D}^\pm_{\psi K}\equiv\frac{1}{2}\left[
{\cal A}_{\rm CP}^{\rm dir}(B_d\to J/\psi K_{\rm S})\pm
{\cal A}_{\rm CP}^{\rm dir}(B^\pm\to J/\psi K^\pm)\right].
\end{equation}
As discussed in detail in Refs.~\cite{RF-Phys-Rep,FM-BpsiK}, 
the observables ${\cal B}_{\psi K}$ 
and ${\cal D}^-_{\psi K}$ are sensitive to NP in the $I=1$ isospin sector, 
whereas a non-vanishing value of ${\cal D}^+_{\psi K}$ would signal NP in the
$I=0$ isospin sector. Moreover, the NP contributions with $I=1$ are expected
to be dynamically suppressed with respect to the $I=0$ case because of their 
flavour structure. The most recent $B$-factory results yield
\begin{equation}\label{B-Dpm-PsiK-res}
\hspace*{-0.7truecm} {\cal B}_{\psi K} =-0.035\pm0.037,\quad
{\cal D}^-_{\psi K}=0.010\pm0.023, \quad
{\cal D}^+_{\psi K}=0.017\pm0.023.
\end{equation}
Consequently, NP effects of ${\cal O}(10\%)$ in the $I=1$ sector of the 
$B\to J/\psi K$ decay amplitudes are already disfavoured by the
data for ${\cal B}_{\psi K}$ and ${\cal D}^-_{\psi K}$. However, since a 
non-vanishing value of ${\cal D}^+_{\psi K}$ requires also a large CP-conserving 
strong phase, this observable still leaves room for sizeable  $I=0$ NP 
contributions.

\subsubsection{A Closer Look at New-Physics Effects}
Thanks to the new Belle result listed in (\ref{s2b-psiK-exp}), the average for
$(\sin2\beta)_{\psi K_{\rm S}}$ went down by about $1 \sigma$, which 
was a somewhat surprising development of the summer of 2005. Consequently, 
the comparison of (\ref{s2b-average}) with the CKM fits in the
$\bar\rho$--$\bar\eta$ plane does no longer look ``perfect", as we saw 
in Fig.~\ref{fig:UTfits}. Let us have a closer look at this feature.
If we use $\gamma$ determined from non-leptonic $B\to D^{(*)} K^{(*)}$ 
tree modes and $R_b$ from semileptonic decays, we may calculate
the ``true" value of $\beta$ with the help of the relations
\begin{equation}
\sin\beta=\frac{R_b\sin\gamma}{\sqrt{1-2R_b\cos\gamma+R_b^2}}\,, \quad
\cos\beta=\frac{1-R_b\cos\gamma}{\sqrt{1-2R_b\cos\gamma+R_b^2}},
\end{equation}
which follow from the unitarity of the CKM matrix; the UTfit value 
\begin{equation}\label{gamma-tree}
\gamma=(65\pm20)^\circ
\end{equation}
in (\ref{gam-DK}) and the inclusive and exclusive values of $R_b$ in 
(\ref{Rb}) yield
\begin{equation}\label{beta-true}
\beta_{\rm incl} = (26.7\pm 1.9)^{\circ},\quad 
\beta_{\rm excl} = (22.9\pm3.8)^\circ,
\end{equation}
which can be converted into
\begin{equation}
\sin2\beta|_{\rm incl} = 0.80\pm0.04,\quad 
\sin2\beta|_{\rm excl} = 0.71\pm0.09.
\end{equation}
Consequently, we find
\begin{equation}\label{S-psi-K}
{\cal S}_{\psi K}\equiv (\sin 2\beta)_{\psi K_{\rm S}}-\sin 2\beta=
\left\{\begin{array}{ll}
-0.11\pm0.05 & \mbox{(incl)}\\
-0.02\pm0.10 & \mbox{(excl),}
\end{array}
\right.
\end{equation}
and see nicely the discrepancy arising for the inclusive determination of
$|V_{ub}|$. As discussed in detail in Ref.~\cite{BF-DMs}, $R_b$ is actually
the key parameter for this possible discrepancy with the SM, whereas 
the situation is remarkably stable with respect to $\gamma$. There are two 
limiting cases of this possible discrepancy with the KM mechanism of CP violation: 
\begin{itemize}
\item NP contributions to the $B\to J/\psi K$ decay amplitudes;
\item NP effects entering through $B^0_d$--$\bar B^0_d$ mixing. 
\end{itemize}

\begin{figure}[t]
\centerline{
\begin{tabular}{ll}
   {\small(a)} & {\small(b)} \\
   \includegraphics[width=6.4truecm]{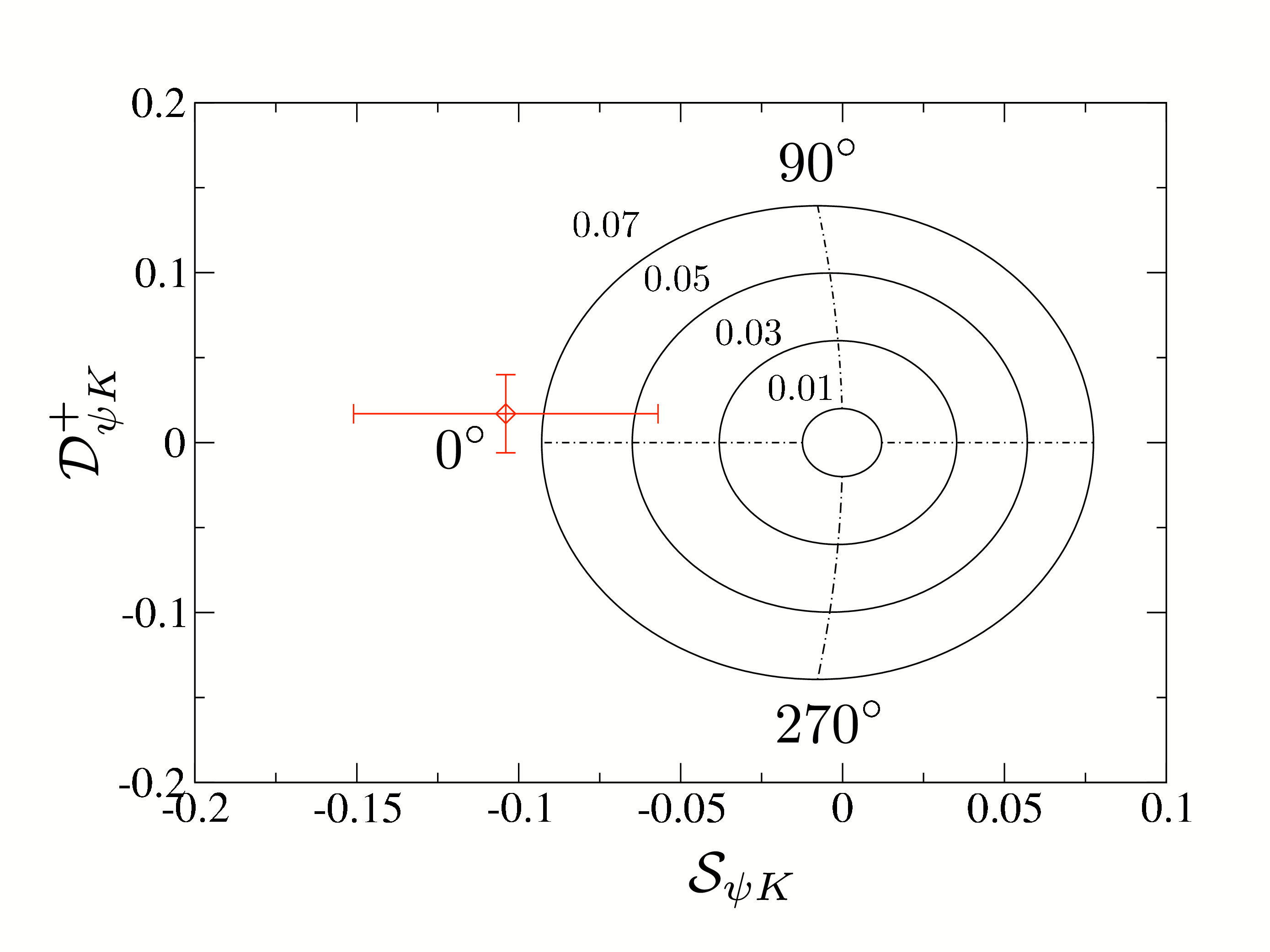}  &
   \includegraphics[width=6.4truecm]{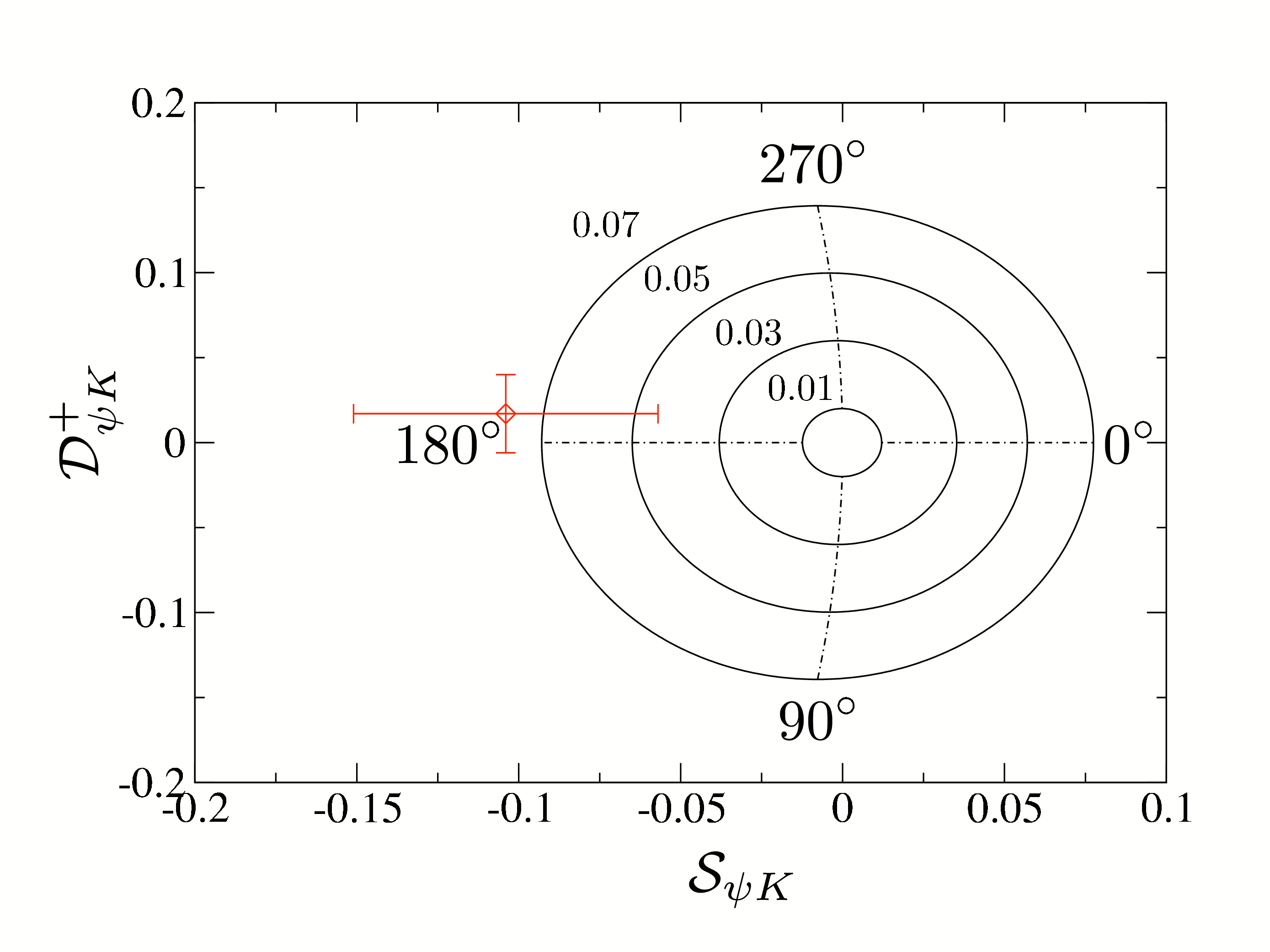}
 \end{tabular}}
 \vspace*{-0.4truecm}
\caption{The situation in the ${\cal S}_{\psi K}$--${\cal D}^+_{\psi K}$ plane for
NP contributions to the $B\to J/\psi K$ decay amplitudes in the $I=0$ isospin 
sector for NP phases $\phi_0=-90^\circ$ (a) and $\phi_0=+90^\circ$ (b). The 
diamonds with the error bars represent the averages of the current data (for the
inclusive value of (\ref{S-psi-K})), whereas 
the numbers correspond to the values of $\Delta_0$ and $v_0$.}\label{fig:Plot-BpsiK}
\end{figure}

Let us first illustrate 
the former case. As the NP effects in the $I=1$ sector are expected to be 
dynamically suppressed, we consider only NP in the $I=0$ isospin sector,
which implies ${\cal B}_{\psi K} ={\cal D}^-_{\psi K}=0$, in accordance with
(\ref{B-Dpm-PsiK-res}). To simplify the discussion, we assume that there is 
effectively only a single NP contribution of this kind, so that we may write
\begin{equation}\label{ApsiK-NP}
A(B^0_d\to J/\psi K^0)=A_0\left[1+v_0e^{i(\Delta_0+\phi_0)}\right]=A(B^+\to J/\psi K^+).
\end{equation}
Here $v_0$ and the CP-conserving strong phase $\Delta_0$ are hadronic parameters,
whereas $\phi_0$ denotes a CP-violating phase originating beyond the SM. 
An interesting specific scenario falling into this category arises if the NP effects
enter through EW penguins. This kind of NP has recently received a lot of attention 
in the context of the $B\to\pi K$ puzzle, which we shall discuss in 
Section~\ref{sec:BpiK-puzzle}. Also within the SM, where $\phi_0$ vanishes, 
EW penguins have a sizeable impact on the $B\to J/\psi K$ system \cite{RF-EWP-rev}.
Using factorization, the following estimate can be obtained \cite{BFRS3}:
\begin{equation}\label{v-SM}
\left. v_0e^{i\Delta_0}\right|_{\rm fact}^{\rm SM}\approx -0.03.
\end{equation}
In Figs.~\ref{fig:Plot-BpsiK} (a) and (b), we consider the inclusive value of
(\ref{S-psi-K}), and show the situation in the
${\cal S}_{\psi K}$--${\cal D}^+_{\psi K}$ plane for $\phi_0=-90^\circ$
and $\phi_0=+90^\circ$, respectively. The contours correspond to 
different values of $v_0$, and are obtained by varying $\Delta_0$ between 
$0^\circ$ and $360^\circ$; the experimental data are represented by the diamonds
with the error bars. Since factorization gives $\Delta_0=180^\circ$, as can be
seen in (\ref{v-SM}), the case of $\phi_0=-90^\circ$ is disfavoured. On the
other hand, in the case of $\phi_0=+90^\circ$, the experimental region can straightforwardly be reached for $\Delta_0$ not differing too much from the
factorization result, although an enhancement of $v_0$ by a factor of 
${\cal O}(3)$ with respect to the SM estimate in (\ref{v-SM}), which suffers from 
large uncertainties, would simultaneously be required in order to reach the central 
experimental value. Consequently, NP contributions to the EW penguin sector 
could, in principle, be at the origin of the possible discrepancy indicated by the
inclusive value of (\ref{S-psi-K}). This scenario should be carefully monitored in
the future.

Another explanation of  (\ref{S-psi-K}) is provided by CP-violating NP contributions to 
$B^0_d$--$\bar B^0_d$ mixing, which affect the corresponding mixing phase as 
in (\ref{phiq-NP}), so that
\begin{equation}
\phi_d=2\beta+\phi_d^{\rm NP}. 
\end{equation}
Assuming that the NP contributions to the $B\to J/\psi K$ 
amplitudes are negligible,  (\ref{s2b-average}) implies
\begin{equation}\label{phid-exp}
\phi_d=(43.4\pm2.5)^\circ \quad\lor\quad (136.6\pm2.5)^\circ.
\end{equation}
Here the latter solution would be in dramatic conflict with the CKM fits, and
would require a large NP contribution to $B^0_d$--$\bar B^0_d$ 
mixing \cite{FIM,FlMa}. Both solutions can be distinguished through the 
measurement of the sign of $\cos\phi_d$, where a positive value would 
select the SM-like branch. Using an angular analysis of the 
$B_d\to J/\psi[\to\ell^+\ell^-] K^\ast[\to\pi^0K_{\rm S}]$ decay products,
the BaBar collaboration finds \cite{babar-c2b}
\begin{equation}
\cos\phi_d =2.72^{+0.50}_{-0.79} \pm 0.27,
\end{equation}
thereby favouring the solution around $\phi_d=43^\circ$. Interestingly, this 
picture emerges also from the first data for CP-violating effects in 
$B_d\to D^{(*)\pm}\pi^\mp$ modes \cite{RF-gam-ca}, and an analysis of 
the $B\to\pi\pi,\pi K$ system \cite{BFRS3}, although in an indirect manner.
Recently, a new method has been proposed, which makes use of 
the interference pattern in $D\to K_{\rm S}\pi^+\pi^-$ decays emerging
from $B_d\to D\pi^0$ and similar decays \cite{bo-ge}. The results of
this method are also consistent with the SM, so that a negative value
of $\cos\phi_d$ is now ruled out with greater than 95\% confidence 
\cite{WG-sum}.

\begin{figure}[t]
$$\epsfxsize=0.38\textwidth\epsffile{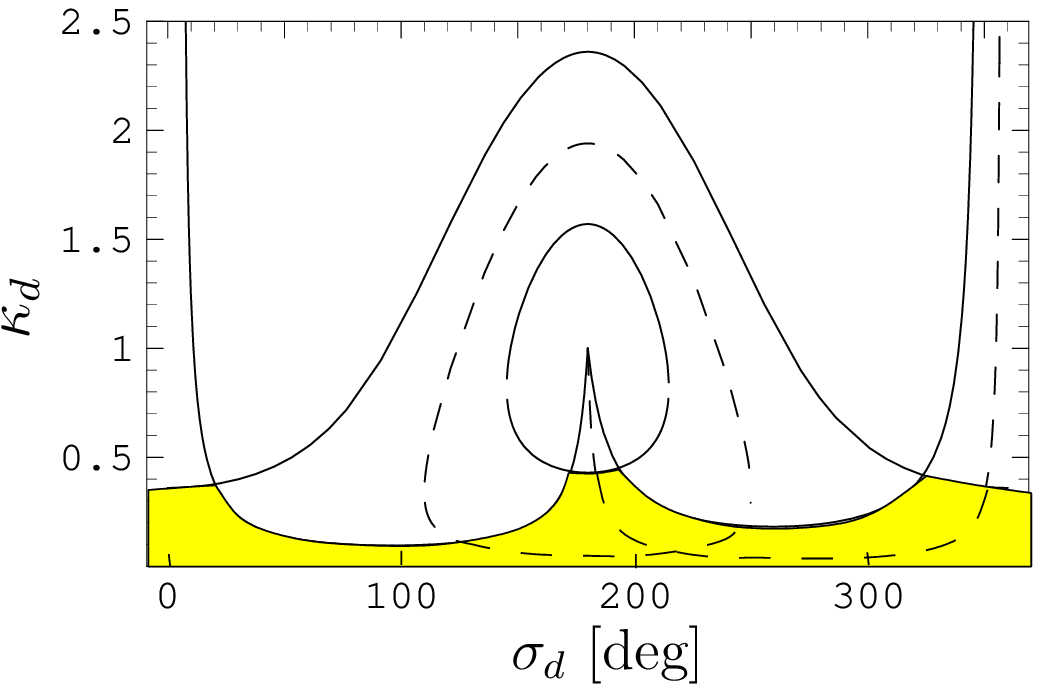}\quad
\epsfxsize=0.38\textwidth\epsffile{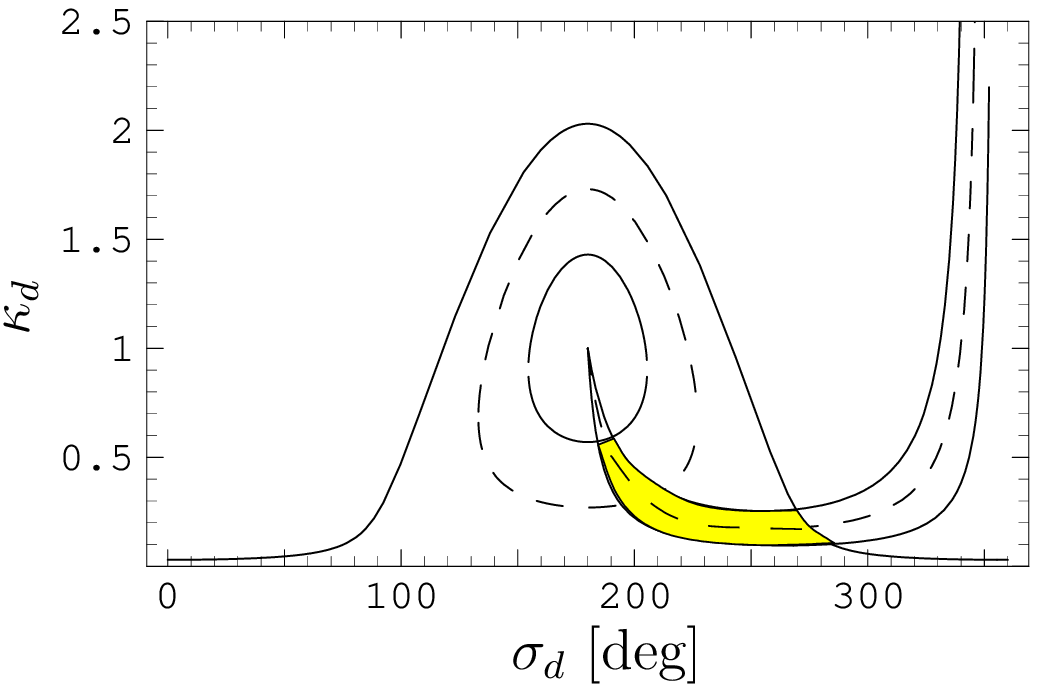}
$$
 \vspace*{-1truecm}
\caption[]{Left panel: allowed region (yellow/grey) in the $\sigma_d$--$\kappa_d$
  plane in a scenario with the JLQCD lattice results (\ref{JLQCD}) and 
  $\left.\phi^{\rm NP}_d\right|_{\rm excl}$. Dashed lines: central values of $\rho_d$ 
  and $\phi^{\rm NP}_d$, solid lines: $\pm 1\,\sigma$. Right panel: ditto for the 
 scenario with the (HP+JL)QCD   lattice results
  (\ref{HPQCD}) and  $\left.\phi^{\rm NP}_d\right|_{\rm incl}$. 
}\label{fig:res-k-sig-d}
\end{figure}

Using the ``true" values of $\beta$ in (\ref{beta-true}), the value of
$\phi_d=(43.4\pm2.5)^\circ$ implies
\begin{equation}\label{phiNPd-num}
\left.\phi^{\rm NP}_d\right|_{\rm incl} = -(10.1\pm 4.6)^\circ\,,\qquad
\left.\phi^{\rm NP}_d\right|_{\rm excl} = -(2.5\pm 8.0)^\circ\,;
\end{equation}
results of $\phi_d^{\rm NP}\approx-10^\circ$ were also recently obtained in 
Refs.~\cite{BFRS-5,UTfit-NP}. The contours in Fig.~\ref{fig:kappa-phi} allow us now
to convert these numbers into constraints in the $\sigma_d$--$\kappa_d$ 
plane. Further constraints can be obtained through the experimental value of 
$\Delta M_d$ in (\ref{DMd-exp}) with the help of the contours in
Fig.~\ref{fig:kappa-rho}, where $\rho_d$ is introduced in (\ref{rhoq-def}).
In addition to  hadronic parameters, the SM prediction of $\Delta M_d$ involves 
also the CKM factor $|V_{td}^\ast V_{tb}|$, which can -- if we use the unitarity of the 
CKM matrix -- be expressed as
\begin{equation}\label{CKM-Bd}
|V_{td}^\ast V_{tb}|=|V_{cb}|\lambda\sqrt{1-2R_b\cos\gamma+R_b^2}.
\end{equation}
The values in (\ref{Rb}) and (\ref{gamma-tree}), as well as the relevant 
lattice parameters in (\ref{JLQCD}) and (\ref{HPQCD}) yield then
\begin{eqnarray}
\left.\rho_d\right|_{\rm JLQCD} &=&
  0.97\pm0.33^{-0.17}_{+0.26}\label{rhod-JLQCD}\\
\left.\rho_d\right|_{\rm  (HP+JL)QCD} &=& 
0.75\pm0.25\pm0.16,\label{rhod-HPJL}
\end{eqnarray}
where the first and second errors are due to $\gamma$ (and a small extent to
$R_b$) and $f_{B_d} \hat B_{B_d}^{1/2}$, respectively \cite{BF-DMs}. 
These results are compatible with the SM value $\rho_d=1$, but suffer from 
considerable uncertainties. In Fig.~\ref{fig:res-k-sig-d}, we finally show the
situation in the $\sigma_d$--$\kappa_d$ plane. We see that the information
about the CP-violating phase $\phi_d$ has a dramatic impact, reducing
the allowed NP parameter space significantly. 

The possibility of having a non-zero value of (\ref{S-psi-K}) could of course just 
be due to a statistical fluctuation. However, should it be confirmed, 
it could be due to CP-violating NP contributions to the $B^0_d\to J/\psi K_{\rm S}$ 
decay amplitude or to $B^0_d$--$\bar B^0_d$ mixing, as we just saw.
A tool to distinguish between these avenues is provided by decays 
of the kind $B_d\to D\pi^0, D\rho^0, ...$, which are pure ``tree" decays, i.e.\
they do {\it not} receive any penguin contributions. If the neutral $D$ mesons
are observed through their decays into CP eigenstates $D_\pm$, these decays
allow extremely clean determinations of the ``true" value of $\sin\phi_d$ 
\cite{RF-BdDpi0}, as we shall discuss in more detail in Subsection~\ref{ssec:BsDsK}. 
In view of (\ref{S-psi-K}), this would be very interesting, so that detailed 
feasibility studies for the exploration of the $B_d\to D\pi^0, D\rho^0, ...$\ modes 
at a super-$B$ factory are strongly encouraged.

\begin{figure}
\centerline{
 \includegraphics[width=5.5truecm]{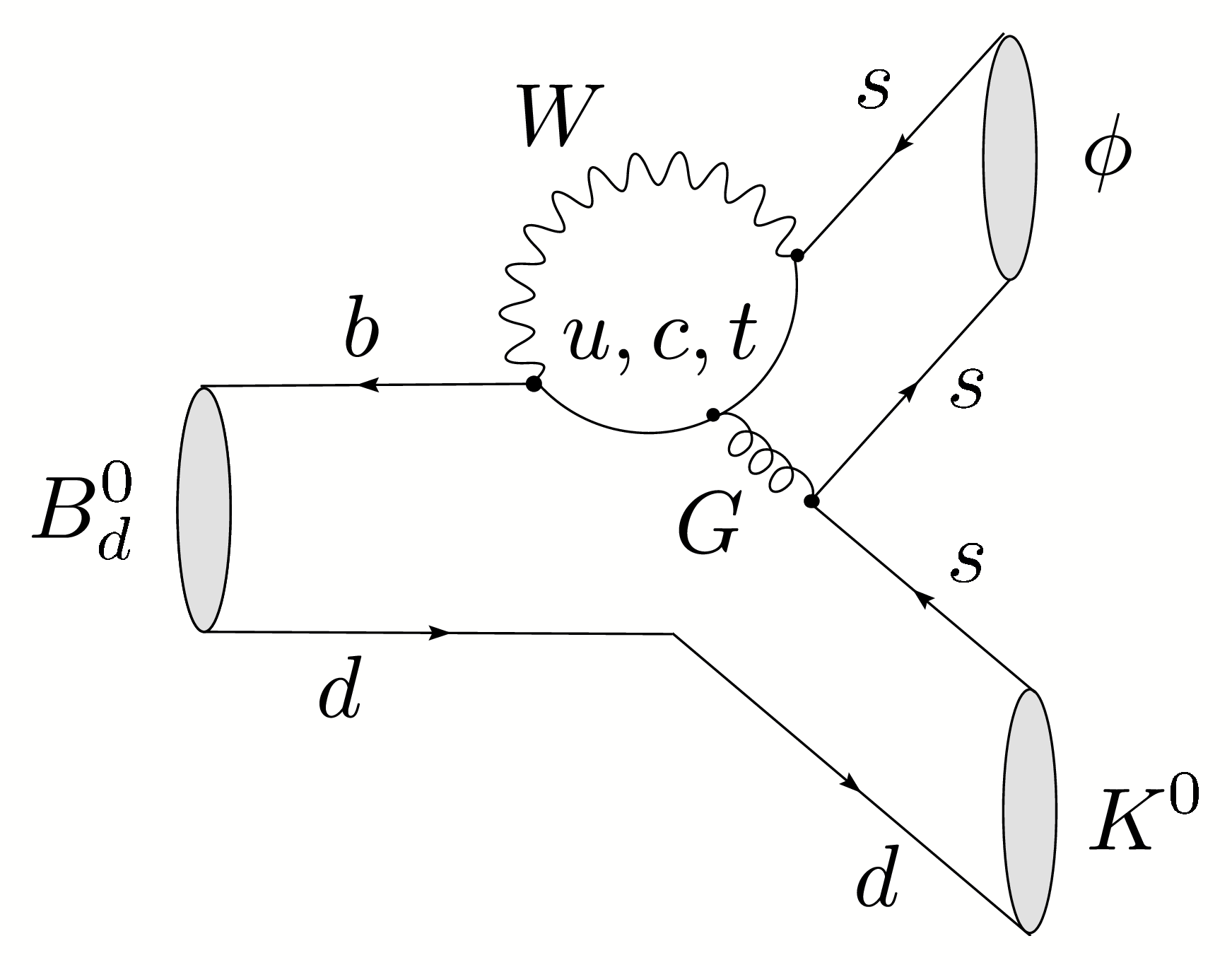} 
 }
 \vspace*{-0.3truecm}
\caption{Feynman diagrams contributing to $B^0_d\to \phi K^0$ 
decays.}\label{fig:BphiK-diag}
\end{figure}

\boldmath
\subsection{$B^0_d\to \phi K_{\rm S}$}\label{ssec:BphiK}
\unboldmath
Another important probe for the testing of the KM mechanism is 
offered by $B_d^0\to \phi K_{\rm S}$, which is a 
decay into a CP-odd final state. As can be seen in Fig.~\ref{fig:BphiK-diag},
it originates from $\bar b\to \bar s s \bar s$ transitions and is, therefore, a 
pure penguin mode. This decay is described by the low-energy effective 
Hamiltonian in (\ref{e4}) with $r=s$, where the current--current operators 
may only contribute through penguin-like contractions, which describe the
penguin topologies with internal up- and charm-quark exchanges. The dominant
r\^ole is played by the QCD penguin operators \cite{BphiK-old}. However,
thanks to the large top-quark mass, EW penguins have a sizeable impact as 
well \cite{RF-EWP,DH-PhiK}. In the SM, we may write
\begin{equation}\label{B0phiK0-ampl}
A(B_d^0\to \phi K_{\rm S})=\lambda_u^{(s)}\tilde A_{\rm P}^{u'}
+\lambda_c^{(s)}\tilde A_{\rm P}^{c'}+\lambda_t^{(s)}\tilde A_{\rm P}^{t'},
\end{equation}
where we have applied the same notation as in Subsection~\ref{ssec:BpsiK}.
Eliminating the CKM factor $\lambda_t^{(s)}$ with the help of
(\ref{CKM-UT-Rel}) yields
\begin{equation}
A(B_d^0\to \phi K_{\rm S})\propto
\left[1+\lambda^2 b e^{i\Theta}e^{i\gamma}\right],
\end{equation}
where 
\begin{equation}
b e^{i\Theta}\equiv\left(\frac{R_b}{1-\lambda^2}\right)\left[
\frac{\tilde A_{\rm P}^{u'}-\tilde A_{\rm P}^{t'}}{\tilde A_{\rm P}^{c'}-
\tilde A_{\rm P}^{t'}}\right].
\end{equation}
Consequently,  we obtain
\begin{equation}\label{xi-phiKS}
\xi_{\phi K_{\rm S}}^{(d)}=+e^{-i\phi_d}
\left[\frac{1+\lambda^2b e^{i\Theta}e^{-i\gamma}}{1+
\lambda^2b e^{i\Theta}e^{+i\gamma}}\right].
\end{equation}
The theoretical estimates of $b e^{i\Theta}$ 
suffer from large hadronic uncertainties. However, since this parameter enters 
(\ref{xi-phiKS}) in a doubly Cabibbo-suppressed way, we obtain the 
following expressions \cite{RF-EWP-rev}:
\begin{eqnarray}
{\cal A}_{\rm CP}^{\rm dir}(B_d\to \phi K_{\rm S})&=&0+
{\cal O}(\lambda^2)\label{BphiK-rel1}\\
{\cal A}_{\rm CP}^{\rm mix}(B_d\to \phi K_{\rm S})&=&-\sin\phi_d
+{\cal O}(\lambda^2),\label{BphiK-rel2}
\end{eqnarray}
where we made the plausible assumption that $b={\cal O}(1)$. On the other 
hand, the mixing-induced CP asymmetry of 
$B_d\to J/\psi K_{\rm S}$ measures also $-\sin\phi_d$, as we saw in
(\ref{Amix-BdpsiKS}). We arrive therefore at the following 
relation \cite{RF-EWP-rev,growo}:
\begin{equation}\label{Bd-phiKS-SM-rel}
-(\sin2\beta)_{\phi K_{\rm S}}\equiv
{\cal A}_{\rm CP}^{\rm mix}(B_d\to \phi K_{\rm S}) 
={\cal A}_{\rm CP}^{\rm mix}(B_d\to J/\psi K_{\rm S}) + 
{\cal O}(\lambda^2),
\end{equation}
which offers an interesting test of the SM. Since $B_d\to \phi K_{\rm S}$ is 
governed by penguin processes in the SM, this decay may well be affected by 
NP. In fact, if we assume that NP arises generically in the TeV regime, it can be 
shown through field-theoretical estimates that the NP contributions to  
$b\to s\bar s s$ transitions may well lead to sizeable violations of
(\ref{BphiK-rel1}) and (\ref{Bd-phiKS-SM-rel}) \cite{RF-Phys-Rep,FM-BphiK}. Moreover, 
this is also the case for several specific NP scenarios; for examples, see 
Refs.~\cite{CFMS,Ko,GHK,Z-prime-BpiK}.

\begin{figure}
\centerline{
\begin{tabular}{ll}
   {\small(a)} & {\small(b)} \\
   \includegraphics[width=6.8truecm]{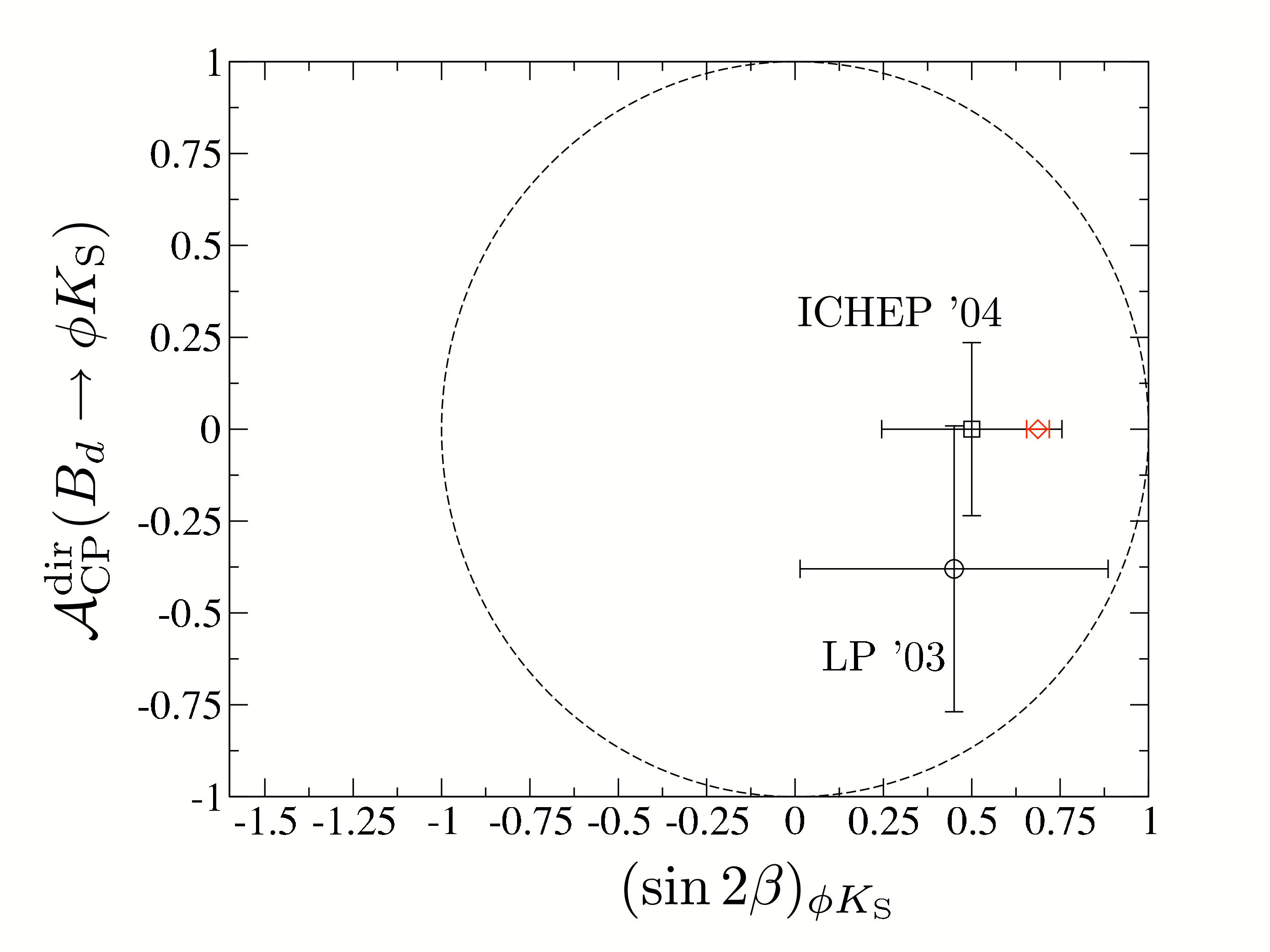}  &
   \includegraphics[width=6.8truecm]{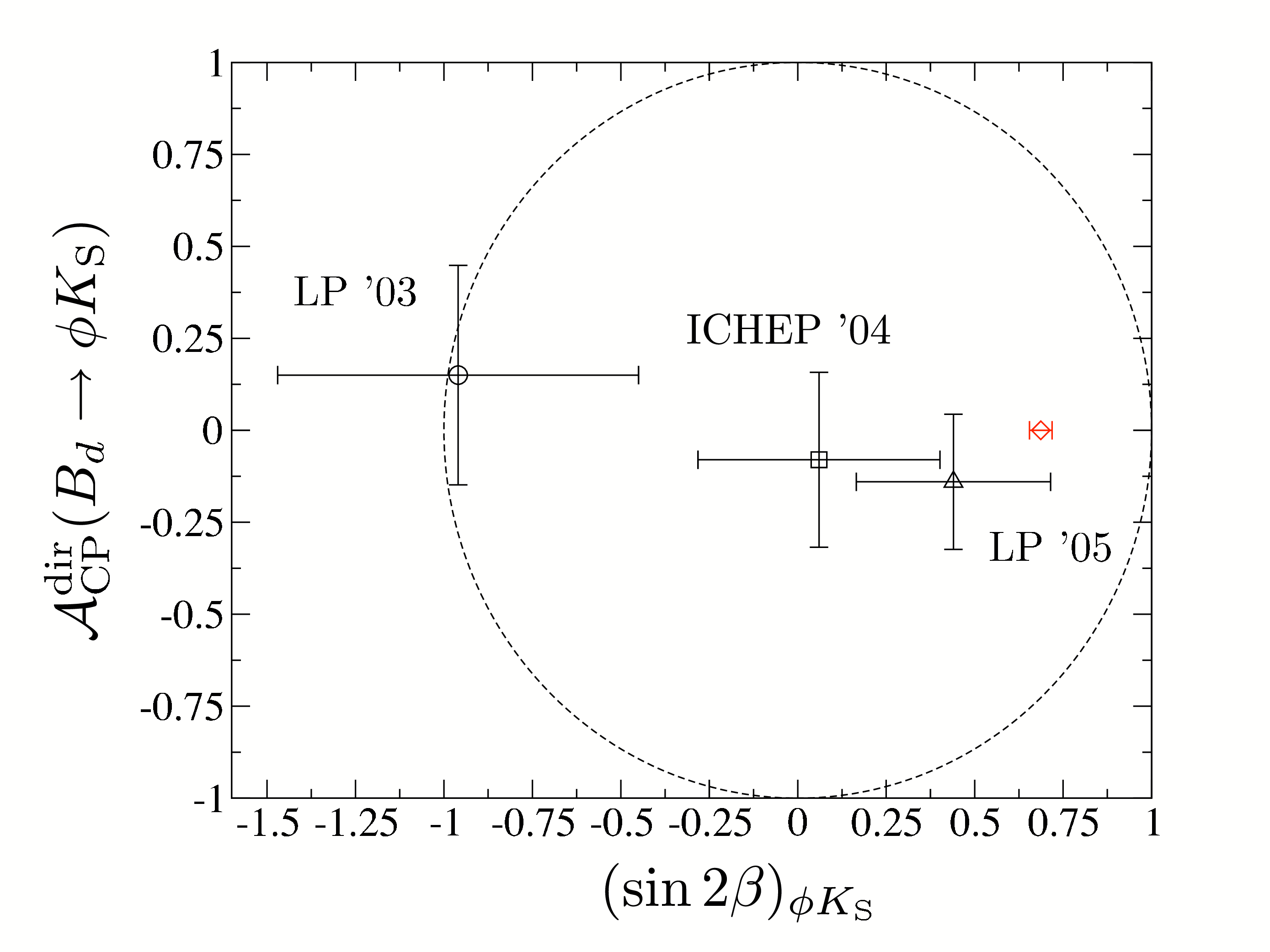}
 \end{tabular}
 }
 \vspace*{-0.4truecm}
   \caption{The time evolution of the BaBar (a)  and Belle (b) data for 
   the CP violation in $B_d\to \phi K_{\rm S}$. The diamonds represent the 
   SM relations (\ref{BphiK-rel1})--(\ref{Bd-phiKS-SM-rel}) with 
   (\ref{s2b-average}).}\label{fig:BphiK-data}
\end{figure}

In Fig.~\ref{fig:BphiK-data}, we show the time evolution of the $B$-factory data
for the measurements of CP violation in $B_d\to\phi K_{\rm S}$, using the results 
reported at the LP~'03 \cite{LP03}, ICHEP~'04 \cite{ICHEP04} and LP~'05 \cite{LP05}
conferences. Because of (\ref{Obs-rel}), the corresponding observables have
to lie inside a circle with radius one around the origin, which is represented by the
dashed lines. The result announced by the Belle collaboration in
2003 led to quite some excitement in the community. Meanwhile, the Babar
\cite{BaBar-Bphi-K} and Belle \cite{Belle-Bphi-K} results are in good agreement 
with each other, yielding the following averages \cite{HFAG}:
\begin{equation}\label{BphiK-av}
{\cal A}_{\rm CP}^{\rm dir}(B_d\to \phi K_{\rm S})=-0.09\pm0.14, \quad
(\sin2\beta)_{\phi K_{\rm S}}=0.47\pm0.19. 
\end{equation}
If  we take (\ref{s2b-average}) into account, we obtain the following result for
the counterpart of (\ref{S-psi-K}):
\begin{equation}\label{S-phi-K}
{\cal S}_{\phi K}\equiv (\sin 2\beta)_{\phi K_{\rm S}}- (\sin 2\beta)_{\psi K_{\rm S}}
=-0.22\pm0.19.
\end{equation}
This number still appears to be somewhat on the lower side, thereby indicating potential 
NP contributions to $b\to s \bar s s$ processes.

Further insights into the origin and the isospin structure of NP contributions 
can be obtained through a combined analysis of the neutral and charged 
$B\to \phi K$ modes with the help of observables 
${\cal B}_{\phi K}$ and ${\cal D}^\pm_{\phi K}$ \cite{FM-BphiK}, which are 
defined in analogy to (\ref{BpsiK}) and (\ref{DpmPsiK}), respectively. The
current experimental results read as follows:
\begin{equation}\label{B-Dpm-PhiK-res}
\hspace*{-0.7truecm} {\cal B}_{\phi K} =0.00\pm0.08,\quad
{\cal D}^-_{\phi K}=-0.03\pm0.07, \quad
{\cal D}^+_{\phi K}=-0.06\pm0.07.
\end{equation}
As in the $B\to J/\psi K$ case, ${\cal B}_{\phi K}$ and ${\cal D}^-_{\phi K}$ probe
NP effects in the $I=1$ sector, which are expected to be dynamically suppressed,
whereas ${\cal D}^+_{\phi K}$ is sensitive to NP in the $I=0$ sector. The latter 
kind of NP could also manifest itself as a non-vanishing value of (\ref{S-phi-K}).

\begin{figure}
\centerline{
\begin{tabular}{ll}
   {\small(a)} & {\small(b)} \\
   \includegraphics[width=6.8truecm]{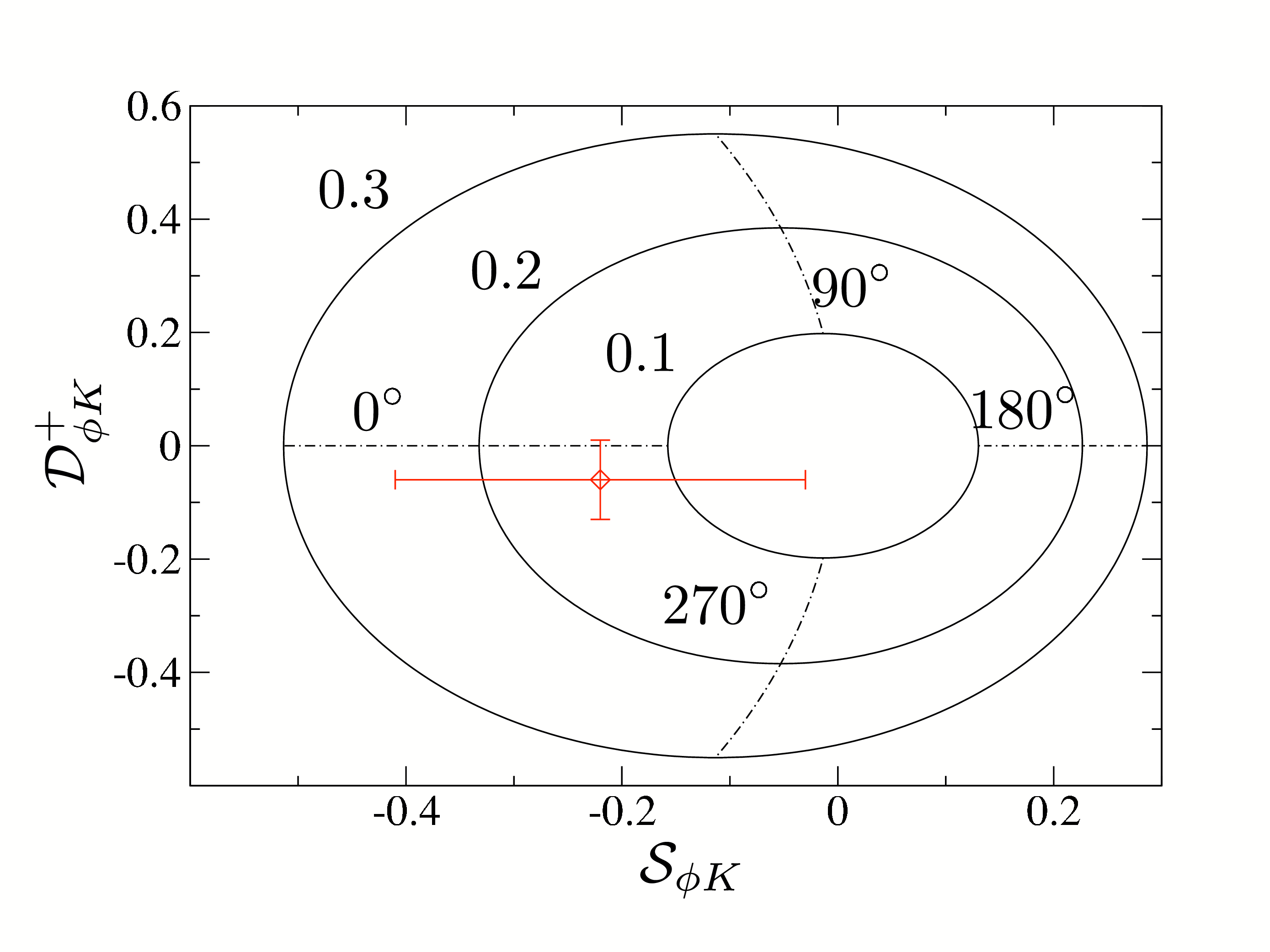}  &
   \includegraphics[width=6.8truecm]{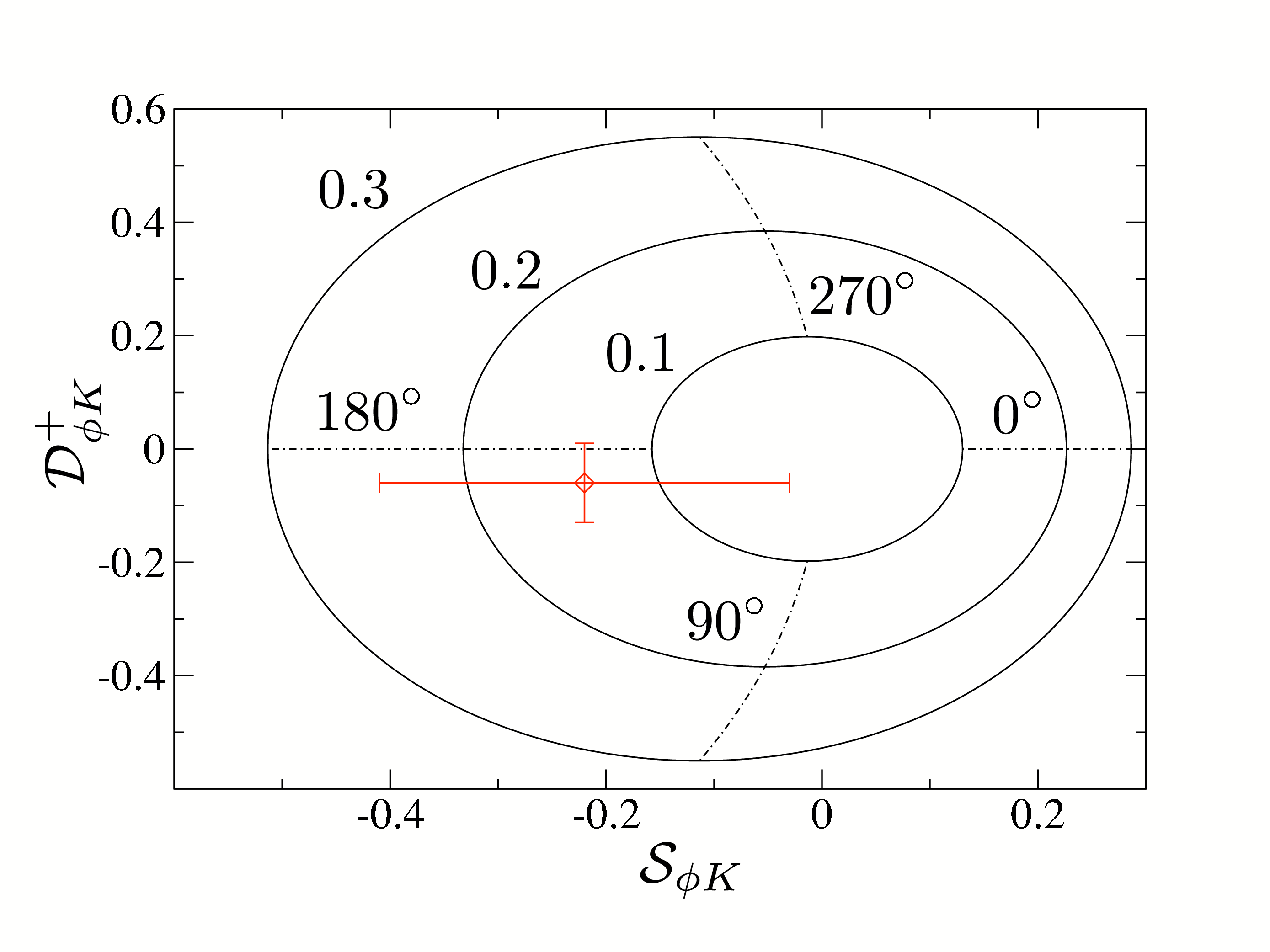}
 \end{tabular}}
 \vspace*{-0.4truecm}
\caption{The situation in the ${\cal S}_{\phi K}$--${\cal D}^+_{\phi K}$ plane for
NP contributions to the $B\to \phi K$ decay amplitudes in the $I=0$ isospin 
sector for NP phases $\phi_0=-90^\circ$ (a) and $\phi_0=+90^\circ$ (b). The 
diamonds with the error bars represent the averages of the current data, whereas 
the numbers correspond to the values of $\tilde\Delta_0$ and 
$\tilde v_0$.}\label{fig:Plot-BphiK}
\end{figure}

In order to illustrate these effects, let us consider again the case where NP enters 
only in the $I=0$ isospin sector. An important example is given by EW penguins, 
which have a significant impact on $B\to\phi K$ decays \cite{RF-EWP}. In analogy
to the discussion in Subsection~\ref{ssec:BpsiK}, we may then write
\begin{equation}\label{AphiK-NP}
A(B^0_d\to \phi K^0)=\tilde A_0\left[1+\tilde v_0e^{i(\tilde \Delta_0+\phi_0)}\right]=
A(B^+ \to \phi K^+),
\end{equation}
which implies ${\cal B}_{\phi K} ={\cal D}^-_{\phi K}=0$, in accordance with 
(\ref{B-Dpm-PhiK-res}). The notation corresponds to the one of (\ref{ApsiK-NP}).  
Using the factorization approach to deal with the QCD and EW penguin contributions, 
we obtain the following estimate in the SM, where the CP-violating NP phase
$\phi_0$ vanishes \cite{BFRS3}:
\begin{equation}\label{v-SM-phiK}
\left.\tilde v_0e^{i\tilde \Delta_0}\right|_{\rm fact}^{\rm SM}\approx -0.2.
\end{equation}
In Figs.~\ref{fig:Plot-BphiK} (a) and (b), we show the situation in the 
${\cal S}_{\phi K}$--${\cal D}^+_{\phi K}$ plane for NP phases $\phi_0=-90^\circ$
and $\phi_0=+90^\circ$, respectively, and various values of $\tilde v_0$; each point
of the contours is parametrized by $\tilde\Delta_0\in[0^\circ,360^\circ]$. We observe 
that the central values of the current experimental data, which are represented by the 
diamonds with the error bars, can straightforwardly be accommodated in this scenario 
in the case of $\phi_0=+90^\circ$ for strong phases satisfying $\cos\tilde\Delta_0<0$, 
as in factorization. Moreover, as can also be seen in Fig.~\ref{fig:Plot-BphiK} (b),
the EW penguin contributions would then have to be suppressed with respect
to the SM estimate, which would be an interesting feature in view of the discussion of
the $B\to \pi K$ puzzle and the rare decay constraints in Section~\ref{sec:BpiK-puzzle}.

It will be interesting to follow the evolution of the $B$-factory data,
and to monitor also similar modes, such as $B^0_d\to \pi^0 K_{\rm S}$ 
\cite{PAPIII} and  $B^0_d\to \eta'K_{\rm S}$ \cite{loso}. For a compilation of 
the corresponding experimental results, see Ref.~\cite{HFAG}; recent 
theoretical papers dealing with these channels can be found in 
Refs.~\cite{BFRS2,BFRS3,BFRS-5,GGR,beneke}. We will return to the CP 
asymmetries of the $B^0_d\to \pi^0 K_{\rm S}$ channel in 
Section~\ref{sec:BpiK-puzzle}.

\boldmath
\subsection{$B^0_d\to \pi^+\pi^-$}\label{ssec:Bpi+pi-}
\begin{figure}
\centerline{
 \includegraphics[width=4.5truecm]{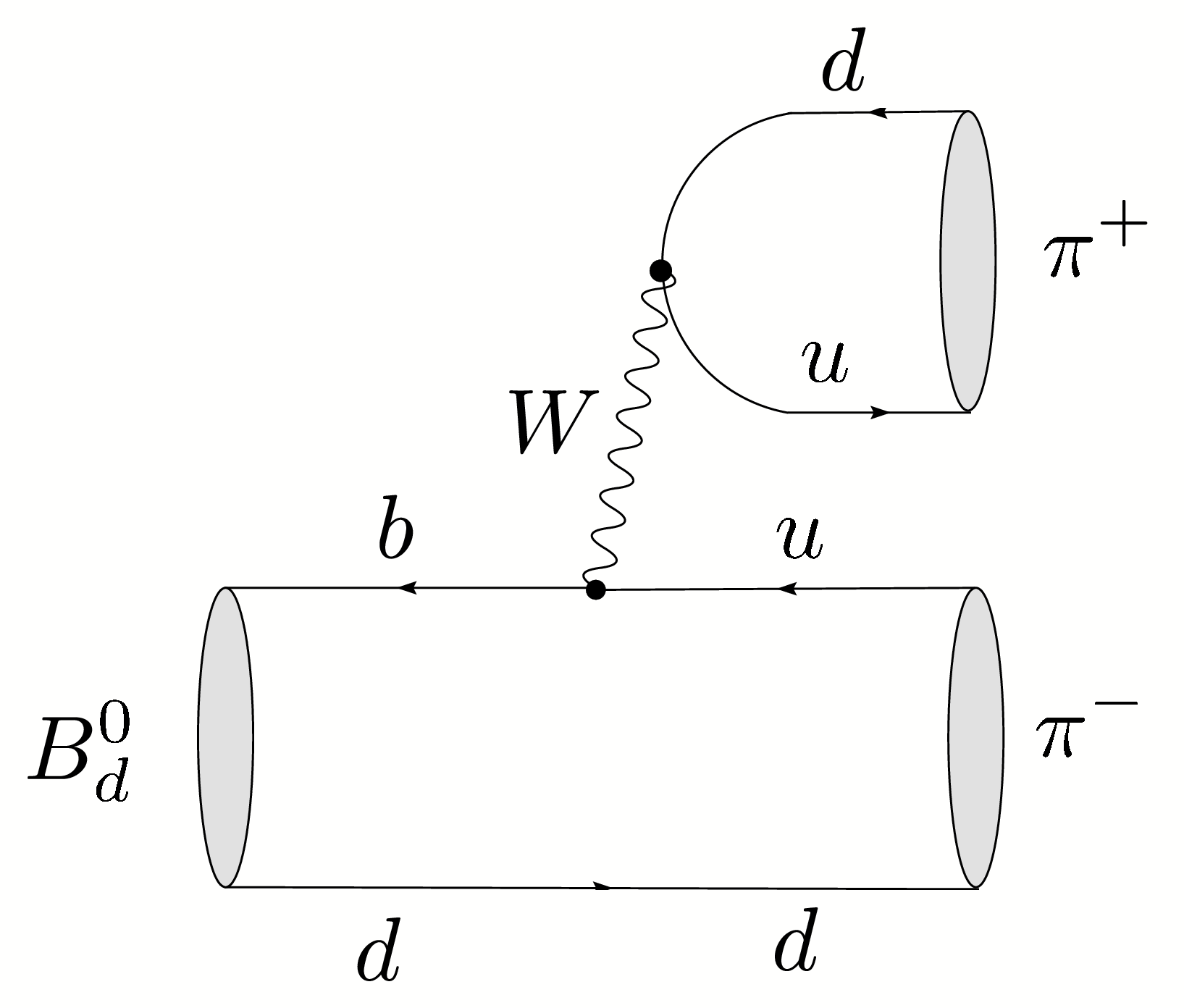}
 \hspace*{0.5truecm}
 \includegraphics[width=5.2truecm]{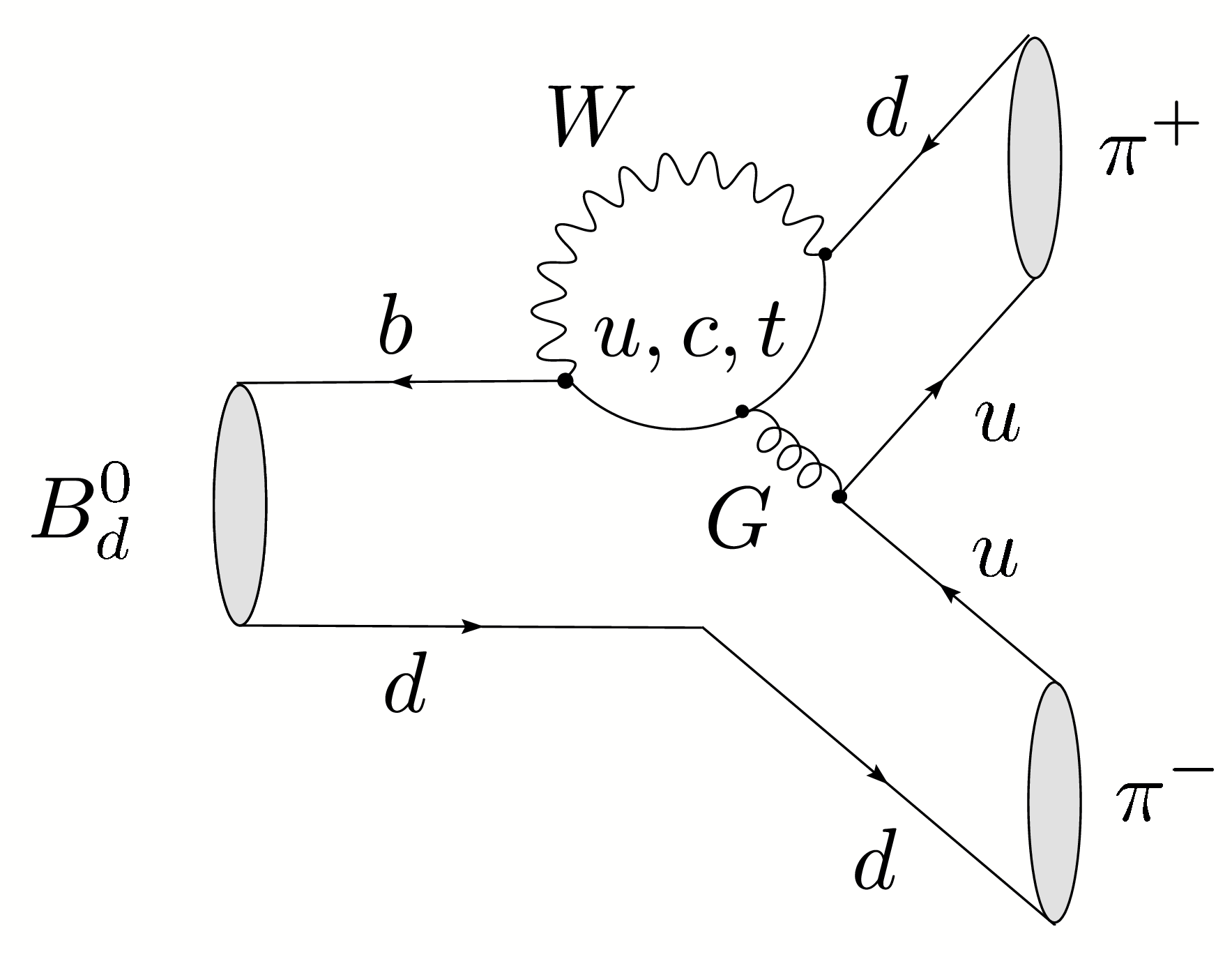}  
 }
 \vspace*{-0.3truecm}
\caption{Feynman diagrams contributing to $B^0_d\to \pi^+\pi^-$ 
decays.}\label{fig:Bpipi-diag}
\end{figure}

\unboldmath
This decay is a transition into a CP eigenstate with eigenvalue $+1$, and 
originates from $\bar b\to\bar u u \bar d$ processes, as can be seen in 
Fig.~\ref{fig:Bpipi-diag}. In analogy to (\ref{Bd-ampl1}) and (\ref{B0phiK0-ampl}), 
its decay amplitude can be written as follows \cite{RF-BsKK}:
\begin{equation}
A(B_d^0\to\pi^+\pi^-)=
\lambda_u^{(d)}\left(A_{\rm T}^{u}+
A_{\rm P}^{u}\right)+\lambda_c^{(d)}A_{\rm P}^{c}+
\lambda_t^{(d)}A_{\rm P}^{t}.
\end{equation}
Using again (\ref{CKM-UT-Rel}) to eliminate the CKM factor 
$\lambda_t^{(d)}=V_{td}V_{tb}^\ast$ and applying once more the 
Wolfenstein parametrization yields
\begin{equation}\label{Bpipi-ampl}
A(B_d^0\to\pi^+\pi^-)={\cal C}\left[e^{i\gamma}-de^{i\theta}\right],
\end{equation}
where the overall normalization ${\cal C}$ and
\begin{equation}\label{D-DEF}
d e^{i\theta}\equiv\frac{1}{R_b}
\left[\frac{A_{\rm P}^{c}-A_{\rm P}^{t}}{A_{\rm T}^{u}+
A_{\rm P}^{u}-A_{\rm P}^{t}}\right]
\end{equation}
are hadronic parameters. 
The formalism discussed in Subsection~\ref{subsec:CPasym} then implies 
\begin{equation}\label{xi-Bdpipi}
\xi_{\pi^+\pi^-}^{(d)}=-e^{-i\phi_d}\left[\frac{e^{-i\gamma}-
d e^{i\theta}}{e^{+i\gamma}-d e^{i\theta}}\right].
\end{equation}
In contrast to the expressions (\ref{xi-BdpsiKS}) and (\ref{xi-phiKS}) 
for the $B_d^0\to J/\psi K_{\rm S}$ and $B^0_d\to\phi K_{\rm S}$ counterparts,
respectively, the hadronic parameter $d e^{i\theta}$, which suffers from large 
theoretical uncertainties, does {\it not} enter  (\ref{xi-Bdpipi}) 
in a doubly Cabibbo-suppressed way. This feature is at the basis of the
famous ``penguin problem'' in $B^0_d\to\pi^+\pi^-$, which was addressed
in many papers (see, for instance, \cite{GL}--\cite{GLSS}). If the penguin 
contributions to this channel were negligible, i.e.\ $d=0$, its CP asymmetries 
were simply given by
\begin{eqnarray}
{\cal A}_{\rm CP}^{\rm dir}(B_d\to\pi^+\pi^-)&=&0 \\
{\cal A}_{\rm CP}^{\rm mix}(B_d\to\pi^+\pi^-)&=&\sin(\phi_d+2\gamma)
\stackrel{\rm SM}{=}\sin(\underbrace{2\beta+2\gamma}_{2\pi-2\alpha})
=-\sin 2\alpha.
\end{eqnarray}
Consequently, ${\cal A}_{\rm CP}^{\rm mix}(B_d\to\pi^+\pi^-)$ would then
allow us to determine $\alpha$. However, in the general case, we obtain 
expressions with the help of (\ref{CPV-OBS}) and (\ref{xi-Bdpipi}) of the form
\begin{eqnarray}
{\cal A}_{\rm CP}^{\rm dir}(B_d\to \pi^+\pi^-)&=&
G_1(d,\theta;\gamma) \label{CP-Bpipi-dir-gen}\\
{\cal A}_{\rm CP}^{\rm mix}(B_d\to \pi^+\pi^-)&=&
G_2(d,\theta;\gamma,\phi_d);\label{CP-Bpipi-mix-gen}
\end{eqnarray}
for explicit formulae, see Ref.~\cite{RF-BsKK}. We observe that actually the 
phases $\phi_d$ and $\gamma$ enter directly in the $B_d\to\pi^+\pi^-$ 
observables, and not $\alpha$. Consequently, since $\phi_d$ can be fixed 
through the mixing-induced CP violation in the ``golden'' mode 
$B_d\to J/\psi K_{\rm S}$, as we have seen in Subsection~\ref{ssec:BpsiK},
we may use $B_d\to\pi^+\pi^-$ to probe $\gamma$.

The current measurements of the $B_d\to\pi^+\pi^-$ CP asymmetries 
are given as follows:
\begin{eqnarray}
{\cal A}_{\rm CP}^{\rm dir}(B_d\to\pi^+\pi^-)
&=&\left\{
\begin{array}{ll}
-0.09\pm0.15\pm0.04 & \mbox{(BaBar \cite{BaBar-Bpipi-05})}\\
-0.56\pm0.12\pm0.06 & \mbox{(Belle \cite{Belle-Bpipi-05})}
\end{array}
\right.\label{Adir-exp}\\
{\cal A}_{\rm CP}^{\rm mix}(B_d\to\pi^+\pi^-)
&=&\left\{
\begin{array}{ll}
+0.30\pm0.17\pm0.03& \mbox{(BaBar \cite{BaBar-Bpipi-05})}\\
+0.67\pm0.16\pm0.06 & \mbox{(Belle \cite{Belle-Bpipi-05}).}
\end{array}
\right.\label{Amix-exp}
\end{eqnarray}
The BaBar and Belle results are still not fully consistent with each other, 
although the experiments are now in better agreement. In Ref.~\cite{HFAG}, 
the following averages were obtained:
\begin{eqnarray}
{\cal A}_{\rm CP}^{\rm dir}(B_d\to\pi^+\pi^-)&=&-0.37\pm0.10
\label{Bpipi-CP-averages}\\
{\cal A}_{\rm CP}^{\rm mix}(B_d\to\pi^+\pi^-)&=&+0.50\pm0.12.
\label{Bpipi-CP-averages2}
\end{eqnarray}
The central values of these averages are remarkably stable
in time. Direct CP violation at this level would require large penguin 
contributions with large CP-conserving strong phases, thereby indicating
large non-factorizable effects. 

This picture is in fact supported by the direct CP violation in $B^0_d\to\pi^-K^+$ 
modes that could be established by the $B$ factories in 
the summer of 2004 \cite{CP-B-dir}. Here the BaBar and Belle results agree 
nicely with each other, yielding the following average \cite{HFAG}:
\begin{equation}\label{AdirBdpimKp-exp}
{\cal A}_{\rm CP}^{\rm dir}(B_d\to\pi^\mp K^\pm)=0.115\pm 0.018.
\end{equation}
The diagrams contributing to $B^0_d\to\pi^-K^+$ can straightforwardly be 
obtained from those in Fig.~\ref{fig:Bpipi-diag} by just replacing the anti-down quark
emerging from the $W$ boson through an anti-strange quark. Consequently, the 
hadronic matrix elements entering $B^0_d\to\pi^+\pi^-$ and $B^0_d\to\pi^-K^+$ can be 
related to one another through the $SU(3)$ flavour symmetry of strong interactions
and the additional assumption that the penguin annihilation and exchange topologies
contributing to $B^0_d\to\pi^+\pi^-$, which have no counterpart in 
$B^0_d\to\pi^-K^+$ and involve the ``spectator" down quark in 
Fig.~\ref{fig:Bpipi-diag}, play actually a negligible r\^ole \cite{RF-Bpipi}. Following 
these lines, we obtain the following relation in the SM:
\begin{equation}\label{H-rel}
\hspace*{-1.7truecm}
H_{\rm BR}\equiv\underbrace{\frac{1}{\epsilon}
\left(\frac{f_K}{f_\pi}\right)^2\left[\frac{\mbox{BR}
(B_d\to\pi^+\pi^-)}{\mbox{BR}(B_d\to\pi^\mp K^\pm)}
\right]}_{\mbox{$7.5\pm 0.7$}} =
\underbrace{-\frac{1}{\epsilon}\left[\frac{{\cal A}_{\rm CP}^{\rm dir}(B_d\to\pi^\mp 
K^\pm)}{{\cal A}_{\rm CP}^{\rm dir}(B_d\to\pi^+\pi^-)}
\right]}_{\mbox{$6.7\pm 2.0$}} \equiv H_{{\cal A}_{\rm CP}^{\rm dir}},
\end{equation}
where 
\begin{equation}\label{eps-def}
\epsilon\equiv\frac{\lambda^2}{1-\lambda^2}=0.053, 
\end{equation}
and the ratio $f_K/f_\pi=160/131$ of the kaon and pion decay constants
defined through
\begin{equation}\label{decay-const-def}
\langle 0|\bar s \gamma_\alpha\gamma_5 u|K^+(k)\rangle=
i f_K k_\alpha, \quad
\langle 0|\bar d \gamma_\alpha\gamma_5 u|\pi^+(k)\rangle=
i f_\pi k_\alpha
\end{equation}
describes
factorizable $SU(3)$-breaking corrections. As usual, the CP-averaged 
branching ratios are defined as
\begin{equation}
\mbox{BR}\equiv\frac{1}{2}\left[\mbox{BR}(B\to f)+
\mbox{BR}(\bar B\to \bar f)\right].
\end{equation}
In (\ref{H-rel}), we have also given the
numerical values following from the data. Consequently, this relation 
is well satisfied within the experimental uncertainties, and does not
show any anomalous behaviour. It supports therefore the SM description
of the $B^0_d\to\pi^-K^+$, $B^0_d\to\pi^+\pi^-$ decay amplitudes,
and our working assumptions listed before (\ref{H-rel}). 

The quantities $H_{\rm BR}$ and $H_{{\cal A}_{\rm CP}^{\rm dir}}$ introduced
in this relation can be written as follows:
\begin{equation}\label{H-fct}
H_{\rm BR} = G_3(d,\theta;\gamma) =
H_{{\cal A}_{\rm CP}^{\rm dir}}.
\end{equation}
If we complement this expression with (\ref{CP-Bpipi-dir-gen}) and
(\ref{CP-Bpipi-mix-gen}), and use (see (\ref{phid-exp}))
\begin{equation}\label{phi-d-det}
\phi_d=(43.4\pm2.5)^\circ,
\end{equation}
we have sufficient information to determine $\gamma$, 
as well as $(d,\theta)$  \cite{RF-BsKK,RF-Bpipi,FleischerMatias}. In using
(\ref{phi-d-det}), we assume that the possible discrepancy with the SM
described by (\ref{S-psi-K}) is only due to NP in $B^0_d$--$\bar B^0_d$ mixing
and not to effects entering through the $B^0_d\to J/\psi K_{\rm S}$ decay amplitude.
As was recently shown in Ref.~\cite{BFRS-5}, the results following from $H_{\rm BR}$ 
and $H_{{\cal A}_{\rm CP}^{\rm dir}}$ give results that are in good agreement with 
one another. Since the avenue offered by $H_{{\cal A}_{\rm CP}^{\rm dir}}$
is cleaner than the one provided by $H_{\rm BR}$, it is preferable to use the former
quantity to determine $\gamma$, yielding the following result \cite{BFRS-5}:
\begin{equation}\label{gamma-det}
\gamma=(73.9^{+5.8}_{-6.5})^\circ.
\end{equation}
Here a second solution around $42^\circ$ was discarded, which can be exclueded 
through an analysis of the whole $B\to\pi\pi,\pi K$ system \cite{BFRS3}. As was recently 
discussed  \cite{BFRS-5} (see also Refs.~\cite{RF-Bpipi,FleischerMatias}), even large 
non-factorizable $SU(3)$-breaking corrections have a remarkably small impact 
on the numerical result in (\ref{gamma-det}). The value of $\gamma$ in 
(\ref{gamma-det}) is somewhat higher than the central values in (\ref{gam-DK}),
but fully consistent within the large errors.  An even larger value in the ballpark of 
$80^\circ$ was recently extracted from the $B\to\pi\pi$ data with the help of SCET 
\cite{gam-SCET,SCET-Bdpi0K0}.

\section{THE \boldmath$B\to\pi K$\unboldmath~PUZZLE AND ITS RELATION 
TO RARE \boldmath$B$\unboldmath~AND
\boldmath$K$\unboldmath~DECAYS}\label{sec:BpiK-puzzle}
\setcounter{equation}{0}
\subsection{Preliminaries}\label{ssec:BpiK-prel}
\begin{figure}
   \centerline{
   \begin{tabular}{lc}
     {\small(a)} & \\
    &  \includegraphics[width=5.2truecm]{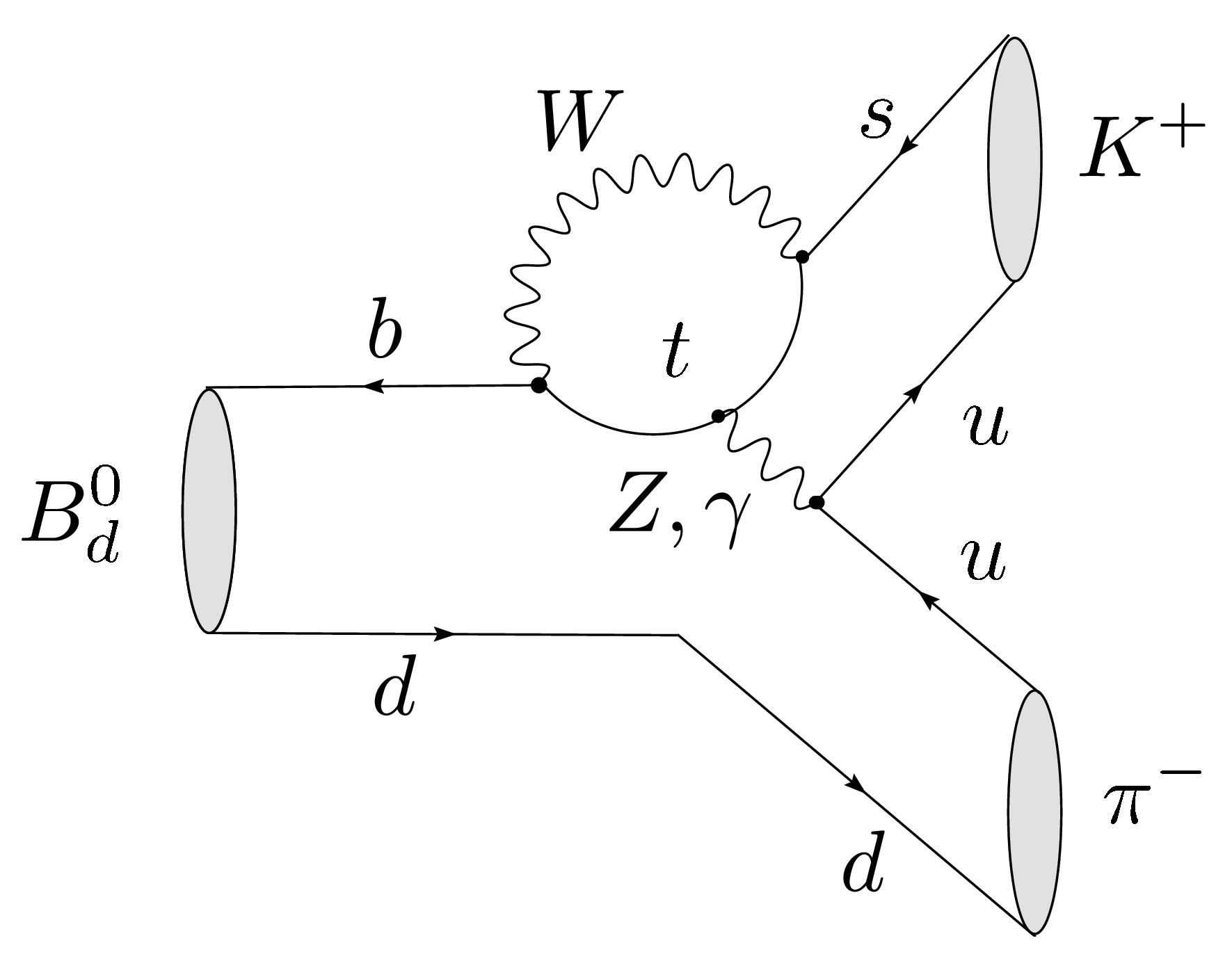}
     \hspace*{0.5truecm}
    \includegraphics[width=5.2truecm]{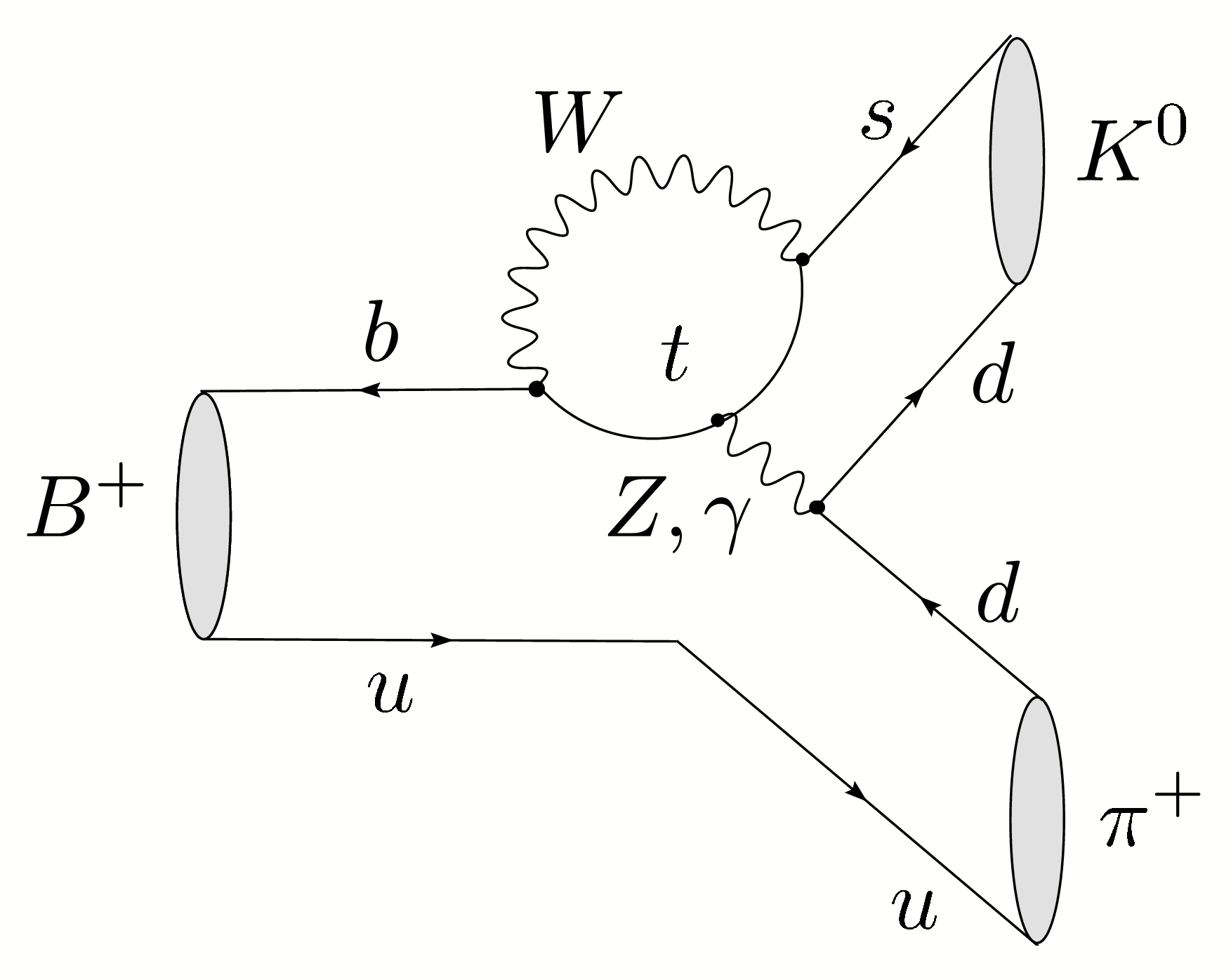} \\
     {\small(b)} & \\
    &     \includegraphics[width=5.2truecm]{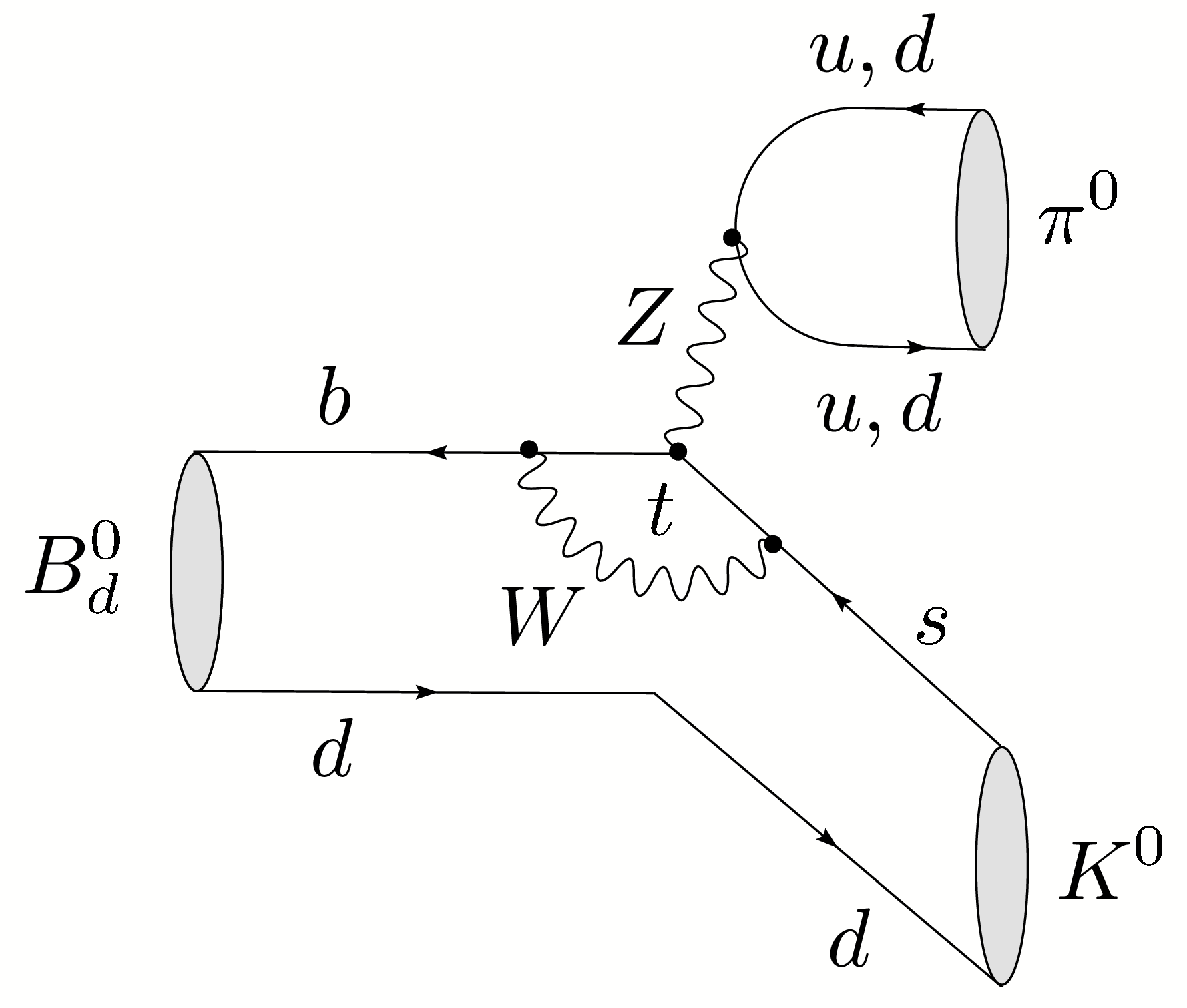} 
     \hspace*{0.5truecm}
    \includegraphics[width=5.2truecm]{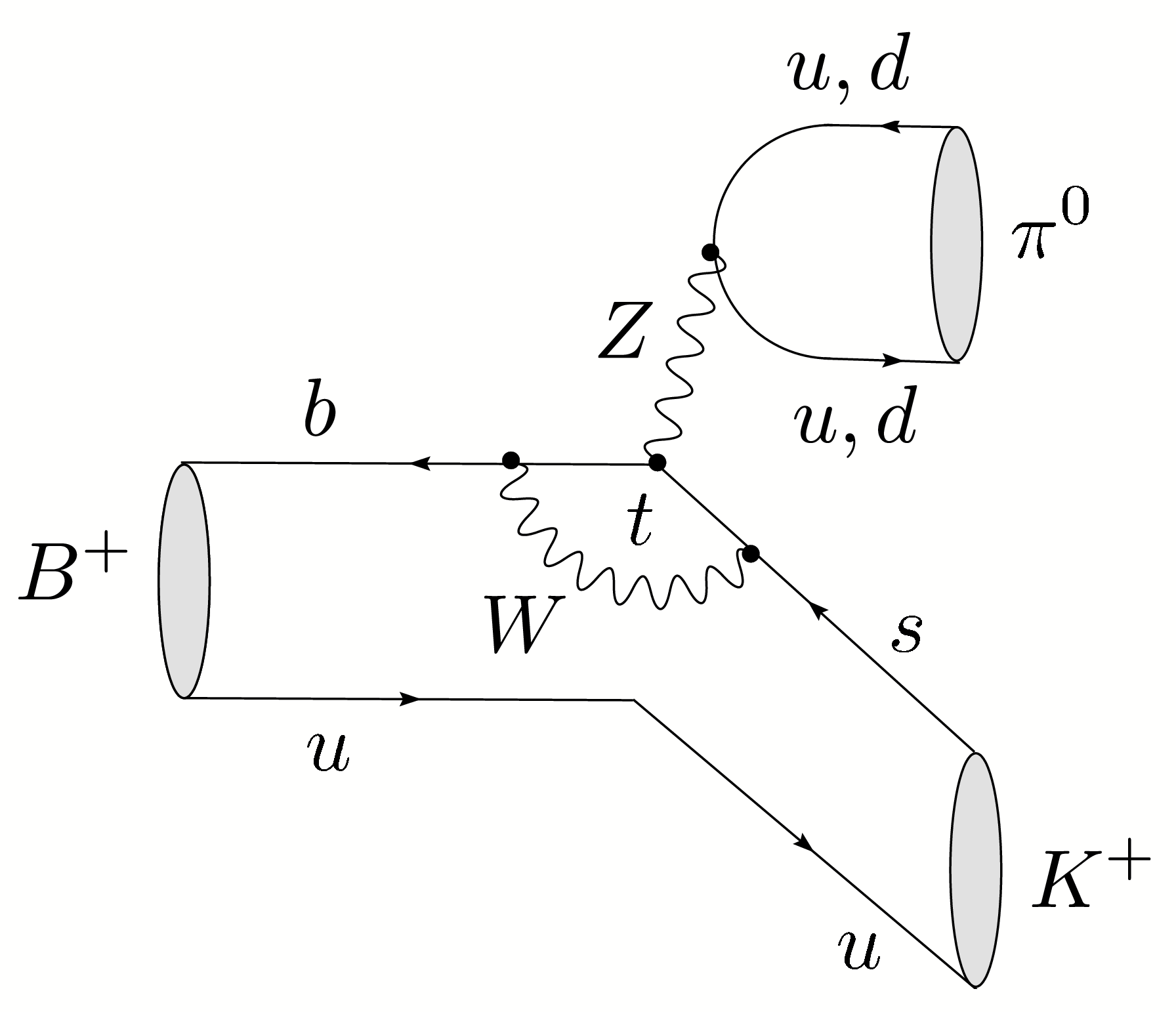} 
     \end{tabular}}
     \caption{Examples of the colour-suppressed (a) and colour-allowed (b) 
     EW penguin contributions to the $B\to\pi K$ system.}\label{fig:BpiK-EWP}
\end{figure}

We made already first contact with a $B\to\pi K$ decay in 
Subsection~\ref{ssec:Bpi+pi-}, the $B^0_d\to\pi^-K^+$ channel. It receives
contributions both from tree and from penguin topologies. Since this decay
originates from a $\bar b\to\bar s$ transition, the tree amplitude is suppressed
by a CKM factor $\lambda^2 R_b\sim 0.02$ with respect to the penguin
amplitude. Consequently, $B^0_d\to\pi^-K^+$ is governed by QCD penguins;
the tree topologies contribute only at the 20\% level to the decay amplitude.
The feature of the dominance of QCD penguins applies to all $B\to\pi K$ modes, 
which can be classified with respect to their EW penguin contributions 
as follows (see Fig.~\ref{fig:BpiK-EWP}):
\begin{itemize}
\item[(a)] In the $B^0_d\to\pi^-K^+$ and $B^+\to\pi^+K^0$ decays, EW penguins 
contribute in colour-suppressed form and are hence expected to play a minor r\^ole.
\item[(b)] In the $B^0_d\to\pi^0K^0$ and $B^+\to\pi^0K^+$ decays, EW penguins 
contribute in colour-allowed form and have therefore a significant impact on the decay 
amplitude, entering at the same order of magnitude as the tree contributions.
\end{itemize}
As we noted above, EW penguins offer an attractive avenue for NP to 
enter non-leptonic $B$ decays, which is also the case for the
$B\to\pi K$ system \cite{FM-BpiK-NP,trojan}. Indeed, the decays of class (b) 
show a puzzling pattern, which may point towards such a NP scenario.
This feature emerged already in 2000 \cite{BF00}, when the CLEO collaboration 
reported the observation of the $B^0_d\to\pi^0K^0$ channel with a surprisingly 
prominent rate \cite{CLEO00}, and is still present in the most recent BaBar and 
Belle data, thereby receiving a lot of attention in the literature (see, for instance, 
Refs.~\cite{Z-prime-BpiK} and \cite{BeNe}--\cite{WZ}).

In the following discussion, we focus on the systematic
strategy to explore the ``$B\to\pi K$ puzzle"  developed in Refs.~\cite{BFRS2,BFRS3}; 
all numerical results refer to the most recent analysis presented in 
Ref.~\cite{BFRS-5}. The logical structure is very simple: the starting point is
given by the values of $\phi_d$ and $\gamma$ in (\ref{phi-d-det})
and (\ref{gamma-det}), respectively, and by the $B\to\pi\pi$ system, 
which allows us to extract a set of hadronic parameters from the data
with the help of the isospin symmetry of strong interactions. Then we make, in
analogy to the determination of $\gamma$ in Subsection~\ref{ssec:Bpi+pi-}, 
the following working hypotheses:
\begin{itemize}
\item[(i)] $SU(3)$ flavour symmetry of strong interactions (but taking factorizable 
$SU(3)$-breaking corrections into account),
\item[(ii)] neglect of penguin annihilation and exchange topologies,
\end{itemize}
which allow us to fix the hadronic $B\to\pi K$ parameters through their $B\to\pi\pi$
counterparts. Interestingly, we may gain confidence in these assumptions through 
internal consistency checks (an example is relation (\ref{H-rel})), which work nicely
within the experimental uncertainties. Having the hadronic $B\to\pi K$ parameters
at hand, we can predict the $B\to \pi K$ observables in the SM. The comparison
of the corresponding picture with the $B$-factory data will then guide us to NP
in the EW penguin sector, involving in particular a large CP-violating NP phase. In the
final step, we explore the interplay of this NP scenario with rare $K$ and $B$
decays.

\boldmath
\subsection{Extracting Hadronic Parameters from the $B\to\pi\pi$ 
System}\label{ssec:Bpipi-hadr}
\unboldmath
In order to fully exploit the information that is provided by the whole $B\to\pi\pi$ 
system, we use -- in addition to the two CP-violating $B^0_d\to\pi^+\pi^-$
observables  -- the following ratios of CP-averaged branching ratios:
\begin{eqnarray}
R_{+-}^{\pi\pi}&\equiv&2\left[\frac{\mbox{BR}(B^+\to\pi^+\pi^0)
+\mbox{BR}(B^-\to\pi^-\pi^0)}{\mbox{BR}(B_d^0\to\pi^+\pi^-)
+\mbox{BR}(\bar B_d^0\to\pi^+\pi^-)}\right]
=2.04\pm0.28
\label{Rpm-def}\\
R_{00}^{\pi\pi}&\equiv&2\left[\frac{\mbox{BR}(B_d^0\to\pi^0\pi^0)+
\mbox{BR}(\bar B_d^0\to\pi^0\pi^0)}{\mbox{BR}(B_d^0\to\pi^+\pi^-)+
\mbox{BR}(\bar B_d^0\to\pi^+\pi^-)}\right]
=0.58\pm0.13.
\end{eqnarray}
The pattern of the experimental numbers in these expressions came as quite 
a surprise, as the central values calculated in QCDF gave 
$R_{+-}^{\pi\pi}=1.24$ and $R_{00}^{\pi\pi}=0.07$ \cite{BeNe}. As discussed in 
detail in Ref.~\cite{BFRS3}, this ``$B\to\pi\pi$ puzzle" can straightforwardly be 
accommodated 
in the SM through large non-factorizable hadronic interference effects, i.e.\ 
does not point towards NP. For recent SCET analyses, 
see Refs.~\cite{SCET-Bdpi0K0,BPRS,FeHu}. 

Using the isospin symmetry of strong interactions, we can write
\begin{equation}\label{Rpipi-gen}
R_{+-}^{\pi\pi}=F_1(d,\theta,x,\Delta;\gamma), \quad
R_{00}^{\pi\pi}=F_2(d,\theta,x,\Delta;\gamma),
\end{equation}
where $xe^{i\Delta}$ is another hadronic parameter, which was introduced
in Refs.~\cite{BFRS2,BFRS3}. Using now, in addition, the CP-violating observables in
(\ref{CP-Bpipi-dir-gen}) and (\ref{CP-Bpipi-mix-gen}), we arrive at the following 
set of haronic parameters:
\begin{equation}\label{Bpipi-par-det}
d=0.52^{+0.09}_{-0.09}, \quad
\theta=(146^{+7.0}_{-7.2})^\circ, \quad
x=0.96^{+0.13}_{-0.14}, \quad
\Delta=-(53^{+18}_{-26})^\circ.
\end{equation}
In the extraction of these quantites, also the EW penguin effects in the 
$B\to\pi\pi$ system are included \cite{BF98,GPY}, although these topologies have a 
tiny impact \cite{PAPIII}. Let us emphasize that the results for the hadronic 
parameters listed above, which are consistent with the picture emerging in the 
analyses of other authors (see, e.g., Refs.~\cite{ALP-Bpipi,CGRS}), 
are essentially clean and serve as a testing ground for 
calculations within QCD-related approaches. For instance, in 
recent QCDF \cite{busa} and PQCD \cite{kesa} analyses, the 
following numbers were obtained:
\begin{equation}
\left.d\right|_{\rm QCDF}=0.29\pm0.09, \quad
\left.\theta\right|_{\rm QCDF}=-\left(171.4\pm14.3\right)^\circ, 
\end{equation}
\begin{equation}
\left.d\right|_{\rm PQCD}=0.23^{+0.07}_{-0.05}, \quad
+139^\circ < \left.\theta\right|_{\rm PQCD} < +148^\circ,
\end{equation}
which depart significantly from the pattern in (\ref{Bpipi-par-det}) that is implied
by the data. 

Finally, we can predict the CP asymmetries of the decay $B_d\to\pi^0\pi^0$:
\begin{equation}\label{ACP-Bdpi0pi0-pred}
{\cal A}_{\rm CP}^{\rm dir}(B_d\to \pi^0\pi^0)=-0.30^{+0.48}_{-0.26}, \quad
{\cal A}_{\rm CP}^{\rm mix}(B_d\to \pi^0\pi^0)=-0.87^{+0.29}_{-0.19}.
\end{equation}
The current experimental value for the direct CP 
asymmetry is given as follows \cite{HFAG}:
\begin{equation}\label{ACP-Bdpi0pi0-exp}
{\cal A}_{\rm CP}^{\rm dir}(B_d\to \pi^0\pi^0)=-0.28^{+0.40}_{-0.39}.
\end{equation}
Consequently, no stringent test of the corresponding prediction 
in (\ref{ACP-Bdpi0pi0-pred}) is provided at this stage, although the 
indicated agreement is encouraging.

\boldmath
\subsection{Analysis of the $B\to\pi K$ System}\label{ssec:BpiK}
\unboldmath
Let us begin the analysis of the $B\to\pi K$ system by having a closer
look at the modes of class (a) introduced above, $B_d\to\pi^\mp K^\pm$
and $B^\pm\to\pi^\pm K$, which are only marginally affected by
EW penguin contributions. We used the banching ratio and direct 
CP asymmetry of the former channel already in the $SU(3)$ relation (\ref{H-rel}),
which is nicely satisfied by the current data, and in the extraction of
$\gamma$ with the help of the CP-violating $B_d\to\pi^+\pi^-$ observables,
yielding the value in  (\ref{gamma-det}). The $B_d\to\pi^\mp K^\pm$
modes provide the CP-violating asymmetry
\begin{equation}\label{ACP-BppipK0}
\hspace*{-1.9truecm}{\cal A}_{\rm CP}^{\rm dir}(B^\pm\to\pi^\pm K)\equiv
\frac{\mbox{BR}(B^+\to\pi^+K^0)-
\mbox{BR}(B^-\to\pi^-\bar K^0)}{\mbox{BR}(B^+\to\pi^+K^0)+
\mbox{BR}(B^-\to\pi^-\bar K^0)} =
0.02 \pm 0.04,
\end{equation}
and enter in the following ratio \cite{FM}:
\begin{equation}\label{R-def}
\hspace*{-0.7truecm}R\equiv\left[\frac{\mbox{BR}(B_d^0\to\pi^- K^+)+
\mbox{BR}(\bar B_d^0\to\pi^+ K^-)}{\mbox{BR}(B^+\to\pi^+ K^0)+
\mbox{BR}(B^-\to\pi^- \bar K^0)}
\right]\frac{\tau_{B^+}}{\tau_{B^0_d}} =
0.86\pm0.06;
\end{equation}
the numerical values refer again to the most recent compilation in \cite{HFAG}. 
The $B^+\to\pi^+ K^0$ channel involves another hadronic parameter,
$\rho_{\rm c}e^{i\theta_{\rm c}}$, which cannot be determined through
the $B\to\pi\pi$ data \cite{BF98,defan,neubert}:
\begin{equation}\label{B+pi+K0}
A(B^+\to\pi^+K^0)=-P'\left[1+\rho_{\rm c}e^{i\theta_{\rm c}}e^{i\gamma}
\right];
\end{equation}
the overall normalization $P'$ cancels in (\ref{ACP-BppipK0}) and
(\ref{R-def}). Usually, it is assumed that the parameter $\rho_{\rm c}e^{i\theta_{\rm c}}$  
can be neglected. In this case, the direct CP asymmetry in (\ref{ACP-BppipK0}) 
vanishes, and $R$ can be calculated through the $B\to\pi\pi$ data with the help 
of the assumptions specified in Subsection~\ref{ssec:BpiK-prel}:
\begin{equation}\label{R-pred-0}
R|_{\rm SM}=0.963^{+0.019}_{-0.022}.
\end{equation}

This numerical result is $1.6 \sigma$ larger than the experimental value
in (\ref{R-def}). As was discussed in detail in Ref.~\cite{BFRS-up}, 
the experimental range for the direct CP asymmetry in (\ref{ACP-BppipK0})  
and the first direct signals for the $B^\pm\to K^\pm K$ decays favour a 
value of $\theta_{\rm c}$ around $0^\circ$. This feature allows us to essentially 
resolve the small discrepancy concerning $R$ for values of $\rho_{\rm c}$ around 
0.05. The remaining small numerical difference between the calculated value of
$R$ and the experimental result, if confirmed by future data, could be due to
(small) colour-suppressed EW penguins, which enter $R$ as well \cite{BFRS3}.
As was recently discussed in Ref.~\cite{BFRS-5}, even large non-factorizable
$SU(3)$-breaking effects would have a small impact on the predicted value 
of $R$. In view of these results, it would not be a surprise to see an increase 
of the experimental value of $R$ in the future.

\begin{figure}
\begin{center}
\includegraphics[width=10cm]{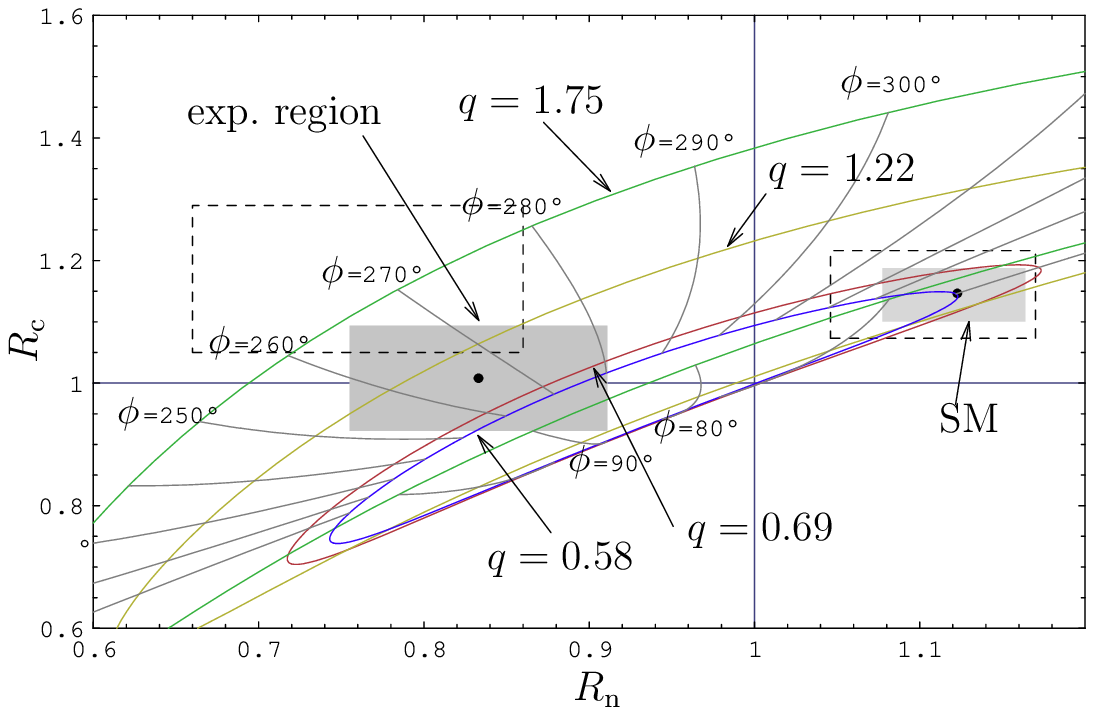}
\end{center}
\vspace*{-0.5truecm}
\caption{The current situation in the $R_{\rm n}$--$R_{\rm c}$ plane: the shaded 
areas indicate the experimental and SM $1 \sigma$ ranges, while the lines show the
theory predictions for the central values of the hadronic parameters
and various values of $q$ with $\phi\in[0^\circ,360^\circ]$.}\label{fig:RnRc}
\end{figure}

Let us now turn to the $B^+\to\pi^0K^+$ and $B^0_d\to\pi^0K^0$ channels,
which are the $B\to\pi K$ modes with significant contributions from EW
penguin topologies. The key observables for the exploration of these modes 
are the following ratios of their CP-averaged branching ratios \cite{BF00,BF98}:
\begin{equation}\label{Rc-def}
R_{\rm c}\equiv2\left[\frac{\mbox{BR}(B^+\to\pi^0K^+)+
\mbox{BR}(B^-\to\pi^0K^-)}{\mbox{BR}(B^+\to\pi^+ K^0)+
\mbox{BR}(B^-\to\pi^- \bar K^0)}\right] =
1.01\pm0.09
\end{equation}
\begin{equation}\label{Rn-def}
R_{\rm n}\equiv\frac{1}{2}\left[
\frac{\mbox{BR}(B_d^0\to\pi^- K^+)+
\mbox{BR}(\bar B_d^0\to\pi^+ K^-)}{\mbox{BR}(B_d^0\to\pi^0K^0)+
\mbox{BR}(\bar B_d^0\to\pi^0\bar K^0)}\right] =
0.83\pm0.08,
\end{equation}
where the overall normalization factors of the decay amplitudes cancel, 
as in (\ref{R-def}). In order to describe the EW penguin effects, both a parameter 
$q$, which measures the strength of the EW penguins with respect to 
tree-like topologies, and a CP-violating phase $\phi$ are introduced. In the SM, this 
phase vanishes, and $q$ can be calculated with the help of the $SU(3)$ flavour 
symmetry, yielding a value of $0.69 \times 0.086/|V_{ub}/V_{cb}|= 0.58$ \cite{NR}.
Following the strategy described above yields the following SM predictions: 
\begin{equation}\label{RncSM}
R_{\rm c}|_{\rm SM}=1.15 \pm 0.05, \quad R_{\rm n}|_{\rm SM}=1.12 \pm 0.05,
\end{equation}
where in particular the value of $R_{\rm n}$ does not agree with the experimental
number, which is a manifestation of the $B\to\pi K$ puzzle. As was recently
discussed in Ref.~\cite{BFRS-5}, the internal consistency checks of the
working assumptions listed in Subsection~\ref{ssec:BpiK-prel} are
currently satisfied at the level of $25\%$, and can be systematically improved 
through better data. A detailed study of the numerical predictions in 
(\ref{RncSM}) (and those given below) shows that their sensitivity on 
non-factorizable $SU(3)$-breaking effects of this order of magnitude is
surprisingly small. Consequently, it is very exciting to speculate that NP
effects in the EW penguin sector, which are described effectively through
$(q,\phi)$, are at the origin of the $B\to\pi K$ puzzle. 
Following Refs.~\cite{BFRS2,BFRS3}, we show the situation in the 
$R_{\rm n}$--$R_{\rm c}$ 
plane in Fig.~\ref{fig:RnRc}, where -- for the convenience of the reader --
also the experimental range and the SM predictions at the time of the 
original analysis of Refs.~\cite{BFRS2,BFRS3} are indicated through the dashed 
rectangles. 
We observe that although
the central values of $R_{\rm n}$ and $R_{\rm c}$ have slightly moved towards
each other, the puzzle is as prominent as ever. The experimental
region can now be reached without an enhancement of $q$, but
a large CP-violating phase $\phi$ of the order of $-90^\circ$ is
still required:
\begin{equation}
\label{q-phi}
q=0.99\,^{+0.66}_{-0.70} ,\quad \phi=-(94\,^{+16}_{-17} )^\circ.
\end{equation}
Interestingly, $\phi$ of the order of $+90^\circ$ can now also bring us rather 
close to the experimental range of $R_{\rm n}$ and $R_{\rm c}$. 

An interesting probe of the NP phase $\phi$ is also provided
by the CP violation in  $B^0_d\to\pi^0 K_{\rm S}$. Within the SM,
the corresponding observables are expected to satisfy the following 
relations \cite{PAPIII}:
\begin{equation}\label{Bdpi0K0-rel}
{\cal A}_{\rm CP}^{\rm dir}(B_d\!\to\!\pi^0 K_{\rm S})\approx 0, \quad
{\cal A}_{\rm CP}^{\rm mix}(B_d\!\to\!\pi^0 K_{\rm S})\approx
{\cal A}_{\rm CP}^{\rm mix}(B_d\!\to\!\psi K_{\rm S}).
\end{equation}
The most recent  Belle \cite{Belle-Bphi-K} and BaBar \cite{BaBar-pi0KS}
measurements of these quantities are in agreement with each other, and 
lead to the following averages \cite{HFAG}:
\begin{eqnarray}
{\cal A}_{\rm CP}^{\rm dir}(B_d\!\to\!\pi^0 K_{\rm S})&=&-0.02\pm0.13\\
{\cal A}_{\rm CP}^{\rm mix}(B_d\!\to\!\pi^0 K_{\rm S})&=&-0.31\pm0.26
\equiv -(\sin2\beta)_{\pi^0K_{\rm S}}.
\end{eqnarray}
Taking (\ref{s2b-average}) into account yields
\begin{equation}\label{DS}
\Delta S \equiv (\sin2\beta)_{\pi^0K_{\rm S}}-
(\sin2\beta)_{\psi K_{\rm S}} =
-0.38\pm 0.26,
\end{equation}
which may indicate a sizeable deviation of the
experimentally measured value of $(\sin2\beta)_{\pi^0K_{\rm S}}$ from
$(\sin2\beta)_{\psi K_{\rm S}}$, and is therefore one of the recent hot topics. 
Since the strategy developed in Refs.~\cite{BFRS2,BFRS3} allows us also to predict the 
CP-violating observables of the $B^0_d\to\pi^0 K_{\rm S}$ channel both within the 
SM and within our scenario of NP, it allows us to address this issue, yielding
\begin{equation}\label{pi0KS-SM}
{\cal A}_{\rm CP}^{\rm dir}(B_d\!\to\!\pi^0 K_{\rm S})|_{\rm SM}=0.06^{+0.09}_{-0.10},
\qquad
\Delta S\vert_{\rm SM}= 0.13\pm0.05, 
\end{equation}
\begin{equation}\label{pi0KS-NP}
{\cal A}_{\rm CP}^{\rm dir}(B_d\!\to\!\pi^0 K_{\rm S})|_{\rm NP}=0.01\,^{+0.14}_{-0.18}, 
\qquad \Delta S\vert_{\rm NP}=0.27\,^{+0.05}_{-0.09},
\end{equation}
where the NP results refer to the EW penguin parameters in (\ref{q-phi}). Consequently,
$\Delta S$ is found to be {\it positive} in the SM. In the literature, values of
$\Delta S\vert_{\rm SM}\sim0.04$--$0.08$ can be found, which were obtained
-- in contrast to (\ref{pi0KS-SM})  -- with the help of dynamical approaches such as QCDF \cite{beneke} and SCET \cite{SCET-Bdpi0K0}. Moreover, 
bounds were derived with the help of the $SU(3)$ flavour symmetry 
\cite{SU3-bounds}. Looking at (\ref{pi0KS-NP}), we see that the modified  
parameters $(q,\phi)$ in (\ref{q-phi}) imply an enhancement of $\Delta S$ with 
respect to the SM case. Consequently, the best values of $(q,\phi)$ that are 
favoured by the measurements of $R_{\rm n,c}$ make the potential 
${\cal A}_{\rm CP}^{\rm mix}(B_d\!\to\!\pi^0 K_{\rm S})$ discrepancy 
even larger than in the SM.

\begin{figure}
\begin{center}
\includegraphics[width=10cm]{AmixAdir-pi0K+0511.epsf}
\end{center}
\vspace*{-0.5truecm}
\caption{The situation in the 
${\cal A}_{\rm CP}^{\rm mix}(B_d\to\pi^0K_{\rm S})$--${\cal A}_{\rm CP}^{\rm dir}
(B^\pm\to\pi^0K^\pm)$ plane:  the shaded regions represent the experimental
and SM $1 \sigma$ ranges, while the lines show the
theory predictions for the central values of the hadronic parameters
and various values of $q$ with 
$\phi\in[0^\circ,360^\circ]$.\label{fig:Adirpi0KS-Amixpi0K+}}
\end{figure}

There is one CP asymmetry of the $B\to\pi K$ system left,  
which is measured as 
\begin{equation}\label{AdirBppi0KP-exp}
{\cal A}_{\rm CP}^{\rm dir}(B^\pm\to\pi^0K^\pm) =
-0.04\pm 0.04.
\end{equation}
In the limit of vanishing colour-suppressed tree and EW penguin topologies,
it is expected to be equal to the direct CP asymmetry of the $B_d\to\pi^\mp K^\pm$
modes. Since the experimental value of the latter asymmetry in 
(\ref{AdirBdpimKp-exp}) does not agree with (\ref{AdirBppi0KP-exp}), the 
direct CP violation in $B^\pm\to\pi^0K^\pm$ has also received
a lot of attention. The lifted colour suppression described by the large value of 
$x$ in (\ref{Bpipi-par-det}) could, in principle, be responsible for a non-vanishing
difference between (\ref{AdirBdpimKp-exp}) and (\ref{AdirBppi0KP-exp}),
\begin{equation}
\Delta A \equiv {\cal A}_{\rm CP}^{\rm dir}(B^\pm\to\pi^0K^\pm)
               -{\cal A}_{\rm CP}^{\rm dir}(B_d\to\pi^\mp K^\pm)\,\stackrel{{\rm exp}}{=} 
                -0.16\pm0.04.  \label{DeltaA}
\end{equation}
However, applying once again the strategy described above yields
\begin{equation}\label{AdirBppi0KP-SM}
{\cal A}_{\rm CP}^{\rm dir}(B^\pm\to\pi^0K^\pm)|_{\rm SM} 
= 0.04\,^{+0.09}_{-0.07},
\end{equation}
so that the SM still prefers a positive value of this CP asymmetry; 
the NP scenario characterized by (\ref{q-phi}) corresponds to 
\begin{equation}\label{AdirBppi0KP-NP}
{\cal A}_{\rm CP}^{\rm dir}(B^\pm\to\pi^0K^\pm)|_{\rm NP}
= 0.09\,^{+0.20}_{-0.16}.
\end{equation}

In view of the large uncertainties, no stringent test is provided at this point.
Nevertheless, it is tempting to play a bit with the CP asymmetries of the
$B^\pm\to\pi^0K^\pm$ and $B_d\to\pi^0K_{\rm S}$ decays. In 
Fig.~\ref{fig:Adirpi0KS-Amixpi0K+}, we show the situation in the 
${\cal A}_{\rm CP}^{\rm mix}(B_d\to\pi^0K_{\rm S})$--${\cal A}_{\rm CP}^{\rm dir}
(B^\pm\to\pi^0K^\pm)$ plane for various values of $q$ with $\phi\in[0^\circ,360^\circ]$.
We see that these observables seem to show a preference for positive values of 
$\phi$ around $+90^\circ$. As we noted above, in this case, we can also get 
rather close to the experimental region in the $R_{\rm n}$--$R_{\rm c}$ plane.
It is now interesting to return to the discussion of the NP effects in the
$B\to\phi K$ system given in Subsection~\ref{ssec:BphiK}. In our scenario of NP 
in the EW penguin sector, we have just to identify the CP-violating phase $\phi_0$ 
in (\ref{AphiK-NP}) with the NP phase $\phi$ \cite{BFRS3}. Unfortunately, we 
cannot determine the hadronic $B\to\phi K$ parameters $\tilde v_0$ and 
$\tilde\Delta_0$ through the $B\to\pi\pi$ data as in the case of the $B\to\pi K$ 
system. However, if we take 
into account that $\tilde\Delta_0=180^\circ$ in factorization and look at 
Fig.~\ref{fig:Plot-BphiK}, we see again that the case of $\phi\sim+90^\circ$ would 
be favoured by the data for ${\cal S}_{\phi K}$. Alternatively, in the case of 
$\phi\sim-90^\circ$, $\tilde\Delta_0\sim 0^\circ$ would be required to 
accommodate a negative value of ${\cal S}_{\phi K}$, which appears unlikely. Interestingly, a similar comment applies to the $B\to J/\psi K$ observables 
shown in Fig.~\ref{fig:Plot-BpsiK}, although here a dramatic enhancement of the 
EW penguin parameter $v_0$ relative to the SM estimate would be simultaneously
needed to reach the central experimental values, in contract to the reduction of 
$\tilde v_0$ in the $B\to\phi K$ case. In view of rare decay constraints, the behaviour 
of the $B\to \phi K$ parameter $\tilde v_0$ appears much more likely, 
thereby supporting the assumption after (\ref{phi-d-det}).

\boldmath
\subsection{The Interplay with Rare $K$ and $B$ Decays and
Future Scenarios}\label{ssec:rareKB}
\unboldmath
In order to explore the implications of the $B\to\pi K$ puzzle for rare 
$K$ and $B$ decays, we
assume that the NP enters the EW penguin sector through 
$Z^0$ penguins with a new CP-violating phase. This scenario was already
considered in the literature, where model-independent analyses and 
studies within SUSY can be found \cite{Z-pen-analyses,BuHi}.
In the strategy discussed here, the short-distance function $C$ characterizing 
the $Z^0$ penguins is determined through the $B\to\pi K$ data \cite{BFRS-I}. 
Performing a renormalization-group analysis yields
\begin{equation}\label{RG}
C(\bar q)= 2.35~ \bar q e^{i\phi} -0.82 \quad\mbox{with}\quad 
\bar q= q \left[\frac{|V_{ub}/V_{cb}|}{0.086}\right].
\end{equation}
Evaluating then the relevant box-diagram contributions in the SM 
and using (\ref{RG}), the short-distance functions
\begin{equation}\label{X-C-rel}
X=2.35~ \bar q e^{i\phi} -0.09 \quad \mbox{and} \quad 
Y=2.35~ \bar q e^{i\phi} -0.64
\end{equation}
can also be calculated, which govern the rare $K$, $B$ decays with $\nu\bar\nu$ 
and $\ell^+\ell^-$ in the final states, respectively. In the SM, we have 
$C=0.79$, $X=1.53$ and $Y=0.98$, with {\it vanishing} CP-violating phases. 
An analysis along these lines shows that the value of $(q,\phi)$ in (\ref{q-phi}), 
which is preferred by the $B\to\pi K$ observables $R_{\rm n,c}$, requires the 
following lower bounds for $X$ and $Y$ \cite{BFRS-5}:
\begin{equation}\label{XY1}
|X|_{\rm min}\approx 
|Y|_{\rm min}\approx 2.2,
\end{equation}
which appear to violate the $95\%$ probability upper bounds
\begin{equation}\label{XY2}
X\le 1.95, \quad Y\le 1.43
\end{equation}
that were recently obtained within the context of MFV \cite{Bobeth:2005ck}. 
Although we have to deal with CP-violating NP phases in our scenario,
which goes therefore beyond the MFV framework, a closer look at 
$B\to X_s \ell^+\ell^-$ shows that the upper 
bound on $|Y|$ in (\ref{XY2}) is difficult to avoid if NP enters only through
EW penguins  and the operator basis is the same as in the SM. A possible
solution to the clash between (\ref{XY1}) and (\ref{XY2}) would be given
by more complicated NP scenarios \cite{BFRS-5}. However, unless a specific 
model is chosen, the predictive power is then significantly reduced. For the
exploration of the NP effects in rare decays, we will therefore not follow
this avenue.

\begin{table}
\vspace{0.4cm}
\begin{center}
\begin{tabular}{|c||c|c|c|c|c|}
\hline
  Quantity & SM & Scen A & Scen B &  Scen C & Experiment
 \\ \hline 
$R_{\rm n}$  & 1.12 &$0.88$ & 1.03 &  1 & $0.83 \pm 0.08$ \\\hline
$R_{\rm c}$  & 1.15 &$0.96$ & 1.13 & 1  & $1.01 \pm 0.09$ \\\hline
${\cal A}_{\rm CP}^{\rm dir}(B^\pm\!\to\!\pi^0 K^\pm) $ &
  0.04 & $0.07$  \rule{0em}{1.05em}& 0.06 & 0.02  &  $-0.04 \pm 0.04$ \\ \hline
${\cal A}_{\rm CP}^{\rm dir}(B_d\!\to\!\pi^0 K_{\rm S})$ & 
  0.06 & $0.04$  \rule{0em}{1.05em}& 0.03  & 0.09 & $-0.02 \pm 0.13$ \\ \hline 
${\cal A}_{\rm CP}^{\rm mix}(B_d\!\to\!\pi^0 K_{\rm S})$ & 
  $-0.82$ & $-0.89$\rule{0em}{1.05em}& $-0.91$ & $-0.70$ &  $-0.31 \pm 0.26$ \\ \hline
$\Delta S$ & 0.13& 0.21& 0.22& 0.01& $-0.38\pm0.26$ \\ \hline
$\Delta A$ & $-0.07$& $-0.04$& $-0.05$& $-0.09$& $-0.16\pm0.04$ \\ \hline
\end{tabular}
\caption{\label{Scentab1} The $B\to\pi K$ observables for the 
 three scenarios introduced in the text. }
\end{center}
\end{table}

\begin{table}
\begin{center}
\begin{tabular}{|c||c|c|c|c|c|}
\hline
  Decay & \quad SM \quad &   Scen A &   Scen B &   Scen C &
  \parbox{2.3cm}{\rule{0em}{1em}Exp. bound \\(90\% {\rm C.L.})}
 \\ \hline
$\mbox{BR}(K^+ \to \pi^+ \nu \bar\nu)/10^{-11}$  &   
 $ 9.3$ & $2.7 $ &  $8.3 $ & $8.4 $ &  $(14.7^{+13.0}_{-8.9}) $\rule{0em}{1.05em} \\ \hline
$\mbox{BR}(K_{\rm L} \to \pi^0 \nu \bar \nu)/10^{-11}$  &
 $ 4.4$ &  $ 11.6$ &  $27.9$ & $7.2$ &  $ < 2.9 \times10^{4} $ \\ \hline
$\mbox{BR}(K_{\rm L} \to \pi^0  e^+ e^-)/10^{-11}$ &  
 $ 3.6$ & $4.6$  &    $7.1$ &  $4.9 $&   $<28$ \\ \hline
$\mbox{BR}(B \to X_s \nu \bar\nu)/10^{-5}$  &  
 $3.6$  &  $ 2.8 $&   $4.8$ &  $3.3 $ &  $<64$ \\ \hline
$\mbox{BR}(B_s \to \mu^+ \mu^-)/10^{-9}$  & 
 $3.9$ & $9.2$ & $ 9.1$ &  $7.0 $&  $<1.5\times 10^{2}$ \rule{0em}{1.05em}\\ \hline 
$\mbox{BR}(K_{\rm L} \to \mu^+ \mu^-)_{\rm SD}/10^{-9}$ &  
 $ 0.9$ & $0.9$  &    $0.001$ &  $0.6 $&   $<2.5$ \\ 
\hline
\end{tabular}
\caption{\label{Scentab2} Rare decay branching ratios for the three scenarios 
introduced in the text. The $B_s\to\mu^+\mu^-$ channel will be discussed
in more detail in Subsection~\ref{ssec:Bmumu}.}
\end{center}
\end{table}

Using an only slightly more generous bound on $|Y|$ by imposing 
$\left|Y \right| \leq 1.5$ and taking only those values of (\ref{q-phi}) 
that satisfy the constraint $\left|Y \right|=1.5$ yields
\begin{equation}
\label{q-phi-RD}
q= 0.48 \pm 0.07 ,\quad \phi=-(93 \pm 17 )^\circ,
\end{equation}
corresponding to a modest {\it suppression} of $q$ relative to its
updated SM value of $0.58$. It is interesting to investigate the impact
of various modifications of $(q,\phi)$, which allow us to satisfy the bounds 
in (\ref{XY2}), for the $B\to\pi K$ observables and rare decays. To this
end, three scenarios for the possible future evolution of the measurements
of $R_{\rm n}$ and $R_{\rm c}$ were introduced in Ref.~\cite{BFRS-5}:
\begin{itemize}
\item {\it Scenario A:} $q=0.48$, $\phi = -93^{\circ}$, which is in accordance with 
the currrent rare decay bounds and the $B \to \pi K$ data (see (\ref{q-phi-RD})).
\item {\it Scenario B:} $q=0.66$, $\phi=-50^{\circ}$, which yields an increase
of $R_{\rm n}$ to 1.03, and some interesting effects in rare decays. This could,
for example, happen if radiative corrections to the  $B_d^0\to\pi^- K^+$ branching 
ratio enhance $R_{\rm n}$ \cite{Baracchini:2005wp}, though this alone would 
probably account for only about $5\%$.
\item {\it Scenario C:} here it is assumed that $R_{\rm n}=R_{\rm c}=1$, which
corresponds to $q=0.54$ and $\phi=61^{\circ}$. The {\it positive} sign of 
$\phi$ distinguishes this scenario strongly from the others.
\end{itemize}
The patterns of the observables of the $B\to\pi K$ and rare decays corresponding
to these scenarios are collected in Tables \ref{Scentab1} and \ref{Scentab2},
respectively. We observe that the $K \to \pi \nu \bar \nu$ modes, which are
theoretically very clean (for a recent review, see Ref.~\cite{BSU}), offer a particularly
interesting probe for the different scenarios. Concerning the observables of the 
$B \to \pi K$ system, ${\cal A}_{\rm CP}^{\rm mix}(B_d\!\to\!\pi^0 K_{\rm S})$ 
is very interesting: this CP asymmetry is found to be very large in Scenarios A and B, 
where the NP phase $\phi$ is negative. On the other hand, the positive sign of 
$\phi$ in Scenario C brings ${\cal A}_{\rm CP}^{\rm mix}(B_d\!\to\!\pi^0 K_{\rm S})$ 
closer to the data, in agreement with the features discussed in
Subsection~\ref{ssec:BpiK}. A similar comment applies to the
direct CP asymmetry of $B^\pm\to\pi^0K^\pm$. 

In view of the large uncertainties, unfortunately no definite conclusions on the
presence of NP can be drawn at this stage. However, the possible anomalies
in the $B\to\pi K$ system complemented with the one in $B\to\phi K$ may actually
indicate the effects of a modified EW penguin sector with a large CP-violating
NP phase. As we just saw, rare $K$ and $B$ decays have an impressive power
to reveal such a kind of NP. Let us finally stress that the analysis of the $B\to\pi\pi$
modes, which signals large non-factorizable effects, and the determination of the 
UT angle $\gamma$ described above are not affected by such NP effects. It will 
be interesting to monitor the evolution of the corresponding data with the help
of the strategy discussed above.

\section{ENTERING A NEW TERRITORY: 
\boldmath$b\to d$\unboldmath~PENGUINS}\label{sec:bd-pengs}
\setcounter{equation}{0}
\subsection{Preliminaries}
Another hot topic which emerged recently is the exploration of 
$b\to d$ penguin processes. The non-leptonic decays belonging
to this category, which are mediated by $b\to d \bar s s$ quark transitions 
(see the classification in Subsection~\ref{sec:class}), are now coming 
within experimental reach at the $B$ factories. A similar comment applies 
to the radiative decays originating from $b\to d\gamma$ processes, whereas
$b\to d\ell^+\ell^-$ modes are still far from being accessible. The $B$ factories
are therefore just entering a new territory, which is still essentially unexplored.
Let us now have a closer look at the corresponding processes.

\boldmath
\subsection{A Prominent Example: $B^0_d\to K^0\bar K^0$}
\unboldmath
The Feynman diagrams contributing to this decay can 
be obtained from those for $B^0_d\to\phi K^0$ shown in 
Fig.~\ref{fig:BphiK-diag} by replacing the anti-strange quark emerging from the 
$W$ boson through an anti-down quark. The $B^0_d\to K^0\bar K^0$ 
decay is described by the low-energy effective Hamiltonian in (\ref{e4}) with $r=d$, 
where the current--current operators may only contribute  through penguin-like 
contractions, corresponding to the penguin topologies with internal up- and
charm-quark exchanges. The dominant r\^ole is played by QCD penguins; 
since EW penguins contribute only in colour-suppressed form, they have a minor 
impact on $B^0_d\to K^0\bar K^0$, in contrast to the case of $B^0_d\to\phi K^0$,
where they may also contribute in colour-allowed form. 

If apply the notation 
introduced in Section~\ref{sec:bench}, make again use of the
unitarity of the CKM matrix and apply the Wolfenstein parametrization, 
we may write the $B^0_d\to K^0\bar K^0$ amplitude as follows:
\begin{equation}\label{ampl-BdKK-lamt}
A(B^0_d\to K^0\bar K^0)=\lambda^3A(\tilde A_{\rm P}^t-\tilde A_{\rm P}^c)
\left[1-\rho_{K\!K} e^{i\theta_{K\!K}}e^{i\gamma}\right],
\end{equation}
where 
\begin{equation}\label{rho-KK-def}
\rho_{K\!K} e^{i\theta_{K\!K}}\equiv R_b
\left[\frac{\tilde A_{\rm P}^t-\tilde A_{\rm P}^u}{\tilde A_{\rm P}^t-\tilde A_{\rm P}^c}\right].
\end{equation}
This expression allows us to calculate the CP-violating asymmetries with 
the help of the formulae given in Subsection~\ref{subsec:CPasym},
taking the following form:
\begin{eqnarray}
{\cal A}_{\rm CP}^{\rm dir}(B_d\to K^0\bar K^0)&=&
D_1(\rho_{K\!K},\theta_{K\!K};\gamma) \label{CP-BKK-dir-gen}\\
{\cal A}_{\rm CP}^{\rm mix}(B_d\to K^0\bar K^0)&=&
D_2(\rho_{K\!K},\theta_{K\!K};\gamma,\phi_d).\label{CP-BKK-mix-gen}
\end{eqnarray}

Let us assume, for a moment, that the penguin contributions are dominated 
by top-quark exchanges. In this case, (\ref{rho-KK-def}) simplifies as
\begin{equation}
\rho_{K\!K} e^{i\theta_{K\!K}} \to R_b.
\end{equation}
Since the CP-conserving strong phase $\theta_{K\!K}$ vanishes in this limit,
the direct CP violation in $B^0_d\to K^0\bar K^0$ vanishes, too. Moreover, 
if we take into account that $\phi_d=2\beta$ in the SM and use trigonometrical
relations which can be derived for the UT, we find that also the mixing-induced
CP asymmetry would be zero. These features suggest an interesting test
of the $b\to d$ flavour sector of the SM (see, for instance, Ref.\  \cite{quinn}). 
However, contributions from penguins with internal up- and charm-quark 
exchanges are expected to yield sizeable CP asymmetries in 
$B_d^0\to K^0\bar K^0$ even within the SM, so that the interpretation of these 
effects is much more complicated \cite{RF-BdKK}; these contributions 
contain also possible long-distance rescattering effects \cite{BFM},
which are often referred to as ``GIM" and ``charming" penguins and received
recently a lot of attention \cite{charming}.

\begin{figure}
\vspace*{0.3truecm}
\begin{center}
\includegraphics[width=10.0cm]{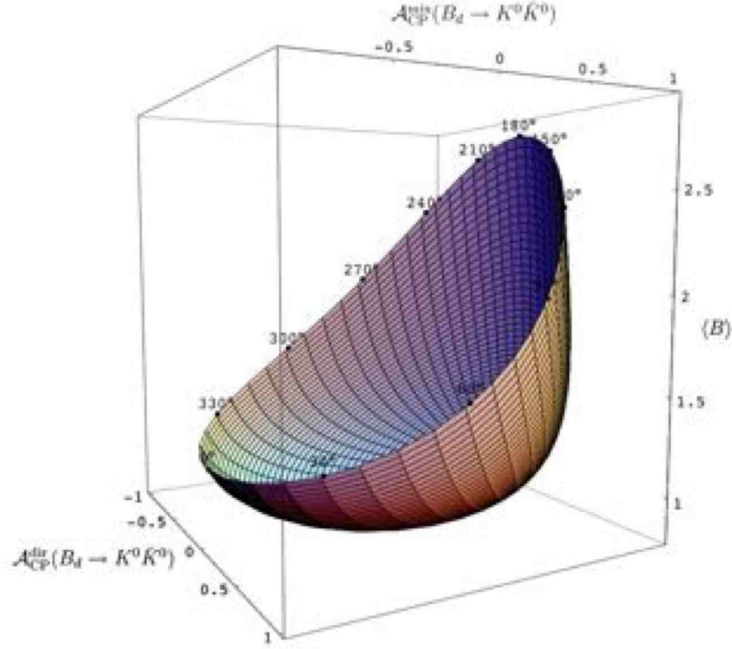}
\end{center}
\vspace*{-0.6truecm}
\caption{Illustration of the surface in the 
${\cal A}_{\rm CP}^{\rm dir}$--${\cal A}_{\rm CP}^{\rm mix}$--$\langle B \rangle$
observable space characterizing the $B^0_d\to K^0\bar K^0$ decay in the SM. 
The intersecting lines on the surface correspond to constant 
values of $\rho_{K\!K}$ and $\theta_{K\!K}$; the numbers on the fringe indicate 
the value of $\theta_{K\!K}$, while the fringe itself is defined by 
$\rho_{K\!K}=1$.}\label{fig:SM-surface}
\end{figure}

Despite this problem, interesting insights can be obtained through the
$B^0_d\to K^0\bar K^0$ observables \cite{FR1}.
By the time the CP-violating asymmetries in (\ref{CP-BKK-dir-gen}) and 
(\ref{CP-BKK-mix-gen}) can be measured, also the angle $\gamma$ of
the UT will be reliably known, in addition to the $B^0_d$--$\bar B^0_d$
mixing phase $\phi_d$. The experimental values of the CP asymmetries
can then be converted into $\rho_{K\!K}$ and $\theta_{K\!K}$, in analogy
to the $B\to\pi\pi$ discussion in Subsection~\ref{ssec:Bpipi-hadr}. Although 
these quantities are interesting to obtain insights into the $B\to\pi K$
parameter $\rho_{\rm c}e^{i\theta_{\rm c}}$ (see (\ref{B+pi+K0}))
through $SU(3)$ arguments, and can be compared  with theoretical predictions,  
for instance, those of QCDF, PQCD or SCET, they do not 
provide -- by themselves -- a test of the SM description of the 
FCNC processes mediating the decay $B^0_d\to K^0\bar K^0$. However, so far, 
we have not yet used the information offered by the CP-averaged branching 
ratio of this channel. It takes the following form:
\begin{equation}\label{BR-BKK-expr}
\mbox{BR}(B_d\to K^0\bar K^0)=\frac{\tau_{B_d}}{16\pi M_{B_d}}
\times \Phi_{KK} \times
|\lambda^3 A \, \tilde A_{\rm P}^{tc}|^2 \langle B \rangle,
\end{equation}
where $\Phi_{KK}$ denotes a two-body phase-space factor, 
$\tilde A_{\rm P}^{tc}\equiv \tilde A_{\rm P}^t-\tilde A_{\rm P}^c$, and
\begin{equation}\label{B-DEF}
\langle B \rangle\equiv 1-2\rho_{K\!K}\cos\theta_{K\!K}
\cos\gamma+\rho_{K\!K}^2.
\end{equation}
If we now use $\phi_d$ and the SM value of $\gamma$, we may characterize
the decay $B^0_d\to K^0\bar K^0$ -- within the SM -- through a surface in 
the observable space of ${\cal A}_{\rm CP}^{\rm dir}$, 
${\cal A}_{\rm CP}^{\rm mix}$ and $\langle B \rangle$. In 
Fig.~\ref{fig:SM-surface}, we show this surface, where each point 
corresponds to a given value of $\rho_{K\!K}$ and $\theta_{K\!K}$. It should 
be emphasized that this surface is {\it theoretically clean} since it 
relies only on the general SM parametrization of $B^0_d\to K^0\bar K^0$. 
Consequently, should future measurements give a value in observable space 
that should {\it not} lie on the SM surface, we would have immediate evidence 
for NP contributions to $\bar b\to \bar d s \bar s$ processes. 

Looking at Fig.~\ref{fig:SM-surface}, we see that $\langle B \rangle$ takes
an absolute minimum. Indeed, if we keep $\rho_{K\!K}$ and $\theta_{K\!K}$
as free parameters in (\ref{B-DEF}), we find
\begin{equation}\label{B-bound}
\langle B \rangle\geq \sin^2\gamma,
\end{equation}
which yields a strong lower bound because of the favourably large value of
$\gamma$. Whereas the direct and mixing-induced CP asymmetries can
be extracted from a time-dependent rate asymmetry (see (\ref{ee6})), 
the determination of $\langle B \rangle$ requires further information to
fix the overall normalization factor involving the penguin amplitude
$\tilde A_{\rm P}^{tc}$. The strategy developed in Refs.~\cite{BFRS2,BFRS3} offers the 
following two avenues, using data for
\begin{itemize}
\item[i)] $B\to\pi\pi$ decays, i.e.\ $b\to d$ transitions, implying the following
lower bound:
\begin{equation}\label{BdKK-bound1}
\mbox{BR}(B_d\to K^0\bar K^0)_{\rm min}= 
\Xi^K_\pi\times\left(1.39\,^{+1.54}_{-0.95}\right) \times 10^{-6},
\end{equation}
\item[ii)] $B\to\pi K$ decays, i.e.\ $b\to s$ transitions, which are complemented 
by the $B\to\pi\pi$ system to determine a small correction, implying the following
lower bound:
\begin{equation}\label{BdKK-bound2}
\mbox{BR}(B_d\to K^0\bar K^0)_{\rm min}= 
\Xi^K_\pi\times\left(1.36\,^{+0.18}_{-0.21}\right) \times 10^{-6}.
\end{equation}
\end{itemize}
Here factorizable $SU(3)$-breaking corrections are included, 
as is made explicit through
\begin{equation}\label{Xi-K-pi}
\Xi^K_\pi=\left[\frac{f_0^K}{0.331}\frac{0.258}{f_0^\pi}\right]^2,
\end{equation}
where the numerical values for the $B\to K,\pi$ form factors $f_0^{K,\pi}$ 
refer to a recent light-cone sum-rule analysis \cite{Ball}. At the time of the 
derivation of these bounds, the $B$ factories reported an experimental {\it upper} 
bound of $\mbox{BR}(B_d\to K^0\bar K^0)<1.5\times 10^{-6}$ (90\% C.L.). Consequently, the theoretical {\it lower} bounds given above suggested that
the observation of this channel should just be ahead of us. Subsequently, the
first signals were indeed announced, in accordance with (\ref{BdKK-bound1}) and 
(\ref{BdKK-bound2}):
\begin{equation}\label{BdK0K0-data}
\hspace*{-0.5truecm}
\mbox{BR}(B_d\to K^0\bar K^0)=\left\{
\begin{array}{ll}
(1.19^{+0.40}_{-0.35}\pm0.13) \times 10^{-6} & \mbox{(BaBar 
\cite{BaBar-BKK}),}\\
 (0.8\pm0.3\pm0.1) \times 10^{-6} & \mbox{(Belle \cite{Belle-BKK}).}
 \end{array}\right.
\end{equation}
The SM description of $B^0_d\to K^0\bar K^0$ has thus successfully passed its 
first test. However, the experimental errors are still very large, and the next crucial 
step -- a measurement of the CP asymmetries -- is still missing. Using QCDF, 
an analysis of NP effects in this channel was recently performed in the minimal 
supersymmetric standard model \cite{giri-moh}. For further aspects of 
$B^0_d\to K^0\bar K^0$, the reader is referred to Ref.~\cite{FR1}.

\boldmath
\subsection{Radiative $b\to d$ Penguin Decays: 
$\bar B\to\rho\gamma$}\label{ssec:radiative}
\unboldmath
Another important tool to explore $b\to d$ penguins is
provided by $\bar B\to\rho\gamma$ modes.  In the SM, these decays 
are described by a Hamiltonian with the following 
structure \cite{B-LH98}:
\begin{equation}\label{Ham-bdgam}
{\cal H}_{\rm eff}^{b\to d\gamma}=\frac{G_{\rm F}}{\sqrt{2}}
\sum_{j=u,c} \! V_{jd}^\ast V_{jb}\left[\sum_{k=1}^{2}C_k Q_k^{jd}\!+\!
\sum_{k=3}^{8}C_k Q_k^{d}\right].
\end{equation}
Here the $Q_{1,2}^{jd}$ denote the current--current operators, whereas the 
$Q_{3\ldots 6}^{d}$ are the QCD penguin operators, which govern the
decay $\bar B^0_d\to  K^0\bar K^0$ together with the
penguin-like contractions of $Q_{1,2}^{cd}$ and $Q_{1,2}^{ud}$. In contrast 
to these four-quark operators,
\begin{equation}
Q_{7,8}^{d}=\frac{1}{8\pi^2}m_b\bar d_i \sigma^{\mu\nu}(1+\gamma_5)
\left\{e b_i F_{\mu\nu} ,\, g_{\rm s}T^a_{ij}b_j G^a_{\mu\nu} \right\}
\end{equation}
are electro- and chromomagnetic penguin operators. 
The most important contributions to $\bar B\to\rho\gamma$
originate from $Q_{1,2}^{jd}$ and $Q_{7,8}^{d}$, 
whereas the QCD penguin operators play only a minor r\^ole, in contrast 
to $\bar B^0_d\to K^0\bar K^0$. If we use again the
unitarity of the CKM matrix and apply the Wolfenstein parametrization,
we may write
\begin{equation}\label{Ampl-Brhogam}
A(\bar B \to \rho\gamma)=c_\rho \lambda^3 A {\cal P}_{tc}^{\rho\gamma}
\left[1-\rho_{\rho\gamma}e^{i\theta_{\rho\gamma}}e^{-i\gamma}\right],
\end{equation}
where $c_\rho=1/\sqrt{2}$ and 1 for $\rho=\rho^0$ and $\rho^\pm$,
respectively, ${\cal P}_{tc}^{\rho\gamma}\equiv
{\cal P}_t^{\rho\gamma}-{\cal P}_c^{\rho\gamma}$, and
\begin{equation}
\rho_{\rho\gamma}e^{i\theta_{\rho\gamma}}\equiv R_b\left[
\frac{{\cal P}_t^{\rho\gamma}-
{\cal P}_u^{\rho\gamma}}{{\cal P}_t^{\rho\gamma}-
{\cal P}_c^{\rho\gamma}}\right].
\end{equation}
Here we follow our previous notation, i.e.\ the ${\cal P}_j^{\rho\gamma}$ 
are strong amplitudes with the following interpretation: 
${\cal P}_u^{\rho\gamma}$ and ${\cal P}_c^{\rho\gamma}$ refer to the matrix 
elements of $\sum_{k=1}^{2}C_k Q_k^{ud}$ and $\sum_{k=1}^{2}C_k Q_k^{cd}$, 
respectively, whereas ${\cal P}_t^{\rho\gamma}$ corresponds to 
$-\sum_{k=3}^{8}C_k Q_k^{d}$. Consequently, ${\cal P}_u^{\rho\gamma}$
and ${\cal P}_c^{\rho\gamma}$ describe the penguin topologies with
internal up- and charm-quark exchanges, respectively, whereas 
${\cal P}_t^{\rho\gamma}$ corresponds to the penguins with the top
quark running in the loop. Let us note that 
(\ref{Ampl-Brhogam}) refers to a given photon helicity. However, 
the $b$ quarks couple predominantly to left-handed photons in 
the SM, so that the right-handed amplitude is usually neglected \cite{GP}; 
we shall return to this point below. Comparing (\ref{Ampl-Brhogam}) with 
(\ref{ampl-BdKK-lamt}), we observe that the structure of both amplitudes is 
the same. In analogy to $\rho_{K\!K} e^{i\theta_{K\!K}}$, 
$\rho_{\rho\gamma}e^{i\theta_{\rho\gamma}}$ may also be affected by 
long-distance effects, which represent a key uncertainty of 
$\bar B\to\rho\gamma$ decays \cite{LHC-Book,GP}. 

If we replace all down quarks in (\ref{Ham-bdgam}) by strange quarks, we obtain the 
Hamiltonian for $b\to s\gamma$ processes, which are already well established 
experimentally \cite{HFAG}:
\begin{eqnarray}
\mbox{BR}(B^\pm\to K^{\ast\pm}\gamma)&=&(40.3\pm2.6)\times 
10^{-6}\label{BR-charged}\\
\mbox{BR}(B_d^0\to K^{\ast0}\gamma)&=&(40.1\pm2.0)\times 
10^{-6}.\label{BR-neutral}
\end{eqnarray}
In analogy to (\ref{Ampl-Brhogam}), we may write
\begin{equation}\label{Ampl-BKastgam}
A(\bar B \!\to\! K^\ast \!\gamma)\!=-\!
\frac{\lambda^3 \! A {\cal P}_{tc}^{K^\ast\!\gamma}}{\sqrt{\epsilon}} \!
\left[1\!+\!\epsilon\rho_{K\!^\ast\!\gamma}e^{i\theta_{K\!^\ast\!\gamma}}
e^{-i\!\gamma}\right]\!,
\end{equation}
where $\epsilon$ was introduced in (\ref{eps-def}). Thanks to the smallness
of $\epsilon$, the parameter 
$\rho_{K\!^\ast\gamma}e^{i\theta_{K\!^\ast\gamma}}$ 
plays an essentially negligible r\^ole for the $\bar B \to K^\ast \gamma$ 
transitions.

Let us have a look at the charged decays $B^\pm \to \rho^{\pm} \gamma$ 
and $B^\pm \to K^{\ast\pm} \gamma$ first. If we consider their 
CP-averaged branching ratios, we obtain
\begin{equation}\label{rare-ratio}
\frac{\mbox{BR}(B^\pm \to \rho^{\pm} 
\gamma)}{\mbox{BR}(B^\pm \to K^{\ast\pm} \gamma)}=\epsilon
\left[\frac{\Phi_{\rho\gamma}}{\Phi_{K\!^\ast\gamma}}\right]
\left|\frac{{\cal P}_{tc}^{\rho\gamma}}{{\cal P}_{tc}^{K\!^\ast\gamma}}
\right|^2 H^{\rho\gamma}_{K\!^\ast\gamma},
\end{equation}
where $\Phi_{\rho\gamma}$ and $\Phi_{K\!^\ast\gamma}$ denote phase-space 
factors, and 
\begin{equation}
H^{\rho\gamma}_{K\!^\ast\gamma}\equiv
\frac{1-2\rho_{\rho\gamma}\cos\theta_{\rho\gamma}\cos\gamma+
\rho_{\rho\gamma}^2}{1+2\epsilon\rho_{K\!^\ast\gamma}
\cos\theta_{K\!^\ast\gamma}
\cos\gamma+\epsilon^2\rho_{K\!^\ast\gamma}^2}.
\end{equation}
Since $B^\pm \to \rho^{\pm} \gamma$ and $B^\pm \to K^{\ast\pm} \gamma$ 
are related through the interchange of all down and strange quarks, 
the $U$-spin flavour symmetry of strong interactions allows us to relate 
the corresponding hadronic amplitudes to each other; the $U$-spin
symmetry is an $SU(2)$ subgroup of the full $SU(3)_{\rm F}$ flavour-symmetry
group, which relates down and strange quarks in the same manner as the 
conventional strong isospin symmetry relates down and up quarks. Following 
these lines, we obtain
\begin{equation}\label{U-spin1}
|{\cal P}_{tc}^{\rho\gamma}|=|{\cal P}_{tc}^{K\!^\ast\gamma}|
\end{equation}
\begin{equation}\label{U-spin2}
\rho_{\rho\gamma}e^{i\theta_{\rho\gamma}}=
\rho_{K\!^\ast\gamma}e^{i\theta_{K\!^\ast\gamma}}\equiv
\rho e^{i\theta}.
\end{equation}
Although we may determine the ratio of the penguin amplitudes 
$|{\cal P}_{tc}|$ in (\ref{rare-ratio}) with the help of  (\ref{U-spin1}) -- up to 
$SU(3)$-breaking effects to be discussed below -- we are still left
with the dependence on $\rho$ and $\theta$. However, keeping $\rho$ 
and $\theta$ as free parameters, it can be shown that
$H^{\rho\gamma}_{K\!^\ast\gamma}$ satisfies the following relation \cite{FR2}:
\begin{equation}\label{H-bound}
H^{\rho\gamma}_{K\!^\ast\gamma}\geq \left[1-2\epsilon
\cos^2\gamma+{\cal O}(\epsilon^2)\right]\sin^2\gamma,
\end{equation}
where the term linear in $\epsilon$ gives a shift of about $1.9\%$. 

Concerning possible $SU(3)$-breaking effects to (\ref{U-spin2}), they 
may only enter this tiny correction and are negligible for our analysis. 
On the other hand, the $SU(3)$-breaking corrections to (\ref{U-spin1}) 
have a sizeable impact. Following Refs.~\cite{ALP-rare,BoBu}, we write
\begin{equation}\label{SU3-break-rare}
\left[\frac{\Phi_{\rho\gamma}}{\Phi_{K\!^\ast\gamma}}\right]
\left|\frac{{\cal P}_{tc}^{\rho\gamma}}{{\cal P}_{tc}^{K\!^\ast\gamma}}
\right|^2=\left[\frac{M_B^2-M_\rho^2}{M_B^2-M_{K^\ast}^2}\right]^3
\zeta^2,
\end{equation}
where $\zeta=F_\rho/F_{K^\ast}$ is the $SU(3)$-breaking ratio of the
$B^\pm\to\rho^\pm\gamma$ and $B^\pm\to K^{\ast\pm}\gamma$ form factors; a 
light-cone sum-rule analysis gives $\zeta^{-1}=1.31\pm0.13$ \cite{Ball-Braun}.
Consequently, (\ref{H-bound}) and (\ref{SU3-break-rare}) allow us to convert 
the measured $B^\pm\to K^{\ast\pm}\gamma$ branching ratio (\ref{BR-charged}) 
into a {\it lower} SM bound for 
$\mbox{BR}(B^\pm\to\rho^\pm\gamma)$ with the help of (\ref{rare-ratio}) \cite{FR2}:
\begin{equation}\label{Brhogam-char}
\mbox{BR}(B^\pm\to \rho^\pm\gamma)_{\rm min}=\left(1.02\,^{+0.27}_{-0.23}
\right)\times10^{-6}.
\end{equation}

A similar kind of reasoning holds also for the $U$-spin 
pairs $B^\pm\to K^\pm K, \pi^\pm K$ and $B^\pm\to K^\pm K^\ast, \pi^\pm K^\ast$,
where the following lower bounds can be derived \cite{FR2}:
\begin{eqnarray}
\mbox{BR}(B^\pm\!\to\! K^\pm K)_{\rm min} \!\!&=&\!\! \Xi^K_\pi\!\times\!
\left(1.69\,^{+0.21}_{-0.24}\right)\!\times\! 10^{-6}\label{BKK-char}\\
\mbox{BR}(B^\pm\!\to\! K^\pm K^\ast)_{\rm min} \!\!&=&\!\! \Xi^K_\pi\!\times\!
\left(0.68\,^{+0.11}_{-0.13}
\right)\!\times \! 10^{-6},\label{BpiKast}
\end{eqnarray}
with $\Xi^K_\pi$ given in (\ref{Xi-K-pi}). Thanks to the most recent
$B$-factory data, we have now also evidence for $B^\pm\to K^\pm K$
decays:
\begin{equation}
\mbox{BR}(B^\pm\!\to\! K^\pm K)=
\left\{\begin{array}{ll}
(1.5\pm0.5\pm0.1)\times 10^{-6} & \mbox{(BaBar \cite{BaBar-BKK})}\\
(1.0\pm0.4\pm0.1)\times 10^{-6} & \mbox{(Belle \cite{Belle-BKK}),}
\end{array} \right.
\end{equation}
whereas the upper limit of $5.3\times 10^{-6}$ for $B^\pm\to K^\pm K^\ast$
still leaves a lot of space. Obviously, we may also consider the
$B^\pm\to K^{\ast\pm} K, \rho^\pm K$ system \cite{FR2}. However,
since currently only the upper bound 
$\mbox{BR}(B^\pm\to \rho^\pm K)<48\times 10^{-6}$ is available, 
we cannot yet give a number for the lower bound on 
$\mbox{BR}(B^\pm\to K^{\ast\pm} K)$. Experimental analyses of
these modes are strongly encouraged.

Let us now turn to $\bar B^0_d\to\rho^0\gamma$, which receives 
contributions from exchange and penguin annihilation topologies that are 
not present in 
$\bar B^0_d\to \bar K^{\ast0}\gamma$; in the case of $B^\pm\to\rho^\pm\gamma$ 
and $B^\pm\to K^{\ast\pm}\gamma$, which are related by the $U$-spin symmetry, 
there is a one-to-one correspondence of topologies. Making the 
plausible assumption that the topologies involving the spectator quarks play 
a minor r\^ole, and taking the factor of $c_{\rho^0}=1/\sqrt{2}$ in 
(\ref{Ampl-Brhogam}) into account, the counterpart of (\ref{Brhogam-char}) 
is given by 
\begin{equation}\label{Brhogam-neut}
\mbox{BR}(B_d\to \rho^0\gamma)_{\rm min}=\left(0.51\,^{+0.13}_{-0.11}
\right)\times10^{-6}.
\end{equation}

At the time of the derivation of the {\it lower} bounds for the
$B\to\rho\gamma$ branching ratios given above, the following experimental {\it upper}
bounds ($90\%$ C.L.) were available:
\begin{equation}\label{Brho-gam-char-EXP}
\mbox{BR}(B^\pm\to \rho^\pm\gamma)<\left\{\begin{array}{ll}
1.8\times 10^{-6} & \mbox{(BaBar \cite{Babar-Brhogamma-bound})}\\
2.2\times 10^{-6} & \mbox{(Belle \cite{Belle-Brhogamma-bound})}\\
\end{array} \right.
\end{equation}
\begin{equation}\label{Brho-gam-neut-EXP}
\mbox{BR}(B_d\to \rho^0\gamma)<\left\{\begin{array}{ll}
0.4\times 10^{-6} & \mbox{(BaBar \cite{Babar-Brhogamma-bound})}\\
0.8\times 10^{-6} & \mbox{(Belle \cite{Belle-Brhogamma-bound}).}
\end{array} \right.
\end{equation}
Consequently, it was expected that the $\bar B\to\rho\gamma$ modes should 
soon be discovered at the $B$ factories \cite{FR2}. Indeed, the Belle 
collaboration reported recently the first observation of $b\to d\gamma$ processes
\cite{Belle-bdgam-obs}:
\begin{eqnarray}
\mbox{BR}(B^\pm\to \rho^\pm\gamma)&=&\left(0.55^{+0.43+0.12}_{-0.37-0.11}\right)
\times 10^{-6}
\label{Belle-Brhogam-p}\\
\mbox{BR}(B_d\to \rho^0\gamma)&=&\left(1.17^{+0.35+0.09}_{-0.31-0.08}\right)
\times 10^{-6}
\label{Belle-Brhogam-n}\\
\mbox{BR}(B\to(\rho,\omega)\gamma)&=&
\left(1.34^{+0.34+0.14}_{-0.31-0.10}\right)\times 10^{-6},
\end{eqnarray}
which was one of the hot topics of the 2005 summer conferences \cite{Belle-press}.
These measurements still suffer from large uncertainties, and the pattern of the 
central values of (\ref{Belle-Brhogam-p}) and (\ref{Belle-Brhogam-n}) would be in
conflict with the expectation following from the isospin symmetry. It will be interesting
to follow the evolution of the data. The next important conceptual step would be the measurement of the corresponding CP-violating observables, though
 this is still
in the distant future.

An alternative avenue to confront the data for the $B\to \rho\gamma$
branching ratios with the SM is provided by converting them into information
on the side $R_t$ of the UT. To this end, the authors of Refs.~\cite{ALP-rare,BoBu}
use also (\ref{SU3-break-rare}), and calculate the CP-conserving (complex) 
parameter $\delta a$ entering 
$\rho_{\rho\gamma}e^{i\theta_{\rho\gamma}}=R_b\left[1+\delta a\right]$
in the QCDF approach. The corresponding result, which favours a small impact 
of $\delta a$, takes leading and next-to-leading order QCD corrections into 
account and holds to leading order in the heavy-quark limit \cite{BoBu}. 
In view of the remarks about possible long-distance effects made above and the 
$B$-factory data for the $B\to\pi\pi$ system, which indicate large corrections 
to the QCDF picture for non-leptonic $B$ decays into two light pseudoscalar 
mesons (see Subsection~\ref{ssec:Bpipi-hadr}), it is, however, not obvious that 
the impact of $\delta a$ is actually small. The advantage of the bound
following from (\ref{H-bound}) is that it is  -- by construction -- {\it not} affected 
by $\rho_{\rho\gamma}e^{i\theta_{\rho\gamma}}$ at all.

\boldmath
\subsection{General Lower Bounds for $b\to d$ Penguin Processes}
\unboldmath
Interestingly, the bounds discussed above are actually 
realizations of a general, model-independent bound that can be derived
in the SM for $b\to d$ penguin processes \cite{FR2}. If we consider such
a decay, $\bar B \to \bar f_d$, we may -- in analogy to (\ref{ampl-BdKK-lamt}) 
and (\ref{Ampl-Brhogam}) -- write
\begin{equation}
A(\bar B \to \bar f_d)= A^{(0)}_d
\left[1-\varrho_de^{i\theta_d}e^{-i\gamma}\right],
\end{equation}
so that the CP-averaged amplitude square is given as follows:
\begin{equation}
\langle|A(B \to f_d)|^2\rangle=|A^{(0)}_d|^2
\left[1-2\varrho_d\cos\theta_d\cos\gamma+\varrho_d^2\right].
\end{equation}
In general, $\varrho_d$ and $\theta_d$ depend on the point in phase space
considered. Consequently, the expression
\begin{equation}
\mbox{BR}(B \to f_d)=\tau_B\left[\sum_{\rm Pol}
\int \!\! d \, {\rm PS} \, \langle|A(B \to f_d)|^2\rangle \right]
\end{equation}
for the CP-averaged branching ratio, where the sum runs over possible
polarization configurations of $f_d$, does {\it not} factorize into 
$|A^{(0)}_d|^2$ and $[1-2\varrho_d\cos\theta_d\cos\gamma+\varrho_d^2]$ as 
in the case of the two-body decays considered above. However, if we 
keep $\varrho_d$ and $\theta_d$ as free, ``unknown'' parameters at any 
given point in phase space, we obtain
\begin{equation}
\langle|A(B \to f_d)|^2\rangle\geq|A^{(0)}_d|^2 \sin^2\gamma,
\end{equation}
which implies
\begin{equation}
\mbox{BR}(B \to f_d)\geq\tau_B\left[\sum_{\rm Pol}
\int \!\! d \, {\rm PS} \, |A^{(0)}_d|^2 \right]\sin^2\gamma.
\end{equation}

In order to deal with the term in square brackets, we use a $b\to s$ 
penguin decay $\bar B \to \bar f_s$, which is the counterpart of $\bar B \to \bar f_d$ 
in that the corresponding CP-conserving strong amplitudes can be related
to one another through the $SU(3)$ flavour symmetry. In analogy to 
(\ref{Ampl-BKastgam}), we may then write
\begin{equation}
A(\bar B \to \bar f_s)= - \frac{A^{(0)}_s}{\sqrt{\epsilon}}
\left[1+\epsilon\varrho_s e^{i\theta_s}e^{-i\gamma}\right].
\end{equation}
If we neglect the term proportional to $\epsilon$ in the square bracket, 
we arrive at
\begin{equation}\label{general-bound}
\frac{\mbox{BR}(B \to f_d)}{\mbox{BR}(B \to f_s)}
\geq \epsilon \left[\frac{\sum_{\rm Pol}\int \! d \, {\rm PS} \, 
|A^{(0)}_d|^2 }{\sum_{\rm Pol}\int \! d \, {\rm PS} \, |A^{(0)}_s|^2 }
\right]\sin^2\gamma.
\end{equation}
Apart from the tiny $\epsilon$ correction, which gave a shift of about
$1.9\%$ in (\ref{H-bound}), (\ref{general-bound}) is valid
exactly in the SM. If we now apply the $SU(3)$ flavour symmetry, we obtain
\begin{equation}\label{SU3-limit}
\frac{\sum_{\rm Pol}\int \! d \, {\rm PS} \, 
|A^{(0)}_d|^2 }{\sum_{\rm Pol}\int \! d \, {\rm PS} \, |A^{(0)}_s|^2 }
\stackrel{SU(3)_{\rm F}}{\longrightarrow} 1.
\end{equation}
Since $\sin^2\gamma$ is favourably large in the SM and the decay
$\bar B \to \bar f_s$ will be measured before its $b\to d$ 
counterpart  -- simply because of the CKM enhancement -- 
(\ref{general-bound}) provides strong lower bounds for 
$\mbox{BR}(B \to f_d)$. 

It is instructive to return briefly to $B\to\rho\gamma$. If we look at 
(\ref{general-bound}), we observe immediately that the assumption that 
these modes are governed by a single photon helicity is no longer 
required. Consequently, (\ref{Brhogam-char}) and (\ref{Brhogam-neut}) 
are actually very robust with respect to this issue, which may only affect 
the $SU(3)$-breaking corrections to a small extend. This feature is interesting
in view of the recent discussion in \cite{GGLP}, where the photon polarization
in $B\to \rho\gamma$ and $B\to K^\ast \gamma$ decays was critically analyzed. 

We can now also derive a bound for the  
$B^\pm\to K^{\ast\pm}K^{\ast}, \rho^\pm K^\ast$ system, where 
we have to sum in (\ref{general-bound}) over three polarization configurations
of the vector mesons. The analysis of the $SU(3)$-breaking corrections is
more involved than in the case of the decays considered above, and the 
emerging lower bound of 
$\mbox{BR}(B^\pm\to K^{\ast\pm} K^\ast)_{\rm min}\sim0.6\times 10^{-6}$
is still very far from the experimental upper bound of $71\times 10^{-6}$.
Interestingly, the theoretical lower bound would be reduced by $\sim 0.6$ in 
the strict $SU(3)$ limit, i.e.\ would be more conservative \cite{FR2}. A similar 
comment applies to (\ref{BdKK-bound1}), (\ref{BdKK-bound2}) and  
(\ref{BKK-char}), (\ref{BpiKast}). On the other hand, the 
$B\to\rho\gamma$ bounds in (\ref{Brhogam-char}) and 
(\ref{Brhogam-neut}) would be enhanced by $\sim 1.7$ in this case.
However, here the theoretical situation is more favourable since we 
have not to rely on the factorization hypothesis to deal with the 
$SU(3)$-breaking effects as in the case of the non-leptonic decays. 

Let us finally come to another application of (\ref{general-bound}), which 
is offered by decays of the kind $\bar B\to \pi \ell^+\ell^-$ and 
$\bar B\to \rho \ell^+\ell^-$. It is
well known that the $\rho_d$ terms complicate the interpretation of
the corresponding data considerably \cite{LHC-Book}; the bound offers
SM tests that are not affected by these contributions. The 
structure of the $b\to d \ell^+\ell^-$ Hamiltonian is similar to 
(\ref{Ham-bdgam}), but involves the additional operators
\begin{equation}
Q_{9,10}=\frac{\alpha}{2\pi}(\bar\ell\ell)_{\rm V\!,\,A}
(\bar d_i b_i)_{\rm V-A}.
\end{equation}
The $b \to s \ell^+\ell^-$ modes $\bar B\to K \ell^+\ell^-$ 
and $\bar B\to K^\ast \ell^+\ell^-$ were already observed at the $B$ 
factories, with branching ratios at the $0.6\times 10^{-6}$ and 
$1.4\times 10^{-6}$ levels \cite{HFAG}, respectively, and received considerable
theoretical attention (see, e.g., \cite{BKll}). For the application
of (\ref{general-bound}), the charged decay combinations
$B^\pm\to \pi^\pm \ell^+\ell^-, K^\pm \ell^+\ell^-$ and
$B^\pm\to \rho^\pm \ell^+\ell^-, K^{\ast\pm} \ell^+\ell^-$ are suited
best since the corresponding decay pairs are related to each other 
through the $U$-spin symmetry \cite{HM}. The numbers given above
suggest
\begin{equation}\label{Bpi-ellell-bounds}
\mbox{BR}(B^\pm\to \pi^\pm \ell^+\ell^-), \quad
\mbox{BR}(B^\pm\to \rho^\pm \ell^+\ell^-)
\mathrel{\hbox{\rlap{\hbox{\lower4pt\hbox{$\sim$}}}\hbox{$>$}}}
10^{-8},
\end{equation}
thereby leaving the exploration of these $b\to d$ penguin decays for the more 
distant future. Detailed studies of the associated $SU(3)$-breaking corrections 
are engouraged. By the time the $B^\pm\to \pi^\pm \ell^+\ell^-$, 
$\rho^\pm \ell^+\ell^-$ modes can be measured, we will
hopefully have a good picture of these effects. 

It will be interesting to confront all of these bounds with experimental data.
In the case of the non-leptonic $B_d\to K^0\bar K^0$, $B^\pm\to K^\pm K$ modes
and their radiative $B\to\rho \gamma$ counterparts, they have already provided a
first successful test of the SM description of the corresponding FCNC processes,
although the uncertainties are still very large in view of the fact that 
we are just at the beginning
of the experimental exploration of these channels. A couple of other non-leptonic
decays of this kind may just be around the corner. It would be exciting if some
bounds were significantly violated through destructive interference between
SM and NP contributions. Since the different decay classes are governed by
different operators, we could actually encounter surprises!

\section{\boldmath$B$\unboldmath-DECAY STUDIES IN THE LHC ERA: FULLY 
EXPLOITING THE \boldmath$B_s$\unboldmath~SYSTEM}\label{sec:LHC}
\setcounter{equation}{0}
\boldmath
\subsection{In Pursuit of New Physics with 
$\Delta M_s$}\label{ssec:Bs-prelim}
\unboldmath
Concerning experimental information about this mass difference, only lower 
bounds were available for many years from the LEP experiments at CERN 
and SLD at SLAC \cite{LEPBOSC}. Since the currently operating $e^+e^-$ $B$ 
factories run at the $\Upsilon(4S)$ resonance, which decays into $B_{u,d}$, 
but not into $B_s$ mesons, the $B_s$ system cannot be explored by the BaBar 
and Belle experiments.\footnote{The asymmetric $e^+e^-$ KEKB collider was 
recently also operated at the $\Upsilon(5S)$ resonance in an engineering run, 
allowing the Belle experiment to take first $B_s$ data \cite{Belle-Y5S}.} 
However, plenty of $B_s$ mesons are produced at 
the Tevatron (and later on will be at the LHC \cite{schneider}), which --
very recently -- allowed the measurement of $\Delta M_s$,
as summarized  in (\ref{D0-range}) and (\ref{CDF-DMs}). 
These new results were one of the hot topics of the spring 2006, and
have already triggered several phenomenological papers 
(see, e.g., \cite{CMNSW}--\cite{GNR}).

As in Section~\ref{sec:NP} and Subsection~\ref{ssec:BpsiK}, 
we shall follow the analysis of
Ref.~\cite{BF-DMs}. In order to describe possible NP effects, we parametrize
them through (\ref{DMq-NP}) and (\ref{phiq-NP}). The relevant CKM factor is
$|V_{ts}^* V_{tb}|$. Using once again the
unitarity of the CKM matrix and including next-to-leading
order terms in the Wolfenstein expansion as given in Ref.~\cite{blo}, we have
\begin{equation}\label{Vts}
\left|\frac{V_{ts}}{V_{cb}}\right|=1-\frac{1}{2}\left(1-2R_b\cos\gamma\right)\lambda^2
+{\cal O}(\lambda^4).
\end{equation}
Consequently, apart from the tiny correction in $\lambda^2$, the  CKM
factor for $\Delta M_s$ is independent of $\gamma$ and $R_b$,
which is an important advantage in comparison with the $B_d$-meson system.
The accuracy of the SM prediction of $\Delta M_s$ is hence  limited by the
hadronic mixing parameter $f_{B_s}\hat{B}_{B_s}^{1/2}$. If we consider
the ratio $\rho_s$ introduced in (\ref{rhoq-def}) and use
the CDF measurement in (\ref{CDF-DMs}), we obtain
\begin{eqnarray}
\left.\rho_s\right|_{\rm JLQCD} &=&
1.08^{+0.03}_{-0.01} \mbox{(exp)} \pm 0.19 \mbox{(th)}\\
\left.\rho_s\right|_{\rm  (HP+JL)QCD} &=& 
0.74^{+0.02}_{-0.01} \mbox{(exp)} \pm 0.18 \mbox{(th)}\,,\label{36}
\end{eqnarray}
where we made the experimental and theoretical errors explicit. 
These numbers are consistent with the
SM case $\rho_s=1$, but suffer from significant theoretical uncertainties, 
which are much larger than the experimental errors. Nevertheless, it is
interesting to note that the (HP+JL)QCD result is $1.5\,\sigma$ below the SM;
a similar pattern arises in (\ref{rhod-JLQCD}) and (\ref{rhod-HPJL}), 
though at the $1\,\sigma$ level. Any more precise statement about the presence 
or absence of NP requires the reduction of theoretical uncertainties.

\begin{figure}[t]
$$\epsfxsize=0.40\textwidth\epsffile{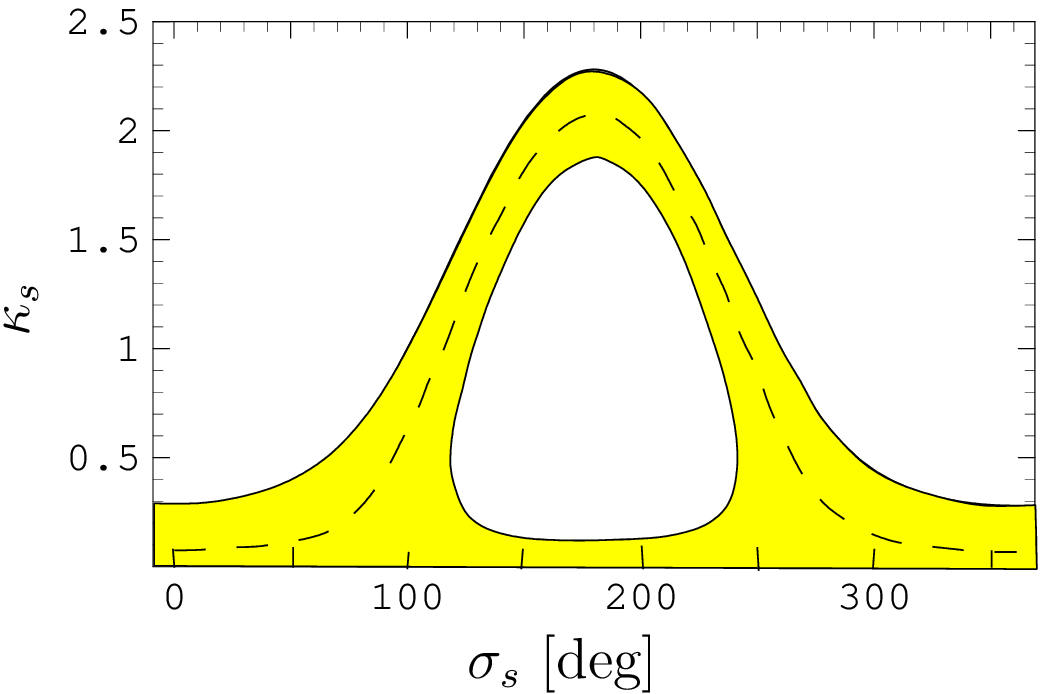}\quad
\epsfxsize=0.40\textwidth\epsffile{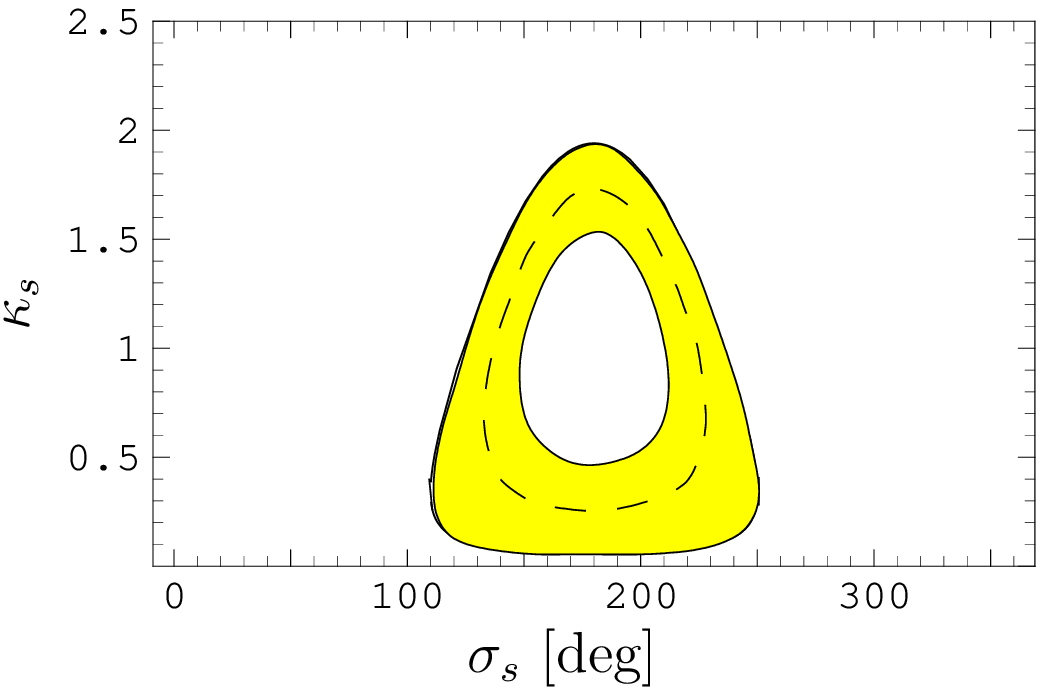}
$$
 \vspace*{-1truecm}
\caption[]{The allowed regions (yellow/grey) in the $\sigma_s$--$\kappa_s$ plane.
Left panel: JLQCD lattice results (\ref{JLQCD}). Right panel: (HP+JL)QCD lattice 
results (\ref{HPQCD}).}\label{fig:MDs-NP}
\end{figure}

In Fig.~\ref{fig:MDs-NP}, we show the constraints in the $\sigma_s$--$\kappa_s$ 
plane, which can be obtained from $\rho_s$ with the help of the contours
shown in Fig.~\ref{fig:kappa-rho}. We see that  upper bounds of $\kappa_s\lsim 2.5$
arise from the measurement of $\Delta M_s$. In the case of (\ref{36}), 
$\sigma_s$ would be constrainted to lie within the range 
$110^\circ\leq\sigma_s\leq250^\circ$. Consequently, the CDF measurement of 
$\Delta M_s$ leaves ample space for the NP parameters $\sigma_s$ and 
$\kappa_s$. As in the case of the $B_d$-meson system discussed in
Subsection~\ref{ssec:BpsiK}, this situation will change significantly 
as soon as information about CP violation in the $B_s$-meson system becomes
available. We shall return to this topic in Subsection~\ref{ssec:BsPsiPhi}.

It is interesting to consider the ratio of $\Delta M_s$ and $\Delta M_d$,
which can be written as follows:
\begin{equation}\label{DMs-DMd-rat}
\frac{\Delta M_s}{\Delta M_d} =  \frac{\rho_s}{\rho_d}
\left|\frac{V_{ts}}{V_{td}}\right|^2 \frac{M_{B_s}}{M_{B_d}}\, \xi^2\,,
\end{equation}
where the hadronic $SU(3)$-breaking parameter $\xi$ is defined  through
\begin{equation}
\xi \equiv 
\frac{f_{B_s}\hat{B}_{B_s}^{1/2}}{f_{B_d}\hat{B}_{B_d}^{1/2}}.
\end{equation}
In the class of NP models with
``minimal flavour violation" (see Section~\ref{sec:NP}, and Ref.~\cite{BBGT} for 
a recent analysis addressing also the $\Delta M_s$ measurement), 
we have $\rho_s/\rho_d=1$, so that (\ref{DMs-DMd-rat}) 
allows the extraction of the CKM factor $|V_{ts}/V_{td}|$, and hence
$|V_{td}|$, as $|V_{ts}|$ is known -- to excellent accuracy -- from (\ref{Vts}). The advantage of this determination lies in the reduced theoretical uncertainty of $\xi$ as compared to $f_{B_d}\hat B_{B_d}^{1/2}$. For the sets of lattice results in
(\ref{JLQCD}) and (\ref{HPQCD}), we have
\begin{eqnarray}
\xi_{\rm JLQCD} & = & 1.14\pm 0.06^{+0.13}_{-0}\label{xi-JLQCD}\\
\xi_{\rm (HP+JL)QCD} & = &
1.210^{+0.047}_{-0.035}.\label{xi-HPJL}
\end{eqnarray}
Using the expression
\begin{equation}\label{Rt-def2}
R_t \equiv \frac{1}{\lambda}\left|\frac{V_{td}}{V_{cb}}\right| 
=  \frac{1}{\lambda}\left|\frac{V_{td}}{V_{ts}}\right| \left[ 
1-\frac{1}{2}\left(1-2R_b\cos\gamma\right)\lambda^2+{\cal O}(\lambda^4)\right],
\end{equation}
we may convert the extracted value of $|V_{ts}/V_{td}|$ into a measurement of the UT
side $R_t$. As we noted in Subsection~\ref{ssec:radiative}, another determination of $R_t$ can, in principle, be obtained from radiative decays, in particular the ratio of 
branching ratios ${\cal B}(B\to (\rho,\omega)\gamma)/{\cal B}(B\to K^*\gamma)$, 
but is presently limited by experimental statistics; see Ref.~\cite{VtdVts} for a 
recent analysis.

\begin{figure}[t] 
$$\epsfxsize=0.40\textwidth\epsffile{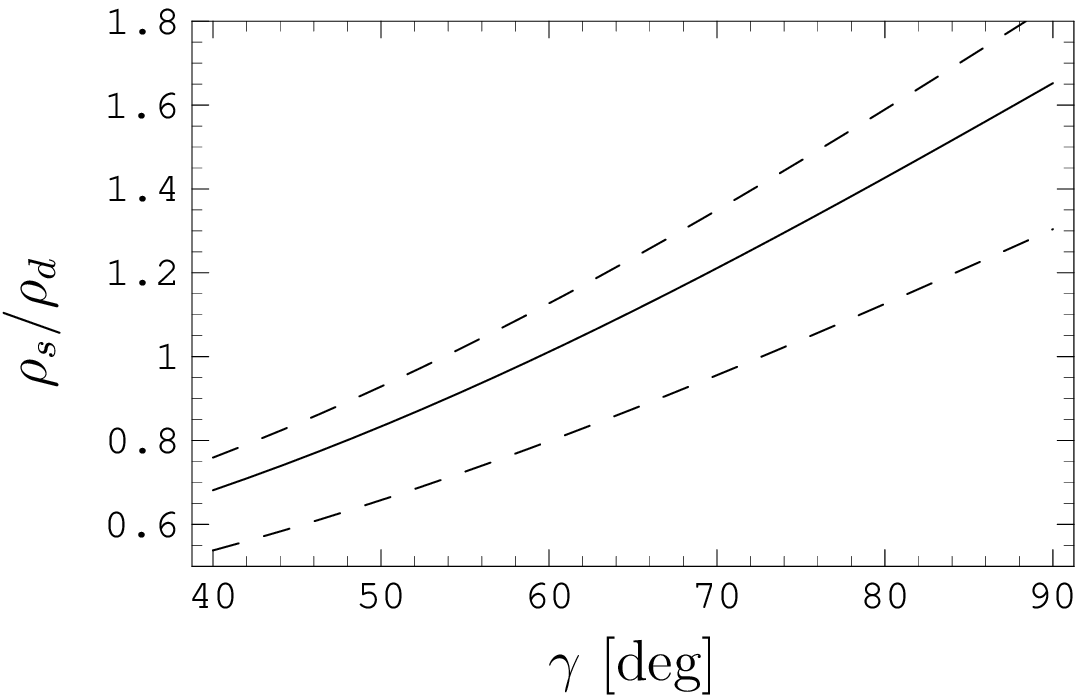}\quad
\epsfxsize=0.40\textwidth\epsffile{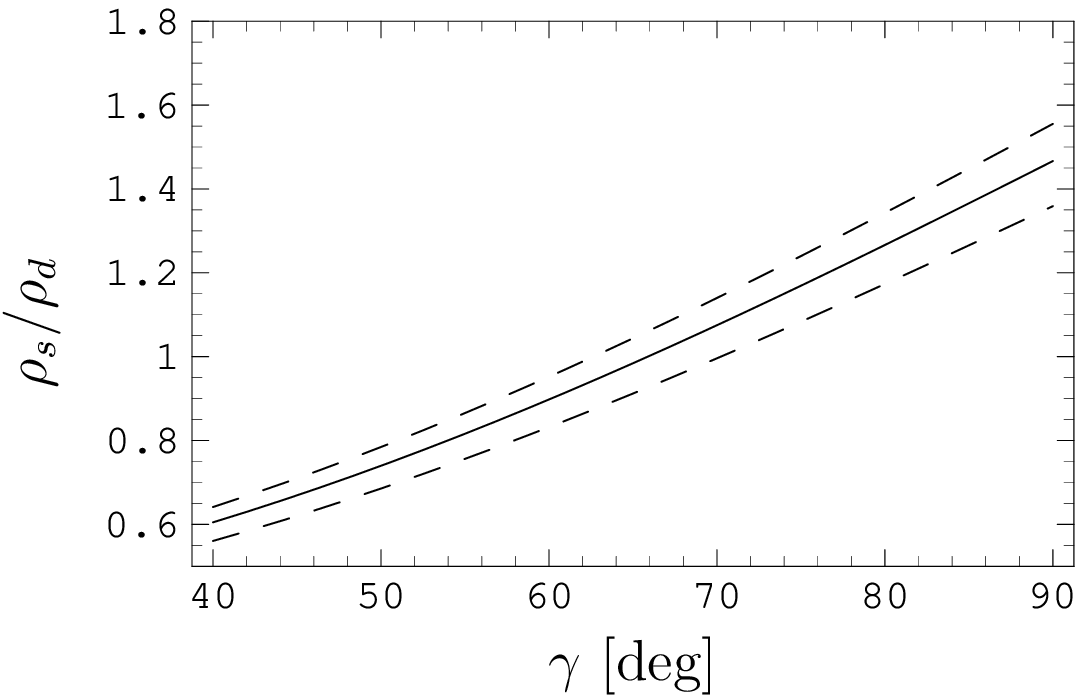}$$
\vspace*{-1cm}
   \caption[]{The dependence of $\rho_s/\rho_d$ on $\gamma$ for the central 
   values of $\Delta M_{d,s}$ in (\ref{DMd-exp}) and (\ref{CDF-DMs}). 
Left panel: JLQCD results (\ref{xi-JLQCD}). 
   Right panel:  (HP+JL)QCD results (\ref{xi-HPJL}). The plots are nearly
   independent of $R_b$.}
   \label{fig:rhos-rhod}
\end{figure}

Alternatively, following Ref.~\cite{BF-DMs}, we may constrain
the ratio $\rho_s/\rho_d$ through the measured value of $\Delta M_s/\Delta M_d$. 
To this end, we express -- in analogy to (\ref{CKM-Bd})  -- the UT side $R_t$ 
in terms of $R_b$ and $\gamma$:
\begin{equation}\label{Rt-expr}
R_t = \sqrt{1- 2 R_b\cos\gamma + R_b^2}\,,
\end{equation}
allowing the determination of $R_t$ through processes that are essentially
unaffected by NP. The resulting value of $R_t$ depends rather strongly on 
$\gamma$, which is the main source of uncertainty. 
Combining then (\ref{DMs-DMd-rat}) and (\ref{Rt-def2}), we obtain the
following expression for $\rho_s/\rho_d$:
\begin{equation}
\frac{\rho_s}{\rho_d}=\lambda^2\left[1-2R_b\cos\gamma+R_b^2\right]
\left[1+(1-2R_b\cos\gamma)\lambda^2+{\cal O}(\lambda^4)\right]
\frac{1}{\xi^2}\frac{M_{B_d}}{M_{B_s}}
\frac{\Delta M_s}{\Delta M_d}\,.
\end{equation}
In Fig.~\ref{fig:rhos-rhod}, we plot this ratio for the central values
of $\Delta M_d$ and $\Delta M_s$ in (\ref{DMd-exp}) and (\ref{CDF-DMs}), 
respectively, as a function of the UT angle $\gamma$ for 
the values of $\xi$ given in (\ref{JLQCD}) and 
(\ref{HPQCD}). We find that the corresponding curves are nearly independent
of $R_b$ and that $\gamma$ is actually the key CKM parameter for the determination
of $\rho_s/\rho_d$. The corresponding numerical values are given by:
\begin{eqnarray}
\left.\frac{\rho_s}{\rho_d}\right|_{\rm JLQCD} &=&
1.11^{+0.02}_{-0.01} \mbox{(exp)} \pm 0.35 (\gamma,R_b)^{+0.12}_{-0.28}(\xi)\,
\label{2006a}\\
\left.\frac{\rho_s}{\rho_d}\right|_{\rm  (HP+JL)QCD} &=& 
0.99^{+0.02}_{-0.01} \mbox{(exp)} \pm 0.31 
(\gamma,R_b)^{+0.06}_{-0.08}(\xi)\,.\label{2006b}
\end{eqnarray}
Because of the large range of allowed values of $\gamma$ in (\ref{gamma-tree}),
this ratio is currently not stringently constrained. This situation should, however,
improve significantly in the LHC era thanks to the impressive determination of 
$\gamma$ to be obtained at the LHCb experiment. In fact,
a statistical accuracy of $\sigma_{\rm stat}(\gamma)\approx 2.5^\circ$ is expected
at LHCb after 5 years of taking data \cite{schneider}. 

Let us introduce a scenario for the year 2010 that is characterized by 
$\gamma=(70\pm5)^\circ$ and the (HP+JL)QCD parameters in (\ref{HPQCD}). 
We then find
\begin{equation}\label{rhos-rhod-2010}
\left.\frac{\rho_s}{\rho_d}\right|_{2010}=1.07 \pm 0.09
(\gamma,R_b)^{+0.06}_{-0.08}(\xi) = 1.07\pm0.12\,,
\end{equation}
where we made the errors arising from the uncertainties of $\gamma$
and $\xi$ explicit, and, in the last step, added them in quadrature. Consequently, 
the hadronic uncertainties and those induced by $\gamma$ would now be of the same 
size, which should provide additional motivation for the lattice
community to reduce the error of $\xi$ even further. Despite the
impressive reduction of uncertainty compared to the 2006 values in
(\ref{2006a}) and (\ref{2006b}), the numerical value 
in (\ref{rhos-rhod-2010}) would still not allow a stringent test of whether
$\rho_s/\rho_d$ equals one: to establish a  $3\,\sigma$ deviation from 1, central 
values of
$\rho_s/\rho_d=1.4$ or 0.7 would be needed. The assumed uncertainty of
$\gamma$ of $5^\circ$ could also turn out to be too pessimistic, in which
case even more progress would be needed from the lattice side to match
the experimental accuracy. 

The result in (\ref{rhos-rhod-2010}) would not necessarily suggest that there is no 
physics beyond the SM. In fact, the central values of $\rho_d=0.69\pm0.16$ and 
$\rho_s=0.74\pm0.18$ would both be smaller than 1, i.e.\ would both 
deviate from the SM picture, although the hadronic uncertainties would again
not allow  us to draw definite conclusions. In order to shed further light on these 
possible NP contributions, the exploration of CP-violating effects in the $B_s$-meson 
system is essential, which can be performed with the help of the ``golden"
decay $B^0_s\to J/\psi \phi$.

\begin{figure}[t]
$$\epsfxsize=0.44\textwidth\epsffile{fig10a.epsf}\quad
\epsfxsize=0.44\textwidth\epsffile{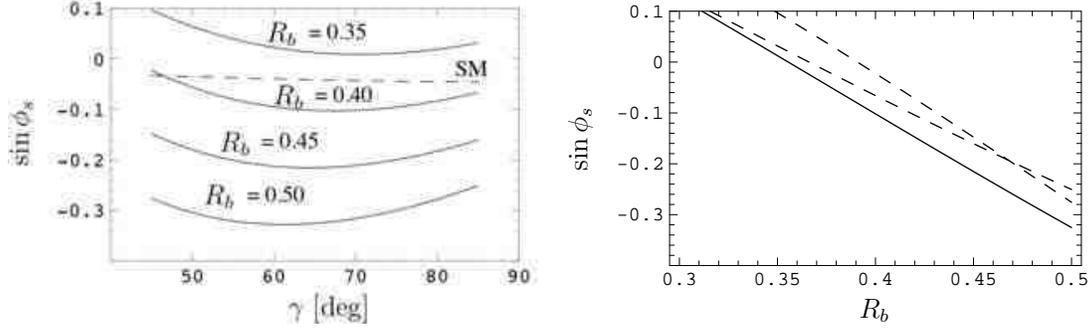}$$
\vspace*{-1cm}
\caption[]{$\sin\phi_s$ for a scenario with flavour-universal NP, i.e.\ $\phi_s^{\rm NP}=
  \phi_d^{\rm NP}$, as specified in Eq.~(\ref{sig-kap-rel}), and $\phi_d=43.4^\circ$.
Left panel: $\sin\phi_s$ as a function of 
$\gamma$ for various values of $R_b$. Right panel: $\sin\phi_s$ 
as a function of $R_b$ for various values of $\gamma$ (solid line:
$\gamma=65^\circ$, dashed lines: $\gamma=(45^\circ,85^\circ)$).
}\label{fig:sinPhis}
\end{figure}

\boldmath
\subsection{$B^0_s\to J/\psi \phi$}\label{ssec:BsPsiPhi}
\unboldmath
As can be seen in Fig.~\ref{fig:BpsiK-diag}, the decay $B^0_s\to J/\psi \phi$
is simply related to $B^0_d\to J/\psi K_{\rm S}$ through a
replacement of the down spectator quark by a strange quark. Consequently,
the structure of the $B^0_s\to J/\psi\phi$ decay amplitude is  
completely analogous to that of (\ref{BdpsiK-ampl2}). On the other hand, the 
final state of  $B^0_s\to J/\psi\phi$ consists of two vector mesons, and is hence
an admixture of different CP eigenstates, which can, however, be disentangled 
through an angular analysis of the $B^0_s\to J/\psi [\to\ell^+\ell^-]\phi[\to K^+K^-]$
decay products \cite{DDF,DDLR}. The corresponding angular distribution 
exhibits tiny direct CP violation, and allows the extraction of
\begin{equation}\label{sinphis}
\sin\phi_s+{\cal O}(\overline{\lambda}^3)=\sin\phi_s+{\cal O}(10^{-3})
\end{equation}
through mixing-induced CP violation.
Since we have $\phi_s=-2\delta\gamma=-2\lambda^2\eta\sim
-2^\circ$ in the SM, the 
determination of this phase from (\ref{sinphis}) is affected by
hadronic uncertainties of ${\cal O}(10\%)$, which may become an issue 
for the LHC era. These uncertainties can be controlled with
the help of flavour-symmetry arguments through the 
$B^0_d\to J/\psi \rho^0$ decay \cite{RF-ang}.

Needless to note, the big hope is that large CP violation
will be found in this channel. Since the CP-violating effects in 
$B^0_s\to J/\psi\phi$ are tiny in the SM, such an observation 
would give us an unambiguous
signal for NP \cite{DFN,NiSi,Branco}. As the situation for NP entering 
through the decay amplitude is similar to $B\to J/\psi K$, we would get 
evidence for CP-violating NP contributions to $B^0_s$--$\bar B^0_s$ mixing, 
and could extract the corresponding sizeable value of $\phi_s$ \cite{DFN}.
Such a scenario may generically arise in the presence of NP with 
$\Lambda_{\rm NP}\sim\mbox{TeV}$ \cite{RF-Phys-Rep}, as well as 
in specific models, including supersymmetric frameworks and models
with extra $Z'$ bosons (see Ref.~\cite{BF-DMs} and references therein).

Thanks to its nice experimental signature, $B^0_s\to J/\psi\phi$ is very accessible
at hadron colliders, and can be fully exploited at the LHC.
After one year of data taking
(which corresponds to
2 $\mbox{fb}^{-1}$), LHCb expects a measurement with the statistical accuracy 
$\sigma_{\rm stat}(\sin\phi_s)\approx 0.031$; adding modes such as 
$B_s\to J/\psi \eta, J/\psi \eta'$ and $\eta_c\phi$, 
$\sigma_{\rm stat}(\sin\phi_s)\approx 0.013$ is expected after
five years \cite{schneider}. Also ATLAS and CMS will contribute to the measurement
of $\sin\phi_s$, expecting uncertainties at the 0.1 level after one year of
data taking, which corresponds to 10 $\mbox{fb}^{-1}$ \cite{smizanska, speer}.
In order to illustrate the impact of NP effects on the quantity
\begin{equation}
\sin\phi_s=\sin(-2\lambda^2R_b\sin\gamma+\phi_s^{\rm NP}),
\end{equation}
let us assume that the NP parameters satisfy the simple relation
\begin{equation}\label{sig-kap-rel}
\sigma_d=\sigma_s,  \quad \kappa_d=\kappa_s, 
\end{equation}
i.e.\ that in particular $\phi_d^{\rm NP}=\phi_s^{\rm NP}$. This scenario would
be supported by (\ref{rhos-rhod-2010}), although it would {\it not} belong to the
class of models with MFV, as new sources of CP violation would be required. 
As we have seen in Subsection~\ref{ssec:BpsiK}, the analysis of the $B^0_d$
data for $R_b^{\rm incl}=0.45$ indicates a small NP phase around
$-10^\circ$ in the $B_d$ system. In the above scenario, that would imply the 
presence of
the same phase in the $B_s$ system, which would interfere constructively 
with the small SM  phase and result in  CP asymmetries at the level of $-20\%$. 
CP-violating effects of that size can easily be detected at the LHC. This 
exercise demonstrates again the great power of the $B_s$-meson system 
to reveal CP-violating NP contributions to $B^0_q$--$\bar B^0_q$ mixing.
The presence of a small NP phase could actually be considerably magnified,
as illustrated in Fig.~\ref{fig:sinPhis}.

\begin{figure}[t] 
$$\epsfxsize=0.40\textwidth\epsffile{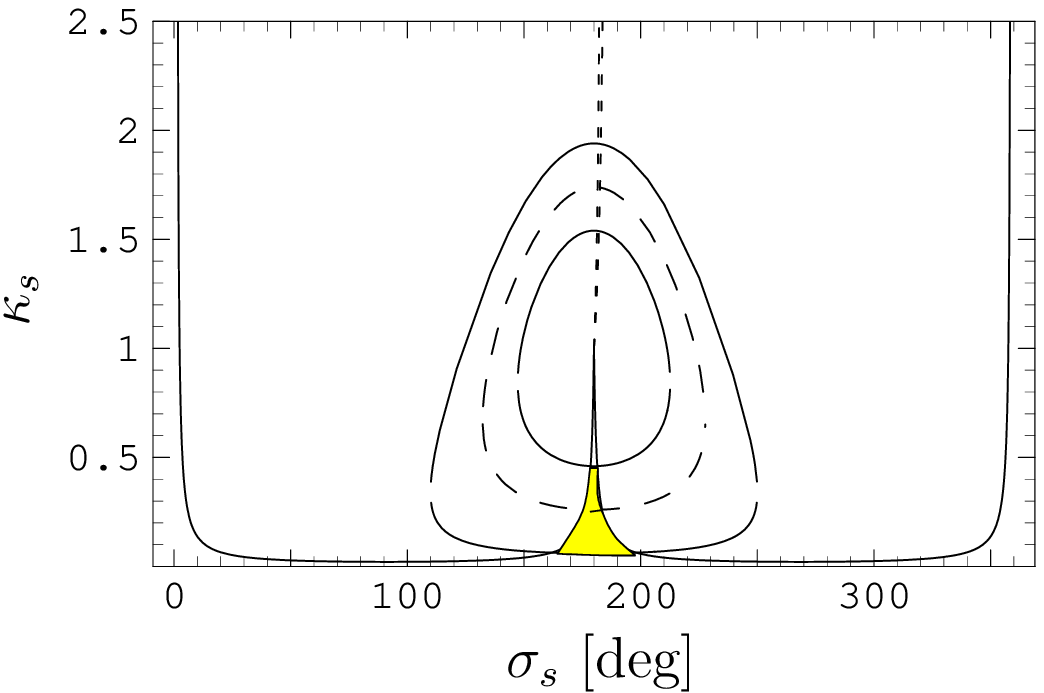}\quad
\epsfxsize=0.40\textwidth\epsffile{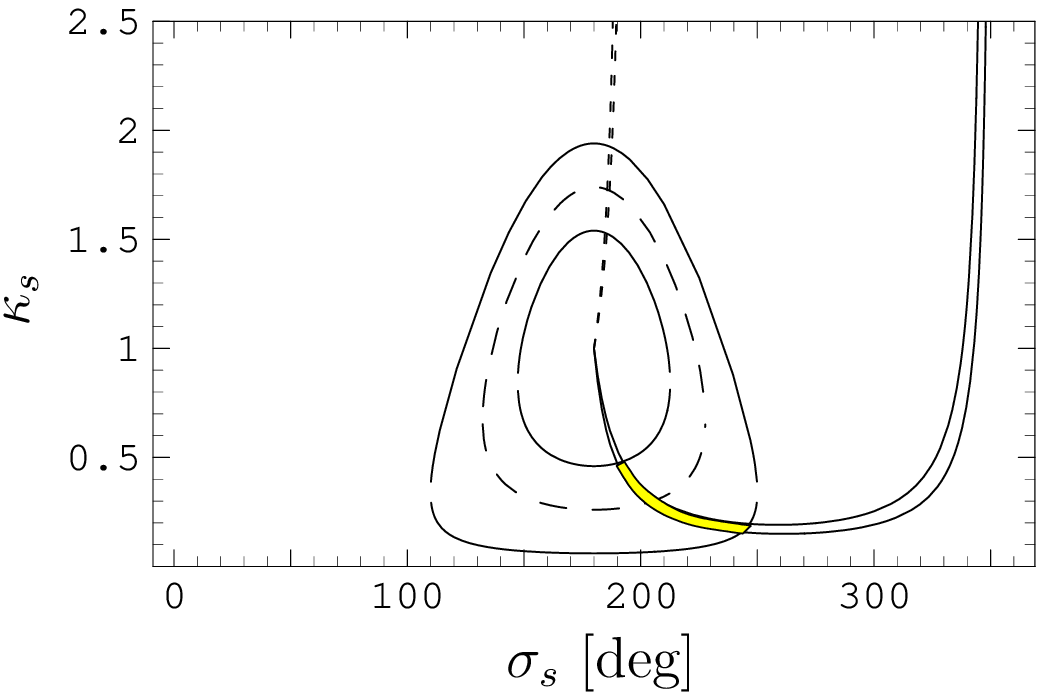}$$
\vspace*{-1cm}
   \caption[]{Combined constraints for the allowed region (yellow/grey) in the 
   $\sigma_s$--$\kappa_s$ plane through $\Delta M_s$ in (\ref{CDF-DMs}) 
   for the (HP+JL)QCD results (\ref{HPQCD}) and CP violation measurements.
   Left panel: the SM scenario $(\sin\phi_s)_{\rm exp}=-0.04\pm0.02$. Right panel: 
   a NP scenario with $(\sin\phi_s)_{\rm exp}=-0.20\pm0.02$. The solid
   lines correspond to $\cos\phi_s>0$, the dotted
   lines to  $\cos\phi_s<0$.}\label{fig:sis-kas-CP}
\end{figure}

Let us finally also discuss the impact of CP violation measurements 
on the allowed region in the $\sigma_s$--$\kappa_s$ plane in our 2010
scenario. To this end, we consider two cases:
\begin{itemize}
\item[i)] $(\sin\phi_s)_{\rm exp}=-0.04\pm0.02$, in accordance with the SM;
\item[ii)] $(\sin\phi_s)_{\rm exp}=-0.20\pm0.02$, in accordance with the NP 
scenario of Fig.~\ref{fig:sinPhis}.
\end{itemize}
The measurement of $\sin\phi_s$ implies a twofold solution for $\phi_s$
and, therefore, also for $\phi_s^{\rm NP}$. However, this ambiguity can 
be resolved through the determination of the sign of $\cos\phi_s$, which
can be fixed through the strategies proposed in Ref.~\cite{DFN}. In 
Fig.~\ref{fig:sis-kas-CP}, we show the situation in the
$\sigma_s$--$\kappa_s$ plane.\footnote{The closed lines agree with those
shown in the right panel of Fig.~\ref{fig:MDs-NP}, as our 2010
scenario is based on the (HP+JL)QCD lattice results.} The dotted lines refer to negative
values of $\cos\phi_s$. Assuming that these are experimentally excluded,
we are left with strongly restricted regions, although $\kappa_s$ could still
take sizeable ranges, with upper bounds $\kappa_s\approx0.5$.
In the SM-like scenario, values of $\sigma_s$ around
$180^\circ$ would arise, i.e.\ a NP contribution with a sign opposite to 
the SM. However, due to the absence of new CP-violating effects, 
the accuracy of lattice results would have to be considerably improved
in order to allow the extraction of a value of  $\kappa_s$ incompatible with 0.
On the other hand, a measurement of $(\sin\phi_s)_{\rm exp}=-0.20\pm0.02$
would give a NP signal at the $10\,\sigma$ level, with $\kappa_s\gsim0.2$.
A determination of  $\kappa_s$ with 10\% uncertainty requires 
$f_{B_s}\hat B_{B_s}^{1/2}$ with 5\% accuracy, i.e.\  the corresponding error 
in (\ref{HPQCD}) has to be reduced by a factor of 2.

Since our discussion  does not refer to a specific model of NP, the 
question arises whether there are actually extensions of the SM that still allow 
large CP-violating NP phases in $B^0_s$--$\bar B^0_s$ mixing. This is in
fact the case, also after the measurement of $\Delta M_s$. In Ref.~\cite{BF-DMs},
where also a comprehensive guide to the relevant literature can be found, 
this exciting feature was illustrated by considering models with an extra $Z'$ boson 
and SUSY scenarios with an approximate alignment of quark and squark masses.

Let us now continue our discussion of the $B_s$-meson system by having
a closer look at other benchmark processes.

\boldmath
\subsection{$B_s\to D_s^\pm K^\mp$ and $B_d\to D^\pm \pi^\mp$}\label{ssec:BsDsK}
\begin{figure}[t]
\centerline{
 \includegraphics[width=5.2truecm]{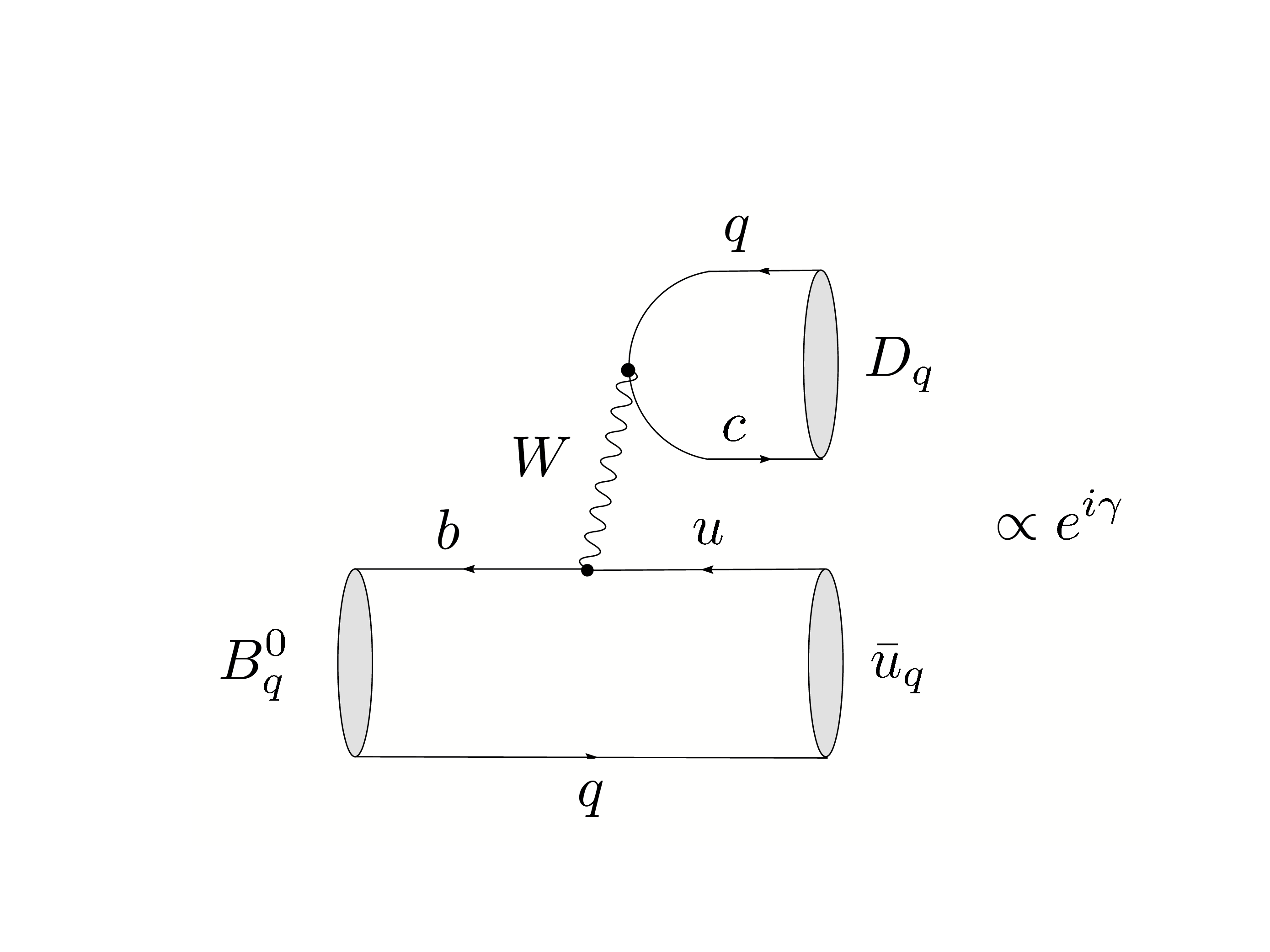}
 \hspace*{0.5truecm}
 \includegraphics[width=5.2truecm]{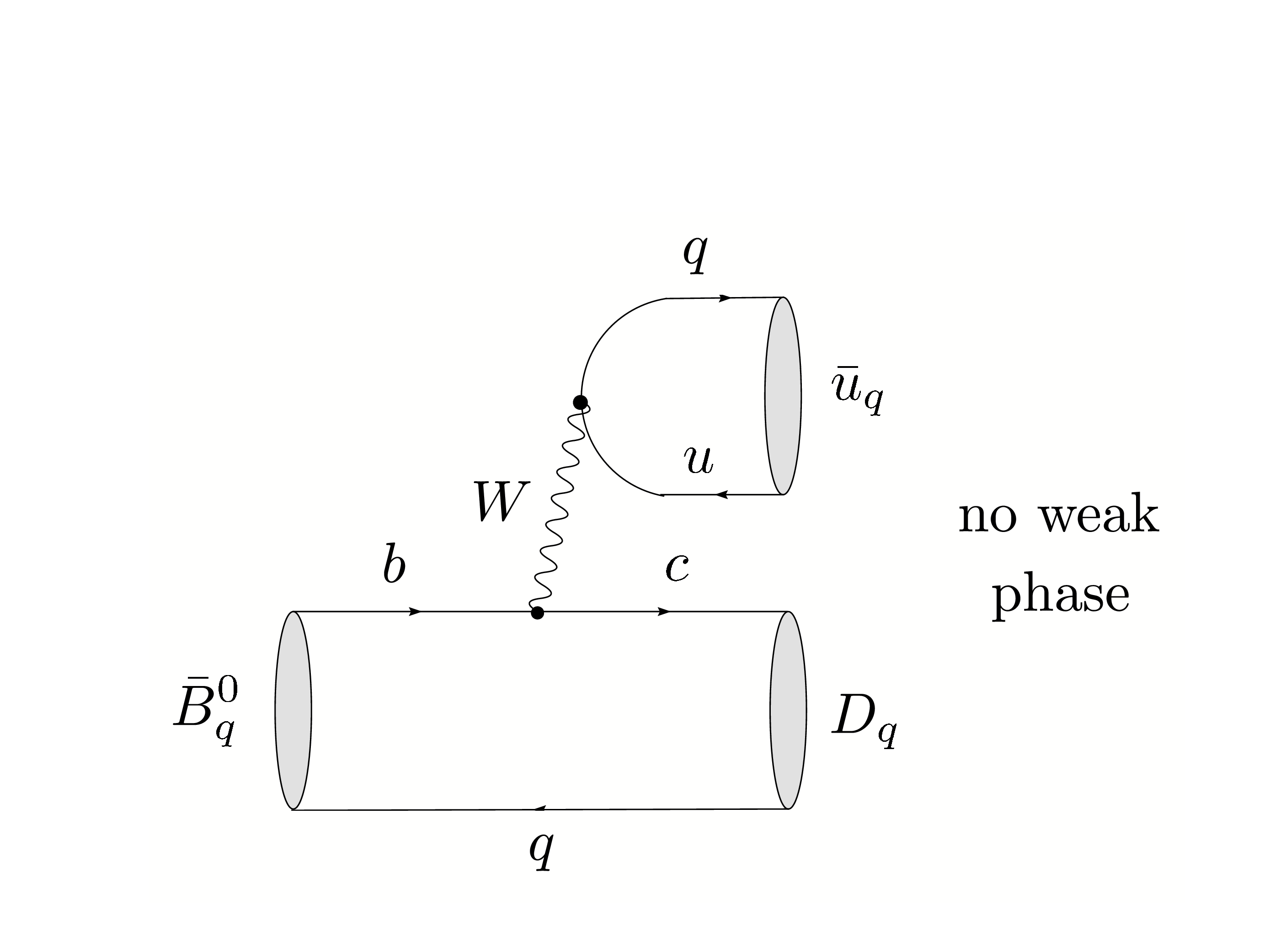}  
 }
 \vspace*{-0.3truecm}
\caption{Feynman diagrams contributing to $B^0_q\to D_q\bar u_q$
and $\bar B^0_q\to D_q \bar u_q$ 
decays.}\label{fig:BqDquq}
\end{figure}

\begin{figure}
\centerline{
 \includegraphics[width=3.0truecm]{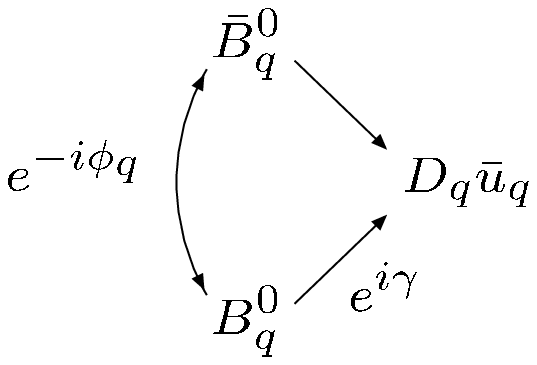} }
 \vspace*{-0.3truecm}
\caption{Interference effects between $B^0_q\to D_q\bar u_q$
and $\bar B^0_q\to D_q\bar u_q$ 
decays.}\label{fig:BqDquq-int}
\end{figure}

\unboldmath
The decays $B_s\to D_s^\pm K^\mp$ \cite{BsDsK} and $B_d\to D^\pm \pi^\mp$
\cite{BdDpi} can be 
treated on the same theoretical basis, and provide new strategies to determine 
$\gamma$ \cite{RF-gam-ca}. Following this paper, we write these modes, which 
are pure ``tree" decays according to the classification of 
Subsection~\ref{sec:class}, generically as $B_q\to D_q \bar u_q$. 
As can be seen from the Feynman diagrams in Fig.~\ref{fig:BqDquq}, their
characteristic feature is that both a $B^0_q$ and a $\bar B^0_q$ meson may decay 
into the same final state $D_q \bar u_q$. Consequently,  as illustrated in 
Fig.~\ref{fig:BqDquq-int}, interference effects between $B^0_q$--$\bar B^0_q$ 
mixing and decay processes arise, which allow us to probe the weak phase 
$\phi_q+\gamma$ through measurements of the corresponding time-dependent
decay rates. 

In the case of $q=s$, i.e.\ $D_s\in\{D_s^+, D_s^{\ast+}, ...\}$ and 
$u_s\in\{K^+, K^{\ast+}, ...\}$, these interference effects are governed 
by a hadronic parameter $X_s e^{i\delta_s}\propto R_b\approx0.4$, where
$R_b\propto |V_{ub}/V_{cb}|$ is the usual UT side, and hence are large. 
On the other hand, for $q=d$, i.e.\ $D_d\in\{D^+, D^{\ast+}, ...\}$ 
and $u_d\in\{\pi^+, \rho^+, ...\}$, the interference effects are described 
by $X_d e^{i\delta_d}\propto -\lambda^2R_b\approx-0.02$, and hence are tiny. 
In the following, we shall only consider $B_q\to D_q \overline{u}_q$ modes, 
where at least one of the $D_q$, $\bar u_q$ states is a pseudoscalar 
meson; otherwise a complicated angular analysis has to be performed.

The time-dependent rate asymmetries of these decays take the same form
as (\ref{ee6}).  It is well known that they allow a {\it theoretically 
clean} determination of $\phi_q+\gamma$, where the ``conventional'' 
approach works as follows \cite{BsDsK,BdDpi}: 
if we measure the observables 
$C(B_q\to D_q\bar u_q)\equiv C_q$ 
and $C(B_q\to \bar D_q u_q)\equiv \overline{C}_q$ provided by the
$\cos(\Delta M_qt)$ pieces, we may determine the following quantities:
\begin{equation}\label{Cpm-def}
\hspace*{-0.9truecm}
\langle C_q\rangle_+\equiv
\frac{1}{2}\left[\overline{C}_q+ C_q\right]=0, \quad
\langle C_q\rangle_-\equiv
\frac{1}{2}\left[\overline{C}_q-C_q\right]=\frac{1-X_q^2}{1+X_q^2},
\end{equation}
where $\langle C_q\rangle_-$ allows us to extract $X_q$. However, to this
end we have to resolve terms entering at the $X_q^2$ level. In the case 
of $q=s$, we have $X_s={\cal O}(R_b)$, implying $X_s^2={\cal O}(0.16)$, so 
that this should actually be possible, though challenging. On the other hand, 
$X_d={\cal O}(-\lambda^2R_b)$ is doubly Cabibbo-suppressed. Although it 
should be possible to resolve terms of ${\cal O}(X_d)$, this will be 
impossible for the vanishingly small $X_d^2={\cal O}(0.0004)$ 
terms, so that other approaches to fix $X_d$ are required
\cite{BdDpi}. For the extraction of $\phi_q+\gamma$, the 
mixing-induced observables $S(B_q\to D_q\bar u_q)\equiv S_q$ and 
$S(B_q\to \bar D_q u_q)\equiv \overline{S}_q$ associated with the
$\sin(\Delta M_qt)$ terms of the time-dependent rate asymmetry must be 
measured. In analogy to (\ref{Cpm-def}), it is convenient to
introduce observable combinations $\langle S_q\rangle_\pm$. Assuming 
that $X_q$ is known, we may consider the quantities
\begin{eqnarray}
s_+&\equiv& (-1)^L
\left[\frac{1+X_q^2}{2 X_q}\right]\langle S_q\rangle_+
=+\cos\delta_q\sin(\phi_q+\gamma)\\
s_-&\equiv&(-1)^L
\left[\frac{1+X_q^2}{2X_q}\right]\langle S_q\rangle_-
=-\sin\delta_q\cos(\phi_q+\gamma),
\end{eqnarray}
which yield
\begin{equation}\label{conv-extr}
\sin^2(\phi_q+\gamma)=\frac{1}{2}\left[(1+s_+^2-s_-^2) \pm
\sqrt{(1+s_+^2-s_-^2)^2-4s_+^2}\right],
\end{equation}
implying an eightfold solution for $\phi_q+\gamma$. If we fix the sign of
$\cos\delta_q$ through factorization,  still a fourfold discrete ambiguity is left,
which is limiting the power for the search of NP significantly.  
Note that this assumption allows us also 
to fix the sign of $\sin(\phi_q+\gamma)$ through $\langle S_q\rangle_+$. 
To this end, the factor $(-1)^L$, where $L$ is the $D_q\bar u_q$ 
angular momentum, has to be properly taken into account. 
This is a crucial issue for the extraction of 
the sign of $\sin(\phi_d+\gamma)$ from $B_d\to D^{\ast\pm}\pi^\mp$ decays.

Let us now discuss new strategies to explore CP violation through 
$B_q\to D_q \bar u_q$ modes, following Ref.~\cite{RF-gam-ca}. 
If $\Delta\Gamma_s$ is sizeable, the ``untagged'' 
rates introduced in (\ref{untagged-rate}) allow us to measure 
${\cal A}_{\rm \Delta\Gamma}(B_s\to D_s\bar u_s)
\equiv {\cal A}_{\rm \Delta\Gamma_s}$ and 
${\cal A}_{\rm \Delta\Gamma}(B_s\to \bar D_s u_s)\equiv 
\overline{{\cal A}}_{\rm \Delta\Gamma_s}$. Introducing, in analogy 
to (\ref{Cpm-def}), observable combinations 
$\langle{\cal A}_{\rm \Delta\Gamma_s}\rangle_\pm$, we may derive the relations
\begin{equation}\label{untagged-extr}
\tan(\phi_s+\gamma)=
-\left[\frac{\langle S_s\rangle_+}{\langle{\cal A}_{\rm \Delta\Gamma_s}
\rangle_+}\right]
=+\left[\frac{\langle{\cal A}_{\rm \Delta\Gamma_s}
\rangle_-}{\langle S_s\rangle_-}\right],
\end{equation}
which allow an {\it unambiguous} extraction of $\phi_s+\gamma$ if we fix
the sign of $\cos\delta_q$ through factorization. 
Another important advantage 
of (\ref{untagged-extr}) is that we do {\it not} have to rely on 
${\cal O}(X_s^2)$ terms, as $\langle S_s\rangle_\pm$ and 
$\langle {\cal A}_{\rm \Delta\Gamma_s}\rangle_\pm$ are proportional to $X_s$.
On the other hand, a sizeable value of $\Delta\Gamma_s$ is of course
needed.

If we keep the hadronic quantities $X_q$ and $\delta_q$  
as ``unknown'', free parameters in the expressions for the
$\langle S_q\rangle_\pm$, we may obtain bounds on $\phi_q+\gamma$ from
\begin{equation}
|\sin(\phi_q+\gamma)|\geq|\langle S_q\rangle_+|, \quad
|\cos(\phi_q+\gamma)|\geq|\langle S_q\rangle_-|.
\end{equation}
If $X_q$ is known, stronger constraints are implied by 
\begin{equation}\label{bounds}
|\sin(\phi_q+\gamma)|\geq|s_+|, \quad
|\cos(\phi_q+\gamma)|\geq|s_-|.
\end{equation}
Once $s_+$ and $s_-$ are known, we may of course determine
$\phi_q+\gamma$ through the ``conventional'' approach, using 
(\ref{conv-extr}). However, the bounds following from 
(\ref{bounds}) provide essentially the same information 
and are much simpler to 
implement. Moreover, as discussed in detail in Ref.~\cite{RF-gam-ca}
for several examples within the SM, the bounds following from the $B_s$ and 
$B_d$ modes may be highly complementary, thereby providing particularly 
narrow, theoretically clean ranges for $\gamma$. 

Let us now further exploit the complementarity between the 
$B_s^0\to D_s^{(\ast)+}K^-$ and $B_d^0\to D^{(\ast)+}\pi^-$ processes.
Looking at the corresponding decay topologies, we see that
these channels are related to each other through an interchange of 
all down and strange quarks. Consequently, applying again the $U$-spin 
symmetry implies $a_s=a_d$ and $\delta_s=\delta_d$, where $a_s\equiv X_s/R_b$ 
and $a_d\equiv -X_d/(\lambda^2 R_b)$ are the ratios of the hadronic matrix elements 
entering $X_s$ and $X_d$, respectively. There are various possibilities 
to implement these relations \cite{RF-gam-ca}. A particularly simple
picture arises if we assume that $a_s=a_d$ {\it and} $\delta_s=\delta_d$, 
which yields
\begin{equation}
\tan\gamma=-\left[\frac{\sin\phi_d-S
\sin\phi_s}{\cos\phi_d-S\cos\phi_s}
\right]\stackrel{\phi_s=0^\circ}{=}
-\left[\frac{\sin\phi_d}{\cos\phi_d-S}\right].
\end{equation}
Here we have introduced
\begin{equation}
S\equiv-R\left[\frac{\langle S_d\rangle_+}{\langle S_s\rangle_+}\right]
\end{equation}
with
\begin{equation}
R\equiv\left(\frac{1-\lambda^2}{\lambda^2}\right)
\left[\frac{1}{1+X_s^2}\right],
\end{equation}
where $R$ can be fixed with the help of untagged $B_s$ rates through
\begin{equation}
R=\left(\frac{f_K}{f_\pi}\right)^2 \left[
\frac{\Gamma(\bar B^0_s \to D_s^{(\ast)+}\pi^-)+
\Gamma(B^0_s\to D_s^{(\ast)-}\pi^+)}{\langle\Gamma(B_s\to D_s^{(\ast)+}K^-)
\rangle+\langle\Gamma(B_s\to D_s^{(\ast)-}K^+)\rangle}\right].
\end{equation}
Alternatively, we can {\it only} assume that $\delta_s=\delta_d$ {\it or} 
that $a_s=a_d$ \cite{RF-gam-ca}. An important feature of this strategy
is that it allow us to extract an {\it unambiguous} value of $\gamma$, 
which is crucial for the search of NP; first studies for LHCb are very promising 
in this respect \cite{wilkinson-CKM}.
Another advantage with respect to the ``conventional'' approach is that 
$X_q^2$ terms have not to be resolved experimentally. In 
particular, $X_d$ does {\it not} have to be fixed, and $X_s$ may only enter 
through a $1+X_s^2$ correction, which can straightforwardly be determined 
through untagged $B_s$ rate measurements. In the most refined implementation 
of this strategy, the measurement of $X_d/X_s$ would only be interesting for 
the inclusion of $U$-spin-breaking corrections in $a_d/a_s$. Moreover, we may 
obtain interesting insights into hadron dynamics and $U$-spin breaking. 

The colour-suppressed counterparts
of the $B_q\to D_q \bar u_q$ modes are also interesting
for the exploration of CP violation. 
In the case of the $B_d\to D K_{\rm S(L)}$, $B_s\to D \eta^{(')}, D \phi$, ...\
modes, the interference effects between $B^0_q$--$\bar B^0_q$ mixing
and decay processes are governed by $x_{f_s}e^{i\delta_{f_s}}\propto R_b$.
If we consider the CP eigenstates $D_\pm$ of the neutral $D$-meson system, 
we obtain additional interference effects at the amplitude level, which involve 
$\gamma$, and may introduce the following ``untagged'' rate asymmetry 
\cite{RF-BdDpi0}:
\begin{equation}
\Gamma_{+-}^{f_s}\equiv
\frac{\langle\Gamma(B_q\to D_+ f_s)\rangle-\langle
\Gamma(B_q\to D_- f_s)\rangle}{\langle\Gamma(B_q\to D_+ f_s)\rangle
+\langle\Gamma(B_q\to D_- f_s)\rangle},
\end{equation}
which allows us to constrain $\gamma$ through the relation
\begin{equation}
|\cos\gamma|\geq |\Gamma_{+-}^{f_s}|. 
\end{equation}
Moreover, if we complement
$\Gamma_{+-}^{f_s}$ with 
\begin{equation}
\langle S_{f_s}\rangle_\pm\equiv \frac{1}{2}\left[S_+^{f_s}\pm S_-^{f_s}\right],
\end{equation}
where $S_\pm^{f_s}\equiv {\cal A}_{\rm CP}^{\rm mix}(B_q\to D_\pm f_s)$,
we may derive the following simple but {\it exact} relation:
\begin{equation}
\tan\gamma\cos\phi_q=
\left[\frac{\eta_{f_s} \langle S_{f_s}
\rangle_+}{\Gamma_{+-}^{f_s}}\right]+\left[\eta_{f_s}\langle S_{f_s}\rangle_--
\sin\phi_q\right],
\end{equation}
with $\eta_{f_s}\equiv(-1)^L\eta_{\rm CP}^{f_s}$. This expression allows 
a conceptually simple, theoretically clean and essentially unambiguous 
determination of $\gamma$ \cite{RF-BdDpi0}. Since the interference effects are 
governed by the tiny parameter $x_{f_d}e^{i\delta_{f_d}}\propto -\lambda^2R_b$
in the case of $B_s\to D_\pm K_{\rm S(L)}$, 
$B_d\to D_\pm \pi^0, D_\pm \rho^0, ...$, these modes are not as interesting 
for the extraction of $\gamma$. However, they provide the relation
\begin{equation}
\eta_{f_d}\langle S_{f_d}\rangle_-=\sin\phi_q + {\cal O}(x_{f_d}^2)
=\sin\phi_q + {\cal O}(4\times 10^{-4}),
\end{equation}
allowing very interesting determinations of $\phi_q$ with theoretical 
accuracies one order of magnitude higher than those of
the conventional  $B^0_d\to J/\psi K_{\rm S}$ and $B^0_s\to J/\psi \phi$
approaches~\cite{RF-BdDpi0}. As we pointed out in Subsection~\ref{ssec:BpsiK},
these measurements would be very interesting in view of the new world
average of $(\sin2\beta)_{\psi K_{\rm S}}$.

\boldmath
\subsection{$B^0_s\to K^+K^-$ and $B^0_d\to\pi^+\pi^-$}
\begin{figure}[t]
   \centering
   \includegraphics[width=7.4truecm]{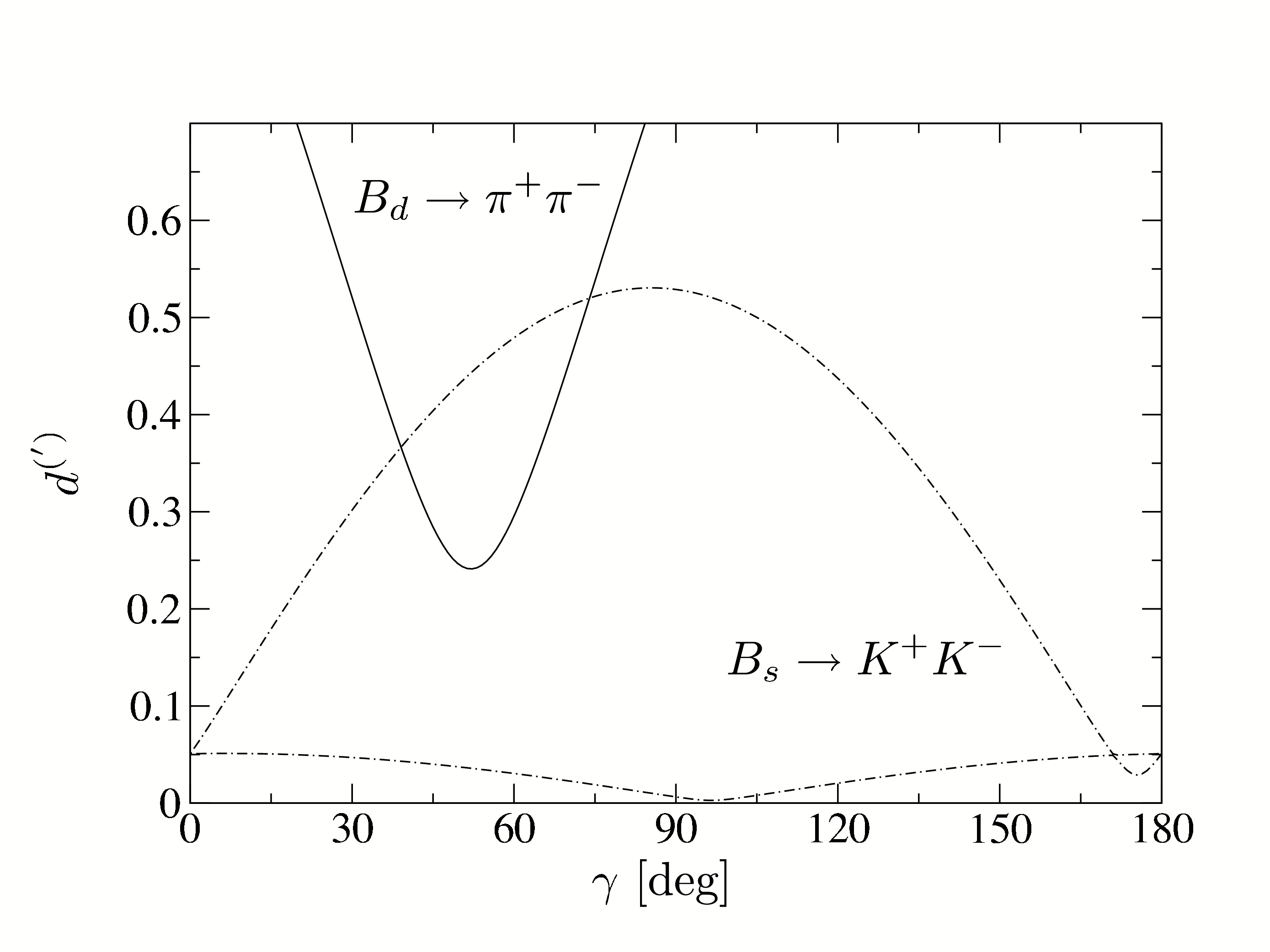} 
   \vspace*{-0.6truecm}
   \caption{The contours in the $\gamma$--$d^{(')}$ plane for an example with
   $d=d'=0.52$, $\theta=\theta'=146^\circ$, $\phi_d=43.4^\circ$, $\phi_s=-2^\circ$,
   $\gamma=74^\circ$, which corresponds to the CP asymmetries
   ${\cal A}_{\rm CP}^{\rm dir}(B_d\to\pi^+\pi^-)=-0.37$ and 
   ${\cal A}_{\rm CP}^{\rm mix}(B_d\to\pi^+\pi^-)=+0.50$
   (see Subsections~\ref{ssec:Bpi+pi-} and \ref{ssec:Bpipi-hadr}), as well as
   ${\cal A}_{\rm CP}^{\rm dir}(B_s\to K^+K^-)=+0.12$ and
   ${\cal A}_{\rm CP}^{\rm mix}(B_s\to K^+K^-)=-0.19$.}\label{fig:Bs-Bd-contours}
\end{figure}

\unboldmath
The decay $B^0_s\to K^+K^-$ is a $\bar b \to \bar s$ transition, and
involves tree and penguin amplitudes, as the $B^0_d\to\pi^+\pi^-$ mode 
\cite{RF-BsKK}. However, because of the different CKM structure, the latter 
topologies play actually the dominant r\^ole in the $B^0_s\to K^+K^-$ channel. 
In analogy to (\ref{Bpipi-ampl}), we may write
\begin{equation}\label{BsKK-ampl}
A(B_s^0\to K^+K^-)=\sqrt{\epsilon} \,\, {\cal C}'
\left[e^{i\gamma}+\frac{1}{\epsilon}\,d'e^{i\theta'}\right],
\end{equation}
where $\epsilon$ was introduced in (\ref{eps-def}), and 
the CP-conserving hadronic parameters ${\cal C}'$ and $d'e^{i\theta'}$ 
correspond to ${\cal C}$ and $de^{i\theta}$, respectively. The corresponding
observables take then the following generic form:
\begin{eqnarray}
{\cal A}_{\rm CP}^{\rm dir}(B_s\to K^+K^-)&=&
G_1'(d',\theta';\gamma) \label{CP-BsKK-dir-gen}\\
{\cal A}_{\rm CP}^{\rm mix}(B_s\to K^+K^-)&=&
G_2'(d',\theta';\gamma,\phi_s),\label{CP-BsKK-mix-gen}
\end{eqnarray}
in analogy to the expressions for the CP-violating $B^0_d\to\pi^+\pi^-$
asymmetries in (\ref{CP-Bpipi-dir-gen}) and (\ref{CP-Bpipi-mix-gen}). 
Since $\phi_d=(43.4\pm 2.5)^\circ$ is already known (see
Subsection~\ref{ssec:BpsiK}) and $\phi_s$ is negligibly small
in the SM -- or can be determined through $B^0_s\to J/\psi \phi$ should CP-violating
NP contributions to $B^0_s$--$\bar B^0_s$ mixing make it sizeable -- 
we may convert the measured values of 
${\cal A}_{\rm CP}^{\rm dir}(B_d\to \pi^+\pi^-)$, 
${\cal A}_{\rm CP}^{\rm mix}(B_d\to \pi^+\pi^-)$ and
${\cal A}_{\rm CP}^{\rm dir}(B_s\to K^+K^-)$, 
${\cal A}_{\rm CP}^{\rm mix}(B_s\to K^+K^-)$ into {\it theoretically clean}
contours in the $\gamma$--$d$ and $\gamma$--$d'$ planes, respectively.
In Fig.~\ref{fig:Bs-Bd-contours}, we show these contours for an example,
which corresponds to the central values of (\ref{Bpipi-CP-averages}) and 
(\ref{Bpipi-CP-averages2}) with the hadronic parameters $(d,\theta)$
in (\ref{Bpipi-par-det}). 

As can be seen in Fig.~\ref{fig:Bpipi-diag}, the decay 
$B^0_d\to\pi^+\pi^-$ is actually related to $B^0_s\to K^+K^-$ through the interchange 
of {\it all} down and strange quarks. Consequently, each decay topology contributing
to $B^0_d\to\pi^+\pi^-$ has a counterpart in $B^0_s\to K^+K^-$, and
the corresponding hadronic parameters can be related to each other
with the help of the $U$-spin flavour symmetry of strong interactions,
implying the following relations \cite{RF-BsKK}:
\begin{equation}\label{U-spin-rel}
d'=d, \quad \theta'=\theta.
\end{equation}
Applying the former, we may extract $\gamma$ and $d$ through the 
intersections of the theoretically clean $\gamma$--$d$ and $\gamma$--$d'$ 
contours. As discussed in Ref.~\cite{RF-BsKK}, it is also possible to resolve 
straightforwardly the 
twofold ambiguity for $(\gamma,d)$ arising in Fig.~\ref{fig:Bs-Bd-contours},
thereby leaving us with the ``true" solution of $\gamma=74^\circ$ in
this example. Moreover, we may determine $\theta$ and $\theta'$, which
allow an interesting internal consistency check of the second $U$-spin relation 
in (\ref{U-spin-rel}). An alternative avenue is provided if we eliminate $d$ and 
$d'$ through the CP-violating $B_d\to\pi^+\pi^-$ and $B_s\to K^+K^-$
observables, respectively, and extract then these parameters and $\gamma$ 
through the $U$-spin relation $\theta'=\theta$.

\begin{figure}
   \centering
   \includegraphics[width=6.6truecm]{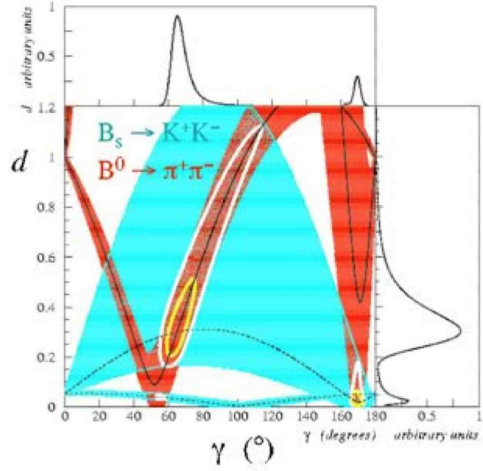} 
   \vspace*{-0.6truecm}
   \caption{Experimental LHCb feasibility study for the contours in the
   $\gamma$--$d^{(')}$ plane, as discussed in 
   Ref.~\cite{LHCb-analyses}.}\label{fig:Bs-Bd-contours-LHCb}
\end{figure}

As illustrated in Fig.~\ref{fig:Bs-Bd-contours-LHCb},
this strategy is very promising from an experimental point of view
for the LHCb experiment, where an accuracy for $\gamma$ of a few degrees
can be achieved \cite{LHC-Book,schneider,LHCb-analyses}. As far as 
possible $U$-spin-breaking 
corrections to $d'=d$ are concerned, they enter the determination of $\gamma$ 
through a relative shift of the $\gamma$--$d$ and $\gamma$--$d'$ contours; 
their impact on the extracted value of $\gamma$ therefore depends on the form 
of these curves, which is fixed through the measured observables. In the examples discussed in Refs.~\cite{RF-Phys-Rep,RF-BsKK}, as well as in the one
 shown in 
Fig.~\ref{fig:Bs-Bd-contours}, the extracted value of $\gamma$ would be very 
stable under such effects. Let us also note that the $U$-spin
relations in (\ref{U-spin-rel}) are particularly robust since they involve only
ratios of hadronic amplitudes, where all $SU(3)$-breaking decay constants
and form factors cancel in factorization and also chirally enhanced terms
would not lead to  $U$-spin-breaking corrections \cite{RF-BsKK}. 
On the other hand, the ratio $|{\cal C}'/{\cal C}|$, which equals 1 in the strict 
$U$-spin limit and enters the $U$-spin relation
\begin{equation}
\frac{{\cal A}_{\rm CP}^{\rm mix}
(B_s\to K^+K^-)}{{\cal A}_{\rm CP}^{\rm dir}(B_d\to\pi^+\pi^-)}=
-\left|\frac{{\cal C}'}{{\cal C}}\right|^2
\left[\frac{\mbox{BR}(B_d\to\pi^+\pi^-)}{\mbox{BR}(B_s\to K^+K^-)}\right]
\frac{\tau_{B_s}}{\tau_{B_d}},
\end{equation}
is affected by $U$-spin-breaking effects within factorization. An 
estimate of the corresponding form factors was recently performed
in Ref.~\cite{KMM} with the help of QCD sum rules, which is an important 
ingredient for a SM prediction of the CP-averaged $B_s\to K^+K^-$ branching 
ratio \cite{BFRS3}. Following these lines, the prediction
\begin{equation}\label{BsKK-pred}
\mbox{BR}(B_s\to K^+K^-)=(35\pm7)\times 10^{-6}
\end{equation}
was obtained in Refs.~\cite{BFRS3,BFRS-up} from the CP-averaged 
$B_d\to\pi^\mp K^\pm$ branching ratio. On the other hand, the CDF collaboration 
announced recently the observation of the $B_s\to K^+K^-$ channel, with the following 
branching ratio \cite{CDF-BsKK}:
\begin{equation}
\mbox{BR}(B_s\to K^+K^-)=(33\pm5.7\pm6.7)\times 10^{-6},
\end{equation}
which is in excellent accordance  with (\ref{BsKK-pred}).
For other recent analyses of the $B_s\to K^+K^-$ decay, see Refs.~\cite{safir,BLMV}.

In addition to the $B_s\to K^+K^-$, $B_d\to\pi^+\pi^-$ and
$B_s\to D_s^\pm K^\mp$, $B_d\to D^\pm \pi^\mp$ strategies discussed
above, also other $U$-spin methods for the extraction of $\gamma$ were 
proposed, using 
$B_{s(d)}\to J/\psi K_{\rm S}$ or $B_{d(s)}\to D_{d(s)}^+D_{d(s)}^-$ 
\cite{RF-BdsPsiK}, $B_{d(s)}\to K^{0(*)}\bar K^{0(*)}$ \cite{RF-Phys-Rep,RF-ang}, 
$B_{(s)}\to \pi K$ \cite{GR-BspiK}, or $B_{s(d)}\to J/\psi \eta$
modes \cite{skands}. In a very recent paper \cite{SoSu}, also two-body decays 
of charged $B$ mesons were considered.

\boldmath
\subsection{$B^0_s\to\mu^+\mu^-$ and $B^0_d\to\mu^+\mu^-$}\label{ssec:Bmumu}
\begin{figure}[t]
\centerline{
 \includegraphics[width=4.6truecm]{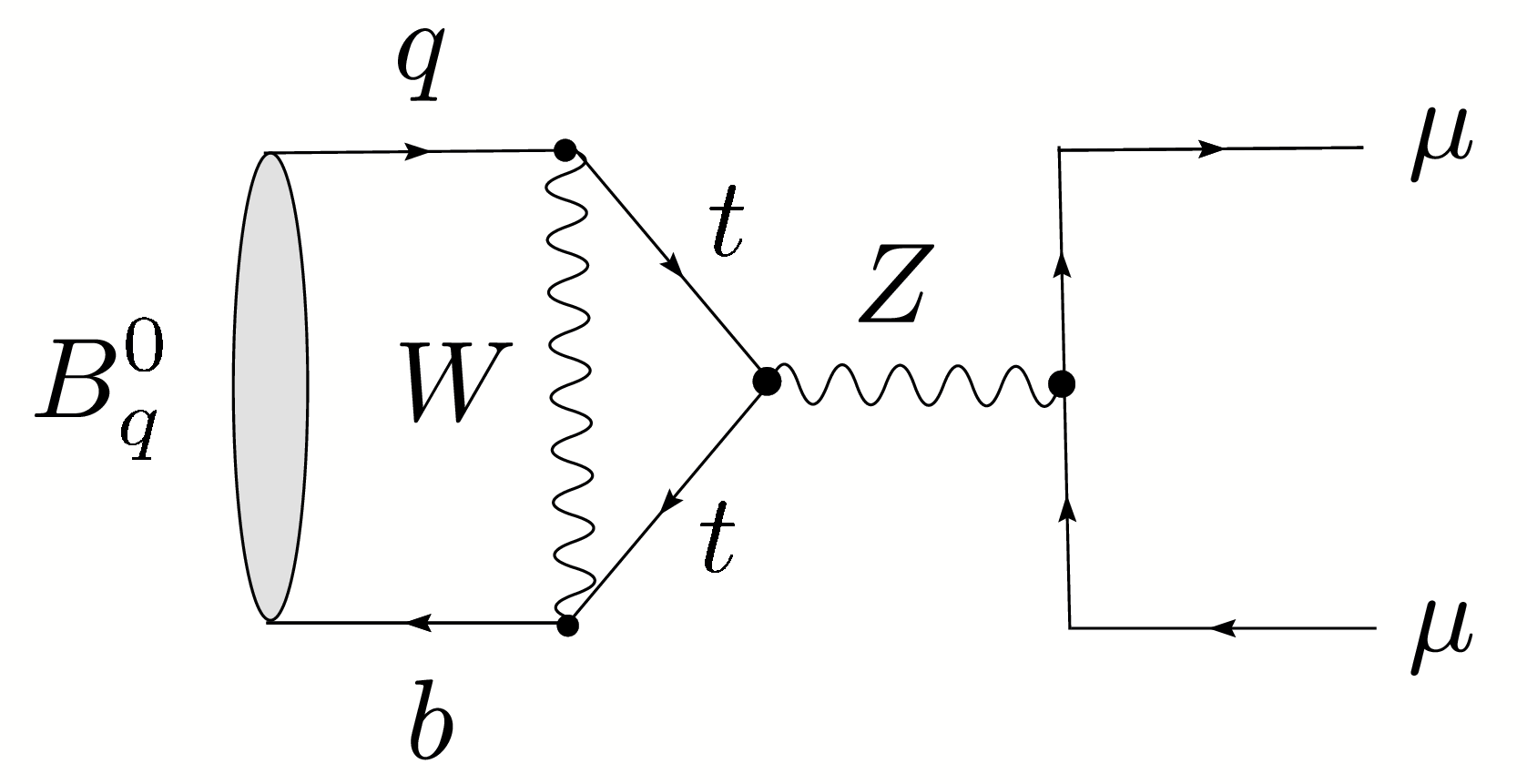}
 \hspace*{0.5truecm}
 \includegraphics[width=4.3truecm]{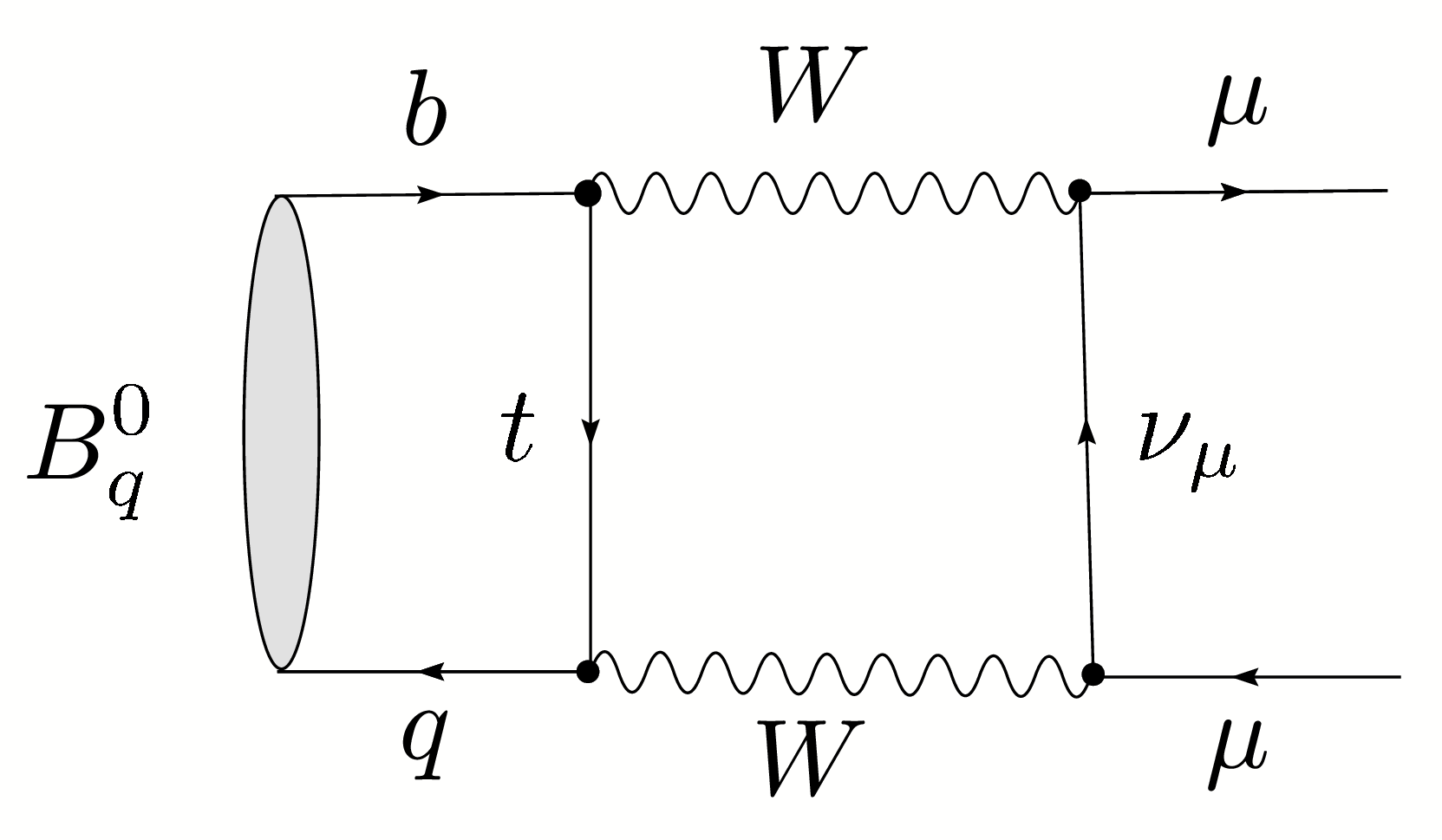}  
 }
\caption{Feynman diagrams contributing to 
$B^0_q\to \mu^+\mu^-$ ($q\in\{s,d\}$).}\label{fig:Bqmumu}
\end{figure}

\unboldmath
Let us finally have a closer look at the rare decay $B^0_s\to\mu^+\mu^-$,
which we encountered already briefly in Subsection~\ref{ssec:rareKB}. 
As can be seen in Fig.~\ref{fig:Bqmumu}, this decay and its $B_d$-meson
counterpart $B^0_d\to\mu^+\mu^-$ originate from $Z^0$-penguin and
box diagrams in the SM. The corresponding low-energy effective Hamiltonian 
is given as follows \cite{B-LH98}:
\begin{equation}\label{Heff-Bmumu}
{\cal H}_{\rm eff}=-\frac{G_{\rm F}}{\sqrt{2}}\left[
\frac{\alpha}{2\pi\sin^2\Theta_{\rm W}}\right]
V_{tb}^\ast V_{tq} \eta_Y Y_0(x_t)(\bar b q)_{\rm V-A}(\bar\mu\mu)_{\rm V-A} 
\,+\, {\rm h.c.},
\end{equation}
where $\alpha$ denotes the QED coupling and $\Theta_{\rm W}$ is the
Weinberg angle. The short-distance physics is described by 
$Y(x_t)\equiv\eta_Y Y_0(x_t)$, where $\eta_Y=1.012$ is a perturbative 
QCD correction \cite{BB-Bmumu}--\cite{MiU}, and the Inami--Lim function
$Y_0(x_t)$ describes the top-quark mass dependence. We observe that
only the matrix element $\langle 0| (\bar b q)_{\rm V-A}|B^0_q\rangle$ 
is required. Since here the vector-current piece vanishes, as
the $B^0_q$ is a pseudoscalar meson, this matrix element is simply
given by the decay constant $f_{B_q}$. 
Consequently, we arrive at a very favourable 
situation with respect to the hadronic matrix elements. Since, moreover, 
NLO QCD corrections were calculated, and long-distance contributions are 
expected to play a negligible r\^ole \cite{BB-Bmumu}, the $B^0_q\to\mu^+\mu^-$ 
modes belong to the cleanest rare $B$ decays. The SM branching ratios
can then be written in the following compact form \cite{Brev01}:
\begin{eqnarray}
\lefteqn{\mbox{BR}( B_s \to \mu^+ \mu^-) = 4.1 \times 10^{-9}}\nonumber\\
&&\qquad\qquad\times \left[\frac{f_{B_s}}{0.24 \, \mbox{GeV}} \right]^2 \left[
\frac{|V_{ts}|}{0.040} \right]^2 \left[
\frac{\tau_{B_s}}{1.5 \, \mbox{ps}} \right] \left[ \frac{m_t}{167 
\, \mbox{GeV} } \right]^{3.12}\label{BR-Bsmumu}\\
\lefteqn{\mbox{BR}(B_d \to \mu^+ \mu^-) = 1.1 \times 10^{-10}}\nonumber\\
&&\qquad\qquad\times\left[ \frac{f_{B_d}}{0.20 \, \mbox{GeV}} \right]^2 \left[
\frac{|V_{td}|}{0.008} \right]^2
\left[ \frac{\tau_{B_d}}{1.5 \, \mbox{ps}} \right] \left[
\frac{m_t}{167 \, \mbox{GeV} } \right]^{3.12}.\label{BR-Bdmumu}
\end{eqnarray}
The most recent upper bounds (95\% C.L.) from the CDF collaboration
read as follows
\cite{CDF-Bmumu}:
\begin{equation}\label{Bmumu-exp-CDF}
\mbox{BR}(B_s\to\mu^+\mu^-)<1.0\times10^{-7}, \quad
\mbox{BR}(B_d\to\mu^+\mu^-)<3.0 \times10^{-8},
\end{equation}
while the D0 collaboration finds the following (95\% C.L.) upper limit \cite{D0-Bmumu}:
\begin{equation}\label{Bmumu-exp-D0}
\mbox{BR}(B_s\to\mu^+\mu^-)<3.7\times10^{-7}.
\end{equation}

Using again relation (\ref{Rt-def2}) and neglecting the tiny corrections entering
at the $\lambda^2$ level, we find that the measurement of the ratio
\begin{equation}\label{RT1-rare}
\frac{\mbox{BR}(B_d\to\mu^+\mu^-)}{\mbox{BR}(B_s\to\mu^+\mu^-)}=
\left[\frac{\tau_{B_d}}{\tau_{B_s}}\right]
\left[\frac{M_{B_d}}{M_{B_s}}\right]
\left[\frac{f_{B_d}}{f_{B_s}}\right]^2
\left|\frac{V_{td}}{V_{ts}}\right|^2
\end{equation}
would allow an extraction of the UT side $R_t$. Since the short-distance
function $Y$ cancels, this determination does not only work in the SM,
but also in the NP scenarios with MFV \cite{buras-MFV}. This
strategy is complementary to that offered by the ratio $\Delta M_s/\Delta M_d$
discussed in the context of (\ref{DMs-DMd-rat}). If we look at this
expression  in the MFV case, where $\rho_s/\rho_d=1$, 
and  (\ref{RT1-rare}), we see that the following relation is implied \cite{Buras-rel}:
\begin{equation}\label{Bmumu-DM-rel}
\frac{\mbox{BR}(B_s\to\mu^+\mu^-)}{\mbox{BR}(B_d\to\mu^+\mu^-)}=
\left[\frac{\tau_{B_s}}{\tau_{B_d}}\right]
\left[\frac{\hat B_{B_d}}{\hat B_{B_s}}\right]
\left[\frac{\Delta M_s}{\Delta M_d}\right],
\end{equation}
which holds again in the context of MFV models, including the SM. 
Here the advantage is that the dependence on $(f_{B_d}/f_{B_s})^2$ cancels. 
Moreover, we may also use the data for the mass differences
$\Delta M_q$ to reduce the hadronic uncertainties 
of the SM predictions of the $B_q\to\mu^+\mu^-$ branching ratios 
\cite{Buras-rel}:
\begin{eqnarray}
\mbox{BR}(B_s \to \mu^+ \mu^-) &=& (3.35 \pm 0.32)\times\times 
10^{-9}\label{Bsmumu-pred}\\
\mbox{BR}(B_d\to \mu^+ \mu^-) &=& (1.03 \pm 0.09)\times 10^{-10},
\end{eqnarray}
where (\ref{Bsmumu-pred}) is another application of the recent $\Delta M_s$ 
measurement at the Tevatron \cite{BBGT}.

The current experimental upper bounds in (\ref{Bmumu-exp-CDF}) 
and  (\ref{Bmumu-exp-D0}) are still about two
orders of magnitude away from these numbers. 
Consequently, should the $B_q \to \mu^+ \mu^-$ decays 
be governed by their SM contributions, we could only 
hope to observe them at the LHC \cite{LHC-Book}.
On the other hand, since the $B_q \to \mu^+ \mu^-$ transitions originate from 
FCNC processes, they are sensitive probes of NP. In particular, 
the branching ratios may be dramatically enhanced in specific NP (SUSY) 
scenarios, as was recently reviewed in Ref.~\cite{buras-NP}. Should this 
actually be the case, these decays may already be seen at run II of the 
Tevatron, and the $e^+e^-$ $B$ factories could observe $B_d\to \mu^+ \mu^-$. 
Let us finally emphasize that the experimental bounds on 
$B_s\to\mu^+\mu^-$ can also be converted into bounds on NP parameters
in specific scenarios. In the context of the constrained minimal 
supersymmetric extension 
of the SM (CMSSM) with universal scalar masses, such constraints were
recently critically  discussed by the authors of Ref.~\cite{EOS}.

\section{CONCLUSIONS AND OUTLOOK}\label{sec:concl}
\setcounter{equation}{0}
CP violation is now well established in the $B$-meson system, thereby 
complementing the neutral $K$-meson system, where this phenomenon
was discovered more than 40 years ago. The data of the $e^+e^-$ $B$ factories 
have provided valuable insights into the physics of strong and weak interactions. Concerning the former aspect, which is sometimes only considered as 
a
by-product, the data give us important evidence for large non-factorizable 
effects in non-leptonic $B$-decays, so that the challenge for a reliable theoretical 
description within dynamical QCD approaches remains, despite interesting 
recent progress. As far as the latter aspect is concerned, the description of
CP violation through the KM mechanism has successfully passed its first
experimental tests, in particular through the comparison between the 
measurement of $\sin 2\beta$ with the help of $B^0_d\to J/\psi K_{\rm S}$ and 
the CKM fits. However, the most recent average for $(\sin2\beta)_{\psi K_{\rm S}}$ 
is now somewhat on the lower side, and there are a couple of puzzles in the
$B$-factory data. It will be very interesting to monitor these effects, which
could be first hints for physics beyond the SM, as the data improve. Moreover,
it is crucial to refine the corresponding theoretical analyses further, to have a 
critical look at the underlying working assumptions and to check them through
independent tests, and to explore correlations with other flavour probes. 

Despite this impressive progress, there are still regions of the 
$B$-physics landscape left that are essentially unexplored. 
For instance, $b\to d$ penguin processes are now entering the 
stage, since lower bounds for the corresponding branching ratios 
that can be derived in the SM turn out to be very close to
the corresponding experimental upper limits. Indeed, we have now
evidence for the $B_d\to K^0\bar K^0$ and $B^\pm\to K^\pm K$ channels,
and the first signals for the radiative $B\to\rho\gamma$ transitions
were reported, representing one of the hot topics of the summer of 2005. 
These modes have now to be explored in much more detail, and several other
decays are waiting to be observed. 

Another very interesting aspect of future studies is the $B_s$-meson system.
Although the mass difference $\Delta M_s$ could eventually be measured
in the spring of 2006 at the Tevatron, many features of $B_s$ physics
are still essentially unexplored. Concerning the measurement of $\Delta M_s$,
NP may actually be hiding in this quantity, but is currently obscured by parameter
uncertainties. The somking-gun signal for NP in $B^0_s$--$\bar B^0_s$
mixing would be the observation of sizeable CP violation in $B^0_s\to J/\psi \phi$
and similar decays. Since there are various specific extensions of the
SM where such effects arise (also when taking the $\Delta M_s$ constraints into
account), we may hope that the LHC will detect them. Moreover, the $B_s$-meson
system allows several determinations of the angle $\gamma$ of the UT in an 
essentially unambiguous way, which are another key ingredient for the search
of NP, and offers further tests of the SM through strongly suppressed rare decays. 
After new results from run II of the Tevatron, the promising physics potential of
the $B_s$-meson system can be fully exploited at the LHC, in particular by 
the LHCb experiment. 

These studies can nicely be complemented through the kaon system, which 
governed the stage of CP violation for more than 35 years. The future lies now 
on rare decays, in particular on the $K^+\to\pi^+\nu\bar\nu$ and
$K_{\rm L}\to\pi^0\nu\bar\nu$ modes; there is a proposal to measure 
the former channel at the CERN SPS, and efforts to explore the latter 
at KEK/J-PARC in Japan. Furthermore, flavour physics offers several other
exciting topics. Important examples are top-quark physics, the $D$-meson 
system, the anomalous magnetic moment of the muon, electric dipole moments
and the flavour violation in the charged lepton and neutrino sectors. 

The established neutrino oscillations as well as the evidence for dark matter and the baryon asymmetry of the Universe tell us that the SM is incomplete, and specific 
extensions contain usually also new sources of flavour and CP violation, which 
may manifest themselves at the flavour factories. Fortunately, the LHC is expected 
to go into operation in the autumn of 2007. This new accelerator will provide 
insights  into electroweak symmetry breaking and, hopefully, also give us direct 
evidence for physics beyond the SM through the production and subsequent 
decays of NP particles in the ATLAS and CMS detectors. It is obvious that there 
should be a very fruitful interplay between these ``direct" studies of NP, and the 
``indirect" information provided by flavour physics.\footnote{This topic is currently 
addressed in detail within a CERN workshop: {\tt http://cern.ch/flavlhc}.}
I have no doubt that an exciting future is ahead of us!

\vskip1cm
\noindent

\section*{ACKNOWLEDGEMENTS}

I would like to thank the students for their interest in my lectures,
the discussion leaders for their efforts to complement them in the
discussion sessions, and the local organizers for hosting this exciting
school in Kitzb\"uhel. I am also grateful to my collaborators for the fun 
we had working on many of the topics addressed in these lectures.

\end{document}